\documentclass[12pt,a4paper,openright,twoside]{book}
\pdfoutput=1

\usepackage[utf8]{inputenc}
\usepackage[french,english]{babel}
\usepackage[T1]{fontenc}
\usepackage{newtxtext}
\usepackage{csquotes}

\usepackage{amsmath}
\usepackage{amsfonts}
\usepackage{amssymb}
\usepackage{mathtools}
\usepackage{newtxmath}
\usepackage{bbm}
\usepackage{stmaryrd} 
\SetSymbolFont{stmry}{bold}{U}{stmry}{m}{n}

\usepackage{graphicx}
\usepackage{geometry}
\usepackage[usenames,dvipsnames]{xcolor}

\usepackage[font=footnotesize, labelfont=bf, labelsep=endash]{caption}
\usepackage[subrefformat=parens, listofformat=subparens]{subfig}
\usepackage[export]{adjustbox}  
\usepackage{standalone}		
\usepackage{tikz}
\usepackage{tikz-3dplot}
\usetikzlibrary{backgrounds,plotmarks}
\usetikzlibrary{arrows.meta}
\usetikzlibrary{patterns}
\usetikzlibrary{decorations.pathmorphing}
\usetikzlibrary{calc}
\tdplotsetmaincoords{60}{110}

\usepackage{varioref}
\usepackage{cleveref}
\crefname{paragraph}{\S}{\S\S}
\crefname{chapter}{chapter}{chapters}
\crefname{app}{appendix}{appendices}

\usepackage[backend=biber, doi=false, url=false, giveninits=true, isbn=false, maxbibnames=99, sorting=none, sortcites=true, date=year, labeldate=year, style=nature]{biblatex}
\usepackage{appendix}

\usepackage{makecell} 		
\usepackage[version=3]{mhchem}  
\usepackage{siunitx}  		
\usepackage{pdfpages}

\newcommand{\ecoli}{\textit{E. coli} }

\newcommand{\rhs}{r.h.s. }
\newcommand{\pdf}{p.d.f. }

\newcommand{\chipseq}{ChIP-seq }
\newcommand{\chipchip}{ChIP-chip }
\newcommand{\hic}{Hi-C }

\newcolumntype{L}[1]{>{\raggedright\let\newline\\\arraybackslash\hspace{0pt}}m{#1}}

\newcommand{\proba}[1]{\mathrm{Pr} \left(#1\right)}
\newcommand{\ud}[1]{\mathrm{d} #1 \,}
\newcommand{\uD}[1]{\mathrm{D} \left[#1\right] \,}
\renewcommand{\vec}[1]{\mathbf{#1}}

\AtBeginEnvironment{subappendices}{%
\chapter*{Appendix}
\addcontentsline{toc}{chapter}{Appendices}
}

\makeatletter
\newcommand*{\toccontents}{\@starttoc{toc}}
\makeatother

\providecommand{\keywords}[1]{\textbf{\textit{Keywords:}} #1}
\providecommand{\motsclefs}[1]{\textbf{\textit{Mots-clefs:}} #1}

\DeclareFieldFormat{labelnumberwidth}{\mkbibbrackets{#1}} 
\AtEveryBibitem{%
    \clearfield{month}%
    \clearfield{day}%
    \clearfield{issue}}

\usepackage[acronym,toc,nonumberlist]{glossaries}
\makeglossaries

\newacronym{afm}{AFM}{Atomic-force microscopy}
\newacronym{bd}{BD}{Brownian dynamics}
\newacronym{bp}{bp}{\glslink{bpdef}{base pair}}
\newacronym{ccc}{CCC}{\glslink{cccdef}{chromosome conformation capture experiment}}
\newacronym{3c}{3C}{\glslink{cccdef}{chromosome conformation capture experiment}}
\newacronym{chipseq}{chIP-seq}{\glslink{chipseqdef}{ChIP-sequencing experiment}}
\newacronym{em}{EM}{electron microscopy}
\newacronym{hic}{Hi-C}{\glslink{hicdef}{CCC with high-throughput DNA sequencing}}
\newacronym{fene}{FENE}{finitely-extensible non-linear elastic potential}
\newacronym{fish}{FISH}{\glslink{fishdef}{Fluorescence \textit{in situ} hybridization}}
\newacronym{gem}{GEM}{Gaussian effective model}
\newacronym{pdf}{p.d.f.}{probability distribution function}
\newacronym{iid}{i.i.d.}{independently and identically distributed (random variables)}
\newacronym{lammps}{LAMMPS}{Large-scale Atomic/Molecular Massively Parallel Simulator}
\newacronym{md}{MD}{molecular dynamics}
\newacronym{naps}{NAP}{nucleoid-associated protein}
\newacronym{palm}{PALM}{\glslink{superresolution}{photo-activated localization microscopy}}
\newacronym{pcr}{PCR}{\glslink{pcrdef}{polymerase chain reaction}}
\newacronym{pwm}{PWM}{position weight matrix}
\newacronym{rnap}{RNAP}{\glslink{polymerase}{RNA polymerase}}
\newacronym{SAW}{SAW}{self-avoiding walk}
\newacronym{storm}{STORM}{\glslink{superresolution}{stochastic optical reconstruction microscopy}}
\newacronym{rpa}{RPA}{\glslink{rpadef}{random-phase approximation}}
\newacronym{tf}{TF}{\glslink{tfdef}{transcription factor}}
\newacronym{tfbs}{TFBS}{\glslink{tfbsdef}{transcription factor binding site}}
\newacronym{wlc}{WLC}{\glslink{wlcdef}{worm-like chain polymer}}

\newglossaryentry{bpdef}
{
   name=base pair,
   description={Two nucleotides in a DNA (or RNA) molecule that are paired by hydrogen bonds}
}

\newglossaryentry{cccdef}
{
   name=chromosome conformation capture experiment,
   description={Set of experimental techniques that have been developed in order to identify interactions between genomic location (loci) \textit{in vivo}. Namely, these techniques count the number of contacts between different loci on the genome, resulting from the particular folding of the chromosome}
}

\newglossaryentry{chipseqdef}
{
   name=ChIP-sequencing,
   description={ChIP-seq assays combine chromatin immunoprecipitation (ChIP) with massively parallel DNA sequencing to identify the binding sites of DNA-associated proteins. After post-processing experimental data, it gives access to the density of binding of a protein as a function of the genomic coordinate}
}

\newglossaryentry{hicdef}
{
  name=Hi-C,
  description={Experimental technique combining \glslink{cccdef}{chromosome conformation capture} and high-throughput DNA sequencing in order to obtain a high-resolution map of the contacts between chromosomal locations}
}

\newglossaryentry{pcrdef}
{
   name=polymerase chain reaction,
   description={Technique used to amplify a target DNA sequence. It relies on denaturation-hybridization-elongation cycles and the usage of a polymerase resistant to high temperatures. This technique enables the production of millions of copies of an initial DNA sequence in just a few hours}
}

\newglossaryentry{tfdef}
{
   name=transcription factor,
   description={Term loosely applied to any protein that can bind to DNA in order to alter (or regulate) the expression of a gene}
}

\newglossaryentry{tfbsdef}
{
   name=transcripton factor binding site,
   description={DNA sequence to which a transcription factor binds to in order to regulate the transcription of a gene}
}

\newglossaryentry{rpadef}
{
  name=random phase approximation,
   description={In the context of polymer field theories, it is a stability analysis of the saddle-point solution for the polymer density. Namely, the Hamiltonian is expanded to second order around the saddle-point solution in order to obtain the quadratic fluctuations. A Fourier mode analysis can reveal if modulations in the polymer density can trigger instabilities, which are associated to a microphase separation. It is also called the Gaussian fluctuations analysis}
}

\newglossaryentry{wlcdef}
{
   name=worm-like chain,
   description={Model used to describe a polymer with bending rigidity. A key parameter of the model is the persistence length, that characterizes the contour distance above which the polymer loses the memory of its orientation. Also known as the Kratky-Porod model}
}

\newglossaryentry{chromosomefolding}
{
  name=chromosome folding,
  description={The chromosome is a long object, and as such can adopt many three-dimensional configurations. One such configuration defines its folding, or architecture}
}

\newglossaryentry{centraldogma}
{
  name=central dogma,
  description={The fundamental principle that genetic information flows from DNA to RNA to protein}
}

\newglossaryentry{cytosol}
{
  name=cytosol,
  description={Intra-cellular space, excluding the organelles}
}

\newglossaryentry{horizontalgenetransfer}
{
  name=horizontal gene transfer,
  description={Process through which DNA is passed from one organism to another. This contrasts with vertical gene transfer which refers to the inheritance of genes from parent to progeny}
}

\newglossaryentry{hydrogenbond}
{
  name=hydrogen bond,
  description={A weak chemical bond between an electronegative atom such as nitrogen or oxygen and a hydrogen atom bound to another electronegative atom}
}

\newglossaryentry{nucleoid}
{
  name=nucleoid,
  description={In prokaryotes which do not have a nucleus, the chromosome is confined to an area near the center of the cell called the nucleoid}
}

\newglossaryentry{nucleus}
{
  name=nucleus,
  description={In eukaryotes, the nucleus contains the chromosomes, and is separated of the rest of the cell by a membrane}
}

\newglossaryentry{monomer}
{
  name=monomer,
  description={Fundamental unit of a polymer}
}

\newglossaryentry{polymer}
{
  name=polymer,
  description={Large molecule made of the repetitive assembly of monomers held by chemical bonds. The number of monomers in the polymer is called the polymerization index}
}

\newglossaryentry{polymerase}
{
  name=polymerase,
  description={General term for a protein that can assemble monomers together to form a polymer. It catalizes the process of polymerisation. For instance RNA polymerase is the enzyme responsible for transcribing coding sequences of genes into RNA}
}

\newglossaryentry{promoter}
{
  name=promoter,
  description={DNA region where RNA polymerase binds in order to initiate transcription}
}

\newglossaryentry{restriction enzyme}
{
  name=restriction enzyme,
  description={Enzyme that can cleave the DNA molecule at specific sites (restriction sites) corresponding to a specific short sequence of nucleotides}
}

\newglossaryentry{chromosome}
{
  name=chromosome (or chromatin),
  description={Term regrouping the DNA molecule with the structuring proteins that are bound to it}
}

\newglossaryentry{bindingenergy}
{
  name=binding energy,
  description={Energy determining the strength of the chemical linkage (or affinity) between two constituents. For instance the binding energy of a protein with DNA}
}

\newglossaryentry{divalent}
{
  name=divalent protein,
  description={Protein with two functional domains enabling the binding to two DNA sites}
}

\newglossaryentry{genome}
{
  name=genome,
  description={It is the unique sequence of nucleotides which embeds the genetic information of one individual}
}

\newglossaryentry{tslevel}
{
  name=transcription level,
  description={The transcription level of one gene can be measured in at least two ways. A direct measure of the transcription level consists in measuring the number of RNA transcripts of a given gene in the cell. This is usually achieved by extracting all messengers RNA from a cell and sequencing them in order to count the transcripts of the gene of interest. An indirect measure of the transcription level is to insert downstream of the gene of interest a reporter gene which encodes a fluorescent protein (such as GFP). The intensity of the fluorescence can be related to the transcription level of the gene of interest}
}

\newglossaryentry{coregulatedgenes}
{
  name=co-regulated genes,
  description={Genes whose expressions are regulated by the same regulators}
}

\newglossaryentry{epigenetics}
{
  name=epigenetics,
  description={Ensemble of biological processes resulting in genetic effects which are not encoded in the DNA sequence. Such effects may result from external factors that affect how cells express genes. For example, DNA methylation can alter how a gene is expressed, yet it does not involve a change in the nucleotide sequence}
}

\newglossaryentry{fishdef}
{
  name=Fluorescence \textit{in situ} hybridization,
  description={Technique that uses fluorescent probes that bind specific DNA (or RNA) sequences by base pair complementarity. Fluorescence microscopy can then be used to detect and localize these specific sequences. In particular, regions of the chromosome can be localized in the cellular compartment, which helps in defining the spatial-temporal patterns of gene expression}
}

\newglossaryentry{superresolution}
{
  name=super-resolution microscopy,
  description={Fluorescence microscopy imaging methods that allow to obtain a resolution beyond the diffraction limit. They rely on the stochastic activation of each fluorophore in the sample from a non-emissive state (or off-state) to an emissive state (or on-state). This ensures that for each image, only a small fraction of the fluorophores (those in the on-state) is emitting photons. This results in very few overlaps between the fluorophore sources and leads to an increased resolution. Transitions from the on-state to the off-state occur through reversible switching in \glslink{storm}{STORM} whereas in \glslink{palm}{PALM} the phenomenon of photobleaching is exploited}
}

\newglossaryentry{xenosilencing}
{
  name=xeno-silencing,
  description={Repression of the transcription of genes in foreign DNA sequences acquired by horizontal transfer}
}

\graphicspath{{figures/}}
\makeatletter
\providecommand*{\input@path}{}
\g@addto@macro\input@path{{figures/}}  
\makeatother

\begin{document}
\frontmatter
\includepdf{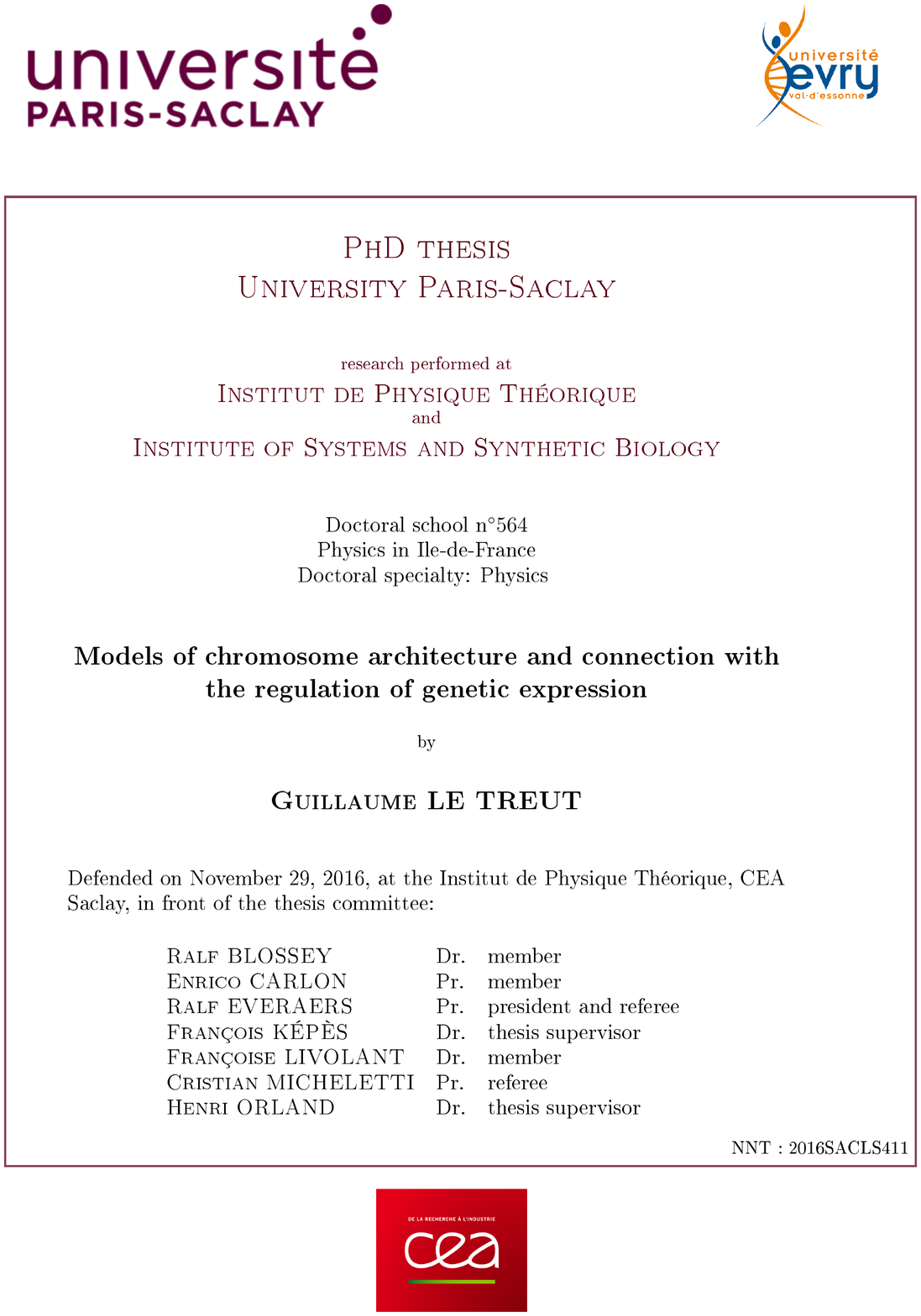}
{\null\thispagestyle{empty}\newpage}
\includepdf{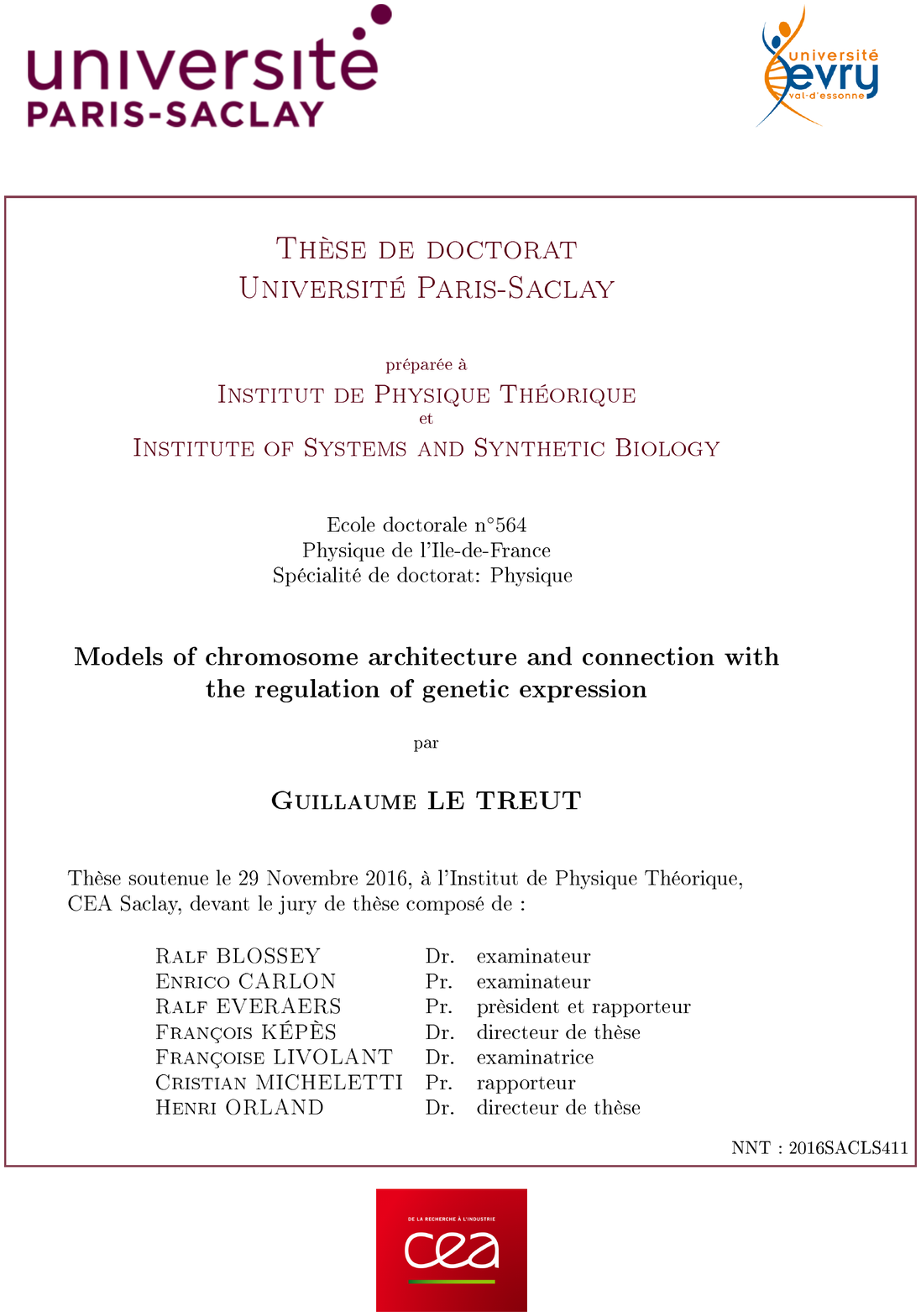}
{\null\thispagestyle{empty}\newpage}
\chapter*{Remerciements}
Je tiens \`a remercier tous ceux qui m'ont soutenu au cours de ces trois ann\'ees.

En premier lieu, merci \`a mes deux directeurs de th\`ese pour m'avoir accept\'e comme \'etudiant. Je remercie Fran\c{c}ois pour avoir cru en mon d\'esir de contribuer \`a la science malgr\'e mon parcours pour le moins atypique. C'est gr\^ace \`a son soutien que j'ai pu m'embarquer dans ce projet de recherche. Je lui en suis tr\`es reconnaissant. J'esp\`ere ne pas l'avoir amen\'e \`a regretter son choix; en ce qui me concerne je ne le regrette pas. Je remercie \'egalement Henri pour m'avoir abreuv\'e de ses connaissances, en physique bien s\^ur, mais pas seulement. Je me tiens pour particuli\`erement chanceux d'avoir pu le c\^otoyer. Nos discussions \'etaient sans doute l'occasion pour lui de s'inqui\'eter de mes lacunes, mais quant \`a moi je les attendais avec impatience. Je rentrais dans ce bureau comme l'on sauterais dans un puits de savoir, s\^ur d'y d\'ecouvrir de nouveaux pans de la physique statistique sur lesquels je plancherais avec enthousiasme ou perplexit\'e sur le chemin du retour. Par ailleurs, beaucoup des r\'esultats pr\'esent\'es dans cette th\`ese seraient sans doute rest\'es des probl\`emes sans son aide pr\'ecieuse. Pour tout cela, je lui suis tr\`es reconnaissant.

Le r\^ole de rapporteur est fastidieux et chronophage. Je tiens donc \`a remercier tout sp\'ecialement Messieurs Everaers et Micheletti d'avoir accept\'e cette t\^ache et de s'en \^etre acquitt\'es dans le temps imparti. J'appr\'ecie le regard critique qu'ils ont jet\'e sur mon travail.

L'Institut de Physique Th\'eorique r\'eunit dans un m\^eme endroit une recherche en physique th\'eorique \`a la pointe, des conditions de travail id\'eales, et des personnes formidables. En particulier, je tiens \`a saluer l'\'equipe administrative qui s'est occup\'ee de toutes mes requ\^etes avec une diligence et avec une facilit\'e d\'econcertantes. Je remercie Patrick, Laurent, Lo\"ic, Laure et Anne. Plus particuli\`erement, je remercie Sylvie pour sa grande gentillesse. Outre son efficacit\'e et son sens de l'anticipation, cela toujours \'et\'e un plaisir de venir la voir. Je remercie \'egalement Michel Bauer de m'avoir conc\'ed\'e une extension de deux mois pour finir ma th\`ese.

Je remercie Kirone d'avoir accept\'e d'\^etre mon parrain. En plus d'\^etre un grand chercheur, sa disponibilit\'e, sa s\'er\'enit\'e et sa capacit\'e d'\'ecoute m'ont \'et\'e salutaires lors des quelques phases de doutes que j'ai travers\'ees.

Je remercie les diff\'erentes g\'en\'erations de th\'esards de l'IPhT pour les grandes tabl\'ees conviviales du d\'ejeuner, les d\^iners organis\'es \`a l'improviste et nos verres occasionnels. Des anciens: Beno\^it, Ekaterina, J\'er\^ome, Antoine, Hanna; \`a ceux de ma g\'en\'eration: Christophe, Xiangyu; puis \`a ceux arriv\'es plus r\'ecemment: Thibault, Christian, S\'everin, Santiago, Romain, Ricardo.

Merci \`a Fr\'ed\'eric d'\^etre venu doubler les effectifs de notre \'equipe de recherche pendant une ann\'ee. Je garde un agr\'eable souvenir de nos dicussions aussi bien \`a l'institut qu'en dehors. Je lui sais gr\'e de m'avoir fait relativiser les difficult\'es rencontr\'ees et de m'avoir encourag\'e \`a pers\'ev\'erer. Je le remercie \'egalement de m'avoir incit\'e \`a adopter une hygi\`ene de vie saine en me faisant d\'ecouvrir notamment ``D\'ed\'e la frite''.

Je remercie toutes les personnes de l'\'equipe MEGA \`a l'iSSB pour avoir entrepris mon instruction en biologie depuis un niveau que nous conviendrons de ne pas divulguer. Plus particuli\`erement, un grand merci \`a Costas, Thibaut, Steff, Julie, Fran\c{c}ois, Brian et Laurent pour la qualit\'e des \'echanges que j'ai pu avoir avec eux, scientifiques certes, mais surtout pour tous les autres. Je salue \'egalement Sylvie, Bernadette et Dominique pour m'avoir assist\'e dans la compl\'etion des diverses formalit\'es administratives dont j'ai d\^u m'acquitter.

Je remercie l'IDEX Paris-Saclay d'avoir financ\'e ma th\`ese.

Je remercie tous mes amis pour les bons moments pass\'es en leur compagnie. Le maintien de mon \'equilibre mental et de mon int\'egrit\'e sociale doit beaucoup aux s\'eances dominicales d'Ultimate. Je pense notamment \`a Antoine, Fady, Hugo, Nizar, Pierre, Simon, Souhad, Charlotte et Juliette; mais aussi aux amiti\'es qui perdurent depuis l'enfance: Rapha\"el, Pauline, Carine, Florian, Arnaud et Adrien.

Je remercie ma famille pour avoir toujours fait confiance dans mes choix, et plus fondamentalement pour leur soutien in\'ebranlable et leur pr\'esence rass\'er\'enante. Je remercie notamment mon p\`ere pour m'avoir donn\'e l'envie de r\'eussir et le go\^ut de l'ambition. Et ma m\`ere, inconditionnellement bienveillante, qui envisage toute chose avec une simplicit\'e et un bon sens qui n'a rien de commun, qui sait tout faire et dont l'exemple me rappelle sans cesse que tout est r\'ealisable au prix d'un effort. Je remercie \'egalement mes soeurs, Claire qui continue de chasser patiemment toutes mes vilaines habitudes, et Camille sur qui je peux toujours compter pour m'apprendre une chor\'egraphie.

Enfin, je remercie Ma\"eva de m'avoir soutenu tout du long, pour avoir apport\'e son regard lucide dans mes moments de d\'esarroi et pour avoir c\'el\'ebr\'e avec moi mes moments de r\'eussite.


\cleardoublepage
\chapter*{Preface}
The work presented in this thesis manuscript was realized between fall of 2013 and fall of 2016, at the Institut de Physique Th\'eorique, CEA Saclay and the Institute of Systems and Synthetic Biology, Universit\'e d'\'Evry.

The motivation of this work was to provide physical models for the characterization of chromosome folding (or architecture) and understand the role it plays in regulating the genetic expression. The manuscript is organized as follows.

In \cref{ch:introduction}, I give an introduction to chromosome architecture and the current biological conjectures for its role. I also review standard techniques in Physics to model the chromosome. I conclude this introductory chapter by giving an outline of the work presented in the subsequent chapters.

In \cref{ch:transcription_factories,ch:structure_function,ch:naps,ch:ccc}, I present the results of my research activity during these three years. This resulted in the publication of one research article:
\begin{itemize}
  \item Phase behavior of DNA in the presence of DNA-binding proteins \cite{LeTreut012016}.
\end{itemize}

Besides, Biology can lead to the usage of a specific vocabulary, or acronyms, whose meaning is sometimes not obvious. Although I have attempted to always define such terms before use, a glossary is available at the end of this manuscript, in order to ease the reading.


\cleardoublepage
\chapter*{Abstract}
Increasing evidence suggests that chromosome folding and genetic expression are intimately connected. For example, the co-expression of a large number of genes can benefit from their spatial co-localization in the cellular space. Furthermore, functional structures can result from the particular folding of the chromosome. These can be rather compact bundle-like aggregates that prevent the access to DNA, or in contrast, open coil configurations with several (presumably) globular clusters like transcription factories. Such phenomena have in common to result from the binding of divalent proteins that can bridge regions sometimes far away on the DNA sequence. The physical system consisting of the chromosome interacting with divalent proteins can be very complex. As such, most of the mechanisms responsible for chromosome folding and for the formation of functional structures have remained elusive.

Using methods from statistical physics, we investigated models of chromosome architecture. A common denominator of our approach has been to represent the chromosome as a polymer with bending rigidity and consider its interaction with a solution of DNA-binding proteins. Structures entailed by the binding of such proteins were then characterized at the thermodynamical equilibrium. Furthermore, we complemented theoretical results with Brownian dynamics simulations, allowing to reproduce more of the biological complexity.

The main contributions of this thesis have been: (i) to provide a model for the existence of transcription factories characterized \textit{in vivo} with fluorescence microscopy; (ii) to propose a physical basis for a conjectured regulatory mechanism of the transcription involving the formation of DNA hairpin loops by the H-NS protein as characterized with atomic-force microscopy experiments; (iii) to propose a physical model of the chromosome that reproduces contacts measured in chromosome conformation capture (CCC) experiments. Consequences on the regulation of transcription are discussed in each of these studies.

To model transcription factories, we implemented a Flory-Huggins polymer theory to characterize the equilibrium of DNA chains interacting with non-specific binding proteins. For sufficiently high DNA-protein binding affinity, this system was shown to exhibit a phase separation with a dilute and a dense phase. We also investigated the structure of the dense phase and showed that for stiff DNA chains, the dense phase may undergo a transition from a globular to a crystalline phase. While globular dense phases can be a model for transcription factories, crystalline dense phases may be a model for bundle-like aggregates in stressed bacteria.

To characterize the formation of DNA hairpin loops by the H-NS protein, we showed the existence of a characteristic length for the H-NS binding region, delimiting two regimes. In one regime, DNA hairpin loops are stable whereas in the other they are not. This result was obtained first from a simplified polymer model with implicit interactions, and then confirmed  using Brownian dynamics simulations with explicit proteins.

To model chromosome architecture, we considered a Gaussian chain polymer model of the chromosome and added Gaussian effective interactions to model the effect of divalent proteins. The contact probability for any pair of monomers was computed and yielded an analytical closed-form which can be used in an inverse approach to reconstruct an effective polymer model of the chromosome reproducing contact probabilities measured in CCC experiments.

\vskip 1em
\keywords{statistical physics, polymer physics, Gaussian chain, worm-like chain, Flory-Huggins theory, random phase approximation, DNA phases, structure function, transfer matrix, Brownian dynamics, contact probability, DNA-binding protein, chromosome architecture, chromosome folding, chromosome dynamics, gene co-localization, transcription factory, transcription regulation, chromosome conformation capture.}

\cleardoublepage
\chapter*{R\'esum\'e}
Plusieurs indices sugg\`erent que le repliement du chromosome et la r\'egulation de l'expression g\'en\'etique sont \'etroitement li\'es. Par exemple, la co-expression d'un grand nombre de g\`enes est favoris\'ee par leur rapprochement dans l'espace cellulaire. En outre, le repliement du chromosome permet de faire \'emerger des structures fonctionnelles. Celles-ci peuvent \^etre des amas condens\'es et fibrillaires, interdisant l'acc\`es \`a l'ADN, ou au contraire des configurations plus ouvertes comportant quelques amas globulaires, comme c'est le cas avec les usines de transcription. Bien que dissemblables au premier abord, de telles structures sont rendues possibles par l'existence de prot\'eines bivalentes, capable d'apparier des r\'egions parfois tr\`es \'eloign\'ees sur la s\'equence d'ADN. Le syst\`eme physique ainsi constitu\'e du chromosome et de prot\'eines bivalentes peut \^etre tr\`es complexe. C'est pourquoi les m\'ecanismes r\'egissant le repliement du chromosome sont rest\'es majoritairement incompris.

Nous avons \'etudi\'e des mod\`eles d'architecture du chromosome en utilisant le formalisme de la physique statistique. Notre point de d\'epart est la repr\'esentation du chromosome sous la forme d'un polym\`ere rigide, pouvant interagir avec une solution de prot\'eines liantes. Les structures r\'esultant de ces interactions ont \'et\'e caract\'eris\'ees \`a l'\'equilibre thermodynamique. De plus, nous avons utilis\'e des simulations de dynamique brownienne en compl\'ement des m\'ethodes th\'eoriques, car elles permettent de prendre en consid\'eration une plus grande complexit\'e dans les ph\'enom\`enes biologiques \'etudi\'es.

Les principaux aboutissements de cette th\`ese ont \'et\'e : (i) de fournir un mod\`ele pour l'existence des usines de transcriptions caract\'eris\'ees \textit{in vivo} \`a l'aide de la microscopie par fluorescence ; (ii) de proposer une explication physique pour une conjecture portant sur un m\'ecanisme de r\'egulation de la transcription impliquant la formation de boucles d'ADN en t\^ete d'\'epingle sous l'effet de la prot\'eine H-NS, qui a \'et\'e \'emise suite \`a l'observation de ces boucles au microscope \`a force atomique ; (iii) de proposer un mod\`ele du chromosome qui reproduise les contacts mesur\'es \`a l'aide des techniques Hi-C. Les cons\'equences de ces m\'ecanismes sur la r\'egulation de la transcription ont \'et\'e syst\'ematiquement discut\'ees.

Afin de mod\'eliser les usines de transcription, nous avons consid\'er\'e une th\'eorie de Flory-Huggins pour des cha\^ines d'ADN en interaction avec des prot\'eines liantes. Cela nous a permis de caract\'eriser l'\'equilibre thermodynamique. En particulier, pour une affinit\'e suffisamment forte avec les prot\'eines, l'ADN se condense, ce qui donne lieu a un syst\`eme biphasique comportant une phase dilu\'ee et une phase dense. Nous avons ensuite montr\'e que pour des chaînes rigides, la phase dense peut passer d'une structure globulaire \`a une structure cristalline. Une phase globulaire semble \^etre un bon mod\`ele pour les usines de transcriptions, tandis que les amas fibrillaires s'apparentent davantage \`a une phase cristalline.

Afin de caract\'eriser la formation de boucles d'ADN en t\^etes d'\'epingles sous l'effet de la prot\'eine H-NS, nous avons montr\'e qu'il existe une taille caract\'eristique pour les r\'egions de liaison avec H-NS. Au-dessus, les boucles sont stables tandis qu'en dessous elles sont instables. En utilisant un mod\`ele simplifi\'e de polym\`ere, nous avons obtenu une expression pour cette grandeur, que nous avons ensuite confirm\'ee \`a l'aide de simulations de dynamique Brownienne.

Afin de mod\'eliser le repliement du chromosome, nous avons consid\'er\'e un mod\`ele de polym\`ere Gaussien auquel nous avons ajout\'e des interactions effectives repr\'esentant l'effet de prot\'eines bivalentes. Nous avons alors pu calculer la probabilit\'e de contact entre deux monom\`eres. L'expression obtenue a ensuite \'et\'e utilis\'ee pour r\'esoudre le probl\`eme inverse consistant \`a trouver le mod\`ele effectif qui reproduit les probabilit\'es de contact mesur\'es lors d'exp\'eriences Hi-C.

\vskip 1em
\motsclefs{physique statistique, physique des polym\`eres, cha\^ine gaussienne, cha\^ine de Kratky-Porod, th\'eorie de Flory-Huggins, random phase approximation, phases de l'ADN, fonction de structure, matrice de transfert, dynamique browniennne, probabilit\'e de contact, prot\'eine se liant \`a l'ADN, architecture du chromosome, repliement du chromosome, dynamique du chromosome, co-localisation des g\`enes, usine \`a transcription, r\'egulation de la transcription, chromosome conformation capture.}


\cleardoublepage

\tableofcontents

\mainmatter
\chapter{Introduction}
\label{ch:introduction}
\section{Chromosome architecture and genetic expression}
\subsection{The central dogma of biology}
Life depends on the ability of cells to store, retrieve and translate a set of instructions commonly denoted as the genetic code. This information is stored in the genes, which determine the characteristics of each individual.

Since the beginning of the twentieth century, we know that the genetic code is carried by deoxyribonucleic acid (DNA) molecules, with a simple chemical composition. The realization of X-ray diffraction experiments in the 1950s led Watson and Crick (Nobel prize 1962) to propose the correct model for the molecular structure of DNA \cite{Watson7371953}. Specifically, a DNA molecule consists of two polynucleotide chains (or strands) wounded in a double-helix. Each nucleotide is made of a sugar and of one of the four bases: adenine (A), thymine (T), guanine (G) and cytosine (C). The sugars are covalently linked together and form the DNA ``backbone''. In addition, the two strands are held together by hydrogen bonds between the bases on the different strands, resulting in a double-helical structure with the base pairs (bp) inside. Actually, bases do not pair at random, but by pair complementarity: A with T and G with C.

The complete sequence of base pairs determines the genetic information of each individual. It is called the genome. The corresponding sequence of letters is enormous. For instance, in the \textit{Escherichia coli} bacterium, it contains \num{4.6e6} letters, and more than \num{3.3e9} in humans. For comparison, in the latter case it would take more than \num{1000} books of \num{1000} pages to write down the full sequence. Besides, specific DNA sequences, the genes, are encountered in the genome. Their number ranges from less than a hundred in simple bacteria to several tens of thousands in higher organisms. For example, approximately \num{4600} genes are found in \ecoli and more than \num{30000} in humans.

The genes encode macromolecules such as ribonucleic acids (RNA), or polypeptides which are chains of amino-acids more commonly known as proteins. These macromolecules are responsible for most of the biochemical workings of a cell and can be envisioned as molecular ``tools''. The central dogma of molecular biology states that DNA sequences from genes are first transcribed into RNA. Furthermore, some RNA transcripts known as messenger RNA (mRNA) are then translated into proteins. The protein synthesis relies on a correspondence between the 4-letter nucleotide alphabet of DNA and the 20-letter amino-acid alphabet of proteins.

In short, the DNA sequences of genes can be seen as a message, handled through two essential and successive processes which are transcription and translation. Yet in order to adjust the synthesis of proteins to the cell needs the genetic expression can be regulated.

\subsection{From a classical to a modern view of transcription}
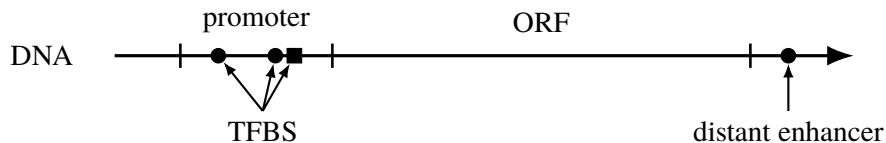
\begin{figure}[!htbp]
  \centering
  \begin{tikzpicture}
\def\dz{1}
\def\rr{0.5em}
\def\tw{0.05em}
\def\th{0.8em}
\def\xa{-4}
\def\xoi{-3.5}
\def\xoii{-2.75}
\def\xoiii{-2.5}
\def\xb{-2}
\def\xc{3.5}
\def\xoiv{4.0}
\def\xlo{-5}
\def\xhi{5}

\tikzset{
  loci/.style = {
    shape = circle,
    align = center,
    draw  = #1,
    text=white,
    fill = black,
    solid,
    inner sep= 0pt,
    minimum size=\rr
  },
  tick/.style = {
    shape = rectangle,
    align = center,
    draw  = #1,
    text=white,
    fill = black,
    solid,
    inner sep= 0pt,
    outer sep= 0pt,
    minimum width=\tw,
    minimum height=\th
  },
}
\small
  \node[label={[label distance=4pt]180:DNA}] (L) at (\xlo,0) {};
  \node (R) at (\xhi,0) {};
  \draw[black, very thick, -{Latex[scale=1.2]}] (L) -- (R);
  \node[tick] (A) at (\xa,0) {};
  \node[tick] (B) at (\xb,0) {};
  \node[tick] (C) at (\xc,0) {};
  \node[loci] (O1) at (\xoi,0) {};
  \node[loci] (O2) at (\xoii,0) {};
  \node[loci,shape=rectangle] (O3) at (\xoiii,0) {};
  \node[loci] (O4) at (\xoiv,0) {};

 \path (A) to node[above,outer sep=5pt]{promoter} (B);
 \path (B) to node[above,outer sep=5pt]{ORF} (C);
 \node (OP) at ({(\xoi+\xoii + \xoiii)/3.},-\dz) {TFBS};
 \node (OE) at (\xoiv,-\dz) {distant enhancer};
 \path[black, thick, -{Latex[scale=0.8]}] (OP.north) edge (O1) edge (O2) edge (O3);
 \path[black, thick, -{Latex[scale=0.8]}] (OE) edge (O4);

\end{tikzpicture}
  \caption{Organisation of a gene under transcriptional regulation. The open reading frame (ORF) encoding for a protein is preceded by a promoter region where several transcription factor binding sites (TFBS) are found. The promoter region comprises one main binding site and several auxiliary binding sites. In eukaryotes, TFBS participating to the gene expression regulation can sometimes be found very far away on the DNA sequence and are called enhancers.}
  \label{fig:operon_sketch}
\end{figure}

\begin{figure}[!htbp]
  \centering
  \includegraphics[width = 1.00 \textwidth]{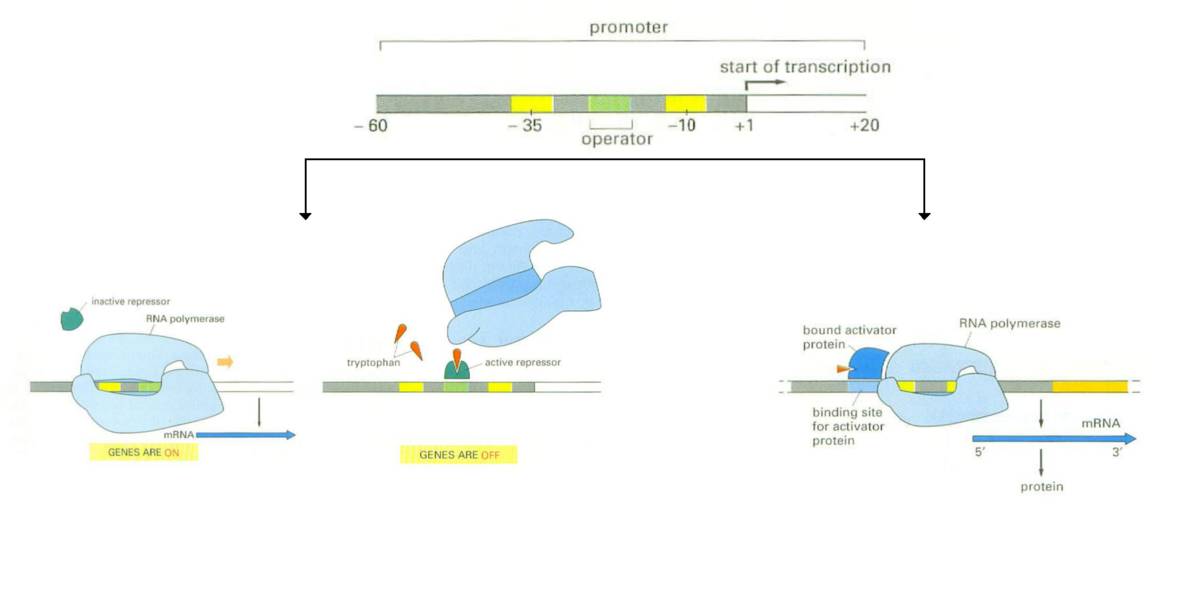}
  \caption{Classical view of the repressor/activator regulation of the transcription \cite{Alberts2013}.}
  \label{fig:intro_transcription_classical}
\end{figure}

In a classical work \cite{Jacob3181961}, Jacob and Monod (Nobel prize 1965) proposed their vision of the operon system in bacteria, which has been extended to the whole living realm and is still nowadays a pillar of molecular biology. Genetic expression is under the control of particular sequences called promoters, found a few tens of base pairs upstream of the protein encoding sequences. Such regions have typically a size of \SI{300}{bp} but sometimes can be even longer. The structural unit constituted of one promoter followed by one or several regulated genes is called an operon. The promoter is of critical importance because it is where the protein responsible for the mRNA synthesis, the RNA polymerase (RNAP), is recruited to initiate the transcription of the downstream gene or operon. The affinity of RNAP with the promoter is therefore an indirect measure for the transcription level and represents a handle for its regulation.
Transcription factors (TFs), that is to say proteins which can regulate the transcription of a gene, can bind to the promoter thanks to the presence of several transcription factor binding sites (TFBS, \cref{fig:operon_sketch}). Importantly the binding of transcription factors to the promoter can alter its affinity with RNAP. When a transcription factors stimulates the transcription it is called an activator (or inducer), and a repressor in the opposite case. At a molecular level, a repressor bound to the promoter area will prevent RNAP binding or obstruct transcription elongation, and an activator bound to the promoter will enhance transcription by recruiting RNAP from the bulk (\cref{fig:intro_transcription_classical}).

\begin{figure}[!htbp]
  \centering
  \includegraphics[width = 0.48 \textwidth]{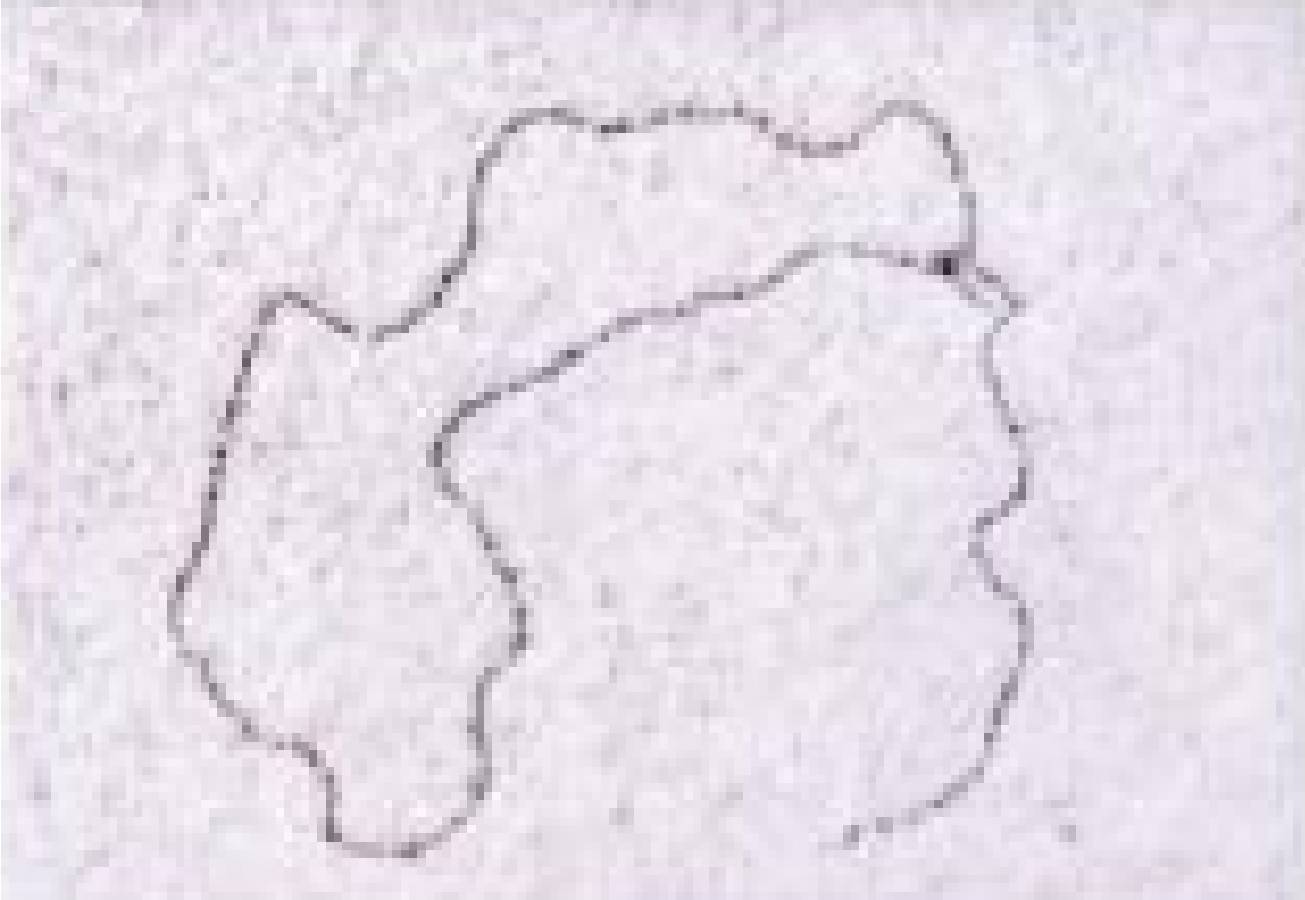}
  \caption{Electronic microscopy image of a repressor system relying on the formation of a \SI{2850}{bp} long loop between the main binding site of a promoter and an auxiliary binding site \cite{Revet1999151}.}
  \label{fig:intro_transcription_loopEM}
\end{figure}

\begin{figure}[!htbp]
  \centering
  \includegraphics[width = 0.60 \textwidth, valign=c]{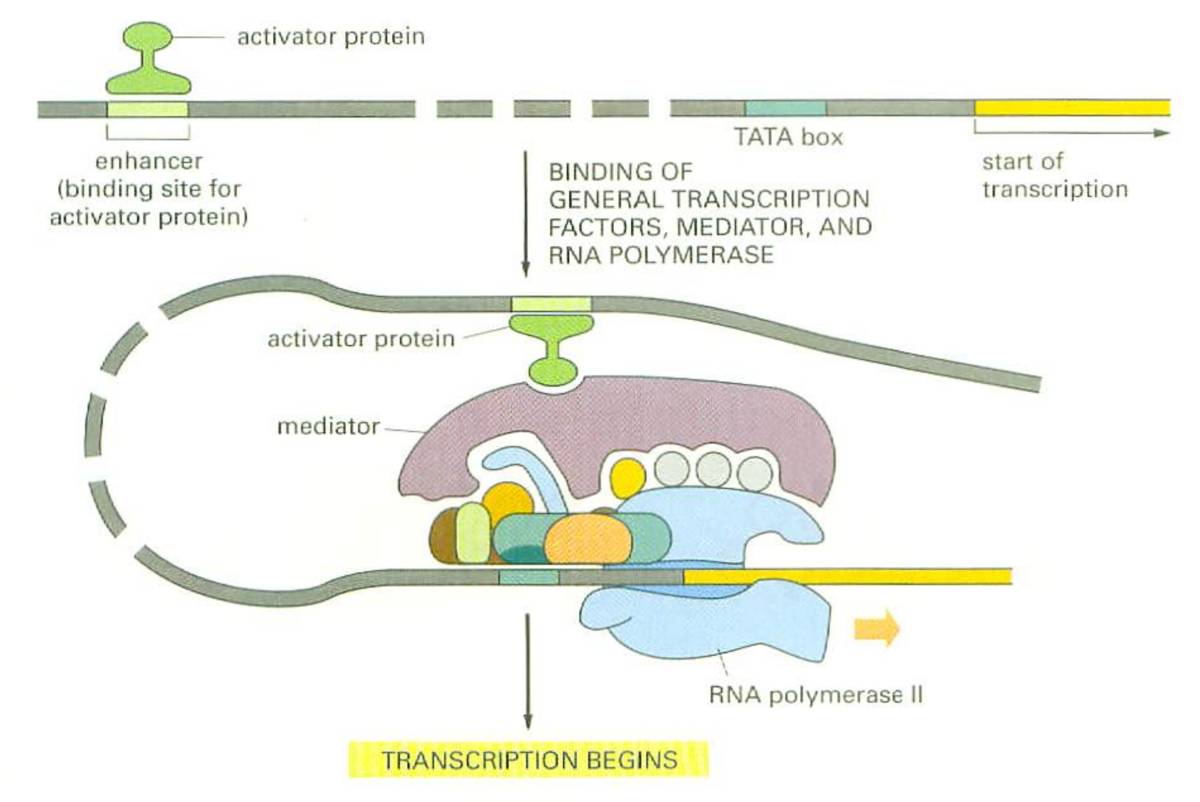}
  \caption{A distant regulatory sequence can interact with the promoter through a looping mechanism to enhance/repress transcription \cite{Alberts2013}.}
  \label{fig:intro_transcription_loop}
\end{figure}

The transcription factor binding sites can be divided into two sets. The binding site with the highest affinity is called the main binding site and is generally found in the promoter region. Others binding sites entailing a weaker binding are called auxiliary binding sites. These are mainly found in the promoter region, but can also be found outside in some cases. The simultaneous binding of a TF with the main and auxiliary binding sites can also participate to the regulation of the transcription. A famous example is the \textit{lac} repressor system in \textit{Escherichia coli}. In this particular case, efficient repression is achieved only when the \textit{lac} repressor binds simultaneously the main and auxiliary binding sites found \SI{401}{bp} from each other in the genome. Actually, many other examples of such regulatory systems have been characterized \cite{Muller-Hill1998,Muller-Hill1998a}. This type of repression involves the formation of DNA loops from tens to a few hundreds base pairs long. An \textit{in vitro} assay has even constructed a synthetic repressor system involving the interaction between the main binding site of a promoter and an auxiliary binding site separated by \SI{2800}{bp} on a plasmid \cite{Revet1999151}. The corresponding DNA loops were observed with electron microscopy imaging and corresponded to the repressed state (\cref{fig:intro_transcription_loopEM}). The existence of distant regulatory elements has now been established for a large number of genes. In eukaryotes specifically, auxiliary binding sites can be found sometimes very far away from the promoter (tens of thousands base pairs), in which case they are called enhancers. When a TF has a low affinity with its main binding, \textit{i.e.} the promoter is weak, the simultaneous binding with an auxiliary binding site may stabilize the binding of the TF to the promoter. When the main and auxiliary sites are not too far from each other, say less than \SI{200}{bp}, the formation of a DNA loop may simply prevent RNAP binding by physically forbidding the access to the promoter. More generally, it is conjectured that the interaction with a remote regulatory site (or CIS element) can favor the formation of a complex comprising DNA and proteins which enhances or represses the transcription (\cref{fig:intro_transcription_loop}).

Let us now consider the problem of a TF diffusing in the nucleoid (in bacteria) or nucleus (in eukaryotes), whose target is the main binding site found in the promoter region of the regulated gene. The typical square displacement of the protein during the time $t$ scales like $\langle x^2 \rangle \sim D t $, where $D$ is the diffusion coefficient. For a protein diffusing in the cytosol, we typically have $D \approx \SI{10}{\mu \meter^{2} . \second^{-1}}$ \cite{Elowitz1999}. Hence the average distance traveled by a protein during \SI{100}{\ms}\index{diffusion time} is approximately \SI{300}{\nm}. If we consider that the typical size of the bacterial nucleoid in \ecoli is \SI{600}{\nm}, then the diffusion time can be quite limiting in regulatory mechanisms of the transcription where proteins have to find their targets on the DNA, scattered within the nucleoid. This is even more critical in eukaryotes, where the size of the nucleus is of several micrometers. The presence of auxiliary binding sites provides an intuitive way to enhance the search process. When a TF is bound to an auxiliary binding sites, it cannot diffuse freely in the cytosol. Instead, it is confined in a sphere whose radius is the contour distance, say $l$, between the main and auxiliary binding sites. Hence the TF only explores a reduced volume in comparison to the whole cellular (or nuclear) compartment, of size $L$. In other words, the apparent concentration of this TF relatively to the promoter is increased by a factor $(L/l)^3$. This is a typical example of local concentration effect.

In summary, it has become clear in the recent decades that the DNA molecule cannot be reduced to a mere ``cookbook'' with a passive role. Instead, it is directly involved in the genetic expression regulation. Specifically, proteins can use the DNA molecule as a scaffold to build complexes or loops that regulate the transcription \cite{Vilar2005136}. This is possible because most TFs are divalent and have several additional binding sites distributed on the genome.

\subsection{Multi-scale description of the chromosome}
In physiological conditions, scores of proteins are bound to the DNA molecule, which consequently is never found ``naked''. The resulting molecule is usually called the chromosome. Although in the classical sense, the chromosome refers to the threadlike structures of condensed DNA observed during mitosis (the process by which a cell becomes two cells), it is nowadays commonly used to designate the double-helical DNA molecule with its structuring proteins. Thus we shall follow this convention from now on. Under the effect of these structuring proteins, the chromosome adopt higher level structures that constitute the chromosome folding, or architecture.

\begin{figure}[!htbp]
  \centering
  \includegraphics[height=0.5 \textwidth, rotate=90]{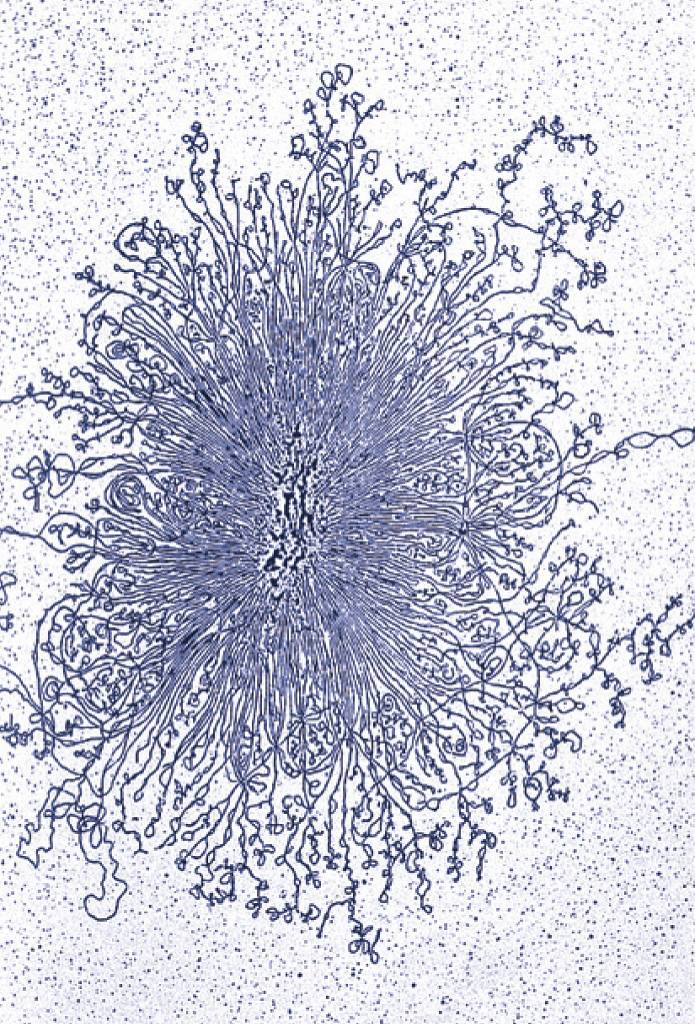}
  \caption{\ecoli chromosome after lysis of the cell walls (at the center) \cite{Wang2013a}.}
  \label{fig:ecoli_chromosome_free}
\end{figure}

In order to fit inside the bacterial cell or the eukaryotic nucleus, the chromosome is compacted nearly \num{e3} times, and this is true in all organisms \cite{Holmes20001322}. In \ecoli for instance, the free chromosome (after lysis of the cell walls) spans a spherical volume with a diameter of approximately \SI{20}{\mu \meter} whereas the length of a bacterium cell is typically of \SI{1}{\mu \meter} (\cref{fig:ecoli_chromosome_free}). Therefore, the chromosome needs to be compacted (or folded) in a multi-scale organisation whose underlying mechanism has remained unclear.

\begin{figure}[!htbp]
  \centering
  \includegraphics[width=1 \textwidth]{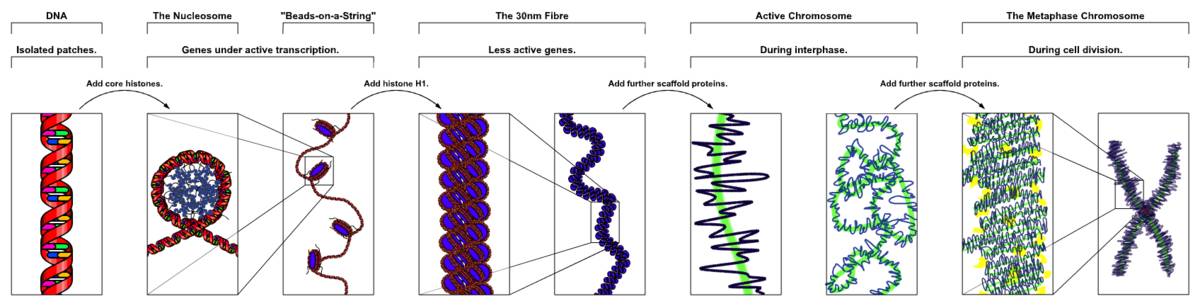}
  \caption{The four levels of folding of the eukaryotic chromosome \cite{wikipedia_chromatin}.}
  \label{fig:chromosome_folding_levels}
\end{figure}

In eukaryotes, there exist four basic levels of folding of the chromosomal chain \cite{Alberts1990,wikipedia_chromatin} (\cref{fig:chromosome_folding_levels}). First, there is the nucleosomal organization enabled by the presence of structuring proteins called histones. Naked DNA wraps around each histone octamer on approximately \SI{147}{bp} to form a nucleosome. Two consecutive nucleosomes are connected by a linker DNA approximately \SI{80}{bp} long. Consequently, the chromosome adopts a "beads-on-string" structure, clearly characterized by \textit{in vitro} assays, where the elementary monomer in the chromosomal chain is the nucleosome. A string of nucleosomes gives rise to the \SI{11}{\nm} fiber. Second, the \SI{30}{\nm} fiber is obtained by coiling the \SI{11}{\nm} fiber in a solenoidal structure with about 6 nucleosomes per turn. The chromosomal fiber is then usually designated as chromatin and has a linear packing fraction $\nu \approx \SI{100}{bp . \nm^{-1}}$ \cite{Langowski2412006,EslamiMossallam1012016,Mergell0119152004,Olins8092003}. Note that actively transcribed chromatin tends to be loosely packed and is called euchromatin whereas chromatin containing non-coding or silent genes tends to adopt more compact conformations (often under the effect of structuring proteins) and is usually called heterochromatin. Yet, the \SI{30}{\nm} fiber is apparently only observed in the interphase nucleus. Therefore, a third level of organization exists, in which the chromosomal chain is organized into domains containing from \num{30} to \SI{100}{\kilo bp} and resulting in a fiber of diameter \num{200}-\SI{300}{\nm}. Finally during mitosis, the ultimate level of compaction consists in an helical folding of the metaphase chromosome, resulting in the well known condensed chromosomes \cite{Delatour1988937,Rattner1985291}. Presumably, structuring proteins are responsible for transitions from one level of organization to the other.

Bacteria lack histones, therefore the primary level of folding is not achieved and the chromosome can be seen as a fiber of diameter \SI{2.5}{\nm} \cite{Langowski2412006}. However, the bacterial DNA is most often circular and negatively super-coiled. This is known to produce plectonemes. In particular, they have been reported to provide a superior level of organization of the chromosome into domains whose size is estimated to $\SIrange{10}{20}{\kilo bp}$ \cite{Cho062008,Brunetti092001}. Yet, we stress that they are not maintained by scaffolding proteins and for this reason can hardly be compared to chromosome folding in eukaryotes.

\subsection{The role of chromosome architecture}
The cooperative binding of hundreds of multivalent TFs producing DNA loops and of structuring proteins on the chromosome can result in sophisticated structures. This kind of global changes can completely redefine the chromosome architecture, and have far reaching consequences on transcription (and presumably other biological processes) that we are just starting to understand.

\begin{figure}[!htbp]
  \centering
  \subfloat[]{\label{fig:intro_structure_model:rosette} \includegraphics[width = 0.40 \textwidth, valign=b]{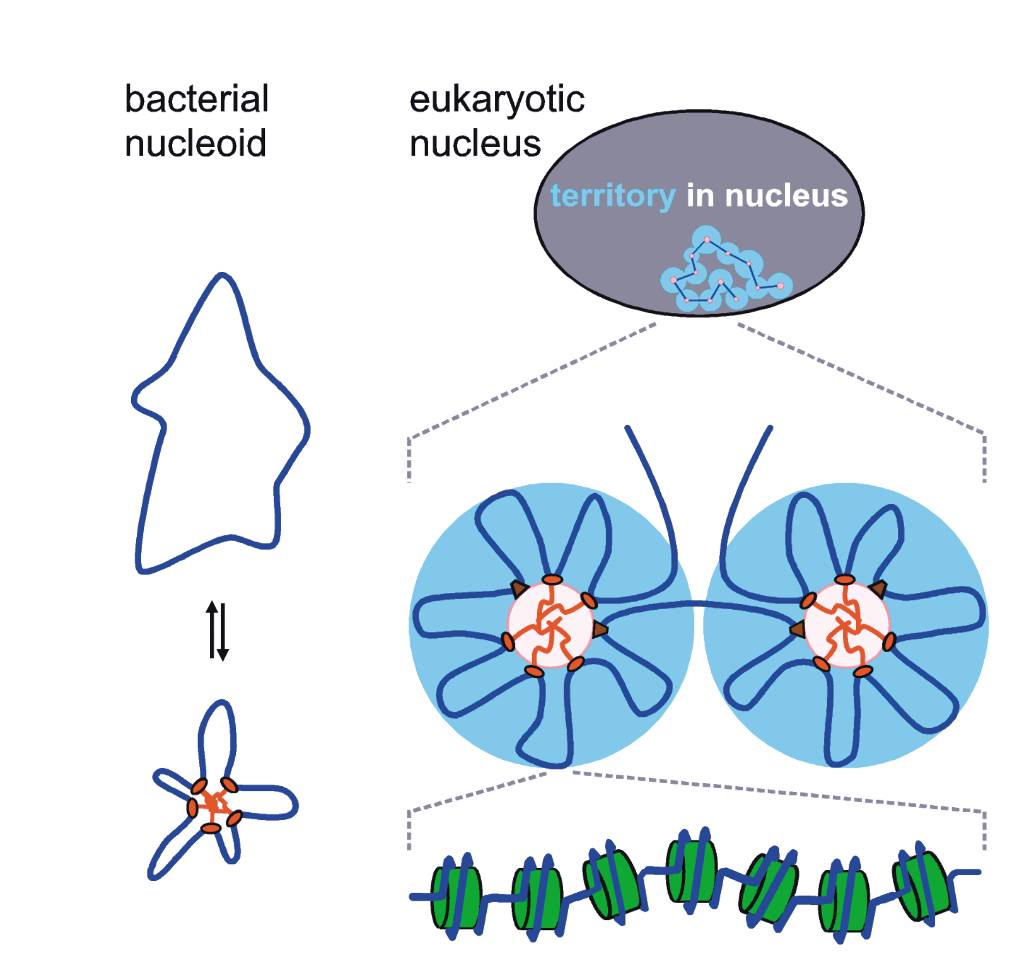}}%
  \quad
  \subfloat[]{\label{fig:intro_structure_model:solenoid} \includegraphics[width = 0.40 \textwidth, valign=b]{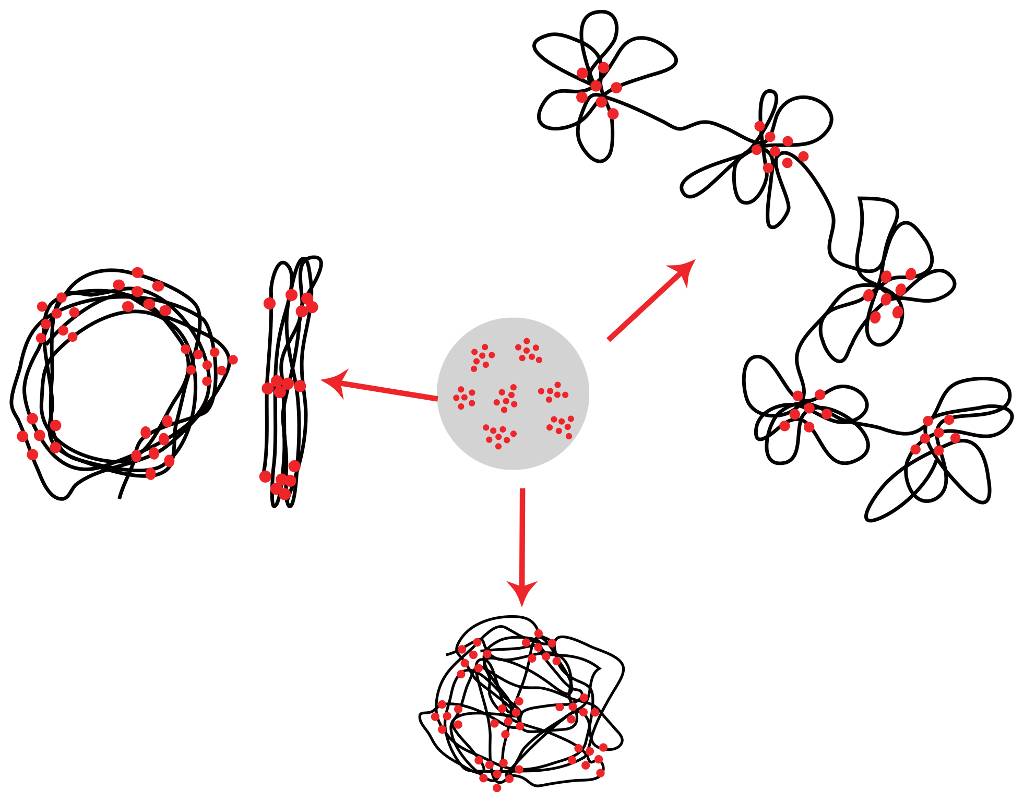}}
  \caption{\protect\subref{fig:intro_structure_model:rosette} Organization of the chromosome in ``rosettes'' by TFs \cite{Cook2002}.\protect\subref{fig:intro_structure_model:solenoid} Organization of the chromosome in ``solenoid'' \cite{Junier2010}.}
  \label{fig:intro_structure_model}
\end{figure}

A modern view of the chromosome is that TFs can form DNA loops resulting in several functional clusters with a ``rosette'' shape \cite{Cook12010,Cook2002} or in a solenoidal topology \cite{Kepes2004,Junier2010} (\cref{fig:intro_structure_model}). It is conjectured that these organizations enable to bring close in space genes whose transcription needs to be synchronized (\cref{fig:intro_structure_model}). In other words, the spatial distribution of genes inside the nucleus/nucleoid matters. This suggests that the genetic expression of a gene will depend on the genes and proteins encountered in its neighborhood. Hence, the specific folding of the chromosome can result in varying transcription levels along the genome, a phenomenon known as context sensitivity which has remained poorly understood.

Functional structures are also encountered in biological processes radically different from genetic regulation. For instance, in \ecoli and \textit{Bacillus subtilis} bacteria, following an exposure to assaults inducing double-strand DNA breaks, the chromosome is reorganized into filamentous bundle-like assemblies maintained by the RecA protein. It is assumed that these structures with quasi-crystalline order can at the same time protect DNA from further damages and enhance DNA repair by limiting the dimensionality of the research in the homologous recombination process, hence justifying the name of ``repairosome'' \cite{FrenkielKrispin2006}. Such ordered states, which have been reproduced \textit{in vitro}, are also encountered in other contexts such as viruses, mitochondrial DNA, stressed bacteria, and induce an inactive state for DNA \cite{Livolant1991117}.

In summary, recent advances in biology are promoting chromosome architecture as a major determinant of the cell physiology. Back to transcription, while the operon system may be seen as the primitive mechanism for genetic expression, it is has become clear that sophisticated regulatory mechanisms require an interplay between chromosome architecture and transcription. Understanding this link is also relevant to several active areas of research including conditional genetic expression, cell differentiation and epigenetics. Yet many unknowns remains, and we are still far from having resolved this phenomenon.

\subsection{Experimental data in biology}
Many important experimental results in biology have been and still continue to be obtained with fluorescence \textit{in situ} hybridization techniques (FISH). In such methods, a fluorescent probe that binds specifically a target DNA (or RNA) sequence by base complementarity is introduced in the cell and monitored with confocal microscopy imaging. This allows the spatio-temporal tracking of a precise location on the chromosome (or locus). For instance, it has been used to investigate spatial organization of transcribed genes \cite{Schoenfelder2010} or to follow the motion of loci during DNA replication \cite{Youngren201471}. However, new technologies as well as new ideas have enabled the steady improvement of experimental techniques in biology. Namely, localization-based super-resolution fluorescence techniques have considerably extended possibilities offered by FISH imaging and enabled to track fluorescent probes at a resolution of a few nanometers, below the diffraction limit. This can be done in two ways. The first is achieved by post-processing images obtained from FISH techniques. Thus the increased resolution does not come from more accurate experimental measurements but from an ingenious data treatment of many consecutive images. The second is achieved in experimental setups implementing stochastic optical reconstruction microscopy or photo-activated localization microscopy (STORM or PALM, Nobel Prize 2014). Alternatively, the tracking of quantum ``dots'' with two-photon adsorption has enabled imaging in living cells at an unprecedented resolution and with less damages caused to the cell.

Other techniques that have revolutionized experimental biology in the last decade or so are those relying on polymerase chain reactions (PCR) combined with high-throughput DNA sequencing. For instance, \chipseq techniques \cite{Bailey20131,Myers20131,Kahramanoglou2011} allow to measure the probability of binding along the genome for a protein of interest at a resolution of a few tens of base-pairs. Similarly, \hic techniques can measure the probability of contact between pairs of DNA sequences on the chromosome and output a contact probability matrix for the whole genome at a resolution of a few \SI{}{\kilo bp} \cite{Imakaev9992012,Marbouty2015,Lieberman-Aiden2009}.

The convergence in technologies now makes it possible to apply each of these techniques not on a population of cells but at the single-cell level. Although they still are at their beginning, single-cell techniques can throw light on stochastic fluctuations from a cell to another one in biological processes including chromosome organization and genetic expression.

\section{Modelling complexity in biology}
Thanks to modern experimental techniques, the spatial structures of the chromosome that we just discussed have been pretty well characterized. Yet understanding the underlying physics, that would pave the way to an era of quantitative predictions, has remained an important challenge. While it is true that all biological processes result from the superimposition of many individuals abiding by the laws of physics, the resulting system can be of a daunting complexity. In particular, problems in biology are characterized by their high-dimensionality, the non unicity of their solutions and the variety of the microscopical players involved (\textit{i.e.} the presence of disorder).

Proteins are ubiquitous in the cell and their many interactions with DNA form a complex system. On a global scale, the chromosome architecture is constrained and shaped by structuring proteins, namely nucleoid-associated proteins (NAPs) in bacteria and histones in eukaryotes. Yet less abundant but dedicated transcription factors can bind to DNA and locally alter the structure the chromosome (\textit{e.g.} by forming loops). Reconciling these two effects into a single physical model is not an easy task and requires a multi-scale approach. For example, there have been models of statistical physics showing how the cooperative binding of transcription factors can result in abrupt transitions leading to the collapse of the chromosome into loops \cite{Barbieri2012} or to an apparent increased affinity of the TF to the gene promoter \cite{Vilar2003}. Methods from statistical physics have been very successful in describing a large variety of complex system phenomena in the twentieth century, yet their application to study biological processes has found many caveats. While they are adapted to describe systems with many but identical constituents, and possibly a source of disorder, difficulties arise in biological systems which involve not a few but tens of protein types, hence these models are rarely tractable.

\begin{figure}[!htbp]
  \centering
  \subfloat[]{\label{fig:intro_territories:chrterr} \includegraphics[width = 0.25 \textwidth, valign=b]{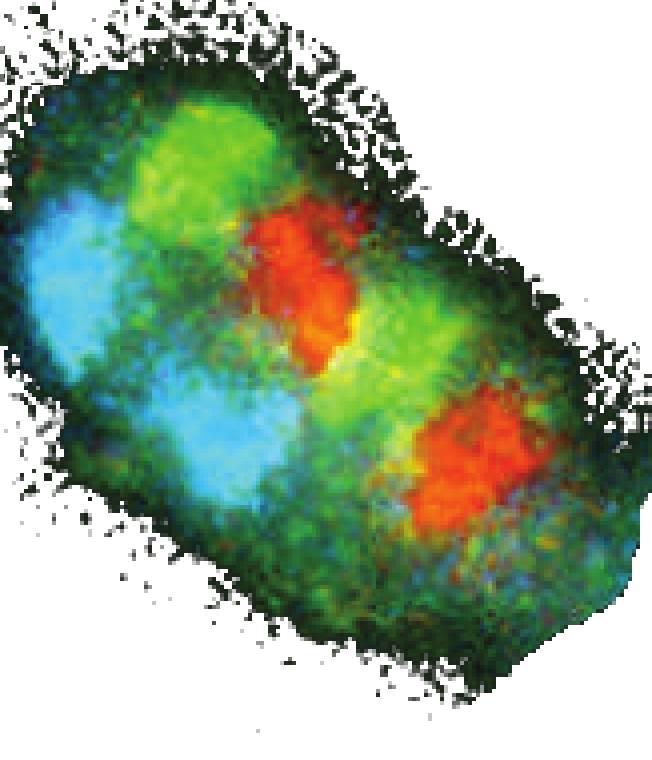}}%
  \quad
  \subfloat[]{\label{fig:intro_territories:crumpled} \includegraphics[width = 0.50 \textwidth, valign=b]{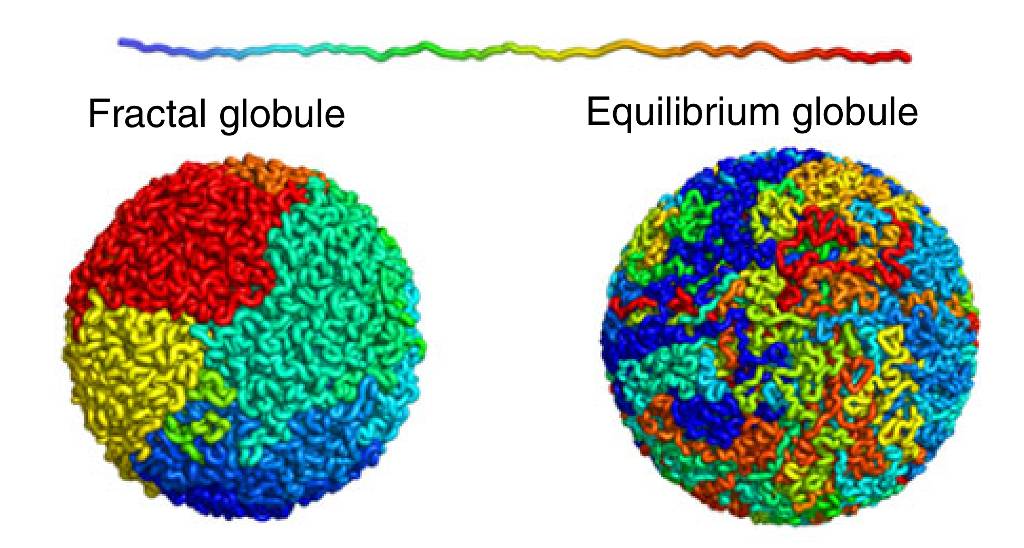}}
  \caption{\protect\subref{fig:intro_territories:chrterr} Fluorescence imaging of the chromosome 12, 14 and 15 in the nucleus of mouse liver cells display an organization into chromosome territories \cite{Meaburn:2007aa}. \protect\subref{fig:intro_territories:crumpled} Crumpled versus equilibrium polymer globules \cite{Mirny2011}.}
  \label{fig:intro_territories}
\end{figure}

For example, the organization of the chromosomes does not comply with predictions from standard polymer physics. Indeed, instead of being entangled, chromosomes remain in separated and non overlapping domains known as chromosome territories \cite{Meaburn:2007aa,Cremer2001292} (\cref{fig:intro_territories}). The configurations adopted by a single chromosome seem best described by the so-called crumpled (or fractal) globule polymer which assumes that strong topological constraints (namely excluded volume) prevents mixing and the equilibrium distribution of the polymer to be reached \cite{Grosbert1993373,Mirny2011}.

It is often hard to know whether a biological process operates at or out of thermal equilibrium. On the one hand the presence of many stationary processes suggests that processes in biology can occur at thermal equilibrium. For example, the transcriptional response to external changes can be achieved within seconds, like the SOS response to stress exposure in \textit{E. coli}. This suggests that for several biological processes, equilibrium, or at least a new stationary state, can be reached quickly. On the other hand, molecular crowding significantly increases the diffusion times, which is also known to result in anomalous diffusion \cite{Bronstein2009018102,Norrelykke2004078102}. Furthermore, consistent with the crumpled globule picture, the equilibration time for chromosomes is expected to count in tens of years, suggesting that the chromosomes in the nucleus are never equilibrated \cite{Rosa2008e1000153}.

A strategy to increase our understanding of biology and still retain a reasonable amount of complexity is to resort to molecular dynamics (MD) simulations. Even then, it is not possible in general to perform molecular dynamics simulations at the atomic resolution and produce trajectories corresponding to time scales compatible with biological times (of the order of seconds). Instead, coarse-grained approaches ignoring molecular details such as sequence effects, the double-helical structure of DNA and modeling the solvent implicitly are preferred. This kind of MD, called Brownian Dynamics (BD), has been broadly used in the past to model the dynamics of the chromosome. It has brought valuable insights on several biological processes, including genes co-localization \cite{Stefano20131}, transcription factories \cite{Brackley36052013}, or the nucleosomal architecture in eukaryotes \cite{Mergell0119152004}, and more generally on chromosome architecture \cite{Rosa2008e1000153,Becker2007,Jost2011,Pichugina2016}. Obviously, BD simulations still rely on several simplifying assumptions that reduce the underlying complexity. In particular, a trade-off must be found between the system size, \textit{i.e.} the number of constituents, and the variety of the interactions, \textit{e.g.} the types of proteins or sequence effects. For instance, a common approach is to consider a generic type of protein with average properties, which represent several protein types at the same time \cite{2016arXiv160102822B,Brackley36052013}. The investigator is then left with several free parameters to fit (or to guess), like binding energies between proteins and DNA, which in general are not known. Because of these limitations, BD simulations cannot be used yet to produce accurate quantitative predictions. However, when they are in qualitative agreement with experimental observations, they can be of precious help to understand the underlying physics and serve as proof of concept for a physical model.

In the next two sections, we spend some time to review some standard results in statistical physics that will be used at different stages of this manuscript. In particular we introduce standard polymer models of the chromosome and the Brownian dynamics framework.

\section{Polymer model of the chromosome}
\label{sec:model_chromosome_description}
\subsection{Beads-on-string polymer}
\begin{figure}[!htbp]
  \centering
  \includegraphics[width=0.30 \textwidth]{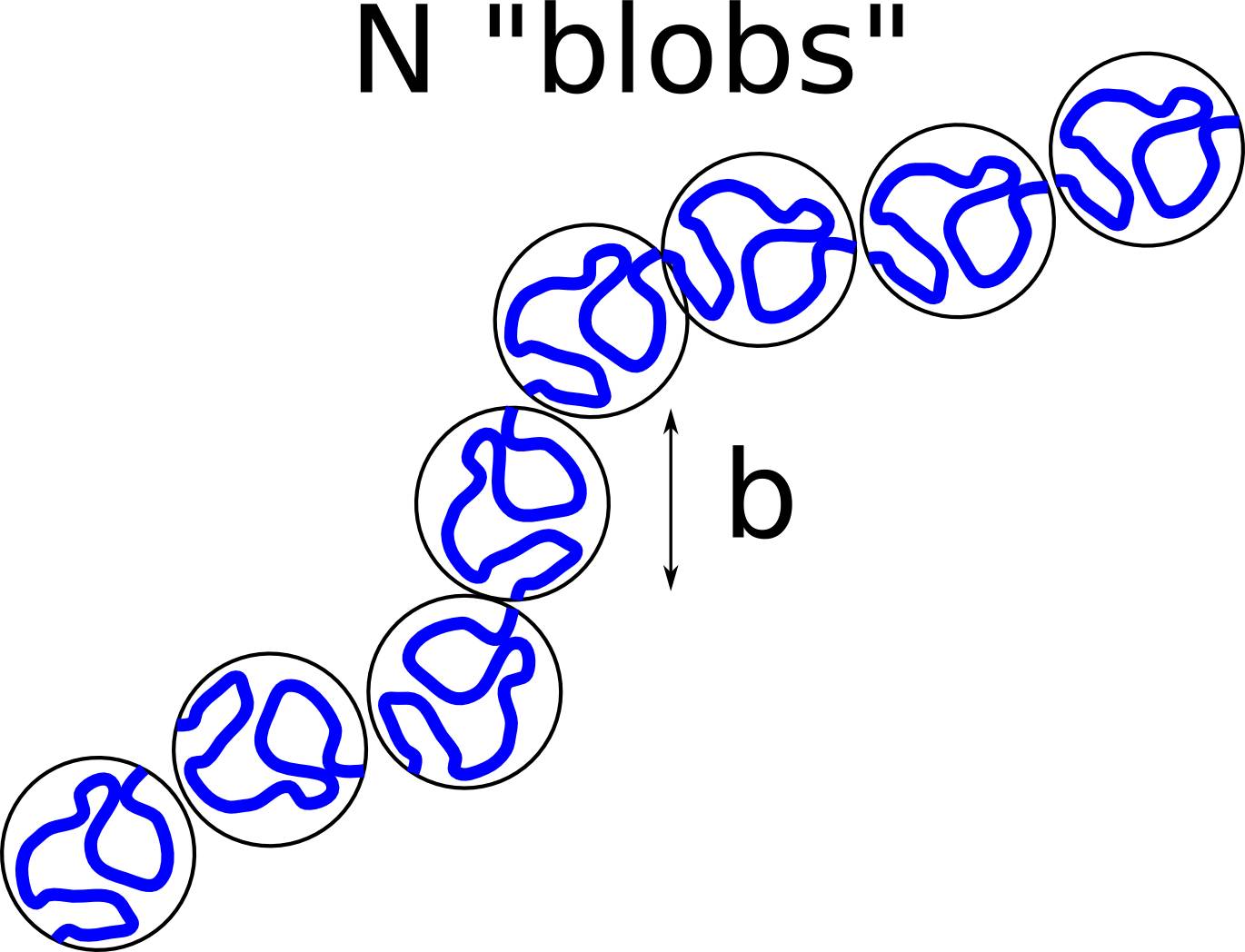}
  \caption{Beads-on-string polymer.}
  \label{fig:polymer_beads_on_string}
\end{figure}

Being a long macro-molecule, the chromosome is commonly modeled as a polymer, consisting of the repetition of structural units called monomers \cite{Bouchiat2000,Peyrard1989}. The chromosome is then divided into $N+1$ ``blobs'' of size $b$, with coordinates $\vec{r}_i$, with $i=0,\dots,N$. This is the so-called beads-on-string polymer of contour length $L=bN$ (\cref{fig:polymer_beads_on_string}). In order to be a consistent model of the reality, $b$ should be equal to the diameter of the chromosome fiber. For eukaryotes, we will consider the \SI{30}{\nm} fiber, and we obtain $b \approx \SI{3000}{bp}$. For bacteria, we will consider the naked DNA with diameter \SI{2.5}{\nm} and we obtain $b \approx \SI{7.3}{bp}$ (where we have used that one base pair has a size of approximately \SI{0.34}{\nm}). In the sequel, we introduce the standard polymer models that are used to model the chromosome. For an exhaustive review on polymers, we refer the interested reader to the classical literature \cite{deGennes1979,Edwards1988,Fredrickson2005}.

\subsection{Gaussian chain}
Assuming that the first monomer is attached to the origin, $\vec{r}_0=0$, the end-to-end vector is defined as:
\begin{align}
  \begin{aligned}
    \vec{R}_e &= \vec{r}_N \\
    &= \sum \limits_{i=1}^N \vec{u}_i,
  \end{aligned}
  \label{eq:polymer_end_to_end}
\end{align}
where $\vec{u}_i=\vec{r}_i - \vec{r}_{i-1}$ is the bond $i$ vector. The expression in \cref{eq:polymer_end_to_end} may be seen as a discrete stochastic process describing the motion of a particle making random jumps $\vec{u}_i$. If we assume that all bonds are independently and identically distributed (i.i.d.) random variables with zero mean and variance $b^2$, we obtain the mean square end-to-end distance:
\begin{equation}
  \langle R_e^2 \rangle = b^2 N,
  \label{eq:polymer_end_to_end_gaussian}
\end{equation}
where $b$, the monomer size, is also called the Kuhn length. Note that for long polymers ($N \gg 1$), we have by the central limit theorem that the probability distribution function (p.d.f.) of $R_e$ converges to a Gaussian distribution. Another useful quantity is the (square) radius of gyration:
\begin{equation}
  R_g^2 = \frac{1}{N+1} \sum \limits_{i=0}^N (\vec{r}_i -\vec{r}_{cm})^2,
  \label{eq:polymer_radius_gyration}
\end{equation}
where $\vec{r}_{cm}$ is the center of mass of the polymer. The radius of gyration gives an account of the spherical volume occupied by the polymer coil, and it has also the advantage of being defined for branched polymers, when the end-to-end vector is not.

\begin{figure}[!htbp]
  \centering
  \includegraphics[width=0.7 \textwidth]{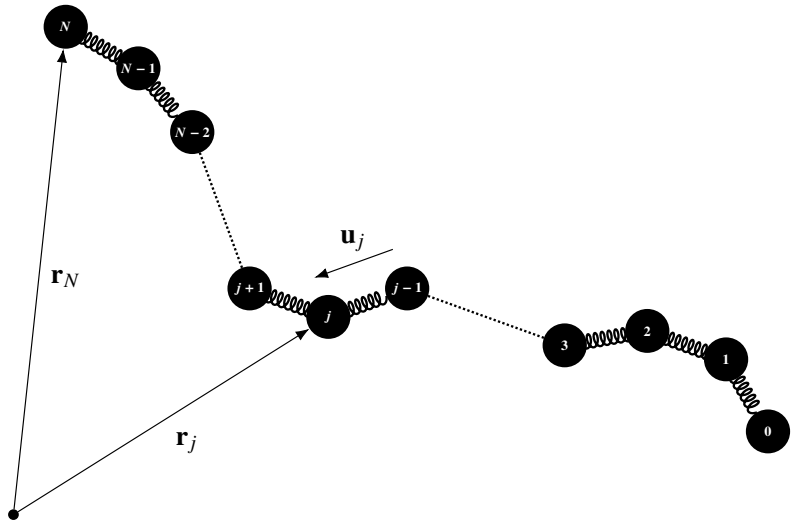}
  \caption{Gaussian chain model.}
  \label{fig:gaussian_chain_model}
\end{figure}

If we assume that all bonds $\vec{u}_i$ have Gaussian distributions, then $\vec{R}_e$ has also a Gaussian distribution. This is the so-called Gaussian chain model, which is equivalent to say that monomers are linked one to another by harmonic springs (\cref{fig:gaussian_chain_model}). The internal energy of the polymer chain is then simply obtained by summing the contributions of each spring:
\begin{equation}
  \beta U_e\left[ \lbrace \vec{r}_i \rbrace \right] = \frac{3}{2 b^2} \sum \limits_{i=1}^{N} (\vec{r}_{i} - \vec{r}_{i-1})^2.
  \label{eq:polymer_gaussian_chain_energy}
\end{equation}

The partition function of the Gaussian chain is then
\begin{align}
  Q_N &= \int \prod \limits_{i=1}^N \ud{^3 \vec{r}_i} \exp{\left( - \beta U_e\left[ \lbrace \vec{r}_i \rbrace \right]\right)},
  \label{eq:polymer_gaussian_chain_partfunc}
\end{align}
and we can compute the characteristic function of the end-to-end distance by Gaussian integral calculus:
\begin{equation}
  \begin{aligned}
    \left\langle \exp{\left( i \vec{k} \cdot \vec{R}_e \right)} \right\rangle &= \frac{1}{Q_N} \int \prod \limits_{i=1}^N \ud{^3 \vec{r}_i} \exp{\left( - \beta U_e\left[ \lbrace \vec{r}_i \rbrace \right] + i \vec{k} \cdot \vec{r}_N \right)} \\
    &= \exp{\left( -\frac{1}{2} b^2 N k^2 \right)},
  \end{aligned}
  \label{eq:polymer_gaussian_chain_characteristicfunc}
\end{equation}
from which we conclude that $\vec{R}_e$ indeed is normally distributed and with second moment as in \cref{eq:polymer_end_to_end_gaussian}. For completeness, note that \cref{eq:polymer_gaussian_chain_energy} can be extended in the continuum limit: $\vec{r}_{i} - \vec{r}_{i-1} \leftarrow \dot{\vec{r}}(s)$. The chain is then determined by the space curve $\vec{r}(s)$, where $s$ is now a continuous variable between $0$ and $N$. The energy of the continuous Gaussian chain reads:
\begin{equation}
  \beta U_e\left[ \vec{r}(s) \right] = \frac{3}{2 b^2} \int \limits_{0}^N \ud{s} \dot{\vec{r}}(s)^2.
  \label{eq:polymer_gaussian_chain_energy_continuous}
\end{equation}

In reality, approximating a polymer to a Gaussian chain is only valid for weak perturbations, and in particular when the end-to-end distance is much smaller than the contour distance: $R_e \ll Nb$. Otherwise, non-linearities in the bonds elasticity may arise. Besides, Gaussian polymers allow the bond distance to fluctuate quite a lot ($\langle u_i^2 \rangle = b^2$). This will be problematic in BD simulations with excluded volume interactions between the monomers because this would result in possible crossings between different bonds. Therefore, for BD implementations, we will prefer to \cref{eq:polymer_gaussian_chain_energy}, the finitely-extensible non-linear elastic potential (FENE):
\begin{equation}
  U_{fene} \left[ \lbrace \vec{r}_i \rbrace \right] = -\frac{3 k_{e} r_0^2}{2 b^2} \sum \limits_{i=1}^{N} \ln{\left(1 - \dfrac{u_i^2}{r_0^2} \right)},
  \label{eq:polymer_gaussian_fene}
\end{equation}
where $r_0$ is a distance above which non-linear effects start to appear in the bonds elasticity and $k_e$ is the rigidity constant of the non-linear spring. Note that for $u_i \ll r_0$ we recover \cref{eq:polymer_gaussian_chain_energy}, \textit{i.e.} a linear spring (with $k_e=1 \, k_B T$). In practical applications, and following the authors who introduced this potential \cite{Kremer50571990}, we will generally take $r_0=1.5 \, b$ and $k_{e}=10 \, k_B T$.

\subsection{Excluded volume and short-range interactions}
Gaussian chains are also known as phantom chains because monomers can overlap. In real polymers however, monomers cannot inter-penetrate, and it is necessary to introduce excluded volume interactions.

In dilute solutions, the size of real chains depends on the quality of the solvent. In good solvent, the end-to-end distance is still expressed as a power law of $N$, as in \cref{eq:polymer_end_to_end}, but with another exponent $\nu$:
\begin{equation}
  R_e \sim b N^\nu
  \label{eq:polymer_real_end_to_end}
\end{equation}
whose value has been well characterized \cite{deGennes1979}. Namely, in three dimensions $\nu \approx 0.588$. This value is well approximated by the Flory exponent $\nu_F = 3/5$. In bad solvent, the chain collapses into a close-packed configuration called globule, in which monomers are in contact. The resulting size of the coil scales like $R_e \sim b N^{1/3}$. In the other limit, for very concentrated solutions, chains behave essentially like ideal chains with size $R_e \sim b N^{1/2}$.

Let us now consider the nucleoid in \ecoli with volume $\SI{0.2}{\mu \cubic \meter}$ and a genome of length $N_g=\SI{4.6e6}{bp}$. Following the description of chromosome organization given in \cref{sec:model_chromosome_description}, we may assume that the chromosome is represented by a beads-on-string polymer with $N=N_g / b$ monomers of size $b = \SI{7.35}{bp} = \SI{2.5}{\nm}$. It follows that the volume occupied by the polymer is approximately $N \pi b^2 /4$, from which we obtain that the chromosome volume fraction in physiological conditions is $\eta \sim \num{e-2}$. By applying similar arguments to an eukaryotic nucleus of size 1-\SI{10}{\mu \meter} with a genome of length $N_g = \SI{e9}{bp}$, and monomers of size $b=\SI{3000}{bp}$ corresponding to the \SI{30}{\nm} fiber packaging, we also obtain a chromosome density of the order of $\eta \sim \numrange{e-3}{e-2}$. Thus, we may consider that the chromosome can be modeled as a polymer in a dilute solution. We will also assume that the cytosol is a good solvent for the chromosome polymer.

Most of DNA-DNA and DNA-protein interactions are in fact Coulombic interactions. Yet ions are present in the cell, giving rise to screened electrostatic interactions. The range of the interactions is given by the Debye-H\"uckel length scale, $r_{DH}$. In physiological conditions, the concentration of salt is $c_0 \approx \SI{0.1}{M}$, giving $r_{DH} \approx \SI{1}{\nm}$ \cite{Zinchenko112005,Kunze20004389}. Since interactions decay exponentially for larger distances and since proteins have a size of the order of the nanometer, the range of the interactions will be typically the size of the objects interacting together.

A commonly used two-parameter empirical form for describing non-bonded interactions between two neutral (but possibly polarized) particles is the Lennard-Jones, or ``6-12'', potential. For a pair of monomers separated by a distance $r$, it reads:
\begin{equation}
  V_{LJ}(r) = 4 \varepsilon \left( \left( \frac{\sigma}{r} \right)^{12} - \left( \frac{\sigma}{r} \right)^{6} \right),
  \label{eq:polymer_lennard_jones_basic}
\end{equation}
where $\varepsilon$ is an energy scale in $k_B T$ and $\sigma$ is the hard core distance. Here, the interaction still decays as a power law of the distance $r$. A standard method to make this interaction short-range, is to introduce a threshold $r^{th}$ such that for distances $r>r^{th}$ the interaction vanishes. Therefore, in practical applications, we will consider the truncated Lennard-Jones potential:
\begin{align}
  U_{ev}(r) =
  \left\lbrace
  \begin{aligned}
    &V_{LJ}(r) - V_{LJ}(r^{th}) & \text{ if } r<r^{th}, \\
    & 0  & \text{ otherwise.}
  \end{aligned}
  \right.
  \label{eq:polymer_lennard_jones_truncated}
\end{align}

\begin{figure}[!hbtp]
  \centering
  \includegraphics[width= 0.6 \textwidth]{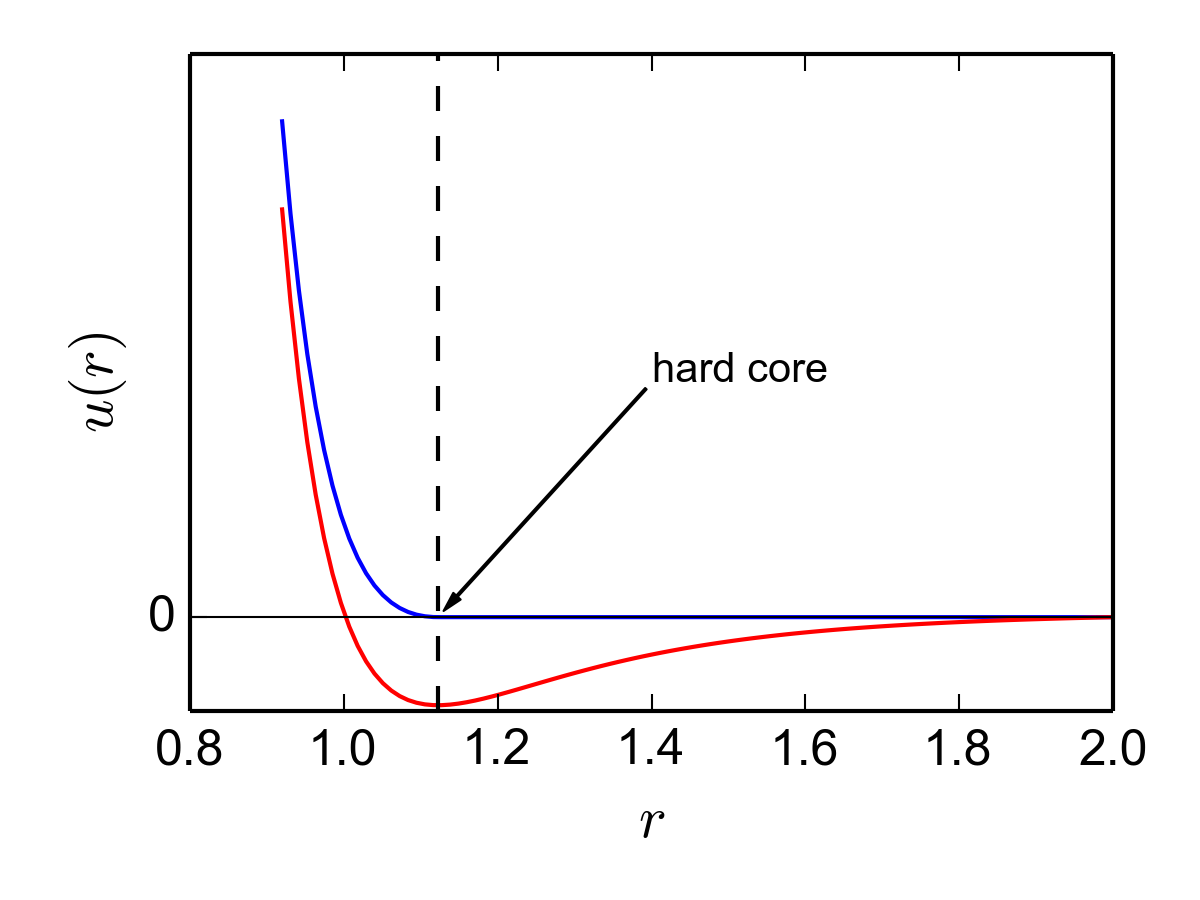}
  \caption{Potentials of interaction with different shapes. The Mayer coefficient $\alpha=\int \mathrm{d}\mathbf{r} \, \left( 1 - \exp{(-u(r))} \right) $ measures the volume excluded for one bead interacting through this potential with a bead attached to the origin. For potentials with an attractive tail (red), $\alpha$ can be negative. When $\alpha < 0$ we say that the potential is attractive, otherwise we say that the potential is repulsive.}
  \label{fig:mayer_coefficient}
\end{figure}

The form in \cref{eq:polymer_lennard_jones_truncated} can be used to describe both repulsive and attractive interactions. Indeed, the repulsive or attractive nature of the interaction depends on the sign of the Mayer coefficient\index{Mayer coefficient}:
\begin{equation}
  \alpha = \int \ud{^3 \vec{r}} \left(1 -  e^{ -\beta U(r)} \right),
  \label{eq:polymer_mayer_coefficient}
\end{equation}
which is the mean-field potential associated to the generic pair potential $U(r)$. When $\alpha>0$, the potential is repulsive, and when $\alpha < 0$ the potential is attractive despite the presence of a hard core (\cref{fig:mayer_coefficient}). Note that $\alpha$ has the dimension of a volume, and can be understood as follows. Let us consider an isolated monomer at the center of a spherical volume equal to $| \alpha |$, that we call the ``volume of influence''. An external monomer penetrating in this volume of influence will tend to be ejected when $\alpha>0$ while it will tend to remain inside when $\alpha < 0$. In the first case, $| \alpha |$ is effectively a volume from which the other monomer is excluded, while in the latter case it defines a ``basin of attraction''.

In practical implementations, we will take $\sigma = b$ and $\varepsilon = 1 \, k_B T$. Furthermore, to model a strict repulsive interaction, we will consider $r^{th}=2^{1/6} \sigma$, resulting in $U_{ev}(r) > 0$ for $r<r^{th}$, and consequently $\alpha > 0$. This choice also ensures that the repulsive force, $- \partial U_{ev} / \partial r$, vanishes precisely for $r=r^{th}$.

\subsection{Bending rigidity}
\label{subsec:polymer_model_bending_rigidity}
In reality, polymer chains are not always flexible and may oppose a resistance to bending. Incidentally, DNA is one of the best characterized examples of polymer with a large bending rigidity. There are different ways to model stiff polymers.

\subsubsection{Worm-like chain}
\begin{figure}[!htbp]
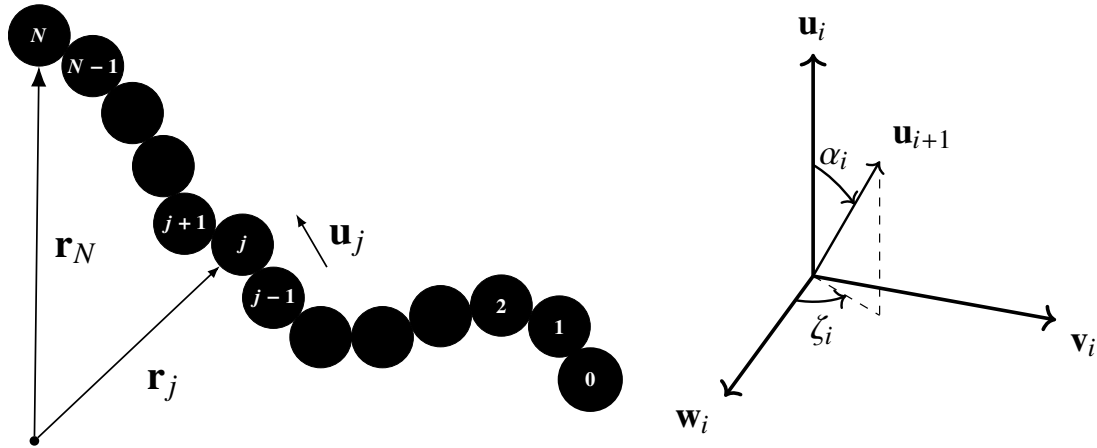

  \centering
  \includegraphics[width=0.55 \textwidth,valign=c]{wlc_chain.tex}
  \quad
  \includegraphics[width=0.4 \textwidth,valign=c]{wlc_spherical_coordinates.tex}
  \caption{Worm-like chain.}
  \label{fig:worm_like_chain}
\end{figure}

\paragraph{Model\newline}
The discrete worm-like chain, originally introduced by Kratky and Porod \cite{Kratky11061949} to describe polymers with bending rigidity, assumes that the bonds $\vec{u}_i$ are of fixed length, $b$ (\cref{fig:worm_like_chain}). For simplicity, we will take $b=1$ in the sequel. We therefore introduce the $N$ spherical coordinates systems $(\vec{w}_i,\vec{v}_i,\vec{u}_i)$ attached to each bond (the zenith is given by the bond $i$ direction). The coordinates of bond $i+1$ in the frame $i$ and the corresponding integration measure read:
\begin{align}
  \vec{u}_{i+1}=
  \begin{pmatrix}
    \sin{\alpha_i} \cos{\zeta_i}\\
    \sin{\alpha_i} \sin{\zeta_i}\\
    \cos{\alpha_i}
  \end{pmatrix}_{\left( \vec{w}_i, \vec{v}_i, \vec{u}_i \right)},
  \qquad
  \mathrm{d}^2 \vec{u}_{i+1} = \sin{\alpha_i} \mathrm{d}\alpha_i \mathrm{d}\zeta_i,
  \label{eq:spherical_coordinate_system_bond}
\end{align}
where $\alpha_i$ (resp. $\zeta_i$) is the polar angle (resp. azimuthal angle) associated to frame $i$. Hence $\alpha_i$ is the angle between bond $i$ and $i+1$. In particular, we have $\vec{u}_{i+1} \cdot \vec{u}_i = \cos \alpha_i$. We will say that the spherical system in \cref{eq:spherical_coordinate_system_bond} characterizes the joint $i$ of the chain.

The Kratky-Porod chain potential is then expressed as:
\begin{equation}
  U_{b}\left[ \left\{ \vec{r}_i \right\} \right] = \beta^{-1} \kappa \sum \limits_{i=1}^{N-1}\left( 1 -  \vec{u}_i \cdot \vec{u}_{i+1} \right),
  \label{eq:wlc_discrete}
\end{equation}
where $\kappa$ is a bending rigidity coefficient expressed in $k_B T$.

\paragraph{Partition function and chain propagator\newline}
The partition function may then be written in a compact form:
\begin{equation}
  Q_N = \int \ud{^2 \vec{u}_N} \ud{^2 \vec{u}_1} T^{N-1}(\vec{u}_N \mid \vec{u}_1 ),
  \label{eq:wlc_discrete_partfunc_formal}
\end{equation}
where we introduced the transfer matrix $T$ with elements:
\begin{equation}
  T\left( \vec{u} \mid \vec{u}' \right) = \exp{\left( -\kappa (1-\vec{u} \cdot \vec{u}') \right)},
  \label{eq:wlc_discrete_transfer_matrix}
\end{equation}
and can be factorized to give an analytical result. To this end, let us introduce the chain propagator $q_{N-1}(\vec{u})$ and the reduced probability function $\Psi_N(\vec{u})$:
\begin{align}
  \begin{aligned}
    & q_{N-1}(\vec{u}) = \int \ud{^2 \vec{u}'} T^{N-1}(\vec{u} \mid \vec{u}'), \\
    & \Psi_{N}(\vec{u}) = \frac{1}{Q_N} q_{N-1}(\vec{u}),
  \end{aligned}
  \label{eq:wlc_discrete_propag_reducedproba}
\end{align}
where we have chosen the underscript $N-1$ for the chain propagator to emphasize that it is expressed as the matrix $T$ to the power $N-1$. Therefore, $q_n(\vec{u})$ is the statistical weight for a polymer with $n$ joints (\textit{i.e.} $n+2$ monomers) to have its last bond (or the first) pointing in the $\vec{u}$ direction. In order to compute $q_n(\vec{u})$, we make use of the change of variable $\vec{u}_{i} \cdot \vec{u}_{i+1} \leftarrow \cos{\alpha}_i$. Using the independence of the joints angles we have:
\begin{align}
  \begin{aligned}
    q_n(\vec{u}) &= \int \ud{^2 \vec{u}_1} \dots \ud{^2 \vec{u}_{n}} \, T \left(\vec{u} \mid \vec{u}_n \right) \dots T \left( \vec{u}_{2} \mid \vec{u}_{1} \right) \\
    &= \prod \limits_{i=1}^{n} \left[ \int \limits_{0}^{2 \pi} \ud{\zeta_i} \int \limits_{0}^{\pi} \ud{\alpha_i} \sin{\alpha_i} \exp{\left( -\kappa(1-\cos{\alpha}_i) \right) } \right] \\
    &= z^n \qquad \text{with} \qquad z= 4 \pi \frac{\exp{\left(-\kappa \right)}}{\kappa}\sinh{\kappa}.
  \end{aligned}
  \label{eq:wlc_discrete_chainprop}
\end{align}

In particular, we get that the reduced probability $\Psi_N(\vec{u})$ is uniform. In other words, the orientation of the final bond is isotropic. Hence we retrieve the rotational invariance allowed in this model. Due to this factorization of the chain propagator, the partition function is trivially expressed as:
\begin{align}
  Q_N &= \int \ud{^2 \vec{u}_{N}} q_{N-1}(\vec{u}_N) = 4 \pi z^{N-1}.
  \label{eq:wlc_discrete_partfunc}
\end{align}

\paragraph{Orientational correlations\newline}
The WLC is also characterized by an exponential decay of the orientational correlations $\langle \vec{u}_{n+1} \cdot \vec{u}_1\rangle$ as a function of the number of joints $n$. In order to briefly review this result, let us now introduce the Green function of the discrete WLC:
\begin{equation}
  G_{n}(\vec{u},\vec{u}') = \frac{1}{z^n} T^n \left(\vec{u} \mid \vec{u}'  \right),
  \label{eq:wlc_discrete_greenfunc}
\end{equation}
where $n$ is the number of joints between the last bond, $\vec{u}_{n+1}=\vec{u}$, and the first bond, $\vec{u}_1 = \vec{u}'$ in a chain with $n+2$ monomers. In order to obtain the orientational correlations, we first compute the thermodynamical average $\langle \cos{\alpha_i} \rangle$ for any joint $i$. A similar computation as in \cref{eq:wlc_discrete_chainprop} yields
\begin{equation}
  \langle \cos{\alpha_i} \rangle = \coth{(\kappa)} - \frac{1}{\kappa}.
  \label{eq:wlc_discrete_cos_avg}
\end{equation}

By expressing $\vec{u}_{n+1}$ in the spherical coordinate system attached to $\vec{u}_n$:
\begin{equation}
  \vec{u}_{n+1} = \sin{\alpha_n} \cos{\zeta_n} \vec{w}_n + \sin{\alpha_n} \sin{\zeta_n} \vec{v}_n + \cos{\alpha_n} \vec{u}_n,
\end{equation}
we see that
\begin{align}
  \begin{aligned}
    \langle \vec{u}_{n+1} \cdot \vec{u}_1 \rangle &= \langle \cos{\alpha_n} \vec{u}_{n} \cdot \vec{u}_1 \rangle + \langle c_1 \cos{\zeta_n} + c_2 \sin{\zeta_n} \rangle \\
  &= \langle \cos{\alpha_n} \rangle \ \langle \vec{u}_{n} \cdot \vec{u}_1 \rangle,
  \end{aligned}
\end{align}
where we have used the independence of consecutive polar and azimuthal angles, and $\langle \cos{\zeta_n} \rangle = \langle \sin{\zeta_n}\rangle = 0$. By recurrence, and by substituting \cref{eq:wlc_discrete_cos_avg}, we obtain the orientational correlations:
\begin{equation}
  \langle \vec{u}_{n+1} \cdot \vec{u}_1 \rangle = \left( \coth{\kappa} - \frac{1}{\kappa} \right)^n \xrightarrow[\kappa \to \infty]{} \exp{\left(-\frac{n}{\kappa}\right)},
  \label{eq:wlc_discrete_correlation_tangents}
\end{equation}
for $\kappa \gg 1$. Therefore, $\kappa$ characterizes the distance (in unit of monomers) above which the chain looses the memory of its orientation. Following standard notations, $\kappa$ is usually called the persistence length and noted $l_p=\kappa$. For moderate values of $\kappa$, we can make use of the two largest eigenvalues ($\lambda_0 > \lambda_1$) of the transfer matrix $T$:
\begin{align}
  \begin{aligned}
    \langle \vec{u}_{n+1} \cdot \vec{u}_1 \rangle &= \int \ud{\vec{u}} \ud{\vec{u}'} \vec{u} \cdot \vec{u}' \, G_n(\vec{u},\vec{u}') \\
    &= \frac{\int \ud{\vec{u}} \ud{\vec{u}'} \vec{u} \cdot \vec{u}' \, T^n \left( \vec{u} \mid \vec{u}' \right)}{\int \ud{\vec{u}} \ud{\vec{u}'} T^n \left( \vec{u} \mid \vec{u}' \right)} \\
    &\sim \left( \dfrac{\lambda_1}{\lambda_0} \right)^{n}
  \end{aligned}.
  \label{eq:wlc_discrete_correlation_tangents_2}
\end{align}

\begin{figure}[!htbp]
  \centering
  \includegraphics[width=0.60 \textwidth]{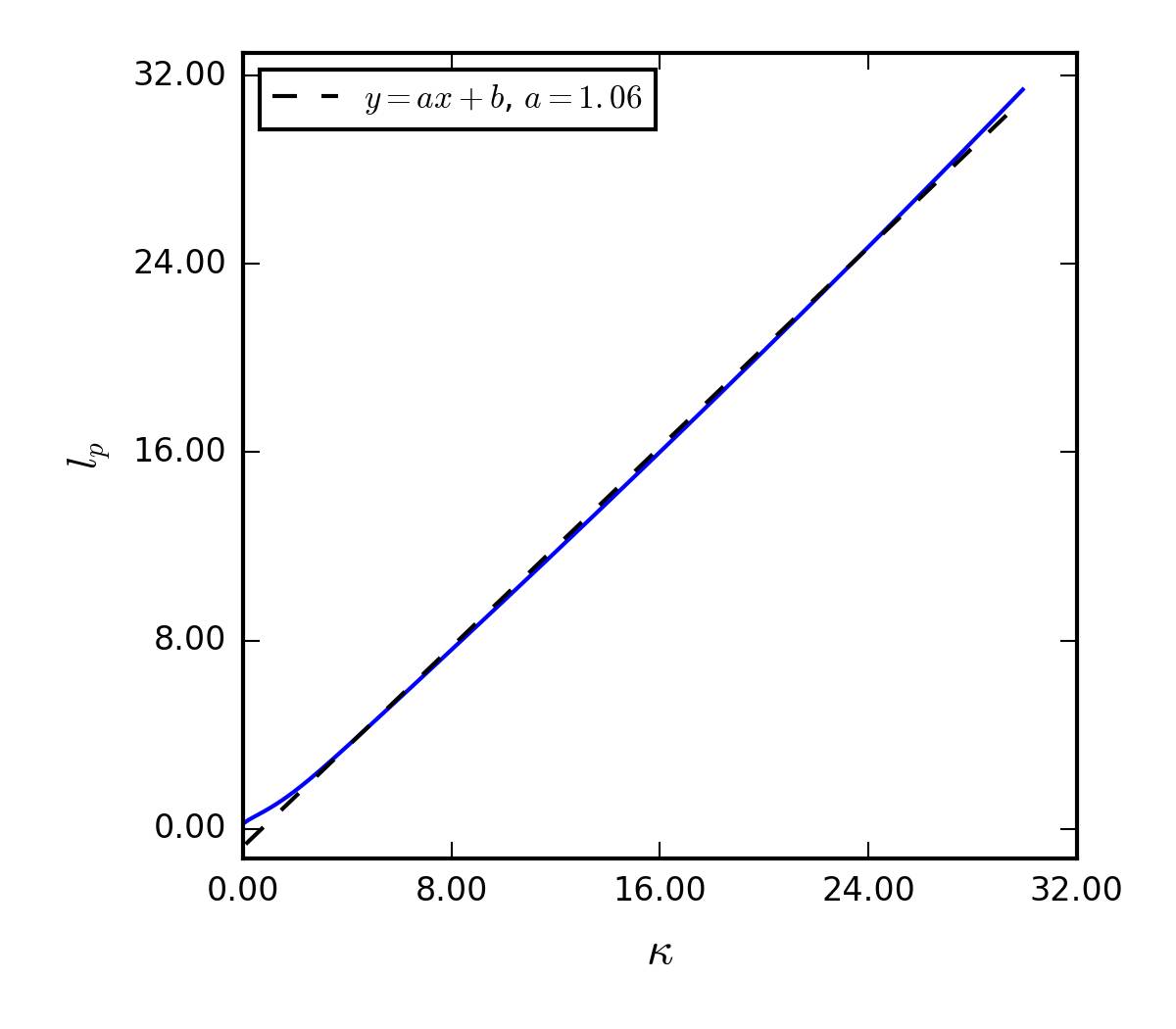}
  \caption{Persistence length computed from the ratio of the two largest eigenvalues of $T$ (\cref{eq:wlc_discrete_correlation_tangents_2}) as $l_p=-1/\log{(\lambda_1 / \lambda_0)}$ for different values of $\kappa$. We have used a discretization of the polar angle $\alpha$ interval $[0,\pi]$ in \num{1000} points (the azimuthal angle is irrelevant and disappear from the integration).}
  \label{fig:persistence_length}
\end{figure}

Therefore, the persistence length is more generally defined as $l_p=-1/\log{(\lambda_1 / \lambda_0)}$. We have computed the persistence length for several values of $\kappa$ (\cref{fig:persistence_length}). Clearly the persistence length quickly converges to the bending rigidity parameter, $l_p \to \kappa$, and \cref{eq:wlc_discrete_correlation_tangents} can be considered as a good approximation in most cases.

Let us note that similarly to the Gaussian chain, the Kratky-Porod potential in \cref{eq:wlc_discrete} can be defined in the continuum limit: $(1 - \vec{u}_i \cdot \vec{u}_{i+1}) \leftarrow \dot{\vec{u}}^2(s)/2$. In that case the correlation of the tangents is always: $\langle \vec{u}(0) \cdot \vec{u}(s) \rangle =\exp{(-s/l_p)}$. The continuous worm-like chain is presented in further details in \cref{app:wlc_continuous}.

\subsubsection{Gaussian chain with curvature penalty}
\label{sec:polymer_models_bending_rigidity_gaussian}
Although for numerical simulations we will generally consider the WLC model of \cref{eq:wlc_discrete}, it is not always adapted to analytical calculations because of the constraint on the bond length: $| \vec{u}_n| = 1$. A simple alternative is to relax this strict constraint and introduce instead a Lagrange multiplier $\lambda$. This trick allows us to have an integration measure for the bonds on the full volume instead of the unit sphere. The partition function for this model reads \cite{Ha199521,Diamant062000,Bhattacharjee1997}:
\begin{equation}
  Q_N = \int \prod \limits_{i=1}^{N} \ud{^3 \vec{u}_i } \exp{\left( -\frac{3}{4}l_p \sum \limits_{i=1}^{N}(\vec{u}_i - \vec{u}_{i-1)})^2 - \lambda \sum \limits_{i=1}^N u_i^2 \right)}
  \label{eq:polymer_semiflexible_partfunc}
\end{equation}
which is a Gaussian integral. Hence it can be computed, and the result in the $N \to + \infty$ limit is:
\begin{equation}
  Q_N=z^N, \qquad z=\left( \frac{4}{3 \pi l_p} \right)^{3/2} \exp{\left( 3 - \sqrt{3 \lambda / l_p} \right)}.
  \label{eq:polymer_semiflexible_partfunc_factorized}
\end{equation}

The Lagrange multiplier can be then determined with a self-consistent argument:
\begin{equation}
  -\frac{1}{N} \frac{\partial \ln{Q_N}}{\partial \lambda} = \langle u_n^2 \rangle = 1 \Leftrightarrow \lambda = \frac{3}{4 l_p}.
  \label{eq:polymer_semiflexible_multiplier}
\end{equation}

Remarkably, this model reproduces the orientational correlations of the WLC, namely
\begin{equation}
\langle \vec{u}_1 \cdot \vec{u}_{n+1} \rangle = \exp{(-n/l_p)}.
\end{equation}

\subsubsection{Persistence length values for DNA}
On the basis of a worm-like chain model, naked DNA has a persistence $l_p=\SI{50}{\nm}$ and the \SI{30}{\nm} fiber has a persistence length $l_p=$ \SIrange{60}{90}{\nm} \cite{Langowski2412006}. Therefore, in our familiar monomer units, we will typically consider in practical applications $l_p=20 \, b$ for the former and $l_p = 3 \, b$ for the latter.

\subsection{Final model of the chromosome}
In summary, by collecting the potentials described in the last paragraphs, the chromosome will be modeled as a beads-on-string polymer with $N+1$ monomers, and with a potential energy given by:
\begin{equation}
  U = U_{fene} + U_b + U_{ev}.
  \label{eq:polymer_total_energy}
\end{equation}

\section{Brownian dynamics}
Brownian dynamics simulations are molecular dynamics simulations in which many molecular details are coarse-grained. In particular, beads in simulations do not represent an atom nor a base pair, but instead a ``blob'' which is the basic entity of a mesoscopic description of the chromosome. Furthermore, the solvent (\textit{i.e.} water molecules, plus salt and ions in solution) are not modeled explicitly. Instead, each bead exchanges energy with a thermal bath at temperature $T$. The classical framework to describe the Brownian motion of a particle is the Langevin equation.

\subsection{The Langevin equation}
Let us consider the motion of a particle with coordinates $x(t)$. The Langevin equation is nothing else than the Newton equation of motion for a particle in a viscous medium plus a stochastic term:
\begin{equation}
  m \ddot{x}(t) = - \gamma \dot{x} - \frac{\partial U}{\partial x}(x(t)) + \gamma \eta(t),
  \label{eq:brownian_motion_langevin_equation}
\end{equation}
in which $m$ is the mass of the particle, $\gamma$ is a damping term and $- \partial U / \partial x$ is the force applied to the particle with $U$ being the potential energy of the particle. These first three terms are deterministic. In addition there is a stochastic term, $\eta(t)$ which represents random collisions with the solvent at temperature $T$. More accurately, $\eta$ is an uncorrelated continuous random process with two first moments:
\begin{equation}
  \langle \eta(t) \rangle = 0, \qquad \langle \eta(t) \eta(t') \rangle = 2 D \delta(t-t'),
  \label{eq:brownian_motion_noise_moments}
\end{equation}
where $D$ is the diffusion coefficient of the particle. It can be shown that in order to sample the Boltzmann equilibrium, $D$ needs to satisfy the Stokes-Einstein relation (see \cref{sec:langevin_boltzmann_sampling}):
\begin{equation}
  D=k_B T / \gamma,
  \label{eq:stokes_einstein}
\end{equation}
where finally from the Stokes' law applied to a bead of diameter $b$ we get $\gamma=3 \pi b \mu$, with $\mu$ being the fluid viscosity.

In order to produce trajectories of polymer dynamics, the Langevin equation \cref{eq:brownian_motion_langevin_equation} is applied to each bead and integrated numerically with the LAMMPS simulation package \cite{Plimpton19951}, which uses a standard velocity Verlet integration scheme \cite{NumericalRecipes2007}. Practically, this requires the choice of an integration time step $dt$. Unless specified otherwise, we will consider in this thesis $dt=\num{e-2}$ when there is no excluded volume interaction, and $dt=\num{e-3}$ otherwise. We also set $\gamma=1$ (in simulation dimensionless units).

\subsection{Mapping to real time}
BD simulations can be used in order to compute equilibrium quantities and validate theoretical predictions. Furthermore, it is possible to map the simulation time to the real time.

Let us write the diffusion coefficient as $D=b^2 / \tau_B$. During the time $\tau_B$, a particle typically travels through a distance $b$, which is its own size. Consequently $\tau_B$ is the natural unit of time for this diffusive process and is called the Brownian time. In BD simulations we take $b=1$ and $D=1$ (in dimensionless units), therefore a unit of simulation time correspond to the Brownian time.

The diffusion coefficient in the bacterial nucleoid was found to be $D=\SI{10}{\mu \meter^{2} . \second^{-1}}$ \cite{Elowitz1999}. Therefore, for $b=\SI{2.5}{\nm}$ we find $\tau_B=\SI{600}{\nano \second}$ and for $b=\SI{30}{\nm}$ we find $\tau_B=\SI{90}{\mu \second}$. Consequently, by performing runs of \num{e5} simulation time units, we can typically produce trajectories corresponding to real times between \SI{10}{\milli \second} and \SI{10}{\second}.

\subsection{A practical detail: relaxation of polymer systems with excluded volume}
In general we will want to start from a random configuration of a self-avoiding polymer. Although we can start from an arbitrary configuration respecting excluded volume constraints, the relaxation to the Boltzmann equilibrium can be very slow. Below is a standard procedure to circumvent this problem and generate quickly an initial configuration for a polymer with excluded volume interactions.

First, perform a relaxation run without excluded volume or short-range attractive interactions. This corresponds to the dynamics of an ideal chain and aims at sampling rapidly a large number of configurations to loose the memory of the initial condition.

Second, perform an intermediate run with few iterations (generally \num{e6} iterations at $dt=\num{e-3}$) with a soft pair potential:
\begin{equation}
  U_{soft}(r)=A\left( 1 + \cos{\left( \frac{\pi r}{r^{th}} \right)} \right),
  \label{eq:polymer_soft_pair}
\end{equation}
where $r^{th}$ is the same cutoff as in the truncated Lennard-Jones potential from \cref{eq:polymer_lennard_jones_truncated}. The magnitude $A$ is progressively increased from 1 to 60 during the run \cite{Kremer50571990}, so that we obtain in the end a configuration with no overlaps between the beads.

Finally, the main run with excluded volume and short-range interactions is performed starting from the configuration without overlaps. Several configurations (generally \num{1000}) are extracted from the resulting trajectory, which sample the Boltzmann ensemble. These configurations can be used to compute equilibrium averages according to the ergodic property of the Boltzmann equilibrium.

\section{Organization of the thesis and personal contributions}
This thesis aims at proposing physical models for some of the functional chromosome architectures characterized or conjectured in biology. In addition, we have sought to understand at a phenomenological level how these structural features can influence the transcription in living cells. Our strategy has been to start from simple physical models that may capture observed features and use methods from statistical physics to obtain analytical results. However, as mentioned previously, such models are not always amenable to analytical solutions. Therefore, we also have used BD simulations in order to complement our studies and sometimes bring unique insights.

In \cref{ch:transcription_factories}, I present the work published in \cite{LeTreut012016} on the modelling of transcription factories. Transcription factories are clusters of DNA and proteins, characterized \textit{in vivo}, from which most of the transcribed RNAs originate. Despite increasing evidences, very little is known about the structure of these clusters, let alone the underlying physical mechanism. At some point during this investigation, and in the context of a polymer field theory, I needed to compute the structure function of a polymer with semi-flexibility. Yet no analytical form is known, and this led me to design a method based on complex transfer matrices, that I present in \cref{ch:structure_function}.

In \cref{ch:naps}, I present a model for a regulatory mechanism of the transcription, based on the formation of DNA hairpin loops by the H-NS structuring protein in \textit{E. coli}. The disruption of these structures by external transcription factors may constitute a way to relieve H-NS mediated repression, although this last mechanism has not been investigated in details in this thesis.

In \cref{ch:ccc}, I propose a method to reconstruct the chromosome architecture from contact matrices obtained with Chromosome Conformation Capture experiments. Namely, the resulting polymer model reproduces the experimental contacts. This achievement constitutes a major improvement compared with other methods proposed in the literature.

\begin{subappendices}

\section{Asymmetrical DNA double-helix}
\begin{figure}[!htbp]
  \centering
  \includegraphics[width= 0.4 \textwidth]{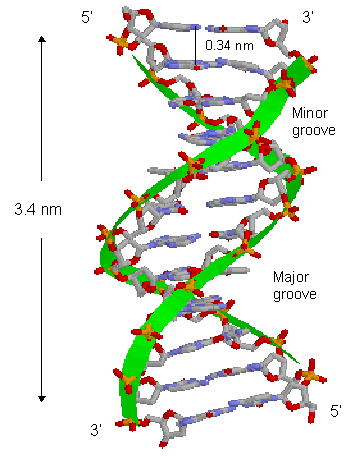}
  \caption{The DNA double-helix is asymmetrical and induced the existence of a major and a minor groove. \cite{wikipedia_dna}}
  \label{fig:dna_structure}
\end{figure}

The four fundamental bases can be divided into purines (A and G) and pyrimidines (T and C). A purine contains a single heterocycle in its chemical composition whereas a pyrimidine contains two of them. This introduces an asymmetry in the DNA double-helix. Namely, the DNA molecule has two asymmetric grooves. One groove is smaller than the other. The larger groove, which is called the major groove, occurs when the backbones are far apart, while the smaller one is called the minor groove and occurs when they are close together (\cref{fig:dna_structure}).

The major and minor grooves expose in a different manner the edges of the bases. As might be expected, the major groove provides an easier access to the bases than the minor groove. Hence the specific binding of proteins to DNA is generally achieved by making contacts with bases through the major groove.

\section{Continuous worm-like chain}
\label[app]{app:wlc_continuous}
\paragraph{Model\newline}
The Kratky-Porod model \cite{Kratky11061949}, also known as Worm-Like Chain (WLC), can be formulated for continuous chains. A polymer of length $L=b N$ is described by a space curve, $\vec{r}(s)$, where $s$ is an arc variable varying continuously from one end at $s=0$, to the other at $s=N$. Here, $b$ represents a unit length, which is typically the length of the smallest building unit of the polymer. In addition, let us introduce $\vec{u}(s)$, the unit tangent vector to the curve $\vec{r}(s)$ at coordinate $s$. The internal energy of the WLC can be written as
\begin{equation}
  \beta U_b \left[ \vec{u}(s) \right] = \frac{\kappa}{2} \int \limits_{0}^N \ud{s} \left( \frac{\mathrm{d} \vec{u}}{\mathrm{d} s} \right)^2
  \label{eq:wlc_continuous}
\end{equation}
where $\beta=(k_B T)^{-1}$ is the inverse temperature and $\kappa$ is a bending rigidity parameter. Therefore, $\beta U_b$ is a functional of $\vec{u}(s)$ which belongs to the unit sphere. Namely, in the spherical coordinate system attached to the z-axis we have:
\begin{align}
  \vec{u} =
  \begin{pmatrix}
    \sin{\theta}\cos{\varphi} \\
    \sin{\theta}\sin{\varphi} \\
    \cos{\theta}
  \end{pmatrix},
  \qquad
  \mathrm{d}^2\vec{u} = \sin{\theta} \mathrm{d}\theta \mathrm{d}\varphi
  .
  \label{eq:spherical_coordinate_system_zaxis}
\end{align}

\paragraph{Partition function and chain propagator\newline}
The partition function reads:
\begin{equation}
  Q_N = \int \uD{\vec{u}(s)} \exp{\left( -\beta U_{0}\left[ \vec{u} \right] \right)} \delta\left( \mid \vec{u}(s) \mid -1 \right).
  \label{eq:wlc_continuous_partfunc}
\end{equation}

Because of the constraint on the norm of $\vec{u}(s)$, computing the integral in \cref{eq:wlc_continuous_partfunc} is rather difficult. We now turn our attention to the Chapman-Kolmogorov (or Schr\"odinger) equation satisfied by $\Psi(\vec{u};s)$, which is the \pdf that the last tangent of a chain of length $s$ is $\vec{u}$. To this end, we introduce the chain propagator $q(\vec{u};s)$, which is the statistical weight that the last (or the first) segment of a chain of length $s$ is $\vec{u}$. More formally:
\begin{equation}
  \begin{aligned}
  &q(\vec{u} ; s) &=&  \int^{\vec{u}(s)=\vec{u}} \uD{\vec{u}(\sigma)} \exp{\left(- \beta U_b\left[ \vec{u}(\sigma) \right] \right)} \delta\left( \mid \vec{u}(\sigma) \mid -1 \right) \\
  &\Psi(\vec{u};s) &=& \frac{1}{Q_N} q(\vec{u};s)
  \end{aligned}.
  \label{eq:wlc_continuous_reducedproba}
\end{equation}

In what follows, we will define the thermodynamical average for any functional $A\left[ \vec{u}(s) \right]$ of the tangent curve as:
\begin{equation}
  \langle A\left[ \vec{u}(s) \right] \rangle = \frac{1}{Q_N} \int \uD{\vec{u}(s)} A\left[ \vec{u}(s) \right] \exp{\left( -\beta \mathbf{U_{b}}\left[ \vec{u} \right] \right)} \delta\left( \mid \vec{u}(s) \mid -1 \right).
\end{equation}

\paragraph{Chapman-Kolmogorov equation\newline}
We can make use of the Markovian structure of the path integral in \cref{eq:wlc_continuous_reducedproba} to split the integration over several sub-chains, connected one to the next. In particular, for small variations in the chain length, $\Delta s$, and small displacements on the unit sphere, $\Delta \vec{u}$ we can write (see \cite{Fredrickson2005}):
\begin{align}
  \begin{aligned}
    \Psi(\vec{u};s+\Delta s) &= \frac{1}{4 \pi} \int \ud{^2(\Delta \vec{u})} \Psi(\vec{u}-\Delta \vec{u} ; s) \exp{\left( -\beta \Delta U_b \right)} \\
    &= \Psi(\vec{u};s) - \langle\Delta \vec{u}\rangle_{\mu} \frac{\partial \Psi}{\partial \vec{u}} + \frac{1}{2} \langle \Delta \vec{u}^2 \rangle_{\mu}  \frac{\partial^2 \Psi}{\partial \vec{u}^2} + o(\langle \Delta \vec{u}^2 \rangle_{\mu})
  \end{aligned}
\end{align}
where the variation in internal energy is written in terms of the displacement $\Delta \vec{u}$ on the unit sphere:
\begin{equation}
\beta \Delta U_b = \frac{\kappa}{2 \Delta s} \Delta \vec{u}^2 = \frac{\kappa}{2 \Delta s} (\Delta \theta^2 + \sin^2{\theta} \Delta \varphi^2),
\label{eq:wlc_continuous_gaussian_weight}
\end{equation}
and the bracket averages are computed from the Gaussian weight $\mu$ such as:
\begin{align}
  \begin{aligned}
    & \mu(\Delta \vec{u}) = \frac{1}{4 \pi} \exp{(-\beta \Delta U_b)} \\
    & \langle\Delta \vec{u}\rangle_{\mu} = 0, \qquad \langle\Delta \vec{u}^2 \rangle_{\mu} = \frac{\Delta s}{\kappa}.
  \end{aligned}
\end{align}

In the limit $\Delta s \to 0$, we obtain the Chapman-Kolmogorov equation:
\begin{equation}
  \frac{\partial \Psi}{\partial s}(\vec{u};s) = \frac{1}{2 \kappa} \frac{\partial^2 \Psi}{\partial \vec{u}^2},
  \label{eq:chapman_kol_wlc}
\end{equation}
where the operator $\partial^2 / \partial \vec{u}^2$ is the Laplacian on the unit sphere:
\begin{equation}
  \frac{\partial^2 \Psi}{\partial \vec{u}^2} = \frac{1}{\sin{\theta}} \frac{\partial}{\partial \theta} \left( \sin{\theta} \frac{\partial \Psi}{\partial \theta} \right) + \frac{1}{\sin^2 \theta} \frac{\partial^2 \Psi}{\partial \varphi^2}.
  \label{eq:laplacian_unit_sphere}
\end{equation}

\paragraph{Orientational correlations\newline}
Let us now introduce the Green function, $G(\vec{u},\vec{u}';s)$, which is the \pdf that a chain of length $s$ has its last segment oriented according to $\vec{u}$, and its first segment oriented according to $\vec{u}'$. Formally, it is defined as
\begin{equation}
  G(\vec{u},\vec{u}';s-s') = \langle \delta(\vec{u}(s) - \vec{u}) \delta(\vec{u}(s')-\vec{u}') \rangle
  \label{eq:wlc_continuous_greenfunc}
\end{equation}
In particular, The Green function is related to the reduced probability function:
\begin{equation}
  \Psi(\vec{u};s) = \frac{1}{4 \pi} \int \ud{^2 \vec{u}'} G(\vec{u}, \vec{u}' ; s-s') \Psi(\vec{u}';s').
  \label{eq:wlc_continuous_greenfunc_relation}
\end{equation}

The Green function is particularly useful for analyzing statistical properties. In particular, let us define the orientational correlation function:
\begin{equation}
  \langle \vec{u}(s) \cdot \vec{u}(0) \rangle = \frac{1}{4 \pi} \int \ud{\vec{u}} \ud{\vec{u}'} G(\vec{u}, \vec{u}';s) \vec{u} \cdot \vec{u}'.
  \label{eq:wlc_continuous_correlation_tangents_interm}
\end{equation}

In order to compute \cref{eq:wlc_continuous_correlation_tangents_interm}, we can use the fact that $G(\vec{u},\vec{u}';s)$ also follows the Chapman-Kolmogorov equation in \cref{eq:chapman_kol_wlc}, with the initial condition $G(\vec{u},\vec{u}' ;0) = \delta(\vec{u} - \vec{u}')$. Then we write:
\begin{align}
  \begin{aligned}
    \frac{\partial  \langle \vec{u}(s) \cdot \vec{u}(0) \rangle}{\partial s} = \frac{1}{4 \pi} \int \ud{\vec{u}} \ud{\vec{u}'} \frac{\partial G(\vec{u}, \vec{u}';s)}{\partial s} \vec{u} \cdot \vec{u}'
  \end{aligned}
\end{align}
and by using \cref{eq:chapman_kol_wlc} and integrating by part (see \cite{Edwards1988}), we obtain:
\begin{equation}
  \langle \vec{u}(s) \cdot \vec{u}(0) \rangle = \exp{\left( - \frac{s}{\kappa} \right)} .
  \label{eq:wlc_continous_correlation_tangents}
\end{equation}

In conclusion, the bending rigidity coefficient is usually referred as the persistence length: $l_p = \kappa$. It characterizes the contour distance over which orientational correlations decay.

\section[Sampling of the Boltzmann equilibrium]{Sampling of the Boltzmann equilibrium by the Langevin equation}
\label[app]{sec:langevin_boltzmann_sampling}
We recall here why the Langevin dynamics in the stationary regime samples the Boltzmann equilibrium. In the over-damped limit $\tau_B \ll m / \gamma $ (light particle or viscous solvent), the acceleration term in \cref{eq:brownian_motion_langevin_equation} can be neglected. Hence we obtain the over-damped Langevin equation:
\begin{equation}
  \dot{x}(t) = - D \beta \frac{\partial U}{\partial x}(x(t)) + \eta(t),
  \label{eq:langevin_overdamped}
\end{equation}
where $\eta(t)$ is an uncorrelated noise with first moments given in \cref{eq:brownian_motion_noise_moments}.

Let us now consider a generic observable of the particle position $f(x)$. According to the It\^o calculation rule, the variations of $f(x)$ along the particle trajectory reads:
\begin{equation}
  \mathrm{d}f(x(t)) = \frac{\partial f}{\partial x} \mathrm{d}x(t) + \frac{1}{2} \frac{\partial^2 f}{\partial x^2} \mathrm{d}x(t)^2.
  \label{eq:langevin_ito_rule}
\end{equation}

We will now obtain the continuity equation for $\rho(x,t)$, which is the probability that the diffusing particle is at position $x$ at time $t$. On one hand, using \cref{eq:langevin_overdamped,eq:langevin_ito_rule}, we have:
\begin{equation}
\begin{aligned}
  \left\langle \frac{\mathrm{d} f(x(t))}{\mathrm{d}t} \right\rangle &= \left\langle \frac{\partial f}{\partial x}(x(t)) \frac{\mathrm{d} x}{\mathrm{d} t}(t) + \frac{1}{2} \frac{\partial^2 f}{\partial x^2}(x(t)) \left( \frac{\mathrm{d} x}{\mathrm{d}t}(t)\right)^2 \right\rangle \\
    &= \left\langle - D \beta \frac{\partial f}{ \partial x}(x(t)) \frac{\partial U}{\partial x}(x(t)) + \frac{\partial^2 f}{\partial x^2}(x(t)) D \right\rangle \\
    &= \int \ud{x} \rho(x,t) \left(- D \beta \frac{\partial f}{\partial x}(x) \frac{\partial U}{\partial x}(x) + \frac{\partial^2 f}{\partial x^2}(x) D   \right) \\
    & = \int \ud{x} f(x) \frac{\partial}{\partial x} \left( D \beta \frac{\partial U}{\partial x}(x) \rho(x,t) + D \frac{\partial \rho}{\partial x}(x,t) \right),
\end{aligned}
  \label{eq:continuity_1}
\end{equation}
where we have used the independence between $x(t)$ and $\eta(t)$, \cref{eq:brownian_motion_noise_moments} and integration by parts to obtain the last line. On the other hand, we have by definition:
\begin{equation}
  \frac{\mathrm{d}}{\mathrm{d} t}\langle f(x(t)) \rangle = \int \ud{x} f(x) \frac{\partial \rho}{\partial t}(x,t).
  \label{eq:continuity_2}
\end{equation}

Therefore, by equating \cref{eq:continuity_1,eq:continuity_2}, we obtain the heat equation:
\begin{equation}
  \begin{aligned}
    & \frac{\partial \rho}{\partial t}(x,t) + \frac{\partial j}{\partial x}(x,t) = 0, \\
    & j(x,t)= - D \beta \frac{\partial U}{\partial x}(x) \rho(x,t) - D \frac{\partial \rho}{\partial x}(x,t),
  \end{aligned}
  \label{eq:heat}
\end{equation}
where $j(x,t)$ is the local density current of particles. The equilibrium is achieved when $j(x,t)=0$, yielding:
\begin{equation}
  \rho_{eq}(x,t) \propto \exp{\left( -\beta U(x) \right)},
  \label{eq:langevin_boltzmann_dist}
\end{equation}
which is the Boltzmann distribution.
\end{subappendices}


\chapter{Modelling of transcription factories}
\label{ch:transcription_factories}
In this chapter, we address the characterization of transcription factories, which are clusters of DNA and proteins where presumably active genes are transcribed. We start by an overview about gene co-regulation, and in particular we introduce recent developments in biology suggesting that the regulation of the expression of genes belonging to a same network entails their co-localization in space. We then introduce transcription factories and discuss what is known about their biological functions.

There are only few physical models for the existence of transcription factories, and still many open questions. Hence, in a first approach, we propose a model grounded in a polymer representation of the chromosome in interaction with a solution of binding proteins, that we call a formal nucleus. In order to characterize the thermodynamical equilibrium of this formal nucleus, a Flory-Huggins free energy model was implemented. We found that depending on the DNA-protein affinity, the DNA chromosome may collapse, resulting in a biphasic regime with a dense and a dilute phase. The dense phase is then a model for transcription factories. Furthermore, we explored the dependence of the collapse on DNA and protein concentrations. In particular, we computed the corresponding phase diagram.

Although the Flory-Huggins theory gives a proof of principle for the existence of clusters of DNA and proteins at equilibrium such as transcription factories, it does not give information on the structure of such a dense phase. By drawing a parallel with an approach based on Hamiltonian paths, used in protein folding, we show with Brownian dynamics simulations that the dense phase has either a molten globule or a crystalline structure, depending on the DNA bending rigidity.

At the end of the Flory-Huggins theory and of the dense phase structure study, we will discuss the biological implications of the results obtained.

\section{Introduction}
\subsection{From co-regulation to co-localization}
Let us consider a network of co-expressed genes\index{co-expressed genes}. By co-expressed we mean that the expression of such genes is coordinated in some way. For instance, co-expressed genes\index{co-regulated genes} might be under the control of the same promoter. More generally, when their expression is directly regulated by the same transcription factor (TF) we say that they are co-regulated. Other cases of co-expression may involve a TF that directly activates the transcription of a gene A, encoding a protein that in turn activates the transcription of a gene B. This example illustrates how transcriptional cascade can occur in network of co-regulated genes\index{gene network}. Such effects may depend on one or just a few TFs, and therefore co-regulated gene networks can entail a broad genetic response to changes coming either from external conditions or from other metabolic pathways. Such transcriptional cascades are not without similarities with cascades occurring in other biological contexts such as for instance kinase pathways in cell signalling.

\begin{figure}[!hbtp]
  \centering
  \subfloat[]{%
    \label{fig:coregulation:a}%
    \includegraphics[width= 0.32 \textwidth]{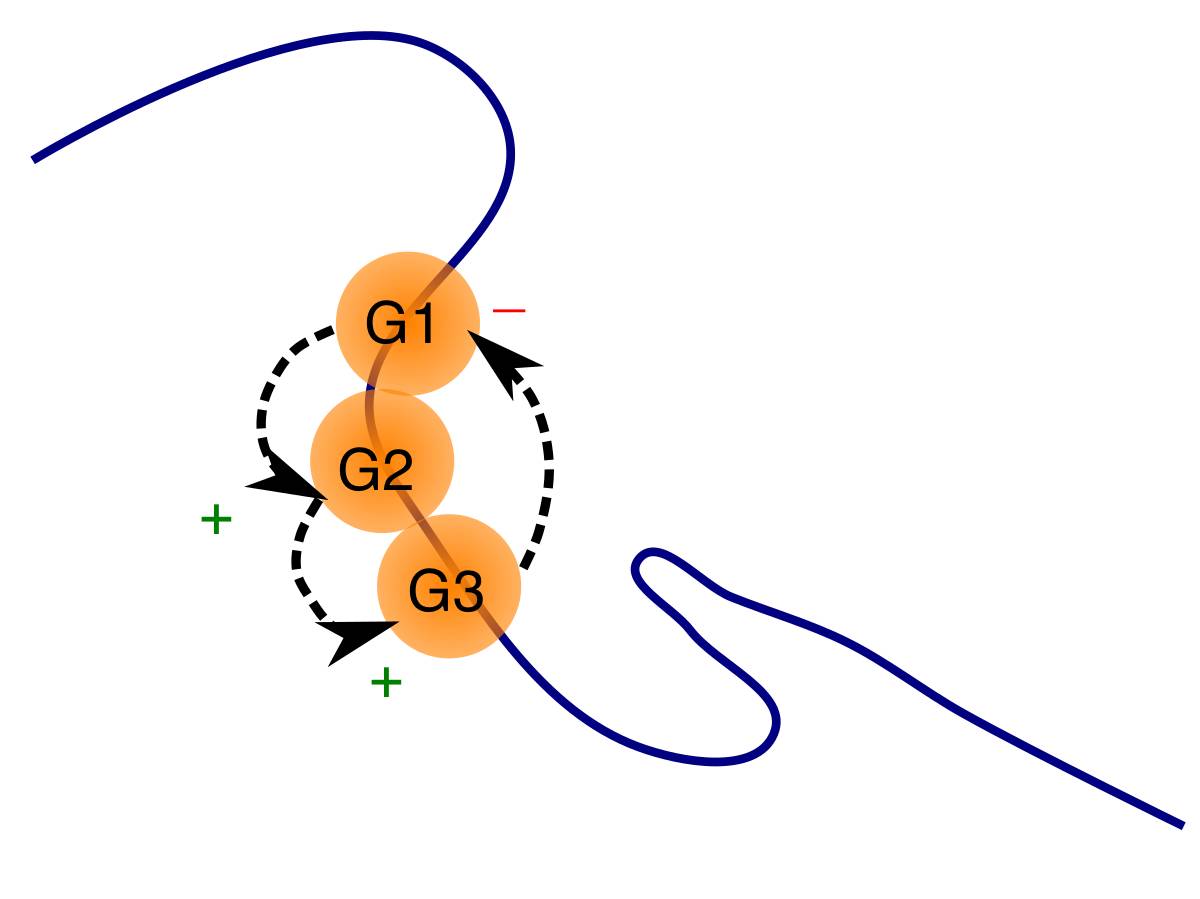}%
  }
  \hspace{0.02 \textwidth}
  \subfloat[]{%
    \label{fig:coregulation:b}%
    \includegraphics[width= 0.32 \textwidth]{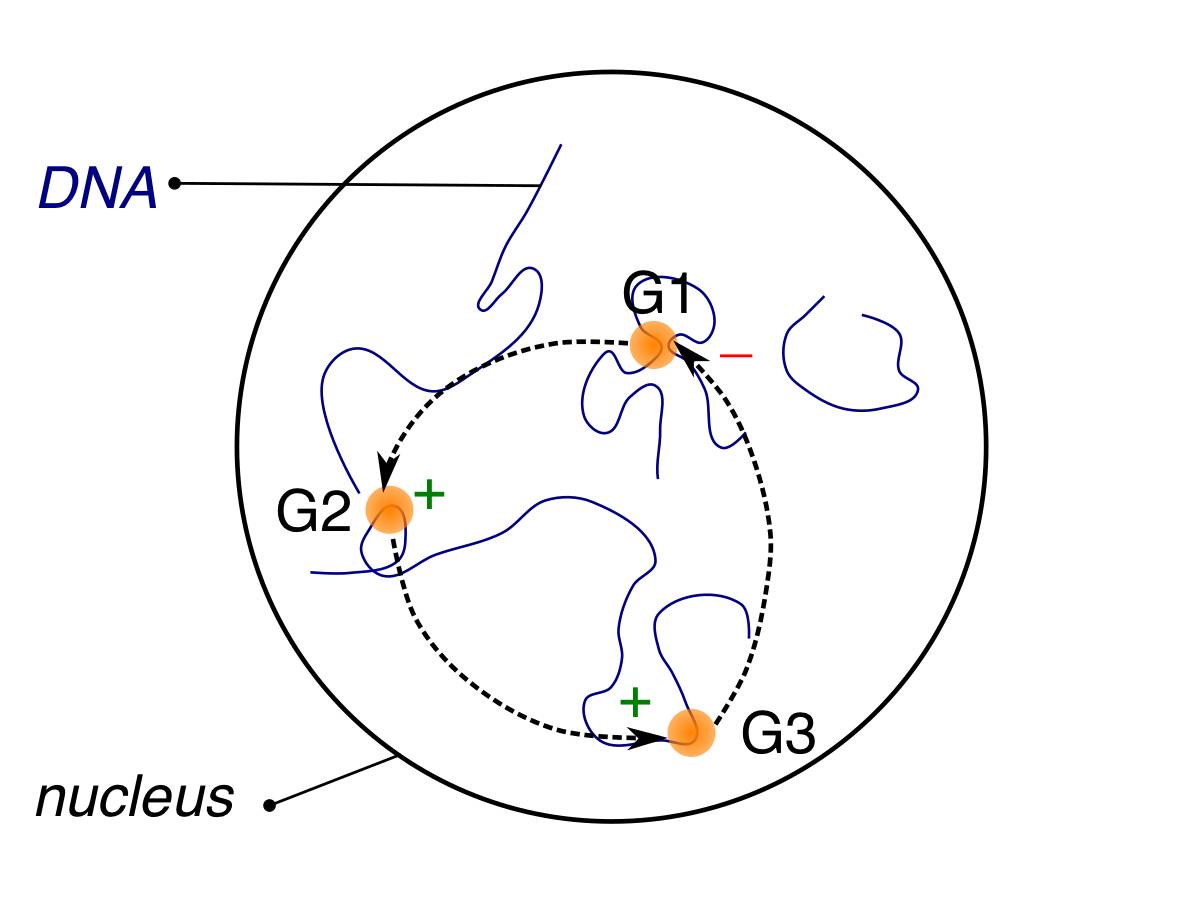}%
  }
  \hspace{0.02 \textwidth}
  \subfloat[]{%
    \label{fig:coregulation:c}%
    \includegraphics[width= 0.30 \textwidth]{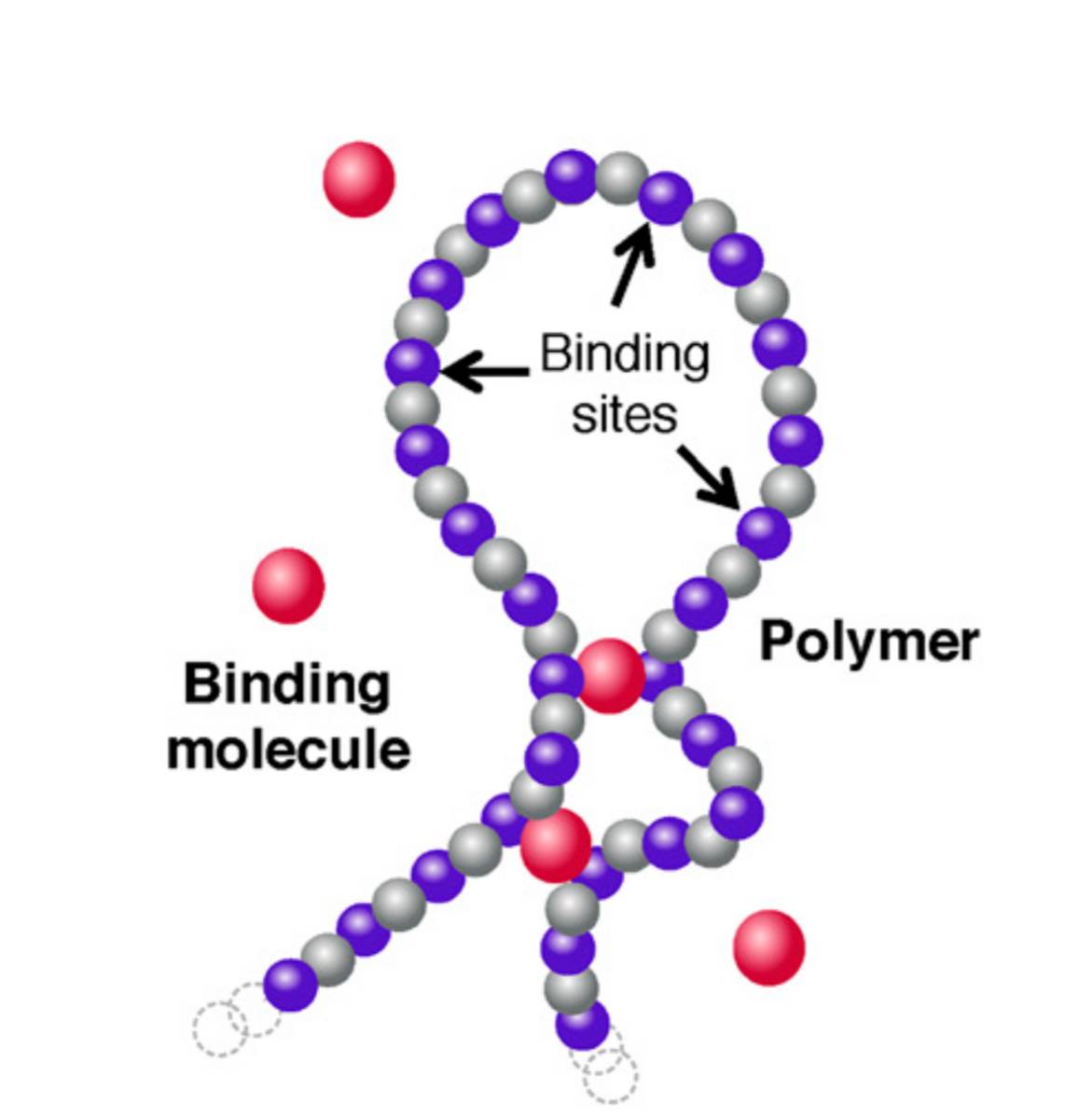}%
  }
  \caption{\protect\subref{fig:coregulation:a} An in-row layout for co-regulated genes enables an efficient regulation of the genes in the network. For instance when a gene produces a protein which down-regulates the expression of the following, or when high local concentration of RNAP results in an increased transcription of one gene and its neighbours. \protect\subref{fig:coregulation:b} The in-row layout is hardly scalable to networks of tens or even hundreds of genes. In eukaryotes, but also in bacteria, several regulation networks involve genes which are located at distant coordinates along the genome sequence. \protect\subref{fig:coregulation:c} Strings and binders switch model \cite{Barbieri2012}.}
  \label{fig:coregulation}
\end{figure}

Whether it is an external TF or a protein encoded by another gene, every protein must diffuse in the nucleoid before reaching its target. In \cref{ch:introduction} we have estimated the time scale to sample the nucleoid/nucleus to be of the order of seconds, which may be a rate-limiting step in transcription regulation. A natural way to overcome diffusion-limited processes is to place co-regulated genes consecutively and next to each other on the DNA sequence (\cref{fig:coregulation:a}). Consequently, the search for the protein target is biased because it is not far from the place where this protein was initially activated, or even produced (note that this last argument does not apply to eukaryotes because proteins are synthesized in the cytoplasm and then imported in the nucleus). In other words, the search time for a protein transiting from one gene to another can be dramatically reduced if these genes are neighbors in space, \textit{i.e.} co-localized.

Although proximity on the genome sequence is one way to achieve co-localization, it is hardly scalable to networks of tens or hundreds of genes, because this would inevitably lead to large genomic distances for some genes of the network, and hence to large spatial distances. Therefore, other mechanisms must exist in order to bring into spatial proximity genes separated by large genomic distances (\cref{fig:coregulation:b}). An influential view is that some TFs have the ability to bind two (or several) sites on the DNA molecules, resulting in an organization of the chromosome into loops \cite{Cook2002,Cook12010,Kepes2004}. Such TFs are said to be divalent\index{divalent transcription factor}, or more generally multivalent if they can bind more than two DNA sequences simultaneously\index{multivalent transcription factors}.

A direct consequence of the binding of divalent TFs is the formation of DNA loops. A case in point is the \textit{lac} operon in \textit{Escherichia coli}, in which repression is achieved when the \textit{lac} repressor binds simultaneously a main site located in the promoter region and an auxiliary sites \SI{401}{bp} away on the sequence \cite{Muller-Hill1998,Muller-Hill1998a}. In this context, the strength of the binding maintaining the DNA loop is directly correlated with the efficiency of the repressor system. This looping mechanism can be envisioned as a mechanical regulatory switch which is turned on and off through the binding of TFs. Let us emphasize that \textit{in vitro} and \textit{in vivo} studies have confirmed the existence of DNA loops, sometimes over long genomic distances \cite{Priest2014,Revet1999151}, suggesting that specific looping can indeed be a key feature of the transcription regulation even in eukaryotes, where enhancers\index{enhancer} can be found several kilo base-pairs away from the promoter\index{DNA looping} \cite{Tolhuis14532002,Schleif1991992,Matthews1231992}.

On the basis of this representation of the chromosome shaped up by divalent TFs, one can think of several physical models. In a stylized view, divalent TFs can be seen as binding spheres able to bind several sites on the DNA sequence (\cref{fig:coregulation:c}). The binding of such a bead at two loci separated by a large genomic distance gives rise to a DNA loop. The superimposition of many such loops not only changes the global chromosome architecture, but has also an impact on transcription, for instance by preventing RNA polymerase (RNAP) to access to the promoter. This has been studied in the so-called strings and binders switch model\index{strings and binders switch model} \cite{Barbieri2012} (\cref{fig:coregulation:c}). Another physical model demonstrated that although it is seemingly more complex than adjusting the affinity of a TF with a given promoter, DNA looping can confer unique and relevant properties to transcription regulation \cite{Vilar2003,Vilar2005136}. In particular, DNA looping leads to an increased effective binding free energy of a TF to its promoter. In other words, the apparent search volume to find the target is reduced and the local concentration of protein is increased\index{local concentration effects}. A consequence of such local concentration effects, which can be envisioned as ``molecular traps'', is to stabilize the protein binding versus global fluctuations of the protein concentration in the cell.

\subsection{Co-localization of genes in transcription factories}
\index{transcription factory}
Co-localization of co-regulated genes has proven to be more than a surmise. It has been confirmed to occur in prokaryotes \cite{Cook2008,Llopis2010} and eukaryotes \cite{Schoenfelder2010,Spilianakis2005} using fluorescence techniques (FISH). In the case of the mouse hemoglobin co-regulated genes (more than 40), a combination of FISH and Chromosome Conformation Capture (3C) techniques demonstrated that these genes tend to co-localize in clusters and exhibited higher contact frequencies than with other non-related genes \cite{Schoenfelder2010}. These clusters were shown to contain of the order of 8 to 10 genes. Associations between co-regulated genes were also shown to happen between different chromosomes \cite{Spilianakis2005}.

In a first series of experiments, it was shown that nascent RNA transcripts are synthesized at discrete foci in the nucleus \cite{Cook2008,Caudron-Herger2015}. Later, it was shown that RNAP itself gathers into clusters instead of being uniformly distributed within the nucleus \cite{Darzacq2013}. These clusters with increased concentration of nascent RNA transcripts and RNAP correspond to areas where active transcription occurs. Hence they were called transcription factories. Although the existence of transcription factories were first obtained on mammalian cells, because their large size is a better fit for fluorescence studies, their existence has also been demonstrated in bacteria \cite{Jin2014}.

In an attempt to connect the co-localization of co-regulated genes with the existence of transcription factories, it was conjectured that by binding to and organizing the chromosome, TFs gather co-regulated genes in transcription factories. In addition to achieve gene co-localization, this may lead to a mutualization of resources\index{transcriptional resources}, such as the availability of RNAP, TFs or epigenetics marks such as methylations. It would also give a more general account of local concentration effects and their role in regulating the transcription. It is very likely that transcription entails a cellular response that will in turn impact chromosome architecture. Incidentally, the life time of transcription factories was found to be of the order of \SI{5}{\second} \cite{Darzacq2013}, suggesting that transcription factories are dynamically re-allocated as a function of the transcriptional state of the cell. In our view, this supports the idea that transcription regulation and chromosome organization are related in some sort of dynamical feedback mechanism which remains to be found.

\begin{figure}[!hbtp]
  \centering
  \subfloat[]{%
    \label{fig:transcription_factories_invivo:a}%
    \includegraphics[width= 0.30 \textwidth]{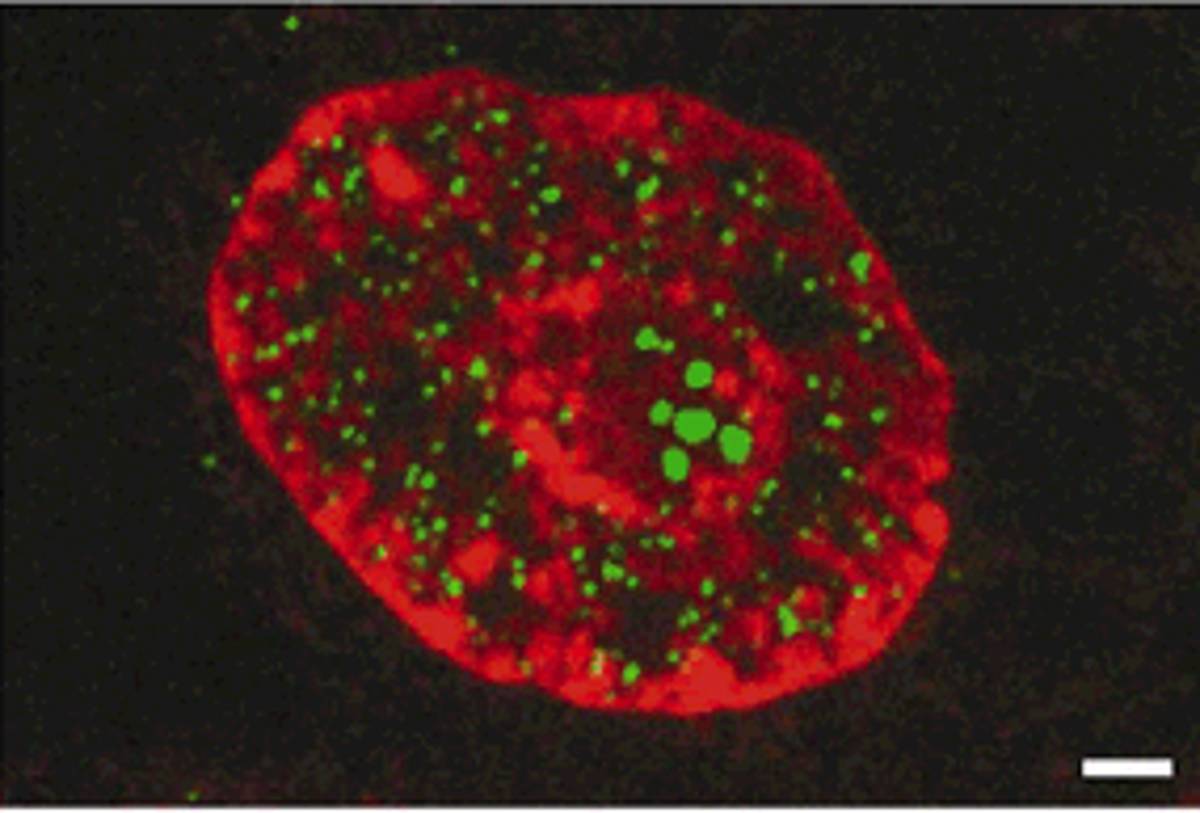}%
  }
  \hspace{0.05 \textwidth}
  \subfloat[]{%
    \label{fig:transcription_factories_invivo:b}%
    \includegraphics[width= 0.25 \textwidth]{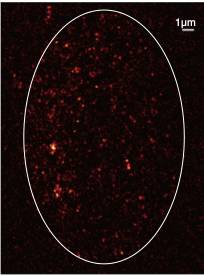}%
  }
  \hspace{0.05 \textwidth}
  \subfloat[]{%
    \label{fig:transcription_factories_invivo:c}%
    \includegraphics[width= 0.25 \textwidth]{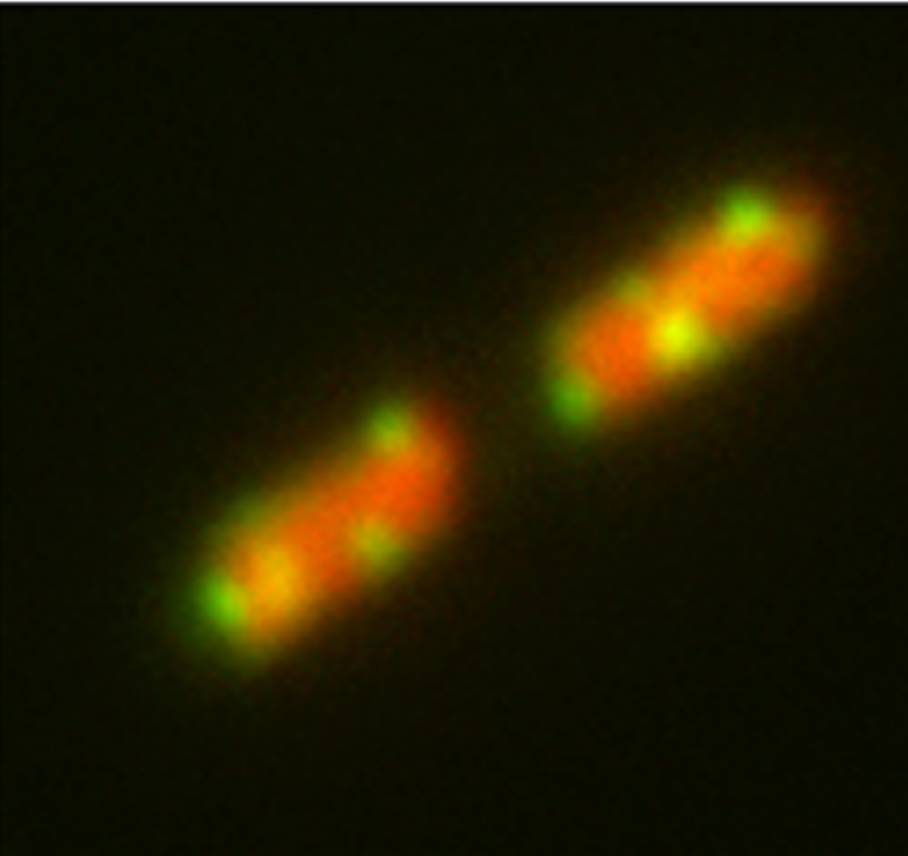}%
  }
  \caption{Transcription factories \textit{in vivo}. \protect\subref{fig:transcription_factories_invivo:a} FISH imaging of nascent RNA transcripts (green) in human HeLa cells \cite{Cook2002}. \protect\subref{fig:transcription_factories_invivo:b} PALM imaging of RNA polymerase (red) in human osteosarcoma cell \cite{Darzacq2013}. \protect\subref{fig:transcription_factories_invivo:c} Fluorescence imaging of RNA polymerase in \ecoli by fusion with green fluorescent proteins \cite{Jin2842006}.}
  \label{fig:transcription_factories_invivo}
\end{figure}

\subsection{Physical origin of transcription factories}
The physical origin of transcription factories has remained controversial. Two questions at least may be formulated. First, one can wonder if the formation of transcription factories constitutes a Boltzmann equilibrium. We have mentioned earlier that the typical time for a protein to sample the bacterial nucleoid is \SI{100}{\ms}\index{diffusion time}, which gives the typical time scale for transcription regulation processes. This figure should be compared with the life time of transcription factories which has been measured to be of the order of seconds or tens of seconds \cite{Darzacq2013}. Thus it seems that transcription factories are reminiscent of an equilibrium phenomenon. Second, the structure of the DNA inside such transcription factories has remained elusive. In particular, it is not clear what is the effective diffusion coefficient of TFs or RNAP inside transcription factories. These remarks have motivated the study of the statistical physics of the DNA interacting with TFs.

\section{Model proposed}
\subsection{Formal nucleus/nucleoid}
\begin{figure}[!hbtp]
  \centering
  \includegraphics[width=0.5 \textwidth]{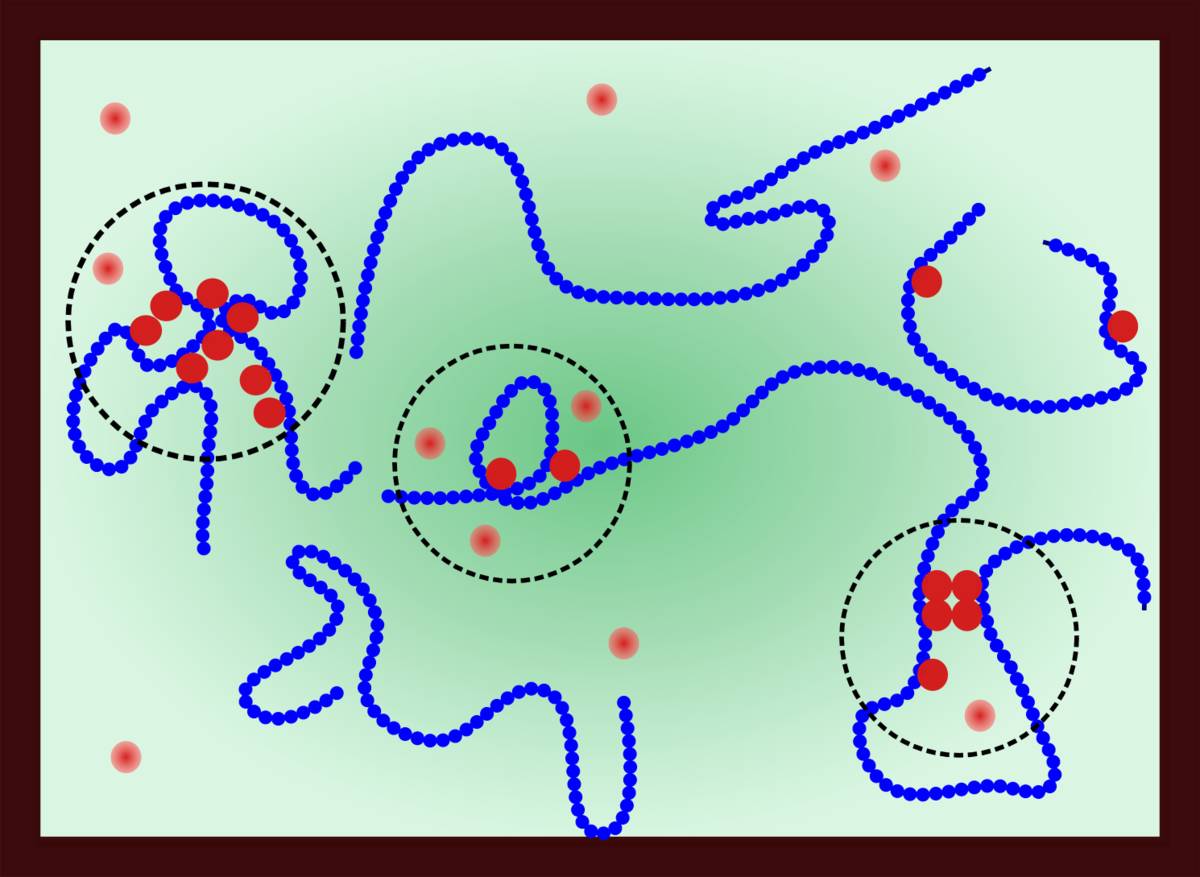}
  \caption{Stylized view of the bacterial cell or nucleus. It contains DNA chains represented as polymers and binding proteins represented as free spheres. The dashed circles denote transcription factories.}
  \label{fig:formal_nucleus}
\end{figure}

We consider a simplified model in which the nucleus (or bacterial nucleoid) is represented by a closed volume $V$ (\cref{fig:formal_nucleus}). In the sequel we will indifferently use the words cells and nucleus, since this model is considered to represent either a bacterial cell or the nucleus of a eukaryotic cell. The double-stranded DNA chains are modeled as $M$ semi-flexible polymer chains of length $b \times N$, where $b$ is the Kuhn length. A DNA monomer is specified by a coordinate $s$ varying from $0$ to $N$, and can interact with $P$ spheres, representing DNA-binding proteins. The typical size of DNA monomer beads and of a protein beads are taken to be equal in this study. In bacteria, $b\approx \SI{2.5}{\nm}$, is the diameter of the naked DNA fiber, and also corresponds to the size of a typical size of a protein. In eukaryotes, $b \approx \SI{30}{\nm}$ is the diameter of the chromatin fiber, therefore a protein bead rather represents a protein complex. Introducing the subscript $D$ for DNA and $P$ for proteins, we shall consider the three pair potentials $u_{DD}(r)$, $u_{PP}(r)$ and $u_{DP}(r)$ for the interactions between two beads separated by a distance $r$. We will also consider that proteins can bind to DNA monomers non-specifically. Although this last assumption is strong, it may be considered as a model for transcription factors with a very large number of targets on the DNA, such as nucleoid-associated proteins (H-NS, FIS or HU in bacteria), or even RNAP itself.

In the sequel, we will assume that DNA monomers experience pure excluded volume interactions with other DNA monomers, and similarly proteins-protein interactions are only repulsive. On the contrary, we will assume that proteins can bind to DNA, and therefore the corresponding interaction potential has an attractive tail. Hence we have the Mayer coefficients\index{Mayer coefficient} (see \vref{eq:polymer_mayer_coefficient}) for each of the three interaction potentials:
\begin{align}
  \begin{aligned}
    \alpha_D &= \int \mathrm{d}\mathbf{r} \, u_{DD}(r) > 0,  \\
    \alpha_P &= \int \mathrm{d}\mathbf{r} \, u_{PP}(r) > 0,  \\
    v &= \int \mathrm{d}\mathbf{r} \, u_{DP}(r) < 0.
  \end{aligned}
\end{align}

\subsection{First principles model}
Starting from the model defined in the last paragraph, we now lay the grounds for a study of the statistical physics of this system. First, we specify the Hamiltonian built from summing over all interactions between microscopical constituents.  Let us consider $M$ polymer chains representing the chromosomes in the nucleus. In this chapter, we will use continuous polymers to model the chromosomes. Thus, every DNA chain $k$ is represented by a space curve $\mathbf{r}_k(s)$ for $k=1,\ldots,M$, giving the spatial coordinates of monomer $s$ along the DNA sequence. In addition, we introduce $P$ proteins with coordinates $\mathbf{R}_i$ for $i=1,\ldots,P$. Using the pair potentials previously introduced, the Hamiltonian for the system reads:
\begin{align}
  \begin{aligned}
    \beta \mathcal{H} \left[\{\mathbf{R}_i\}, \{ \mathbf{r}_k(s) \} \right] &= \dfrac{1}{2} \sum \limits_{i \neq j} u_{PP}(\mathbf{R}_j - \mathbf{R}_i)
    + \dfrac{1}{2} \sum \limits_{k \neq l} \int \limits_0^N \ud{s}\ud{s'}  u_{DD}(\mathbf{r}_l(s') - \mathbf{r}_k(s) ) \\
    &+ \beta \sum \limits_i \sum \limits_k \int \limits_0^N \ud{s} u_{DP}(\mathbf{r}_k(s)- \mathbf{R}_i) \\
    &+ \dfrac{1}{3!}w \sum \limits_{I,I',I''} \delta( \mathbf{R}_{I'} - \mathbf{R}_{I}) \delta(\mathbf{R}_{I''} - \mathbf{R}_{I'})
  \end{aligned} \label{eq:hamiltonian_micro}
\end{align}

The Hamiltonian thus consists in a summation over many bodies interactions. Note that because $u_{DP}(r)$ is a potential with an attractive tail, the system may collapse in a certain range of concentrations of DNA and proteins, if it were not to be compensated by higher order terms. For that reason, we needed to consider an expansion of the Hamiltonian to order three at least, which is a common procedure in polymer physics \cite{DesCloizeaux1990}. In the present study, we have assumed that a Kuhn segment on the DNA and a protein bead have same size $b$, and for that reason, the three-body term is a sum over an index $I$ which can be either a protein or a DNA monomer. The coefficient $w$ represents a penalty in $k_B T$ whenever three such beads collapse on the same coordinates. The prefactor of $1 / 3!$ is here to ensure a standard form of the associated virial expansion of the osmotic pressure. We will come back to this shortly. We also point out that only the DNA-protein interaction is temperature-dependent, as can be seen from the $\beta$ prefactor right in front of the potential $u_{DP}$.

The partition of our formal nucleus therefore reads
\begin{equation}
  Z = \int  \frac{1}{P!} \prod \limits_{i=1}^P \ud{\mathbf{R}_i} \int \frac{1}{M!} \prod \limits_{k=1}^M \uD{\mathbf{r}_k(s)} \exp{\left(-\beta \mathcal{H} \left[\{\mathbf{R}_i\}, \{ \mathbf{r}_k(s) \} \right] - \sum \limits_k \beta U_0\left[ \mathbf{r}_k(s) \right] \right)},
  \label{eq:partfunc_micro}
\end{equation}
where we introduced the internal energy of the DNA chains $\beta U_0$. The functional dependence of the internal energy on the chain configuration depends on the polymer model retained for the chromosome. Specifically, DNA is a rigid biopolymer that can be modeled as a semi-flexible polymer (see \vref{subsec:polymer_model_bending_rigidity}).

\subsection{Field representation}
We now derive the partition function in field representation. This can be used as a starting point for several standard approximations, including the saddle-approximation and the Gaussian fluctuations analysis also know as Random Phase Approximation in the context of polymer physics. In particular, the mean-field theory, that we will introduce in the next section, can be obtained from a saddle-point approximation.

In order to change the integration in \cref{eq:partfunc_micro}, which is performed over the individual coordinates of the constituents to an integration over fields, we make a change of variables by introducing the concentration fields:
\begin{align}
  \rho_P(\mathbf{r}) &= \sum \limits_i \delta(\mathbf{r} - \mathbf{R}_i), \label{eq:field_P} \\
  \rho_D(\mathbf{r}) &= \sum \limits_k \int \limits_0^N \ud{s} \delta(\mathbf{r} - \mathbf{r}_k(s)).
  \label{eq:field_D}
\end{align}

We then introduce the following identity in the partition function in \cref{eq:partfunc_micro}:
\begin{align}
  \begin{aligned}
    1 &= \int \uD{\rho_D(\mathbf{r})} \delta \left(\rho_D(\mathbf{r}) - \sum \limits_k \int \limits_0^N \ud{s} \delta(\mathbf{r} - \mathbf{r}_k(s)) \right) \\
    &= \int \uD{\rho_D(\mathbf{r})} \uD{\varphi_D(\mathbf{r})} \exp{ \left( i \int \ud{\mathbf{r}} \rho_D(\mathbf{r}) \varphi_D(\mathbf{r}) - \sum \limits_k \int \limits_0^N \ud{s} \varphi_D(\mathbf{r}_k(s)) \right)},
  \end{aligned} \label{eq:var_change_field}
\end{align}
where in \cref{eq:var_change_field} we made use of the exponential representation of the delta-functional by introducing an auxiliary field $\varphi_D(\mathbf{r})$. A similar identity can be introduced for the protein concentration field. This leads to the following re-writing for the partition function of the system:
\begin{equation}
  Z = \int \uD{\rho_D} \uD{\varphi_D} \uD{\rho_P} \uD{\varphi_P} \exp{\left( - \beta S\left[\rho_D, \varphi_D, \rho_P, \varphi_P \right] \right) },
  \label{eq:partfunc_field}
\end{equation}
with the action
\begin{align}
  \begin{aligned}
    \beta S &= -i \int \ud{\mathbf{r}} \rho_D(\mathbf{r}) \varphi_D(\mathbf{r}) - i \int \ud{\mathbf{r}} \rho_P(\mathbf{r}) \varphi_P(\mathbf{r}) \\
    & + \int \ud{\mathbf{r}} \ud{\mathbf{r'}} \rho_P(\mathbf{r'}) u_{PP}(\mathbf{r'} - \mathbf{r}) \rho_P(\mathbf{r}) + \int \ud{\mathbf{r}} \ud{\mathbf{r'}} \rho_D(\mathbf{r'}) u_{DD}(\mathbf{r'} - \mathbf{r}) \rho_D(\mathbf{r}) \\
    & + \beta \int \ud{\mathbf{r}} \ud{\mathbf{r'}} \rho_D(\mathbf{r'}) u_{DP}(\mathbf{r'} - \mathbf{r}) \rho_P(\mathbf{r}) \\
    & + \dfrac{1}{3!} w \int \ud{\mathbf{r}} \left( \rho_D(\mathbf{r}) + \rho_P(\mathbf{r}) \right)^3 \\
    & - P \ln{W[i \varphi_P ]} - M \ln{Q[i \varphi_D ]} + P \ln{\dfrac{P}{e}} + M \ln{\dfrac{M}{e}}.
  \end{aligned}  \label{eq:action_field}
\end{align}

An interesting outcome of this re-writing is the separation of the enthalpic and entropic contributions in the action expressed in \cref{eq:action_field}. Namely, the entropy is represented by the last four terms in the action. It depends only on the single-particle partition function of the protein beads, respectively the single-chain partition function of the DNA chains, in the imaginary potential $i\varphi_P(\mathbf{r})$, respectively $i\varphi_D(\mathbf{r})$, whose expressions are given by
\begin{align}
  W[i \varphi_P ] &= \int \ud{\mathbf{R}} \exp{(-i \varphi_P(\mathbf{R}))}, \label{eq:partfunc_singlebead} \\
  Q[i \varphi_D ] &= \int \uD{\mathbf{r}(s)} \exp{\left( -\beta U_0[\mathbf{r}(s)] - i \int \limits_0^N \ud{s} \varphi_D(\mathbf{r}(s))\right)}. \label{eq:partfunc_singlechain}
\end{align}

\section{Flory Huggins theory}
\label{sec:flory_huggins}
In this section, we study the Flory-Huggins theory of the bulk of the nucleus. We will show that the existence of a binding energy between proteins and DNA can lead to the formation of a dense phase that one may identify to transcription factories. As pointed out previously, only the DNA-protein interaction is temperature-dependent. In the sequel, we will often refer to the high, respectively low, temperature regime which corresponds to a weak, respectively strong, DNA-protein attraction.

\subsection{Mean-field free energy}
Intuitively, one might expect that for high temperatures, the attraction between DNA monomers and binding-proteins vanishes. In this regime, the system contains an homogeneous concentration of DNA monomers and proteins. In other terms, we may perform a mean-field approximation and remove the spatial dependence of the concentration fields: $\rho_D(\mathbf{r}) \leftarrow c_D = MN / V$ and $\rho_P(\mathbf{r}) \leftarrow c_P = P / V$. In the context of polymer physics, this is also called the Flory-Huggins theory \cite{deGennes1979}. We will frequently refer to these concentrations as the mean-field solutions. Solving the stationary equations for the action in \cref{eq:action_field} can be done (see \cref{sec:RPA}). If in addition we assume mean-field solutions, we obtain the free energy function per volume unit:
\begin{equation}
  \beta f(c_D,c_P) = \dfrac{1}{2} \alpha_D c_D^2 + \dfrac{1}{2} \alpha_P c_P^2 + \beta v c_D c_P + \dfrac{1}{3!} w (c_D + c_P)^3 + c_P \ln{\dfrac{c_P b^3}{e}} + \dfrac{c_D}{N} \ln{\dfrac{c_D b^3}{Ne}}. \label{eq:free_en_mf}
\end{equation}

Another way to look at this expression is to consider that an excluded volume penalty $\alpha$ is applied whenever two beads are in contact. In the case of the DNA-protein interaction, the effective excluded volume $\beta v$ is negative because the interaction is attractive. The probability to find two beads in contact is proportional to the product of their concentrations. This gives terms of the form $\alpha \times c^2$ contributing to the free energy function. The two logarithmic terms in the free energy accounts for the configurational entropy of the proteins and DNA chains. Briefly, one can see them as contributions of the form $(V/b^3)^P / P! \sim \exp{(V c_P \ln{(c_P b^3/e}))}$ to the Boltzmann weight, where $(V/b^3)$ is the number of accessible configurations for one bead of size $b^3$ distributed uniformly in the volume $V$. We may also give an account for the presence the three-body term without resorting to the microscopical Hamiltonian defined in the last section. For this, one needs to consider the Flory-Huggins theory as the limit of a gas on a lattice, where the enthalpic contributions are the first three terms in the right-and side (r.h.s.) of \cref{eq:free_en_mf}, and the last two terms are the entropy of the particles. Now, in any gas on a lattice representation, one should take into account the entropy of the vacancies, which in our case should be identified to the solvent. If we assume the system to be incompressible, this entropic contribution has the form $(c_0 - c_D -c_P) \ln{((c_0 - c_D -c_P)/e)}$, where $c_0$ is the close-packing concentration. An expansion in powers of $(c_D+c_P)$ truncated at order three yield the three-body term in the free energy.

In the high temperature regime, the DNA-protein interaction term, $\beta v c_D c_P$, in \cref{eq:free_en_mf} vanishes. The free energy is therefore a convex function, making the mean-field solution stable.

\subsection{Spinodal condition}
When the temperature is progressively decreased, the attractive interaction $|\beta v|$ increases, until the mean-field solution is no longer stable. This is the so-called spinodal condition, which delimits the region in which the mean-field solution is stable from the region where it is not. This condition corresponds to an inversion of the curvature of the free energy, which in our case reads
\begin{align}
  \begin{vmatrix}
    \dfrac{\partial^2 f}{\partial c_D^2} & \dfrac{\partial^2 f}{\partial c_D \partial c_P} \\
    \dfrac{\partial^2 f}{\partial c_D \partial c_P} & \dfrac{\partial^2 f}{\partial c_P^2}
  \end{vmatrix}
  = 0, \label{eq:spinodal}
\end{align}
where the array denotes a determinant. For any given choice of the DNA-protein attraction, $\beta v$, the spinodal equation is an implicit equation for a closed curve in the $(c_D,c_P)$ plane. The range of concentrations enclosed within this curve corresponds to unstable mean-field solutions whereas in the outer region are stable mean-field solutions. There is a critical temperature $T^c$ such that for $T>T^c$ there is no solution to the spinodal condition and the mean-field solution is stable, whereas for $T<T^c$, the spinodal condition has solutions. At $T=T^c$, the closed curve of solutions reduces to a single point of coordinates $(c_D^c,c_P^c, T^c)$.

\begin{figure}[!hbtp]
  \centering
  \includegraphics[width= 1 \textwidth]{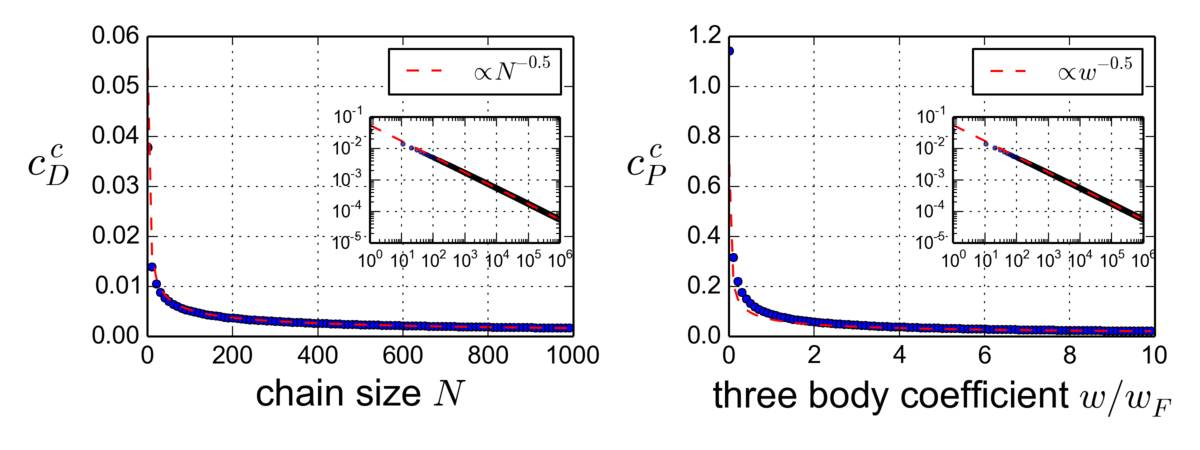}
  \caption{Scaling of the critical concentrations of DNA and proteins as a function of the parameters of the model. $w_F=b^6$ is the value of the three-body repulsive core which naturally arises from a Flory-Huggins theory. The insets display the dependence of the critical concentrations on the parameters in log-log scale, over a broad range of values.}
  \label{fig:critical_scalings}
\end{figure}

Let us point out that \cref{eq:spinodal} is not tractable by hands, so we solve it numerically. As shown in \cref{fig:critical_scalings}, we obtain the following scaling relations for the critical concentrations:
\begin{align}
  \begin{aligned}
    c_D^c &\sim \dfrac{1}{\sqrt{N}} \dfrac{1}{\sqrt{w}}, \\
    c_P^c &\sim 1 / \sqrt{w}.
  \end{aligned} \label{eq:critical_concentrations}
\end{align}

\begin{figure}[!hbtp]
  \centering
  \subfloat[]{%
    \label{fig:flory_section_spinodal}%
    \includegraphics[width= 0.48 \textwidth]{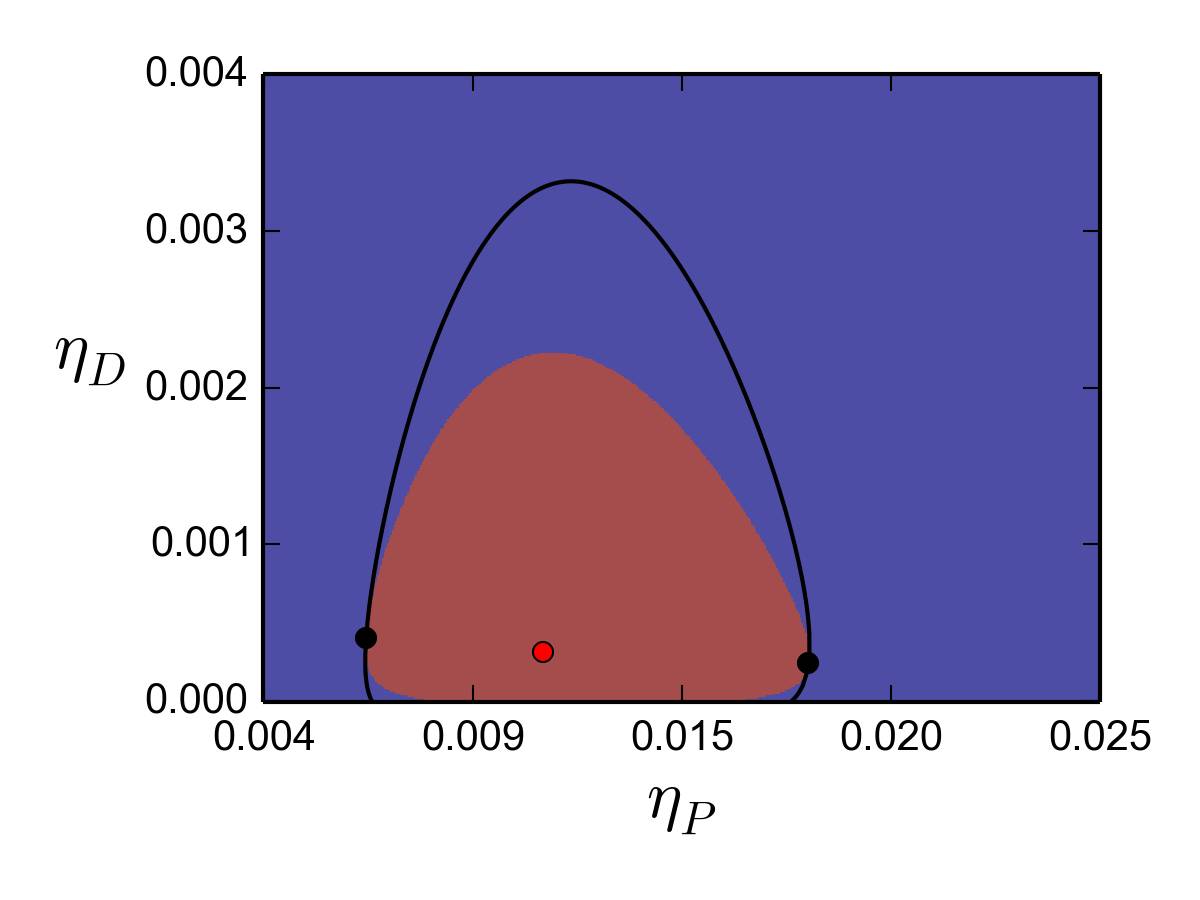}%
  }%
  \quad
  \subfloat[]{%
    \label{fig:flory_section_coexistence}%
    \includegraphics[width= 0.48 \textwidth]{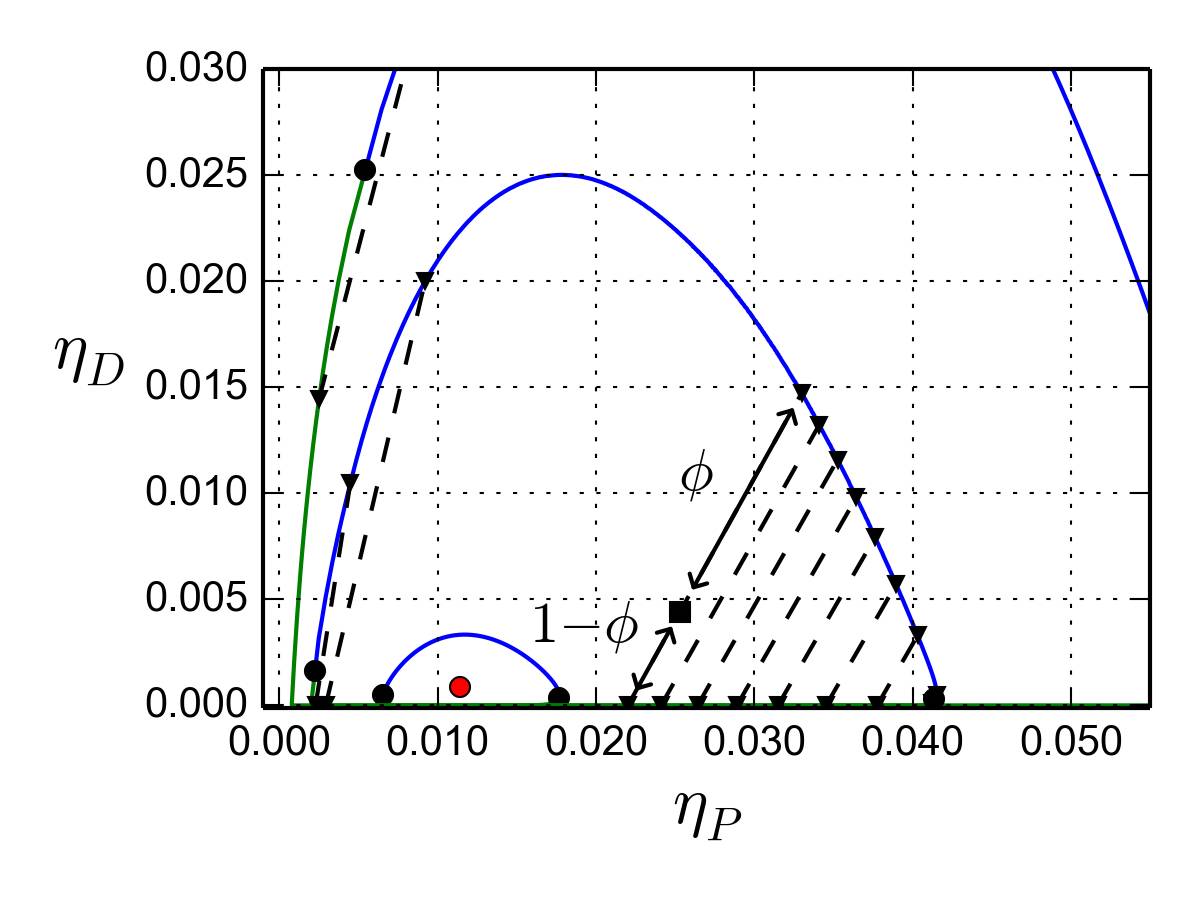}%
  }
  \caption{\protect\subref{fig:flory_section_spinodal} At $t = 0.05$. The binodal, or coexistence, line is the solid black curve. The mean-field solution is unstable in the region colored in red and the system splits into two phases. For concentrations falling into the blue region, the mean-field solution is stable. The black circles are two critical points where the coexistence and the spinodal lines intersect. \protect\subref{fig:flory_section_coexistence} Binodal, or coexistence, lines at $t = 0.05, \, 0.5, \, 1.0$. The coexistence line shrinks toward the tricritical point (red dot) when $t \to 0$. For each curve, the dilute phase is shown in green and the concentrated phase is shown in blue. Coexisting states are connected by tie lines (dotted segments). The volume fraction of each phase is determined (black arrows) according to \cref{eq:maxwell_construct}.}
  \label{fig:flory_section}
\end{figure}

We observe that for $T<T^c$, the spinodal equation consists of an infinite set of doublet pairs $(c_D^1, c_P^1)$-$(c_D^2, c_P^2)$ which together are a parametrization for the spinodal line (\cref{fig:flory_section_spinodal}). However, on each spinodal line are found two double solutions, meaning that each doublet pair $(c_D^1, c_P^1)$ and $(c_D^2, c_P^2)$ merge into a single critical point $(c_D^0, c_P^0)$.

In summary, \cref{eq:spinodal} is an implicit equation for a surface in the $(c_D,c_P,T)$ space, below which the mean-field solutions are unstable. On this surface lies a critical curve where the spinodal condition has a double solution $(c_D^0, c_P^0)$, instead of two different solutions (\cref{fig:flory_phasediag}). At the apex of the critical curve lies the tricritical point $(c_D^c,c_P^c, T^c)$, which is the first point where the mean-field solution becomes unstable when the temperature is decreased.

\subsection{The biphasic regime}
For $T<T^c$, the free energy function in \cref{eq:free_en_mf} has an unstable range of concentrations. In this range, the system splits into two phases $I$ and $II$ which belong to the stable region. The total free energy of the system can be written as the sum of the free energies of the two phases. The equilibrium state in then found by minimizing the total free energy function under the constraints that the total number of particles is conserved and the volume constant. This is most easily done by minimizing the Lagrangian:
\begin{align}
  \begin{aligned}
    \mathcal{L} &= \phi^I f(I) + \phi^{II} f(II) \\
    & - \mu_D \left( \phi^I c_D^{I} + \phi^{II} c_D^{II} \right)
    - \mu_P \left( \phi^I c_P^{I} + \phi^{II} c_P^{II} \right)
    - \Pi \left(\phi^I + \phi^{II} \right),
  \end{aligned}
\end{align}
in which $f(I)$ is a short-hand for $f(c_D^{I},c_P^{I})$ and $\phi^I$ is the volumic fraction of phase $I$. The same notations apply for phase $II$. The Lagrangian multipliers $\mu_D$ and $\mu_P$ have been introduced to conserve the number of DNA monomers and the number of proteins, and are to be identified with chemical potentials. The Lagrangian multiplier $\Pi$ has been introduced to conserve the volume, and is to be identified to the osmotic pressure. The minimization of $\mathcal{L}$ relatively to the variables $c_D^{I}, \, c_P^{I}, c_D^{II}, c_P^{II}, \phi^I, \phi^{II}$ yields the system of equation
\begin{align}
  \left\lbrace
  \begin{aligned}
    & \dfrac{\partial f}{\partial c_D} (c_D^{I}, c_P^{I}) &=& \quad \dfrac{\partial f}{\partial c_D} (c_D^{II}, c_P^{II}) &=& \quad \mu_D \\
    & \dfrac{\partial f}{\partial c_P} (c_D^{I}, c_P^{I}) &=& \quad \dfrac{\partial f}{\partial c_P} (c_D^{II}, c_P^{II}) &=& \quad \mu_P \\
    & f(c_D^I,c_P^I) - \mu_D c_D^I - \mu_P c_P^I &=& \quad f(c_D^{II},c_P^{II}) - \mu_D c_D^{II} - \mu_P c_P^{II} &=& \quad - \Pi
  \end{aligned}
  \right., \label{eq:binodal}
\end{align}
which states that the chemical potentials and the osmotic pressure in the two phases are equal. Note that the writing of the last line can be rewritten more compactly if we introduce the Gibb's free energy, which is the Legendre transform of \cref{eq:free_en_mf}:
\begin{equation}
  g(c_D,c_P) = f(c_D,c_P) - \mu_D c_D - \mu_P c_P,
\end{equation}
and where the chemical potentials depend implicitly on the concentrations through the equilibrium condition:
\begin{align}
  \left\lbrace
  \begin{aligned}
    & \dfrac{\partial g}{\partial c_D} &=& \quad 0 \\
    & \dfrac{\partial g}{\partial c_P} &=& \quad 0 \\
  \end{aligned}
  \right.
  \quad
  \Leftrightarrow
  \quad
  \left\lbrace
  \begin{aligned}
    & \dfrac{\partial f}{\partial c_D} &=& \quad \mu_D \\
    & \dfrac{\partial f}{\partial c_P} &=& \quad \mu_P \\
  \end{aligned}
  \right..
\end{align}

\begin{figure}[!hbtp]
  \centering
  \includegraphics[width= 1 \textwidth]{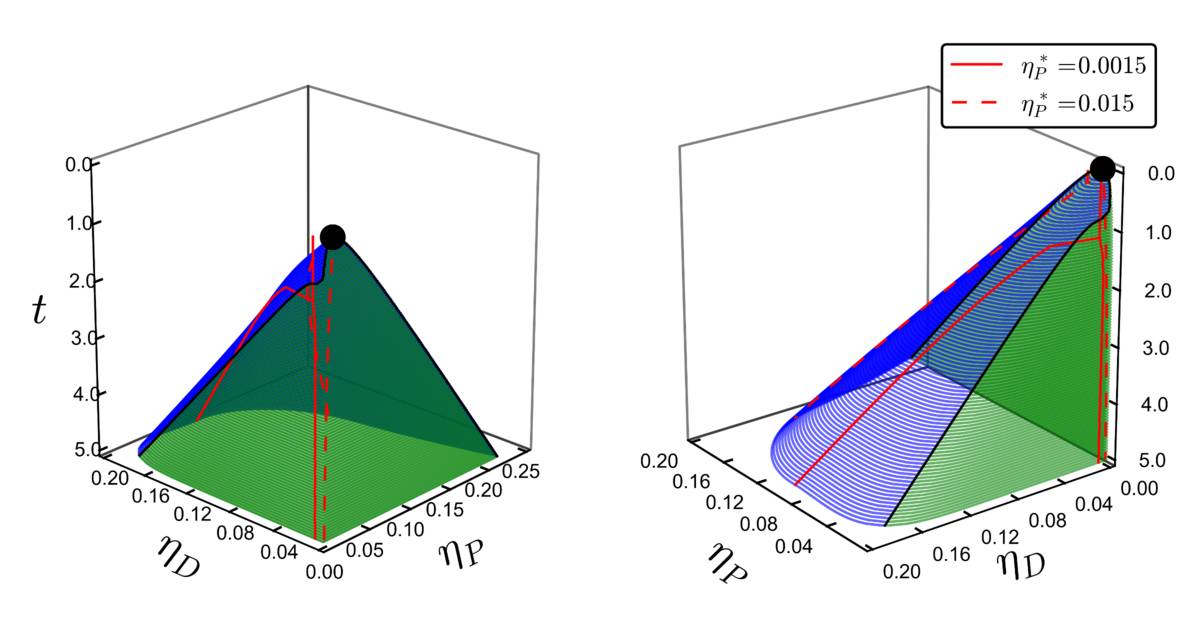}
  \caption{Phase diagram obtained from the resolution of \cref{eq:binodal}. For $T<T^c$, any combination of concentrations lying below the binodal curve is an unstable mean-field solution, which results in the system splitting into two phases $I$ and $II$ which sits on the binodal surface. The dilute phase $I$ (green) has lower concentrations than the dense phase $II$ (blue). The parameter $t$ from the vertical axis is defined in \cref{eq:expansion_param}. In this diagram, we chose to use the volumic densities instead of the concentrations, which are defined as $\eta_D= b^3 c_D$ and $\eta_P= b^3 c_P$. The black line is the critical curve.}
  \label{fig:flory_phasediag}
\end{figure}

The system in \cref{eq:binodal} is a system of 3 equations with the four unknown concentration variables, plus the temperature. Hence it is a parametrization for a surface, called the binodal, which completely determines the phase diagram of the system. We solved this system of equations numerically using a quasi-Newton root finding method \cite{NumericalRecipes2007}. The surface obtained is represented in \cref{fig:flory_phasediag}, where a parameter $t$ has been introduced as an expansion of the DNA-protein interaction around the tricritical point:
\begin{equation}
  \beta v = (\beta v)^c (1 + t). \label{eq:expansion_param}
\end{equation}

At a given temperature though, there is only one solution for the phase separated system. Indeed, the points on the binodal are uniquely determined by the self-consistent relations:
\begin{align}
  \begin{aligned}
    \dfrac{\partial \mathcal{L}}{\partial \mu_D} &= M N / V, \\
    \dfrac{\partial \mathcal{L}}{\partial \mu_P} &= P / V, \\
    \dfrac{\partial \mathcal{L}}{\partial \Pi} &= 1,
  \end{aligned}
\end{align}
which yield
\begin{align}
  \phi
  \begin{pmatrix}
    c_D^I \\
    c_P^I
  \end{pmatrix}
  +
  (1-\phi)
  \begin{pmatrix}
    c_D^{II} \\
    c_P^{II}
  \end{pmatrix}
  =
  \begin{pmatrix}
    c_D = MN / V \\
    c_P = P / V
  \end{pmatrix},
  \label{eq:maxwell_construct}
\end{align}
and completely determine the dilute and the dense phase from the mean-field concentrations $c_D$ and $c_P$. This relation has a very straightforward graphical interpretation and is a generalization of a Maxwell construct (\cref{fig:flory_section_coexistence}). If we imagine changing the concentration $c_D$ or $c_P$, or more generally going along a path in the $(c_D,c_P)$ plane, the system will not phase-separate right away when the path crosses the coexistence line. That is because there is a range of metastable mean-field solutions enclosed within the coexistence curve (\cref{fig:flory_section_spinodal}). Continuing along this fictive path, the mean-field solution will become unstable only when the determinant in \cref{eq:spinodal} becomes negative. Only then will the system split into two phases. Therefore in general, the transition is first order, except at the two critical points where the coexistence line and the spinodal line intersect.

\subsection{Conclusion}
\begin{figure}[!hbtp]
  \centering
  \subfloat[]{%
    \label{fig:MD_snapshots:a}%
    \includegraphics[width= 0.25 \textwidth]{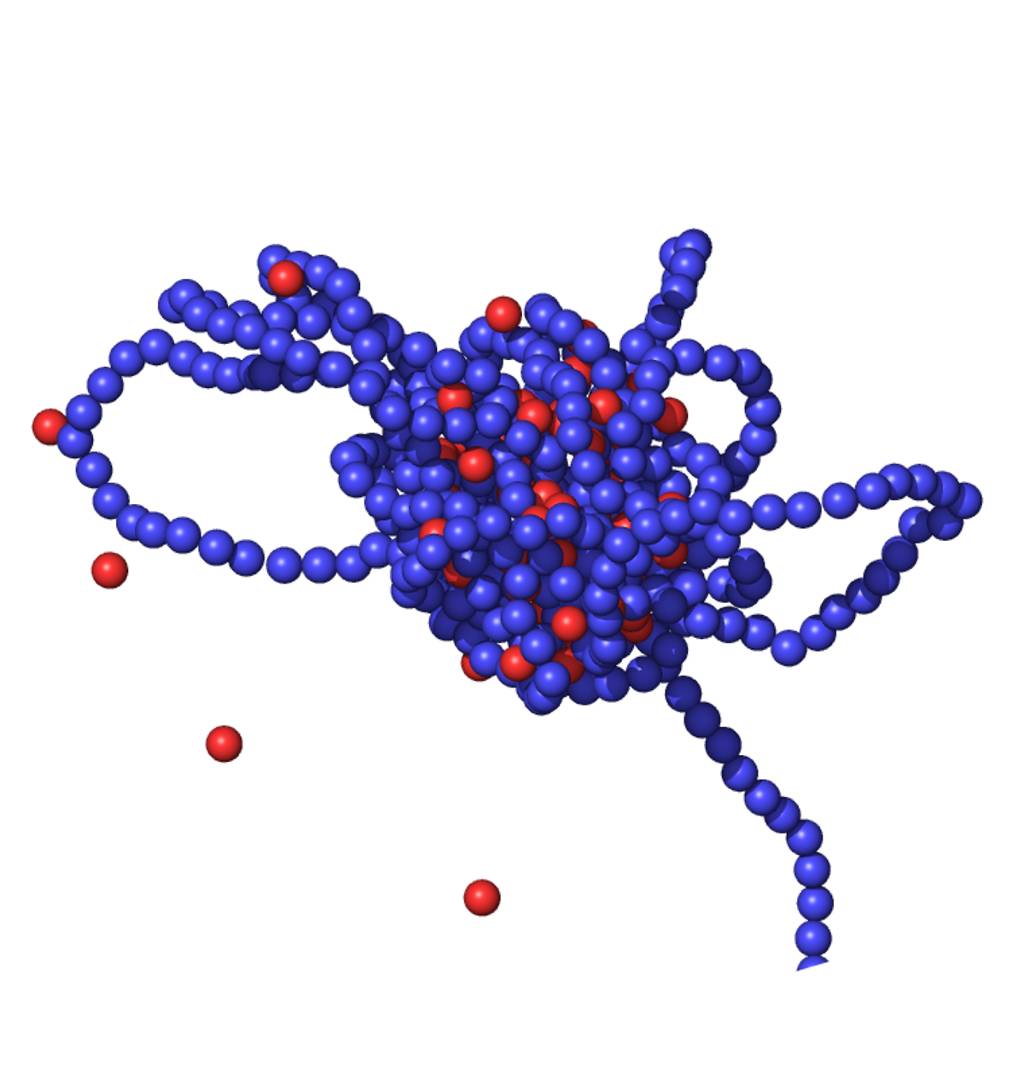}%
  }
  \hspace{0.05 \textwidth}
  \subfloat[]{%
    \label{fig:MD_snapshots:b}%
    \includegraphics[width= 0.25 \textwidth]{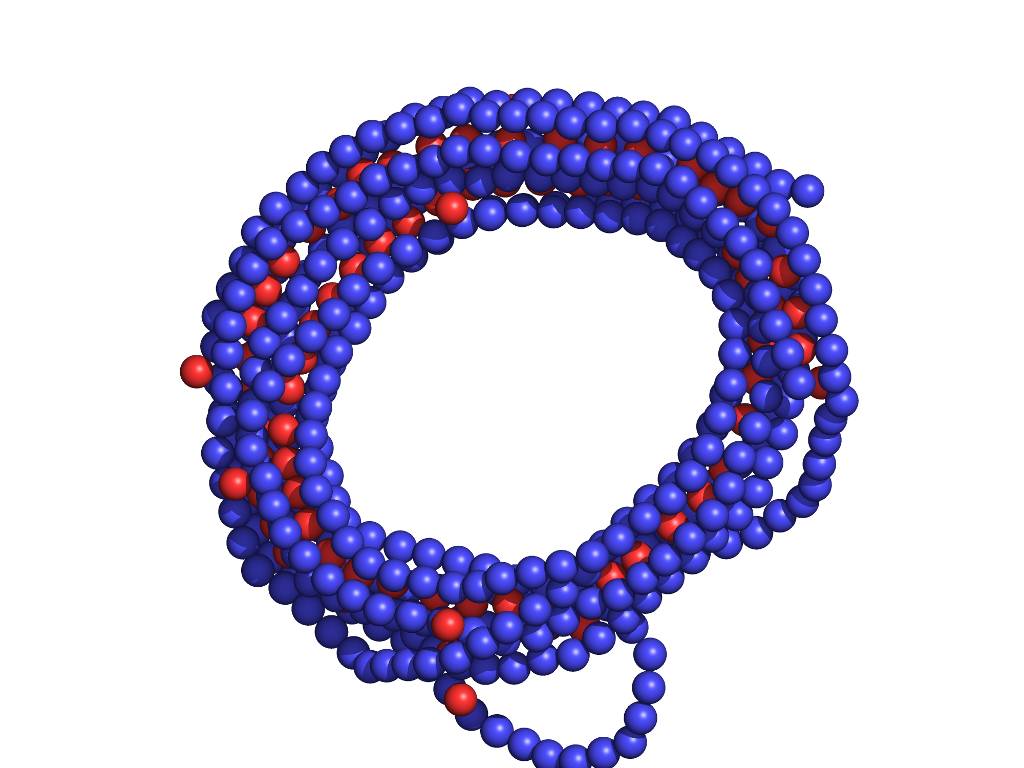}%
  }
  \hspace{0.05 \textwidth}
  \subfloat[]{%
    \label{fig:MD_snapshots:c}%
    \includegraphics[width= 0.30 \textwidth]{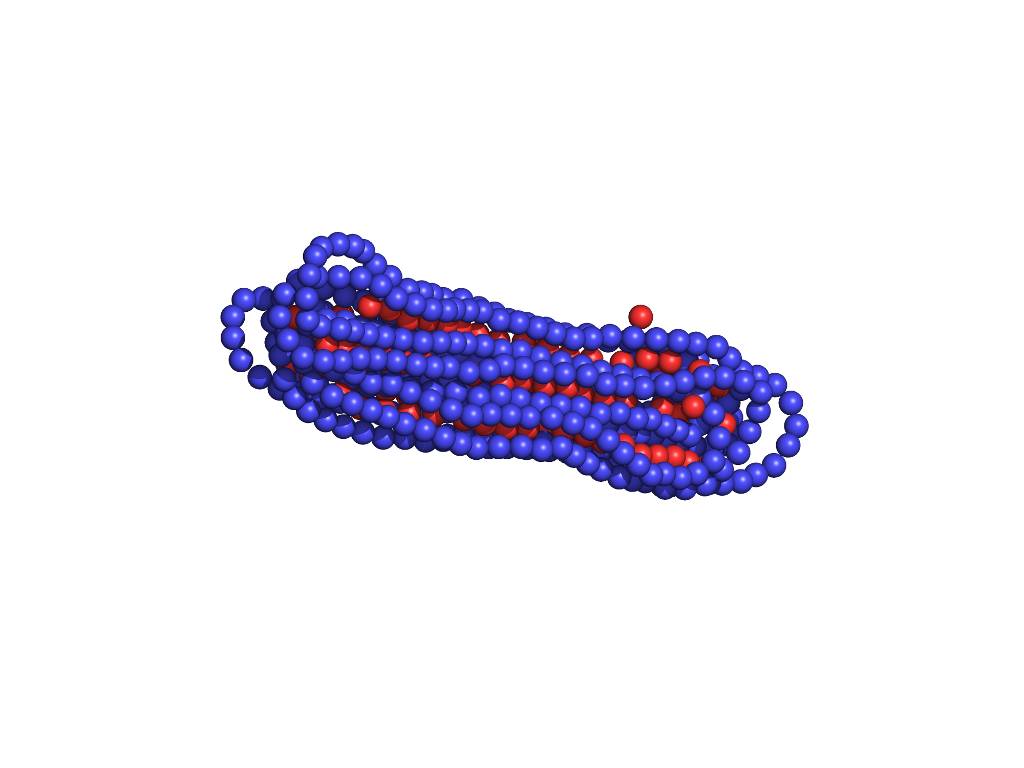}%
  }
  \caption{Snapshots of Brownian dynamics simulations (polymer with $N+1=400$ monomers). For small persistence lengths \protect\subref{fig:MD_snapshots:a} the dense phase is a molten globule while it is crystal-like for stiffer chains \protect\subref{fig:MD_snapshots:b}-\protect\subref{fig:MD_snapshots:c}.}
  \label{fig:MD_snapshots}
\end{figure}

The relevant biological parameter in this approach is the binding interaction between proteins and DNA, which is represented by the mean-field coefficient $\beta v$. It is a function of the affinity of the proteins with DNA and depends on the biochemistry of the interaction. What this approach tells us is that for a generic DNA-binding protein, a phase separation is to be expected if the affinity is such that $\beta v < (\beta v)^c$ and if the concentrations of DNA and proteins fall within the biphasic region. The phase separation gives rise to a dilute phase ($I$) with few DNA and few proteins and a dense phase ($II$) with higher concentrations of DNA and proteins. Because in general $c_D^I < c_D^c \sim 1 / \sqrt{N}$, the concentration of DNA in the dilute phase is very small. Essentially, the DNA chains are collapsed into molten globules which form the dense phase, with few protruding loops which form the dilute phase. This feature is visible in BD simulations (\cref{fig:MD_snapshots}). The phase transition characterized is first order almost everywhere, in agreement with a thermodynamical model for the agglomeration of DNA-looping proteins, based on a description with graph ensembles \cite{SumedhaP110052008}.

\begin{figure}[!hbtp]
  \centering
  \includegraphics[width=1 \textwidth]{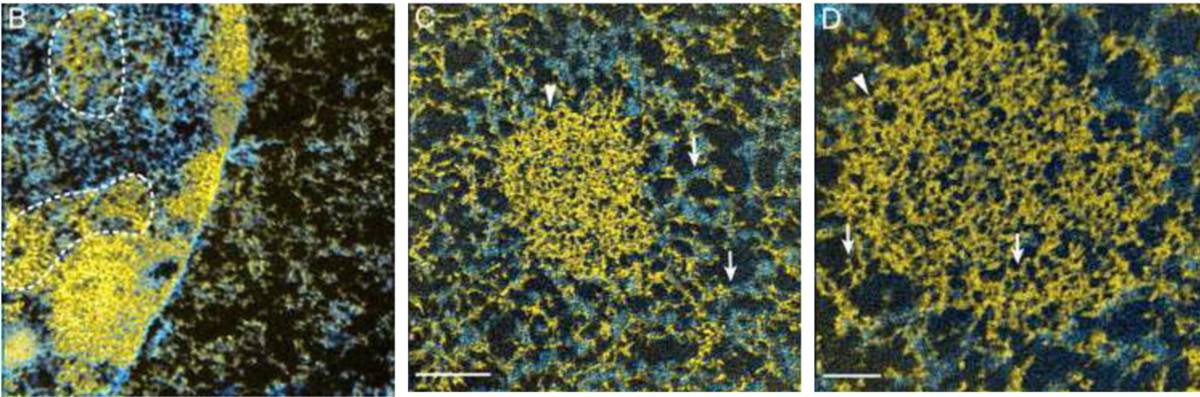}
  \caption{DNA-protein condensates observed in the nucleus in mouse cells with electron microscopy techniques \cite{Rapkin1502012}. The condensates are globular and have DNA (yellow) and proteins (blue), while the rest of the nucleus is filled mostly with proteins.}
  \label{fig:globular_dna_invivo}
\end{figure}

The globular clusters of the dense phase can be considered as a model for transcription factories observed \textit{in vivo}. Of course in reality there are many different TFs in the cell, which together may contribute to the architecture of the chromosome, including the formation of transcription factories. However, here we considered a generic type of protein. This come to say that the effect of one abundant protein prevails on the others in given physiological conditions. Note that we have adopted a coarse-grained approach in which many details of the chromosome organization such as specific DNA loops or protein complexes are embedded in the bead representation of DNA monomers and proteins. However, DNA clusters are indeed observed in the cell (\cref{fig:globular_dna_invivo}). These clusters do not display any internal structure, suggesting that they are globular. This effect can be accounted by the present theory by considering in a general way the effect of all binding proteins on the DNA.

Increasing evidence has suggested that transcription partly proceeds from transcription factories. Although the non-specific hypothesis for the binding of proteins to DNA that we have taken in this model is an over-simplifying assumption of the reality, it is true for instance that RNAP binds widely on the DNA thanks to its $\sigma$-unit. The conclusions reached in this Flory-Huggins theory suggest that a biphasic regime can exist, with a dense phase spanning a volume of size $(1-\phi) V$ and with local concentrations of DNA and RNAP increased with respect to the mean-field ones. Hence, the equilibrium of complexation reactions such as:
$$
\text{ DNA } + \text{ protein } \rightleftarrows RNAP \text{ bound to } DNA
$$
may be shifted towards the formation of complexes and may favour transcription initiation in the transcription factories. This is consistent with some experimental study showing that RNAP clusters are formed during pre-initiation and initiation of transcription \cite{Darzacq2013}. The same authors also proposed that crowding of enzymes, \textit{i.e.} higher local concentrations can aid in rate-limiting steps of gene regulation. From a dynamical standpoint, the confinement of unbound RNAP in a restricted volume can reduce the search time for a promoter. To this extent, it is worth pointing out a study claiming that the promoter search mechanism is indeed dominated by three-dimensional diffusion of RNAP over the monodimensional diffusion (\textit{i.e.} sliding) along DNA \cite{Wang2013}.

Note that the theory described here is somewhat different from more standard polymer-colloid systems treatments. In the latter case, the system consists of a solution of polymer coils and colloids, and the radius of gyration of one single coil and the diameter of the colloidal particles are comparable. In our case, the proteins can hardly be compared to colloidal particles because their size if comparable not to the radius of gyration of the whole chromosome but instead to the size of one monomer.

The presence of ions in solutions (\textit{e.g.} \ce{Ca^2+}, \ce{Cl-} \ce{Mg^2+}) gives rise to screened electrostatic interactions. For objects of nanometric size like proteins, interactions are short-ranged with a range typically given by the size of the objects in question. Let us also point out that at the mean-field level, the effect of ions in solution only arise through an adjustment of the Mayer coefficients $\alpha_D$, $\alpha_P$ and $\beta v$. The DNA excluded volume coefficient $\alpha_D$ will always be positive and accounts for the electrostatic repulsion between negatively charged monomers. The protein excluded volume coefficient $\alpha_P$ will be positive in general for the same reasons, but for proteins able to dimerize it may take negative values.

The present theory is stated with a general formalism. Hence, we think it may also be adapted to the description of the condensation of DNA by other condensing agents such as multivalent ions. As stated above, within the Flory-Huggins theory, the phase transition induced by condensing agents appears to be first order, except at the tricritical point and on the critical lines. Therefore the transition from the swollen to the condensed state should be discontinuous and present hysteresis effects, which was indeed observed \cite{Yoshikawa1995,Widom1980}. Interestingly, the Flory-Huggins theory also predicts another effect. For a fixed temperature and for a DNA concentration taken in a prescribed range, if we start with a small concentration of condensing agents that we progressively increase, there will be a value at which the system splits into two phases. Yet, if we keep adding condensing agents, the system will at some point exit the biphasic regime. This phenomenon called re-entrance has been observed in some experimental work using polyethylene glycol (PEG) \cite{Vasilevskaya1995}.

\section{Structure of the dense phase}
\begin{figure}[!hbtp]
  \centering
  \subfloat[]{%
    \label{fig:rigidity_effect:a}%
    \includegraphics[width= 0.38 \textwidth]{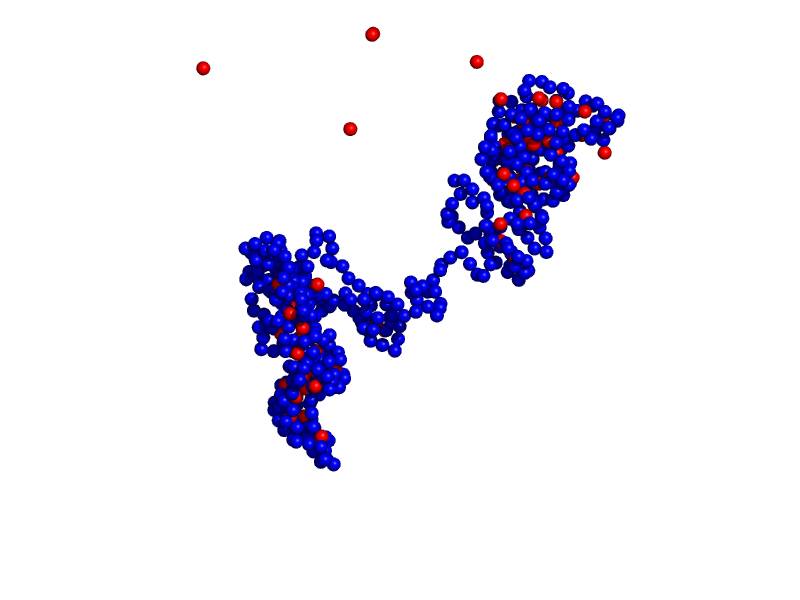}%
  }
  \subfloat[]{%
    \label{fig:rigidity_effect:b}%
    \includegraphics[width= 0.58 \textwidth]{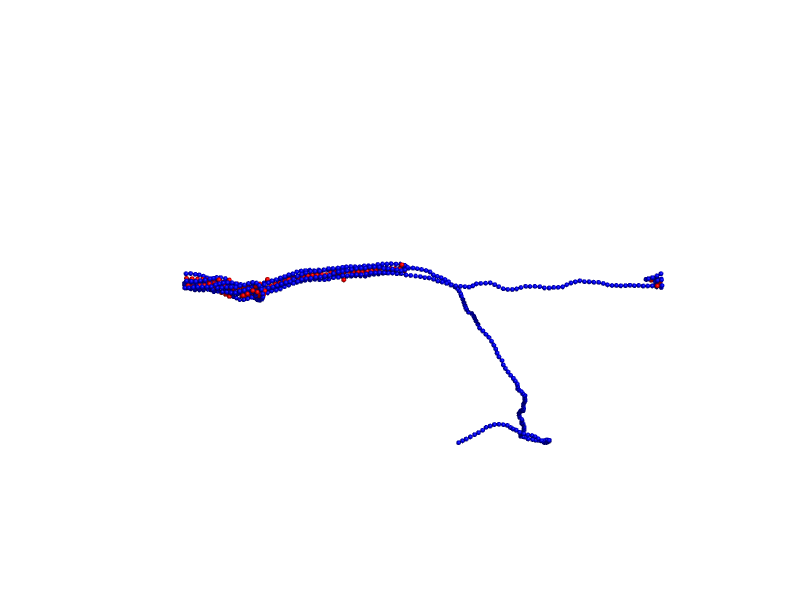}%
  }
  \caption{Two equilibrium configurations of a polymer chain (blue) interacting with binding spheres (red) in the absence of bending rigidity ($l_p=0$) and with strong bending rigidity ($l_p=30$), obtained with BD simulations. We used a truncated Lennard-Jones potential with $\varepsilon=3.0 \, k_B T$ and $r^{th}=2$. In both cases, the system is phase separated. \protect\subref{fig:rigidity_effect:a} For $l_p=0$, the dense phase consists of several globular aggregates distributed in a necklace fashion along the chain. \protect\subref{fig:rigidity_effect:b} For $l_p=30$, the dense phase adopts a tubular structure.}
  \label{fig:rigidity_effect}
\end{figure}

In the last section, the Flory-Huggins theory predicted the existence of a phase separation between two homogeneous phases. However, in the Flory-Huggins theory, the chain structure does not come into play, except through the suppression of the translational entropy of the chains. Namely, the dense phase predicted is a globule, that is to say a melt of collapsed DNA with proteins, regardless of the rigidity of the DNA chains. It turns out that DNA is a rigid biopolymer. It is an example of polyelectrolytes, and as such its bending rigidity depends on the screening effect of salt because of the presence of negative charges along its backbone. In physiological condition, the naked DNA has a persistence length $l_p$ of approximately \SI{150}{bp}. Several studies have highlighted that the bending rigidity of the polymer has an influence on the micro-structure of the dense phase \cite{Brackley36052013,Marenduzzo2015,Orland1996}. The dense phase then adopts stretched configurations which are characterized by the apparition of tube-like or helical structures. This effect is also well characterized through BD simulations (\cref{fig:rigidity_effect}).

\subsection{Random Phase Approximation}
\label{sec:RPA}
A standard way to characterize the effect of the chain structure is to use the Random Phase Approximation (RPA) \cite{deGennes1979}. The method consists in expanding the action in \cref{eq:partfunc_field} to second order around the mean-field solution and checking for the stability of the Gaussian fluctuations. The structure function of the DNA chain then naturally arises as a functional parameter for the stability condition of the Gaussian fluctuations. For the sake of simplicity, we will introduce the notation
\begin{align}
  \mathbf{X}(r) =
  \begin{pmatrix}
    \rho_D (r) \\
    \varphi_D (r) \\
    \rho_P (r) \\
    \varphi_P (r)
  \end{pmatrix},
\end{align}
and write the partition function in \cref{eq:partfunc_field} as
\begin{equation} \label{eq:partfunc_field_simple}
  Z = \int \uD{\mathbf{X}(r)} \exp{\left( - S[\mathbf{X}] \right)},
\end{equation}
where to alleviate notations, we took $\beta=1$.

The saddle-point condition, $\delta S / \delta \mathbf{X}(r) = 0$, yields the Lagrange equations:
\begin{align}
  & \begin{multlined}
    i \varphi_D(\mathbf{r}) - \int \ud{\mathbf{r}'} u_{DD}(\mathbf{r}-\mathbf{r}') \rho_D(\mathbf{r}') - \int \ud{\mathbf{r}'} u_{DP}(\mathbf{r}-\mathbf{r}') \rho_P(\mathbf{r}') \\
    - \dfrac{1}{2} w (\rho_D(\mathbf{r}) + \rho_P(\mathbf{r}))^2
  \end{multlined} &= 0, \\
  & i \rho_D(\mathbf{r}) + M \dfrac{\delta \ln{Q}}{\delta
    \varphi_D(\mathbf{r})}[i \varphi_D] &= 0, \\
    & \begin{multlined}
      i \varphi_P(\mathbf{r}) - \int \ud{\mathbf{r}'} u_{PP}(\mathbf{r}-\mathbf{r}') \rho_P(\mathbf{r}') - \int \ud{\mathbf{r}'} u_{DP}(\mathbf{r}-\mathbf{r}') \rho_D(\mathbf{r}') \\
      - \dfrac{1}{2} w (\rho_D(\mathbf{r}) + \rho_P(\mathbf{r}))^2
    \end{multlined} &= 0, \\
    & i \rho_P(\mathbf{r}) + P \dfrac{\delta \ln{W}}{\delta
      \varphi_P(\mathbf{r})}[i \varphi_P]<++> &= 0.
      \label{eq:saddle_point_eqn}
    \end{align}

    Starting from a guess solution (\textit{e.g.} homogeneous fields), the previous system can be solved iteratively or using continuous steepest descent methods. Such procedure is known as numerical self-consistent field methods in the polymer literature \cite{Fredrickson2005}. Although the convolutions are easily handled in Fourier space, the difficulty lies in the computation of the functional derivative $ \delta \ln{Q} / \delta \varphi_D(\mathbf{r})$. Hence it is a hard computational problem to solve. Based on physical considerations presented in the last section, we may look for the particular homogeneous (mean-field) solution:
    \begin{align}
      \mathbf{X}^* =
      \begin{pmatrix}
        c_D \\
        \phi_D \\
        c_P \\
        \phi_P
      \end{pmatrix}.
    \end{align}

    The resolution of \cref{eq:saddle_point_eqn} is now straightforward. We obtain the following saddle-point approximation for the partition function:
    \begin{equation}
      Z \simeq \exp(-S^*) = \exp\left(- V \beta f(c_D,c_P) \right),
    \end{equation}
    where $\beta f$ is given by \cref{eq:free_en_mf}, with $c_D=MN/V$ and $c_P=P/V$. That is to say we recover the Flory-Huggins theory from the last section. The RPA consists in the analysis of the effect of the thermodynamical fluctuations on the mean-field solution. To this end, let us introduce the vector field $\mathbf{Y} = \mathbf{X} - \mathbf{X}^*$. An expansion of the action in \cref{eq:partfunc_field_simple} to second  order gives
    \begin{equation}
      Z \simeq \int \uD{\mathbf{Y}(\mathbf{r}} \exp\left(-S^* - \dfrac{1}{2} \int \ud{\mathbf{r}} \ud{\mathbf{r}'} \mathbf{Y}(\mathbf{r}) \left. \dfrac{\delta^2 S}{\delta \mathbf{X}(\mathbf{r}) \delta \mathbf{X}(\mathbf{r}')} \right|_{\mathbf{X}=\mathbf{X^*}} \mathbf{Y}(\mathbf{r}') \right),
    \end{equation}
    or in Fourier space
    \begin{equation}
      Z \simeq \int \prod \limits_{\mathbf{k} > 0} \ud{\mathbf{Y}_{\mathbf{k}}} \exp\left(-S^* - \dfrac{1}{V} \sum \limits_{\mathbf{k} > 0} \mathbf{Y}_{\mathbf{k}} \left. \dfrac{\partial^2 S}{\partial \mathbf{X}_{\mathbf{k}} \partial \mathbf{X}_{-\mathbf{k}}} \right|_{\mathbf{X}=\mathbf{X^*}} \mathbf{Y}_{-\mathbf{k}} \right),
      \label{eq:partfunc_rpa}
    \end{equation}
    where the summation in Fourier space is carried out over the first Brillouin zone. The operator in the quadratic form is a $4 \times 4$ matrix. Its matrix elements can be computed (see \cref{app:rpa_matrix_elements}) and the following expression is obtained in Fourier representation:
\begin{align}
  \left. \dfrac{\partial^2 S}{\partial \mathbf{X}_{\mathbf{k}} \partial \mathbf{X}_{-\mathbf{k}}} \right|_{\mathbf{X}=\mathbf{X^*}}
  =
  \begin{pmatrix}
    A_{DD}(\mathbf{k}) & -i & A_{DP}(\mathbf{k}) & 0 \\
    -i & c_D S_N(\mathbf{k}) & 0 & 0 \\
    A_{DP}(\mathbf{k}) & 0 & A_{PP}(\mathbf{k}) & -i \\
    0 & 0 & -i & c_P
  \end{pmatrix},
\end{align}
with
\begin{align}
  \begin{aligned}
    A_{DD}(\mathbf{k}) &= u_{DD}(\mathbf{k}) + w(c_D+c_P)^2, \\
    A_{PP}(\mathbf{k}) &=  u_{PP}(\mathbf{k}) + w(c_D+c_P)^2, \\
    A_{DP}(\mathbf{k}) &=  u_{DP}(\mathbf{k}) + w(c_D+c_P)^2.
  \end{aligned}
\end{align}

Note that the structure of the chain comes into play through the structure factor $S_N(\mathbf{k})$ which appears in one of the matrix elements. The quadratic fluctuations operator is diagonal in Fourier space. Hence the partition function in \cref{eq:partfunc_rpa} is a product of Gaussian integrals. It is well-defined as long as for all $\mathbf{k}$
    \begin{equation}
      \Gamma(\mathbf{k}) = \det{\left( \dfrac{\partial^2 S}{\partial \mathbf{X}_{\mathbf{k}} \partial \mathbf{X}_{-\mathbf{k}}} \right) } > 0.
      \label{eq:rpa_spinodal}
    \end{equation}

    The previous equation is a generalization of the spinodal condition for a Fourier mode $\mathbf{k}$. When the temperature is lowered, the first mode $\mathbf{k}^*$ to become unstable sets the temperature when the mean-field solution is no longer stable. If $k^*=0$, then it is simply the spinodal condition of the Flory-Huggins theory, and at the instability, the system splits into two homogeneous phases as described previously. If $k^* > 0$, then it is the signature of a micro-phase separation. In that case the homogeneous mean-field solution is no longer a stable solution and instead stable solutions will display spatial modulations of typical length $2\pi / k^*$. The determinant in \cref{eq:rpa_spinodal} can be computed easily and gives
    \begin{equation}
      \Gamma(\mathbf{k}) = \left( A_{DD}(\mathbf{k}) + \dfrac{1}{c_D S_N(\mathbf{k})} \right) \left( A_{PP}(\mathbf{k}) + \dfrac{1}{c_P} \right) - A_{DP}(\mathbf{k})^2.
    \end{equation}

    The potentials are essentially contact-like because of the screened interactions (we recall that the Debye-H\"uckel length is $\sim 1 \, nm$ in the cell), thus the Fourier transforms of the interaction potentials have the expressions:
    \begin{align}
      \begin{aligned}
        u_{DD}(\mathbf{k}) &= \alpha_D, \\
        u_{PP}(\mathbf{k}) &= \alpha_P, \\
        u_{DP}(\mathbf{k}) &= \beta v.
      \end{aligned}
    \end{align}

  The structure factor for a Gaussian chain with $N$ Kuhn segments of length $b$ is given by $S_N(\mathbf{k}) = N D(k^2 R_g^2)$, where $R_g^2=b^2 N / 6$ is as usual the radius of gyration of the chain and $D(x) = 2/x^2 (x-1+\exp{(-x)})$ is the Debye function \cite{deGennes1979}. There is no analytical expression available for the structure function for a Worm Like Chain (with persistence $l_p$) as it is the case for the Gaussian chain. To date, to the best of our knowledge, it is still an open problem despite several classical and more recent works \cite{DesCloizeaux1973,Kholodenko1993,Pedersen1996,Spakowitz2004,Zhang2014}. We first used an approximate expression for the structure factor of a semi-flexible chain found by Thirumalai and co-worker \cite{Bhattacharjee1997}, which lead us to propose an alternative method to compute the structure function of a worm-like chain, that we will present in \cref{ch:structure_function}.

\begin{figure}[!ht]
      \centering
      \subfloat[]{%
        \label{fig:phase_transitions_I_II:a}%
        \includegraphics[width= 0.48 \textwidth]{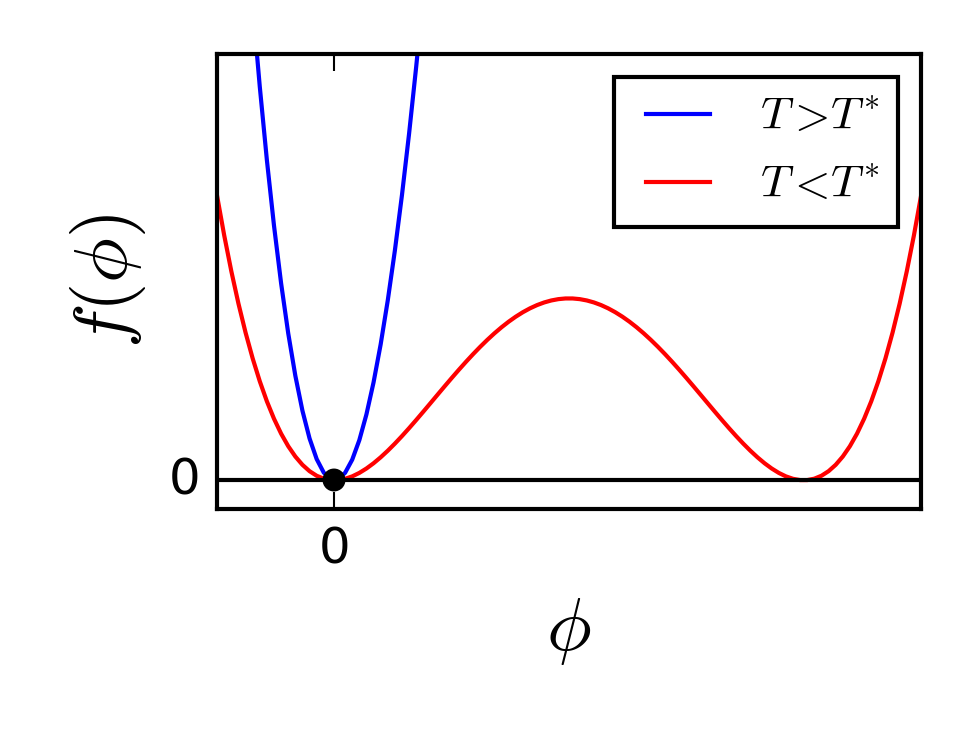}%
      }
      \subfloat[]{%
        \label{fig:phase_transitions_I_II:b}%
        \includegraphics[width= 0.48 \textwidth]{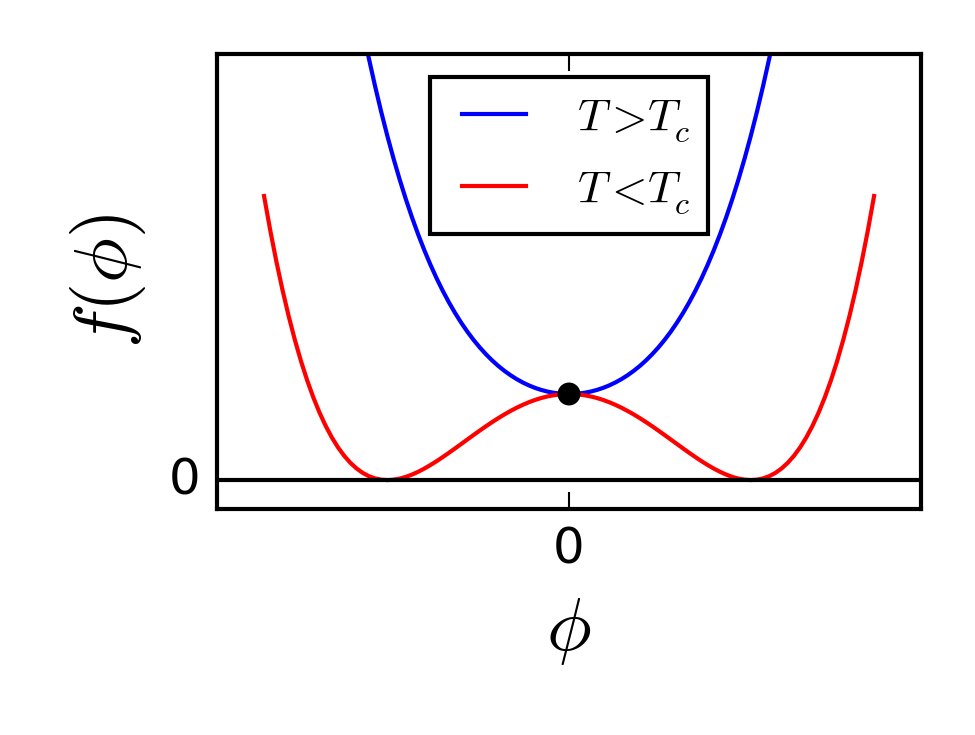}%
      }
      \caption{Comparison between a first order phase transition \protect\subref{fig:phase_transitions_I_II:a} and a second order phase transition \protect\subref{fig:phase_transitions_I_II:b}. A second order phase transition is therefore driven by critical fluctuations at the high temperature equilibrium, which is not the case for a first order phase transition.}
    \end{figure}

 We have monitored the sign of $\Gamma(k)$ as a function of the wave number $k$. However we do not report it here because we have not found any instability arising for a non-zero Fourier mode, that would characterize a dense phase with micro-structure. On second thought, we suspect that the RPA analysis is not well suited for this system because the phase transition is first order in general (it is only second order when crossing the critical line). In order to illustrate this point, let us consider a generic system with a mean-field order parameter $\phi$ and a free energy function $F(\phi)$ displaying a second order phase transition at a critical temperature $T_c$. For $T>T_c$, the free energy has a single minimum $\phi^*$ and $\partial^2 F / \partial \phi^2 (\phi^*) > 0$, \textit{i.e.} the solution is stable. At $T=T_c$, the free energy has still one single minimum $\phi^*$, yet  the curvature at this minimum is null. For $T<T_c$, the former solution $\phi^*$ is unstable, \textit{i.e.} $\partial^2 F / \partial \phi^2 (\phi^*) < 0$, and the free energy has two minima $\phi^I$ and $\phi^{II}$. Similarly to what has been presented above, one may give an approximation of the partition function of this system by integrating over the Gaussian fluctuations:
    \begin{equation}
      Z = \int \ud{y} \exp{\left(-F(\phi^*) - \dfrac{1}{2} \dfrac{\partial F }{\partial \phi^2}(\phi^*) y^2 \right)},
    \end{equation}
    where $y=\phi-\phi^*$ is the difference with the high-temperature solution. When $T<T_c$, $\partial^2 F / \partial \phi^2 (\phi^*) < 0$ and this Gaussian integral is no longer defined. The instability is driven by an inversion of the curvature around the high-temperature solution $\phi^*$ which occurs at $T=T_c$ (\cref{fig:phase_transitions_I_II:a}). Therefore, for a second order transition the critical temperature $T_c$ coincides with an instability of the fluctuations around the high-temperature solution. In that case, we say that the phase transition is driven by critical fluctuations. To the contrary, in the case of a first order transition, the high-temperature solution $\phi^I$ remains stable as we cross the phase transition temperature $T^*$ (\cref{fig:phase_transitions_I_II:b}). For that reason, the integral over the fluctuations around the high-temperature solution remains a well-defined Gaussian integral and is of no use to identify the phase transition temperature $T^*$. In that case, we say that the phase transition is not driven by critical fluctuations. A generalization of this reasoning suggests that RPA will not help in characterizing a micro-phase separation because the transition is not driven by critical fluctuations around the saddle-point solution. Since the RPA did not give any interesting results, we turned to another way to characterize the structure of the dense phase.

\subsection{Lattice model of the dense phase}
\subsubsection{Model}
\begin{figure}[!hbtp]
  \centering
  \includegraphics[width=0.48 \textwidth]{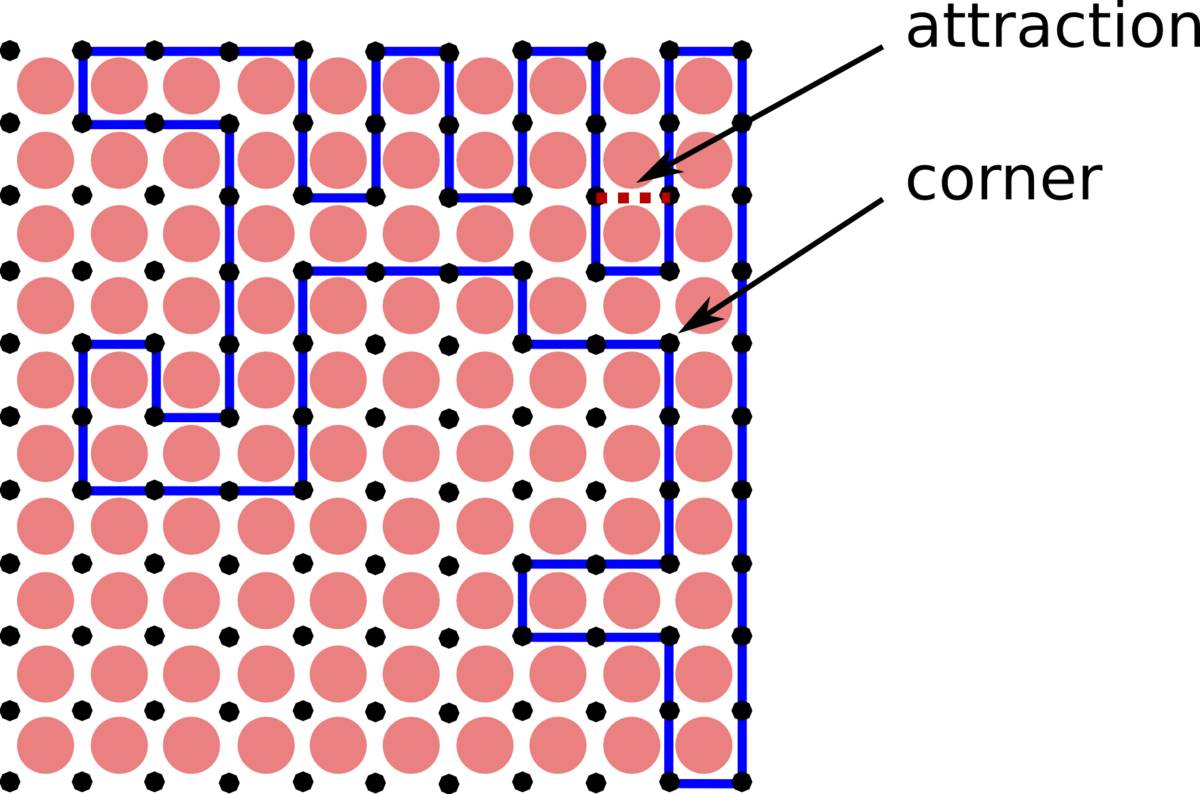}
  \caption{Lattice model for the collapse of a DNA polymer. The attractive interaction mediated by binding proteins (red spheres) is taken into account implicitly through the introduction of an attractive interaction between neighboring sites.}
  \label{fig:lattice_model_low_density}
\end{figure}

Since the RPA is not appropriate to describe the system in the dense phase, we adopt another approach. Because of their attractive interactions with the DNA, the spheres induce an effective attraction between the DNA monomers. Before coming back to a system in a continuous volume at the end of this section, let us turn our attention to a model of a semi-flexible polymer chain on a lattice that was proposed initially to explain the folding of a protein in compact structures \cite{Orland1996,Orland1992,Orland1993} (see \cref{fig:lattice_model_low_density}). An attraction energy $\varepsilon_v$ between non-bonded nearest neighbors is included, which favors compact configurations. A bending energy of the chain is introduced as a corner penalty which favor stretched configurations (or ``helices'' in the protein folding vocabulary). It penalizes corners by an energy $\varepsilon_h$ and thus plays the role of a bending rigidity. As we will see, this term induces an ordering transition between a random (molten) globule where corners are mobile in the bulk, and a crystalline phase, where corners are expelled to the surface of the globule.

In order to explore the equilibrium physics of this system, we write the partition function for this system as
\begin{equation}
  Z_N = \sum \limits_{SAW} \exp{\left( - \beta \varepsilon_h \mathcal{N}_c + \dfrac{1}{2} \beta \varepsilon_v \sum \limits_{\mathbf{r},\mathbf{r}'} n_{\mathbf{r}} \Delta_{\mathbf{r},\mathbf{r}'} n_{\mathbf{r}'} - N \beta \varepsilon_v \right)},
  \label{eq:partfunc_lattice_low_d}
\end{equation}
where the sum is carried out on the self-avoiding walks (SAW) of length $N$, $\mathcal{N}_c$ is the number of corners in the configuration, $\Delta_{\mathbf{r}, \mathbf{r}'}$ is the nearest neighbor operator on the lattice, and $n_{\mathbf{r}}=0,1$ is the occupancy variable of the lattice sites. The first term in the exponent gives a penalty proportional to the number of corners, whereas the second gives a bonus proportional to the number of neighbors pairs. Note the third term which is introduced to cancel off the interactions between consecutive monomers along the path that were already counted in the second term. Using a mean field theory, Orland and colleagues \cite{Orland1996,Orland1992,Orland1993} showed that depending on the attraction energy and the corner penalty, three phases can exist, namely a dilute phase where the polymer is swollen, a condensed phase, which we call a molten globule, where the polymer is collapsed and disordered  and finally a second condensed phase where the polymer is collapsed but with a local crystalline ordering. The phase diagram is described simply in the plane $(\varepsilon_v/T, \varepsilon_h / \varepsilon_v)$. For fixed small $\varepsilon_h$, there is a second-order phase transition at a temperature $T=T_\theta$ between a dilute and a disordered condensed phase, followed by a first-order freezing  transition at $T_F$ between the disordered condensed phase and a locally ordered condensed phase of the polymer. Upon increasing the chain stiffness $\varepsilon_h$, the molten globule region shrinks until it eventually vanishes. Thus, for large stiffness, the polymer goes abruptly from a swollen to a frozen configuration ($T_F > T_\theta$) through a direct first order transition. These theoretical results were readily confirmed and improved by Monte-Carlo simulations \cite{Grassberger2008,Frenkel1998}.

\subsubsection{Counting Hamiltonian paths}
For the sake of completeness we review here in a simple case the methods that were used to obtain the announced results. We consider the extreme case where the polymer has collapsed completely (density $\eta =1$). That is to say, each of the $N$ sites of the lattice are occupied by a DNA monomer. When both $\varepsilon_h=0$ and $\varepsilon_v=0$, the partition function for this system reduces to
\begin{equation}
  Z_N = \sum \limits_{HP} 1,
  \label{eq:partfunc_lattice_hp}
\end{equation}
where the sum is carried out on the Hamiltonian paths (HP) on the lattice, that is to say paths that visit each of the $N$ sites of the lattice once and only once. Hence, the partition function is just the number of HP on the lattice. An equivalence to a field theory is obtained by introducing for each lattice site a $n$-component field $\boldsymbol{\varphi}_\mathbf{r}$, and the corresponding partition function
\begin{equation}
  Q_n = \int \prod \limits_{\mathbf{r}} \ud{\boldsymbol{\varphi}_{\mathbf{r}}} e^{-A_G} \prod \limits_{\mathbf{r}} \left( \dfrac{1}{2} \boldsymbol{\varphi}_{\mathbf{r}}^2 \right),
  \label{eq:partfunc_lattice_hp_field}
\end{equation}
where $A_G$ is a quadratic action given by
\begin{equation}
  A_G = \dfrac{1}{2} \sum \limits_{\mathbf{r},\mathbf{r}'} \boldsymbol{\varphi}_{\mathbf{r}} \Delta_{\mathbf{r},\mathbf{r}'}^{-1} \boldsymbol{\varphi}_{\mathbf{r}'},
\end{equation}
and $\Delta_{\mathbf{r},\mathbf{r}'}$ is as before the nearest neighbor operator on the lattice. The quantity in \cref{eq:partfunc_lattice_hp_field} is (up to a normalization), a simple Gaussian average. This average can be computed using Wick's theorem once it has been pointed out that the elementary contraction reads
\begin{equation}
  \left\langle \boldsymbol{\varphi}_{\mathbf{r}}^u \boldsymbol{\varphi}_{\mathbf{r}'}^v \right\rangle = \delta^{uv} \Delta_{\mathbf{r},\mathbf{r}'}.
\end{equation}

Therefore, only products of fields corresponding to sites which are nearest neighbors will give a non zero average. As a result, we obtain that
\begin{align}
  \begin{aligned}
    \left\langle \prod \limits_{\mathbf{r}} \left( \dfrac{1}{2} \boldsymbol{\varphi}_{\mathbf{r}} \right)^2 \right\rangle  &= \dfrac{1}{2^N} \sum \limits_{\text{all permutations }} \left\langle \boldsymbol{\varphi}_{\mathbf{r}_1} \boldsymbol{\varphi}_{\mathbf{r}_2} \right\rangle \ldots
    \left\langle \boldsymbol{\varphi}_{\mathbf{r}_{2N-1}} \nonumber \boldsymbol{\varphi}_{\mathbf{r}_{2N}} \right\rangle, \\
    &= \sum_{k=1} n^k c_k,
  \end{aligned}
\end{align}
where $c_k$ is the number of graphs on the lattice containing $k$ closed paths which together visit all sites of the lattice once and only once. Eventually, $c_1$ is just the number of Hamiltonian paths on the lattice, and we have the equivalence
\begin{align}
  \sum_{HP} 1 = \lim_{n \to 0} \dfrac{1}{n} \dfrac{\displaystyle \int \prod \limits_{\mathbf{r}} \ud{\boldsymbol{\varphi}_{\mathbf{r}}} e^{-A_G} \prod \limits_{\mathbf{r}} \left( \dfrac{1}{2} \boldsymbol{\varphi}_{\mathbf{r}}^2 \right)}{\displaystyle \int \prod \limits_{\mathbf{r}} \ud{\boldsymbol{\varphi}_{\mathbf{r}}} e^{-A_G} \prod \limits_{\mathbf{r}}}.
  \label{eq:lattice_field_equivalence}
\end{align}

This simple equivalence gives an original way to compute the number of Hamiltonian paths on a lattice. Namely, one can perform a saddle-point approximation on the field partition function in \cref{eq:partfunc_lattice_hp_field}. In that case, an approximation of $Q_n$ is:
\begin{equation}
  Q_n \simeq \int \prod \limits_{\mathbf{r}} \ud{\boldsymbol{\varphi}_{\mathbf{r}}^*} \exp{\left( -N + d\frac{1}{2} \sum_{\mathbf{r}} \ln{\left(\dfrac{1}{2}\boldsymbol{\varphi}_{\mathbf{r}}^*\right)}\right)},
\end{equation}
where the $*$ superscript denotes that the integration is carried out on the saddle-point solutions. One can then go further by making a mean-field approximation, that is to say by considering only saddle-point solutions of norm $\Vert \boldsymbol{\varphi}_{\mathbf{r}}^* \Vert = \varphi$. It follows from the saddle-point equation that $\varphi^2 = 2q$ where $q=2d$ is the coordination number of the lattice. The integration in the last expression is then performed over the $n$-dimensional sphere, whose area is $2 \pi^{n/2} / \Gamma(n/2)$. Eventually, using the equivalence in \cref{eq:lattice_field_equivalence}, one obtains the approximation for the number of Hamiltonian paths on the lattice of $N$ sites:
\begin{equation}
  Z_N \simeq \left( \dfrac{q}{e} \right)^N.
  \label{eq:partfunc_lattice_hp_final}
\end{equation}

\subsubsection{The effect of rigidity}
Still assuming that the polymer has collapsed completely, we can now slightly modify the previous model by introducing a corner penalty $\beta \varepsilon_h \neq 0$. In that case, the partition function for this system reads
\begin{equation}
  Z_N = \sum \limits_{HP} e^{-\beta \varepsilon_h \mathcal{N}_c},
  \label{eq:partfunc_lattice_hp_rigid}
\end{equation}
where as before $\mathcal{N}_c$ counts the number of corners in the HP realization. In close analogy to the previous developments, we seek for an equivalence with a field theory. For each site of the lattice, we introduce a $n$-component field $\boldsymbol{\varphi}_\alpha(\mathbf{r})$ for each of the direction $\alpha=1,\ldots,d$ of the $d$-dimensional lattice. This time we  introduce the partition function
\begin{equation}
  Q_n = \int \prod \limits_{\mathbf{r}} \prod \limits_{\alpha=1}^d \ud{\boldsymbol{\varphi}_\alpha(\mathbf{r})} e^{-A_G} \prod \limits_{\mathbf{r}} \left( \dfrac{1}{2} \sum \limits_{\alpha=1}^d \boldsymbol{\varphi}_\alpha(\mathbf{r})^2 + e^{-\beta \varepsilon_h} \sum \limits_{\alpha < \gamma} \boldsymbol{\varphi}_\beta(\mathbf{r}) \cdot \boldsymbol{\varphi}_\gamma(\mathbf{r}) \right),
  \label{eq:partfunc_lattice_hp_rigid_field}
\end{equation}
where $A_G$ is a quadratic action given by
\begin{equation}
  A_G = \dfrac{1}{2} \sum \limits_{\alpha=1}^d \sum \limits_{\mathbf{r},\mathbf{r}'} \boldsymbol{\varphi}_\alpha(\mathbf{r}) \left[ \Delta_{\mathbf{r},\mathbf{r}'}^{\alpha}\right]^{-1} \boldsymbol{\varphi}_\alpha(\mathbf{r}'),
\end{equation}
and $\Delta_{\mathbf{r},\mathbf{r}'}^\alpha$ is the nearest neighbor operator on the lattice in direction $\alpha$. The quantity in \cref{eq:partfunc_lattice_hp_rigid_field} is again a simple Gaussian average. It can be computed using Wick's theorem with the elementary contraction
\begin{equation}
  \left\langle \boldsymbol{\varphi}_\alpha^u(\mathbf{r}) \boldsymbol{\varphi}_\beta^v(\mathbf{r}') \right\rangle = \delta^{uv} \delta^{\alpha \beta} \Delta_{\mathbf{r},\mathbf{r}'}^\alpha,
\end{equation}
and again, only products of fields corresponding to sites which are nearest neighbors will give a non-zero average. Therefore this average selects closed paths on the lattice. However, there is an important difference with the previous case. While making products of the quantity between parenthesis in \cref{eq:partfunc_lattice_hp_rigid_field}, the first term of this quantity will tend to select nearest neighbors only in the direction $\alpha$. Yet, given a site $\mathbf{r}$ and a nearest neighbor $\mathbf{r}'$, one can choose for the former a term of the form $\boldsymbol{\varphi}_\alpha(\mathbf{r})^2 / 2$, and for the latter a term $\boldsymbol{\varphi}_\alpha(\mathbf{r}') \cdot \boldsymbol{\varphi}_\beta(\mathbf{r}')$, in which case a Boltzmann weight equal to $\exp{(-\beta \varepsilon_h)}$ must be applied. Applying Wick's theorem, one is led to consider products of elementary contractions like $\langle \boldsymbol{\varphi}_\alpha(\mathbf{r}) \cdot \boldsymbol{\varphi}_\alpha(\mathbf{r}') \rangle e^{-\beta \varepsilon_h} \langle \boldsymbol{\varphi}_\beta(\mathbf{r}') \cdot \boldsymbol{\varphi}_\beta(\mathbf{r}'')\rangle$. In summary, like before, the partition function in \cref{eq:partfunc_lattice_hp_rigid_field} generates closed paths on the lattice. Yet this time a Boltzmann weight with a penalty equal to the number of turns times $\beta \varepsilon_h$ is applied to each path. This leads to the following equivalence:
\begin{equation}
  \sum \limits_{HP} e^{-\beta \varepsilon_h \mathcal{N}_c} = \lim_{n \to 0} \dfrac{1}{n} \dfrac{\displaystyle \int \prod \limits_{\mathbf{r}} \prod \limits_{\alpha=1}^d \ud{\boldsymbol{\varphi}_\alpha(\mathbf{r})} e^{-A_G} \prod \limits_{\mathbf{r}} \left( \dfrac{1}{2} \sum \limits_{\alpha=1}^d \boldsymbol{\varphi}_\alpha(\mathbf{r})^2 + e^{-\beta \varepsilon_h} \sum \limits_{\alpha < \gamma} \boldsymbol{\varphi}_\beta(\mathbf{r}) \cdot \boldsymbol{\varphi}_\gamma(\mathbf{r}) \right)}{\displaystyle \int \prod \limits_{\mathbf{r}} \prod \limits_{\alpha=1}^d \ud{\boldsymbol{\varphi}_\alpha(\mathbf{r})} e^{-A_G}}.
\end{equation}

In the same spirit as in the previous case, a saddle-point approximation supplemented by a mean-field approximation yields the approximate expression for the partition function in \cref{eq:partfunc_lattice_hp_rigid}:
\begin{equation}
  Z_N \simeq \left( \dfrac{q(\beta)}{e} \right)^N,
\end{equation}
where
\begin{equation}
  q(\beta) = 2 + 2(d-1) e^{-\beta \varepsilon_h}.
  \label{eq:lattice_effective_q}
\end{equation}

\begin{figure}[!ht]
  \centering
  \includegraphics[width=0.7 \textwidth]{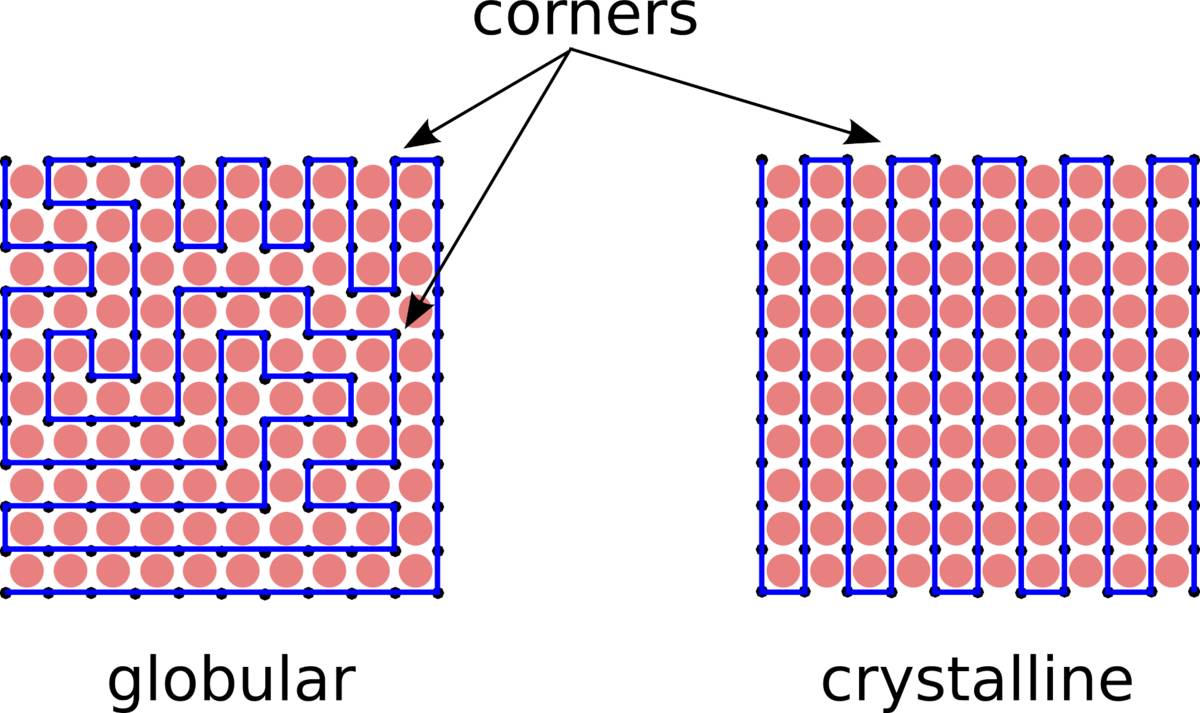}
  \caption{Two realizations of Hamiltonian paths on a cubic lattice. The globular state contains an extensive number of corners whereas the crystalline state contains a non-extensive number of corners (proportional to the surface).}
  \label{fig:lattice_model_globule_crystalline}
\end{figure}

The expression obtained for the partition function is remarkable because it has the same form as \cref{eq:partfunc_lattice_hp_final}, but with an effective lattice coordination number $q(\beta)$. In the limit $T \to \infty$, one recovers the previous case, with $q(0)=2d$. To the contrary, when $T \to 0$, one has $q(\infty)=2$, which corresponds to straight paths where each monomer only "sees" the previous and the next monomer. The free energy $f(T) = -T/N \ln{Z_N}$ has the property to vanish at the temperature $T_F$, defined by $q(\beta_F)=e$. Starting from the high temperatures, $f(T)$ is a decreasing function of the temperature until $T = T^*$, at which point the entropy $S(T^*)= - \partial f / \partial T (T^*) = 0$ and $f(T^*)>0$. When the entropy vanishes, the system freezes in one configuration. Nonetheless this result is obtained within the context of the saddle-point and mean field approximation. A more careful analysis of the partition function in \cref{eq:partfunc_lattice_hp_rigid_field} based on a Schwartz inequality shows that the free energy is bounded: $f(T) \le 0$ \cite{Orland1992}. Consequently, the positivity of the free energy in the range of temperature $T^* < T < T_F$ can only correspond to metastable states. In conclusion, there is a freezing transition at $T=T_F$ which separates a high temperature regime in which the collapsed polymer is a molten globule from a low temperature regime in which the collapsed polymer is crystal-like and has a (quasi) zero entropy. In the crystalline phase, the configurations look like straight paths with corners expelled to the surface of the lattice (\cref{fig:lattice_model_globule_crystalline}). These configurations have been studied previously: they are elongated neck structures or toro{\"i}ds \cite{MacKintosh2004}.

\subsubsection{Phase diagram}
The methods presented in the two previous paragraphs can be generalized to the case with attractive interactions between nearest neighbors ($\beta \varepsilon_v < 0$) and to a lattice with vacancies ($\eta < 1$). Namely, an equivalence with a field theory in the limit $n \to 0$ can be written for the partition function in \cref{eq:partfunc_lattice_low_d}. Similarly to the previous cases, a combination of saddle-point and mean-field approximations yields the free energy per monomer:
\begin{equation}
  f(\eta,T) = - T \ln{\left( \dfrac{q(\beta)}{e} \right)} + T \dfrac{1 - \eta}{\eta} \ln{(1 - \eta)} + \varepsilon_v(1 - d \eta),
  \label{eq:lattice_full_freeenergy}
\end{equation}
where $q(\beta)$ is defined in \cref{eq:lattice_effective_q}. The full derivation is presented in details in \cite{Orland1996}. In the open coil regime $\eta \simeq 0$ and the free energy reduces to $f_0(T) = -T \ln{q} + \varepsilon_v$. Introducing the reduced temperature $t=T / \varepsilon_v$, the equilibrium equation $\partial f / \partial \eta = 0$ yields non-zero solutions $\eta^*$ only when
\begin{equation}
  t < t_\theta = 2d.
\end{equation}

Therefore, there is a second-order transition between an open-coil state and a globule at $t_\theta$. Though, as characterized previously, for $\beta \varepsilon_h \neq 0$, there is a freezing transition at a temperature $t_F$ toward a crystalline phase. For each $\varepsilon_h$ the freezing temperature is obtained by equating the free energy in \cref{eq:lattice_full_freeenergy} with the free energy of a frozen configuration where all the corners, delimiting the non-vacant region, are expelled to the surface:
\begin{equation}
  g = - (d-1) \varepsilon_v.
  \label{eq:lattice_full_freeenergy_frozen}
\end{equation}

\begin{figure}[!hbtp]
  \centering
  \subfloat[]{%
    \label{fig:phase_diagrams_dense_phase:a}%
    \includegraphics[width= 0.44 \textwidth,valign=c]{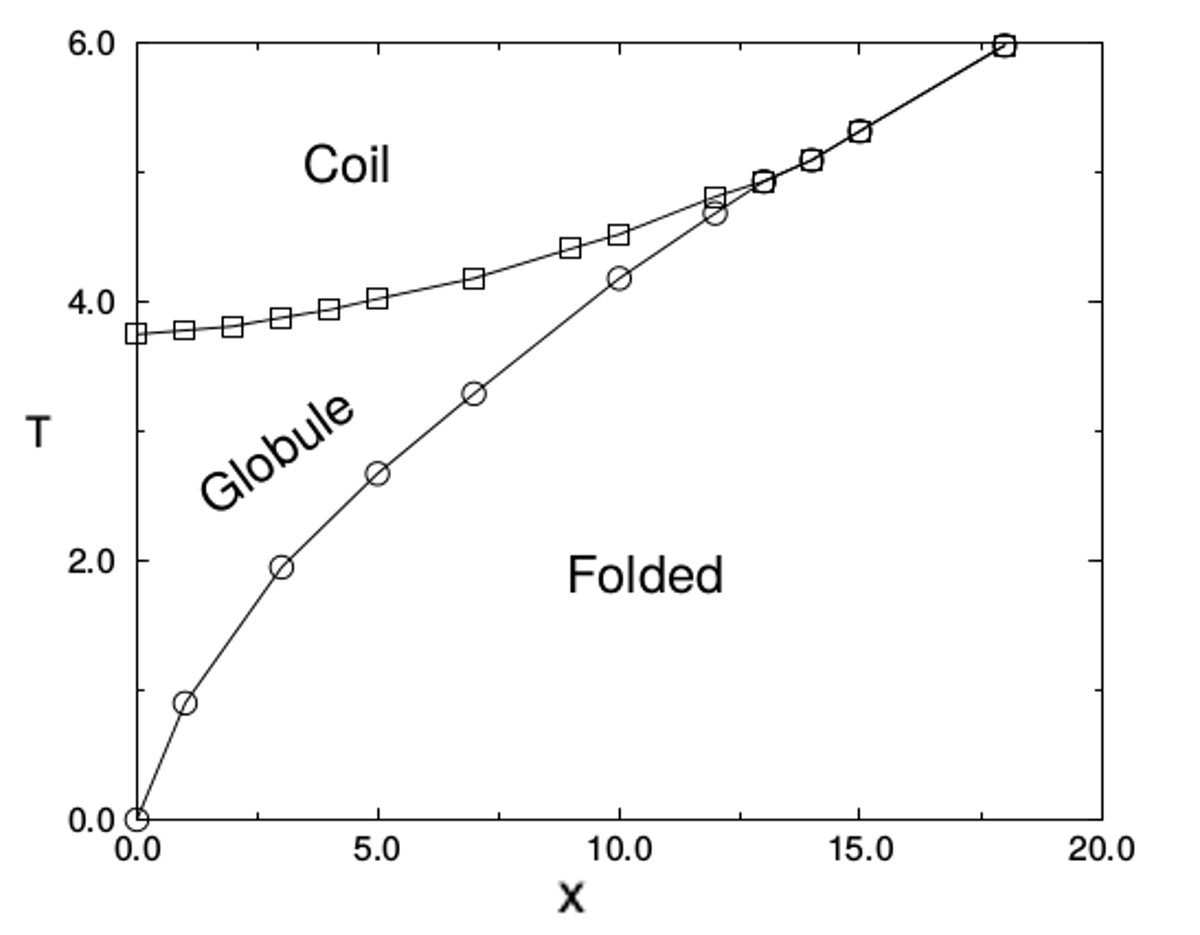}%
  }
  \subfloat[]{%
    \label{fig:phase_diagrams_dense_phase:b}%
    \includegraphics[width= 0.52 \textwidth,valign=c]{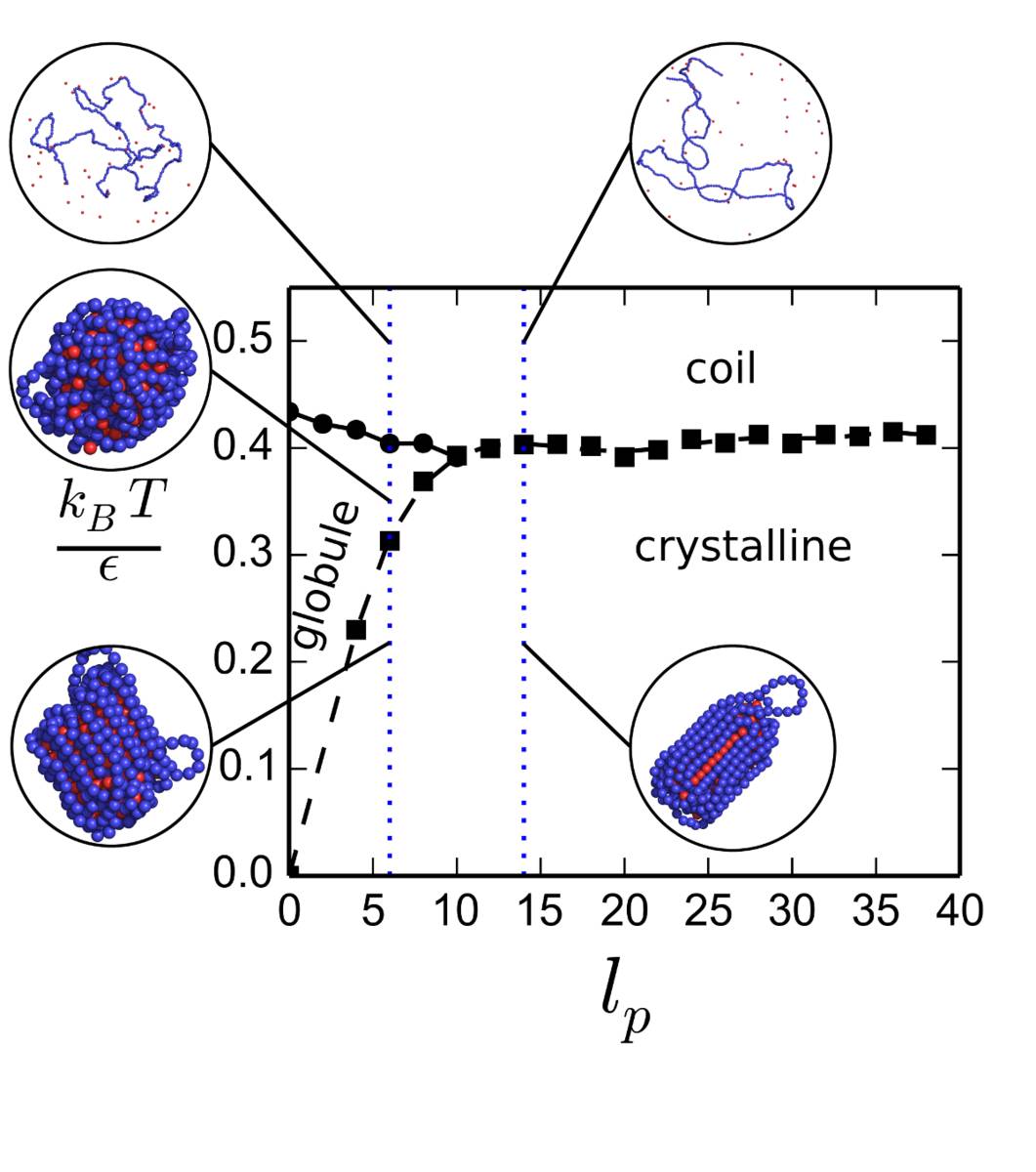}%
  }
  \caption{\protect\subref{fig:phase_diagrams_dense_phase:a} Phase diagram of the collapse of a polymer on a lattice as defined in \cref{eq:partfunc_lattice_low_d}, and computed in \cite{Grassberger2008} from Monte-Carlo simulations. $x=\varepsilon_h / \varepsilon_v$ \protect\subref{fig:phase_diagrams_dense_phase:b} Phase diagram of a polymer interacting with spheres in a continuous volume. The phase diagram was computed using BD simulations with $P=100$ spheres, a polymer with $N+1=400$ monomers, as a function of the persistence length $l_p$ and of the strength of the Lennard-Jones DNA-protein interaction $\varepsilon$.}
  \label{fig:phase_diagrams_dense_phase}
\end{figure}

In conclusion, the phase diagram announced is obtained in the plane $(t,\varepsilon_h / \varepsilon_v)$, in which $t_\theta$ delimits the coil-globule transition and $t_F$ the globule-crystalline freezing transition. This phase diagram was later confirmed and refined with the help of Monte-Carlo simulations \cite{Grassberger2008} (see \cref{fig:phase_diagrams_dense_phase:a}).

\subsection{Phase diagram of the dense phase structure}
  The results obtained in this section have enlightened our understanding of the collapse of a rigid polymer. There are however differences with the model of DNA condensation by binding proteins presented in \cref{sec:flory_huggins}. First, the formal nucleus model is not a lattice model. Second, the attractive interactions between DNA monomers are not implicit but mediated by proteins. In that respect, the collapse depends on the concentration of proteins. Third, the bending rigidity of the DNA is not taken into account by discrete corner penalties but is instead modeled using a Kratky-Porod potential with persistence length $l_p$ (see \vref{subsec:polymer_model_bending_rigidity}).

  Despite these differences, we have found that the phase diagram of the collapse of a semi-flexible polymer interacting explicitly with spheres in an off-lattice volume is very similar to the phase diagram obtained for the collapse of a polymer on a lattice presented above (\cref{fig:phase_diagrams_dense_phase}). Indeed, we performed BD simulations with a polymer chain of $N+1=400$ beads and $P=100$ protein spheres in a cubic volume of size $L=100$ with periodic boundary conditions. Polymer beads and protein spheres were interacting through a truncated Lennard-Jones potential with a well depth given by the energy scale $\varepsilon$ (in $k_B T$). By varying $l_p$ and $\varepsilon$ independently, we explored the phase behaviour of this system (\cref{app:coil_globule_crystal_transitions}).

  There are minor quantitative differences between the lattice/implicit and the off-lattice/explicit cases. Namely the numerical values for the coil-globule and globule-crystal transitions are different. This is due to the difference in the definition of the order parameters, the use of a Lennard-Jones potential for the attractive interaction, the Kratly-Porod model used to take into account the chain bending rigidity, and it is also a consequence of going from a lattice model to a continuous model. Furthermore, the concentration of spheres in solution is smaller than the close packing concentration, making it hardly comparable to an actual solvent. Despite these discrepancies, we can say that the results in both cases are in qualitative agreement.

The main conclusion obtained from the phase diagram of the dense phase is the existence of specific persistence length $l_p^* \simeq 10$ such that:
\begin{itemize}
  \item	for $l_p < l_p^*$, the polymer collapses through a second order coil-globule transition, followed by a first order globule-crystal transition when $\varepsilon$ increases;
  \item	for $l_p > l_p^*$, the coil-globule transition no longer exists and the polymer collapses directly from a coil to a crystalline phase through a first order phase transition.
\end{itemize}

The coil-globule transition is the same as the phase transition depicted using a Flory-Huggins theory in \cref{sec:flory_huggins}. Yet, when the DNA-protein attraction is strong enough, it appears that the dense phase can be crystalline. Besides, for very rigid chains ($l_p > l_p^*$), the coil-globule transition does not exist because it is precluded by the freezing transition. In this case, the results of the Flory-Huggins theory are no longer valid. Snapshots of the coil, globule, and frozen state computed from MD simulations can be seen in \cref{fig:MD_snapshots} and \cref{fig:phase_diagrams_dense_phase:b}.

\subsection{Conclusion}
Let us sum up what has been obtained in this section. We have first noticed that the bending rigidity has an influence on the structure of the DNA-protein condensates. We have tried to use the RPA in order to characterize modulations of the DNA and protein concentrations in the dense phase, which is the signature of the existence of a microphase. Yet, the RPA failed because the collapse transition being generally first order, it is not driven by critical fluctuations. We turned to a theory of polymer collapse on a lattice, which is instructive because it predicts the existence of a crystalline phase for large values of the rigidity parameter $\varepsilon_h$. Using BD simulations, we effectively recovered this result for our formal nucleus model.

DNA condensation has been well characterized in \textit{in vitro} experimental works \cite{Bloomfield1996,Livolant1996,Vasilevskaya1995,Yoshikawa1995,Durand1992,Sergey1995}. Consequently, it is well known that DNA collapses from disperse structures corresponding to swollen coil configurations into ordered, highly condensed states, namely toroids or hexagonal bundles \cite{Sung14212016,Lansac219952016}. Namely, such studies have concluded that during its collapse, DNA undergoes transitions through the following three phases: isotropic fluid, cholesteric and crystalline (hexagonal). This is in agreement with our results, demonstrating that our model for the collapse of DNA by proteins is actually more general that what was intended in the first place.

Although it is premature to draw any clear biological conclusion, it is tempting to discuss at least qualitatively potential effects of the crystalline phase on biological functions. In eukaryotes, nucleosomal organization provides an effective protection against detrimental factors. This organization is absent in prokaryotes, which have a significantly lower ratio of DNA-binding proteins \cite{Kellenberger1993}. However, in harsh environmental conditions (radiations, temperature, oxidizing agents and radicals), several bacteria resort to DNA condensation mechanisms to protect their genome. Maybe the most spectacular case is the appearance of macroscopic DNA aggregates with crystal-like order in starved \ecoli cells. In stressful conditions, the alternative $\sigma^S$ factor is expressed, in response to low temperature, cell surface stress or oxidative shock. This in turn induces the expression of the DNA-binding protein DPS \cite{Almiron1992,Almiron1994}. In starved cells, DPS is the most abundant DNA-binding protein, with approximately \num{20000} DPS proteins per cell. Consequently, DNA is condensed into crystal-like aggregates, which makes it less accessible to damaging factors. Wild-type \ecoli cells starved for three days remain unaffected by a high dose of oxidizing agents whereas mutants lacking DPS lose viability \cite{Almiron1992}. This process is reversible. Interestingly, DPS binds non-specifically to DNA. We speculate that when DPS concentration increases, it induces the DNA collapse, and a dense phase appears. For proteins of the size of DPS ($< \SI{10}{\nm}$), the apparent rigidity of DNA is large ($\approx \SI{50}{\nm}$). Hence we might be in a case where the coil-globule transition is precluded by the freezing transition. Other examples of DNA compaction by non-specific proteins seem to exist. For instance the protein RecA induces the formation of DNA bio-crystals in \ecoli which have an essential role in the DNA repair system \cite{FrenkielKrispin2006}, and the condensation of DNA in crystal-like configurations by spermine and polyamines has also been well characterized \cite{Newton1996}.

Earlier studies have demonstrated that the frozen phase can present various metastable states \cite{Frenkel1998}. In the large $N$ limit ($N$ is the length of one chain), the transition time scale from one to another could be very large, and the system might well never equilibrate within biological time scales. Moreover, the parallel drawn between the Hamiltonian paths theory and the Flory-Huggins theory does not pretend to mathematical rigor. One essential difference is that in our case the attractive interaction between monomers is mediated by spheres. A way to compute more precisely the structure of the dense phase would be to go beyond the homogeneous saddle point approximation leading to the Flory-Huggins theory, for instance by using self-consistent field methods \cite{Edwards1988,Fredrickson2005}, which are quite complex methods in the case of semi-flexible polymers.

\begin{figure}[!hbtp]
  \centering
  \subfloat[]{%
    \label{fig:dna_cystals:a}%
    \includegraphics[width= 0.6 \textwidth]{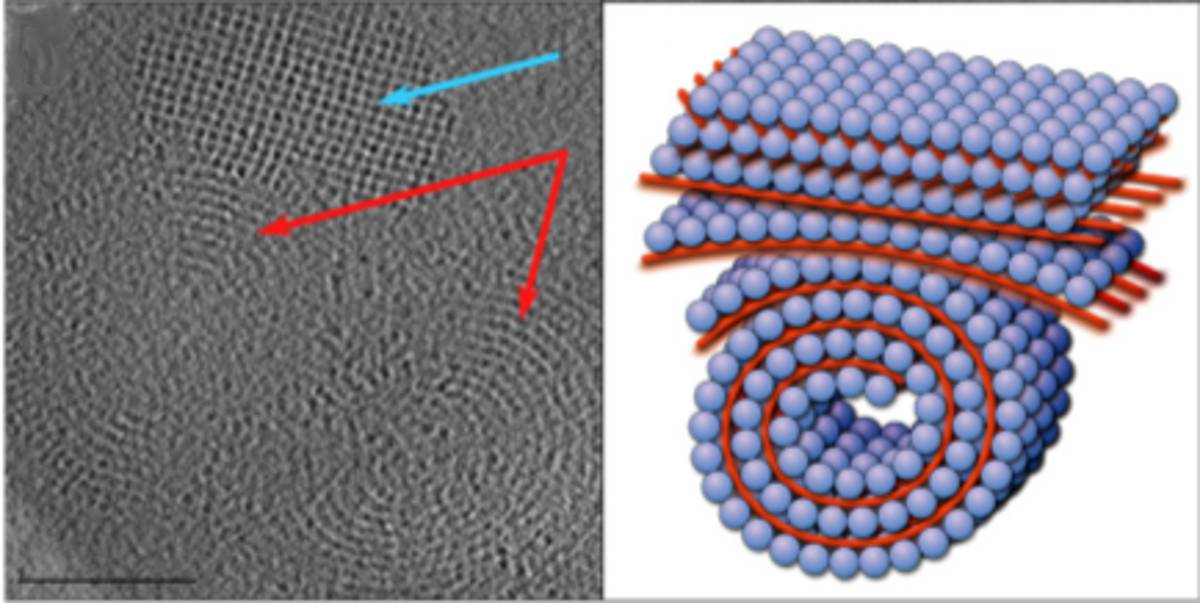}%
  }
  \\
  \subfloat[]{%
    \label{fig:dna_cystals:b}%
    \includegraphics[width= 0.3 \textwidth]{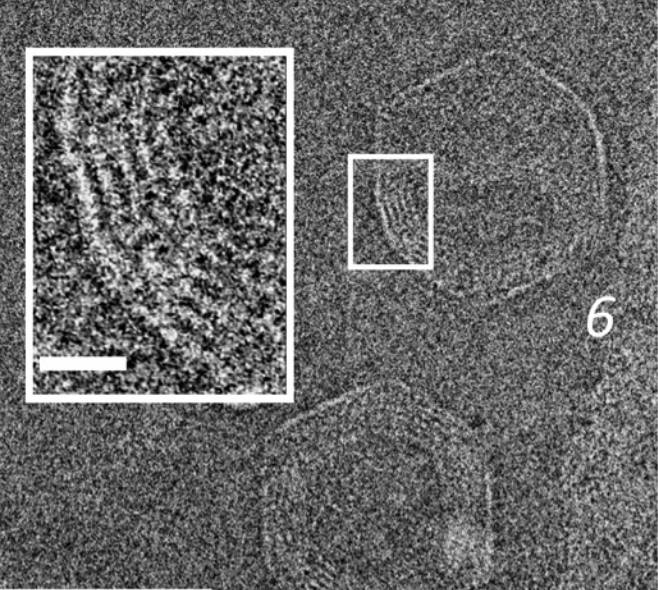}%
  }
  \quad
  \subfloat[]{%
  \label{fig:dna_cystals:c}%
  \includegraphics[width= 0.3 \textwidth]{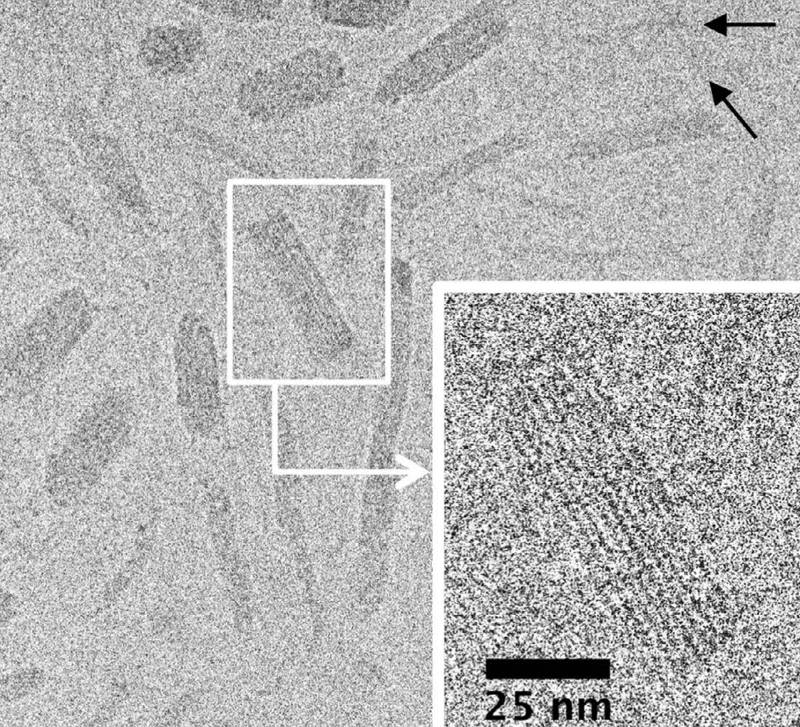}%
}
\caption{\protect\subref{fig:dna_cystals:a} Electron microscope (EM) images of DNA co-crystals in \ecoli \cite{FrenkielKrispin2006}. \protect\subref{fig:dna_cystals:b} CryoEM of DNA condensed inside the T5 bacteriophage in toroidal shapes \cite{Sung14212016}. \protect\subref{fig:dna_cystals:c} DNA-protamines complexes with bundle shapes observed by cryo-imaging with transmission electron microscope \cite{Lansac219952016}.}
\end{figure}

\section{Discussion}
We presented here two complementary frameworks to describe the phase diagram of polymeric fluids induced by binding particles, and applied it to a DNA chain interacting with DNA-binding proteins. Starting from a Flory-Huggins free energy, we first computed the mean-field phase diagram and found that at low temperature (\textit{i.e.} high DNA-protein affinity) a biphasic regime exists, consisting of the coexistence of a dilute phase and a concentrated phase. The dilute phase may correspond to swollen configurations of the DNA whereas the concentrated phase is a model for condensed states of DNA. It turns out that the theory may also apply to DNA condensation by multivalent ions. Second, we addressed the characterization of the dense phase structure and showed that the chain bending rigidity can have dramatic effects. Without bending rigidity, the dense phase has no directional order and is a molten globule. However, when the chain bending rigidity is large enough, there is a freezing transition from the globular to a crystalline phase. Eventually for very rigid chains, the coil-globule transition is precluded by the freezing transition and the phase transition predicted in the Flory-Huggins framework does not occur.

In the cell, the existence of a dense phase could be a good approximation for the transcription factories observed experimentally. It is conjectured that this may increase the rate of success in transcription initiation  by means of protein crowding and by enhancing the promoter search mechanism. Note that at a scale coarse-grained to a few thousand base-pairs (gene scale), the chromosome is flexible and the dense phase has the structure of a molten globule. Conversely, at a scale of a few base-pairs, the apparent rigidity of DNA is much higher. Thus, the DPS protein, which binds non-specifically to DNA, can induce the collapse of the \ecoli chromosome into crystal-like aggregates; the dense phase is then frozen. This is not an efficient state for a searching mechanism. But on the contrary, it is very adequate to protect DNA or to halt transcription.

\begin{subappendices}
\section{Matrix elements of the Gaussian fluctuations}
\label[app]{app:rpa_matrix_elements}
We recall the expression of the action in \cref{eq:action_field}:
\begin{align}
  \begin{aligned}
    \beta S &= -i \int \ud{\mathbf{r}} \rho_D(\mathbf{r}) \varphi_D(\mathbf{r}) - i \int \ud{\mathbf{r}} \rho_P(\mathbf{r}) \varphi_P(\mathbf{r}) \\
    & + \int \ud{\mathbf{r}} \ud{\mathbf{r'}} \rho_P(\mathbf{r'}) u_{PP}(\mathbf{r'} - \mathbf{r}) \rho_P(\mathbf{r}) + \int \ud{\mathbf{r}} \ud{\mathbf{r'}} \rho_D(\mathbf{r'}) u_{DD}(\mathbf{r'} - \mathbf{r}) \rho_D(\mathbf{r}) \\
    & + \beta \int \ud{\mathbf{r}} \ud{\mathbf{r'}} \rho_D(\mathbf{r'}) u_{DP}(\mathbf{r'} - \mathbf{r}) \rho_P(\mathbf{r}) \\
    & + \dfrac{1}{3!} w \int \ud{\mathbf{r}} \left( \rho_D(\mathbf{r}) + \rho_P(\mathbf{r}) \right)^3 \\
    & - P \ln{W[i \varphi_P ]} - M \ln{Q[i \varphi_D ]} + P \ln{\dfrac{P}{e}} + M \ln{\dfrac{M}{e}},
  \end{aligned}
\end{align}
with the single bead and single chain partition functions:
\begin{align}
  \begin{aligned}
    W[i \varphi_P ] &= \int \ud{\mathbf{R}} \exp{(-i \varphi_P(\mathbf{R}))}, \\
    Q[i \varphi_D ] &= \int \uD{\mathbf{r}(s)} \exp{\left( -\beta U_0[\mathbf{r}(s)] - i \int \limits_0^N \ud{s} \varphi_D(\mathbf{r}(s))\right)}.
  \end{aligned} \label{eq:app_partfunc_singlechain}
\end{align}

We now give the first and second order derivative of the action, which are used to find the saddle-point equations and perform the Gaussian fluctuations analysis. For the sake of clarity, we take $\beta=1$ in the sequel. The first order functional derivatives are:
\begin{align}
  \begin{aligned}
    \dfrac{\delta S}{\delta \rho_D(\mathbf{r})} &= -i \varphi_D(\mathbf{r}) + \int \ud{\mathbf{r}'} u_{DD}(\mathbf{r}-\mathbf{r}') \rho_D(\mathbf{r}') + \int \ud{\mathbf{r}'} u_{DP}(\mathbf{r}-\mathbf{r}') \rho_P(\mathbf{r}') \\
    & + \dfrac{1}{2} w (\rho_D(\mathbf{r}) + \rho_P(\mathbf{r}))^2, \\
    \dfrac{\delta S}{\delta \varphi_D(\mathbf{r})} &= -i \rho_D(\mathbf{r}) - M \dfrac{\delta \ln{Q}}{\delta \varphi_D(\mathbf{r})}, \\
    \dfrac{\delta S}{\delta \rho_P(\mathbf{r})} &= -i \varphi_P(\mathbf{r}) + \int \ud{\mathbf{r}'} u_{PP}(\mathbf{r}-\mathbf{r}') \rho_P(\mathbf{r}') + \int \ud{\mathbf{r}'} u_{DP}(\mathbf{r}-\mathbf{r}') \rho_D(\mathbf{r}') \\
    & + \dfrac{1}{2} w (\rho_D(\mathbf{r}) + \rho_P(\mathbf{r}))^2, \\
    \dfrac{\delta S}{\delta \varphi_P(\mathbf{r})} &= -i \rho_P(\mathbf{r}) - P \dfrac{\delta \ln{W}}{\delta \varphi_P(\mathbf{r})},
  \end{aligned} \label{eq:app_dS}
\end{align}
and the second order derivatives are:
\begin{align}
  \begin{aligned}
    \dfrac{\delta^2 S}{\delta \rho_D(\mathbf{r}) \delta \rho_D(\mathbf{r}')} &= u_{DD}(\mathbf{r} - \mathbf{r}') + w(\rho_D(\mathbf{r})+\rho_P(\mathbf{r}))^2 \delta(\mathbf{r} - \mathbf{r}'), \\
    \dfrac{\delta^2 S}{\delta \rho_P(\mathbf{r}) \delta \rho_P(\mathbf{r}')} &= u_{PP}(\mathbf{r} - \mathbf{r}') + w(\rho_D(\mathbf{r})+\rho_P(\mathbf{r}))^2 \delta(\mathbf{r} - \mathbf{r}'),  \\
    \dfrac{\delta^2 S}{\delta \rho_D(\mathbf{r}) \delta \rho_P(\mathbf{r}')} &= u_{DP}(\mathbf{r} - \mathbf{r}') + w(\rho_D(\mathbf{r})+\rho_P(\mathbf{r}))^2 \delta(\mathbf{r} - \mathbf{r}'), \\
    \dfrac{\delta^2 S}{\delta \rho_D(\mathbf{r}) \delta \varphi_D(\mathbf{r}')}  &= -i, \\
    \dfrac{\delta^2 S}{\delta \rho_P(\mathbf{r}) \delta \varphi_P(\mathbf{r}')} &= -i, \\
    \dfrac{\delta^2 S}{\delta \varphi_D(\mathbf{r}) \delta \varphi_D(\mathbf{r}')}  &= -M \dfrac{\delta^2 \ln{Q}}{\delta \varphi_D(\mathbf{r}) \delta \varphi_D(\mathbf{r}')}, \\
    \dfrac{\delta^2 S}{\delta \varphi_P(\mathbf{r}) \delta \varphi_P(\mathbf{r}')}  &= -P \dfrac{\delta^2 \ln{W}}{\delta \varphi_P(\mathbf{r}) \delta \varphi_P(\mathbf{r}')}.
  \end{aligned} \label{eq:app_ddS}
\end{align}

In the RPA analysis of \cref{sec:RPA}, these derivatives are to be computed at the mean-field solution:
\begin{align}
  \begin{aligned}
    \rho_D(\mathbf{r}) &= c_D, \\
    \varphi_D(\mathbf{r}) &= \phi_D,\\
    \rho_P(\mathbf{r}) &= c_P, \\
    \varphi_P(\mathbf{r}) &= \phi_P,
  \end{aligned}
\end{align}
which we denote by the $*$ subscript in what follows. Evaluating the five first equations in \cref{eq:app_ddS} at the mean field saddle-point is easily done, whereas it requires further computations for the two last ones. We obtain:
\begin{align}
  \begin{aligned}
    \left. \dfrac{\delta^2 S}{\delta \varphi_P(\mathbf{r}) \delta \varphi_P(\mathbf{r}')} \right\vert_{*} &= c_P \left( \delta(\mathbf{r}-\mathbf{r}') - \dfrac{1}{V} \right), \\
    \left. \dfrac{\delta^2 S}{\delta \varphi_D(\mathbf{r}) \delta \varphi_D(\mathbf{r}')} \right\vert_{*} &= c_D \left( S_N(\mathbf{r}-\mathbf{r}') - c_D \right),
  \end{aligned}
\end{align}
where $S_N(\mathbf{r} - \mathbf{r}')$ is the structure function of the polymer chain. Its expression follows directly from taking the second order derivative in \cref{eq:app_partfunc_singlechain}:
\begin{align}
  \begin{aligned}
    c_D S_N(\mathbf{r} - \mathbf{r}') &= \left\langle \int \limits_0^N \ud{s} \ud{s'} \delta(\mathbf{r}-\mathbf{r}(s)) \delta(\mathbf{r}'-\mathbf{r}(s')) \right\rangle \\
    &= \dfrac{1}{Q_0} \int \limits_0^N \ud{s} \ud{s'} \int \uD{\mathbf{r}(t)} \delta(\mathbf{r}-\mathbf{r}(s)) \delta(\mathbf{r}'-\mathbf{r}(s')) \exp{\left( -U_0[\mathbf{r}] \right)} \\
    &= \dfrac{1}{Q_0} \int \limits_0^N \ud{s} \ud{s'} \int \ud{\mathbf{r}_N} \ud{\mathbf{r}_0} \langle \mathbf{r}_N \vert e^{-(N-s) \hat{U}_0} \vert \mathbf{r}' \rangle \langle \mathbf{r}' \vert e^{-(s'-s) \hat{U}_0} \vert \mathbf{r} \rangle \langle \mathbf{r} \vert e^{-s \hat{U}_0} \vert \mathbf{r}_0 \rangle, \\
  \end{aligned} \label{eq:app_strucfunc_expression}
\end{align}
where we have introduced the chain propagator:
\begin{align}
  \begin{aligned}
    q(\mathbf{r}'s ; \mathbf{r} 0) &= \int \limits_{\mathbf{r}(0) = \mathbf{r}}^{\mathbf{r}(s) = \mathbf{r}'} \uD{\mathbf{r}(t)} \exp{\left( -U_0[\mathbf{r}] \right)} \\
    &= \langle \mathbf{r}' \vert e^{-s \hat{U}_0} \vert \mathbf{r} \rangle \\
    &= \langle \mathbf{r}' \vert \psi(s) \rangle \text{,} \qquad \vert \psi(0) \rangle = \vert \mathbf{r} \rangle.
  \end{aligned}
\end{align}

A proper choice of normalization results in $\vert \psi (s) \rangle$ to be the probability distribution function for the last monomer. Consequently, we have
\begin{align}
  \begin{aligned}
    &\int \ud{\mathbf{r}'} q(\mathbf{r}'s ; \mathbf{r} 0) &=& \int
    \ud{\mathbf{r}'} \langle \mathbf{r}' \vert \psi(s) \rangle &=& \quad 1, \\
    &\int \ud{\mathbf{r}} \ud{\mathbf{r}'} q(\mathbf{r}'s ; \mathbf{r} 0) &=&
    \int \ud{\mathbf{r}} \ud{\mathbf{r}'} \langle \mathbf{r}' \vert \psi(s)
    \rangle &=& \quad V.
  \end{aligned}
\end{align}

Eventually, the structure function of the chain has the expression:
\begin{equation}
  S_N(\mathbf{r}'-\mathbf{r}) = \dfrac{1}{N} \int \limits_0^N \ud{s} \ud{s'} q(\mathbf{r}'s' ; \mathbf{r} s).
  \label{eq:app_structfunc}
\end{equation}

Let us emphasize that this result is valid only because $S_N(\mathbf{r}'-\mathbf{r})$ can be expressed in \cref{eq:app_strucfunc_expression} as a matrix product of the three subchain propagators $q(\mathbf{r}_N N ; \mathbf{r}' s')$, $q(\mathbf{r}'s' ; \mathbf{r} s)$ and $q(\mathbf{r}s ; \mathbf{r}_0 0)$. More generally, the chain propagator must obeys a Chapman-Kolmogorov equation. For a Gaussian chain it is \cite{deGennes1979,Fredrickson2005}:
\begin{equation}
  \dfrac{\partial q}{\partial s}(\mathbf{r}s ; \mathbf{r}' s') = \dfrac{a^2}{6} \Delta_{\mathbf{r}} q (\mathbf{r}s ; \mathbf{r}' s') - V(\mathbf{r}) q (\mathbf{r}s ; \mathbf{r}' s'),
\end{equation}
where $\Delta_{\mathbf{r}}$ is the Laplacian operator and $V(\mathbf{r})$ is an external field. It should be noted that the \rhs is diagonal in Fourier basis. In the case of the free Gaussian chain, $V(\mathbf{r})=0$, one obtains for the chain propagator:
\begin{align}
  \begin{aligned}
    q(\mathbf{k}, s'-s) 	&= \langle \mathbf{k} \vert e^{-\frac{a^2 \vert s'-s \vert \hat{\mathbf{k}}^2}{6} } \vert \mathbf{k} \rangle \\
    &= \exp{\left( -\dfrac{a^2 k^2 \vert s'-s \vert}{6} \right)}. \\
  \end{aligned}
\end{align}

Plugging this result back into \cref{eq:app_structfunc} gives the structure function of the Gaussian chain:
\begin{equation}
  S_N (\mathbf{k}) = N D(k^2 R_g^2),
\end{equation}
where $R_g^2=a^2 N / 6$ is the radius of gyration of the chain and $D(x) = 2 / x^2 (x + \exp{(-x)} -1 )$ is the Debye function.

\section{Detection of the coil-globule-crystal transitions}
\label[app]{app:coil_globule_crystal_transitions}
In order to detect the coil-globule transition, we monitored the quantity:
\begin{equation}
  q=\dfrac{\log R_g}{\log N},
\end{equation}
where $R_g$ is the radius of gyration of the polymer. For a self-avoiding polymer with scaling law $R_g \sim b N^{\nu}$, $q = \nu + cst / \log N$. In a good solvent, the polymer is swollen with $\nu=0.588$ whereas in a bad solvent it collapses with $\nu=1/3$ . Thus $q$ varies like $\nu$.

In order to detect the coil-crystalline transition, and following \cite{Grassberger2008}, we first defined the quantity
\begin{equation}
 n_\alpha = \sum \mid \mathbf{u}_i \cdot \mathbf{e}_\alpha \mid,
\end{equation}
in which $i$ runs over all the bonds of the polymer, $\mathbf{u}_i$ is the unit vector having the same direction as the bond $i$ and $\mathbf{e}_\alpha$ is the unit vector of the corresponding $\alpha$-axis ($\alpha=x,y,z$). We chose to monitor the quantity:
\begin{equation}
  p=1 - \dfrac{n_{min}}{n_{max}},
\end{equation}
where $n_{min}=\min_{\alpha} (n_\alpha)$ and $n_{max}=\max_{\alpha} (n_\alpha)$. It is clear that for an isotropic configuration, $n_x=n_y=n_z$ resulting in $p=0$. Conversely, for a configuration stretched in one direction, say along the x-axis, $n_x=1$ and $n_y=n_z=0$, resulting in $p=1$. Thus $p$ effectively measures the directional order of the polymer.

\begin{figure}[!htbp]
  \centering
  \includegraphics[width = 1 \textwidth]{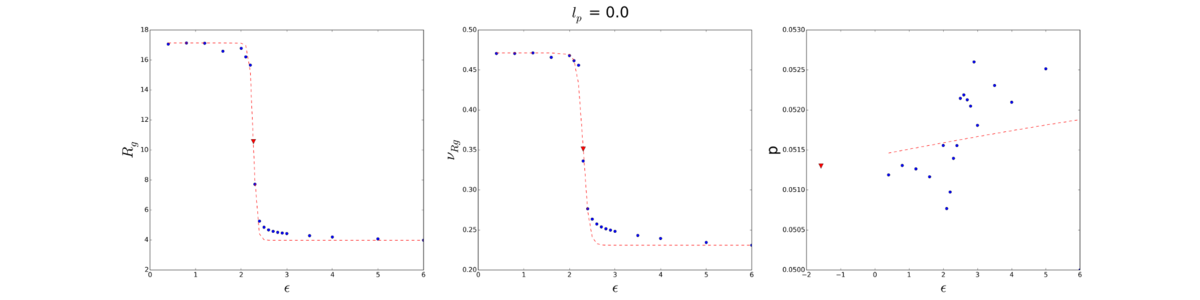}
  \includegraphics[width = 1 \textwidth]{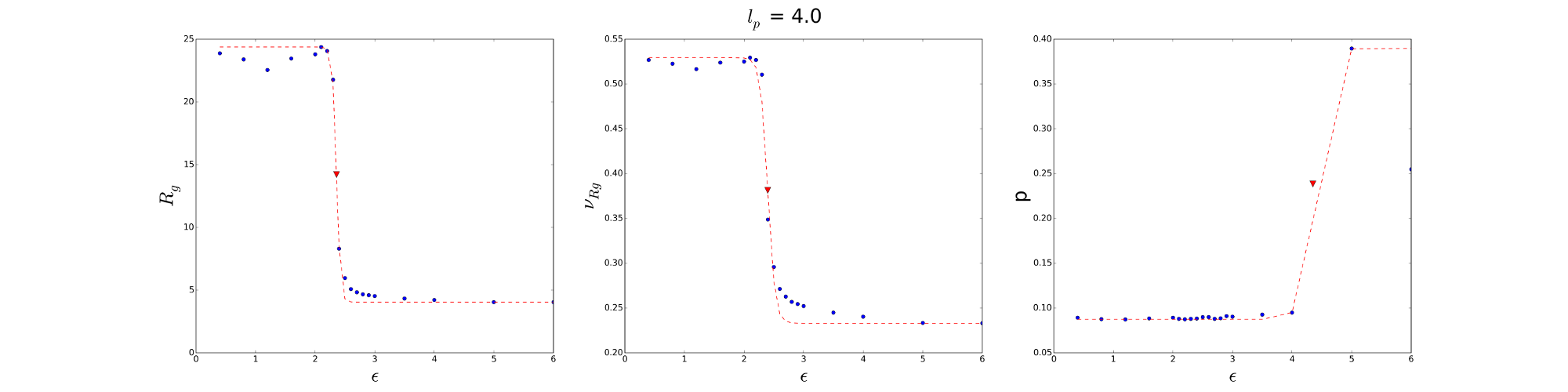}
  \includegraphics[width = 1 \textwidth]{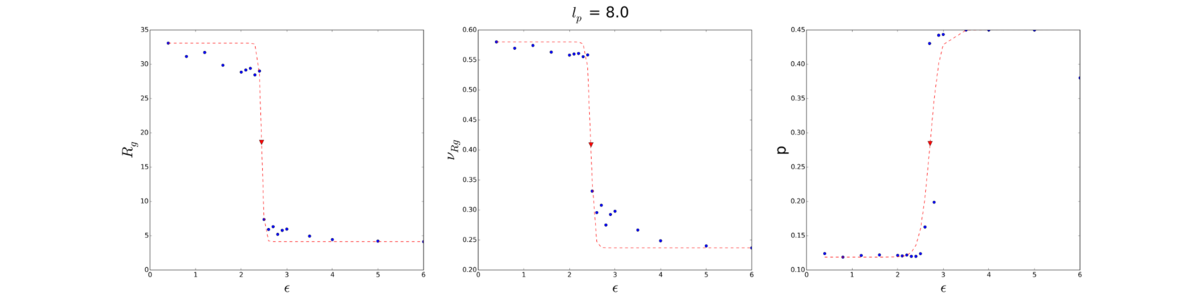}
  \includegraphics[width = 1 \textwidth]{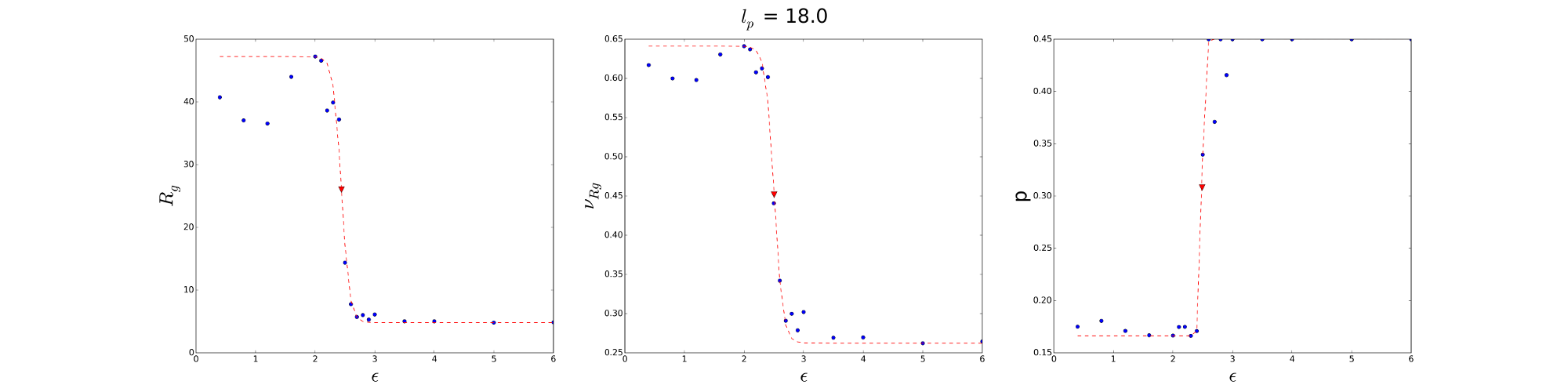}
  \caption{Collapse of a DNA polymer interacting with binding proteins through a Lennard-Jones potential with strength $\varepsilon$, for different persistence length $l_p$. For each $\varepsilon$ the average values $\langle R_g \rangle$, $\langle q \rangle \simeq \nu_{R_g}$ and $\langle p \rangle$ are computed from BD simulations. A fit with a sigmoid function gives access to the coil-globule transition (for $R_g$ and $q$) and the globule-crystal or coil-crystal transition (for $p$).}
  \label{fig:dense_phase_diagram_transition_curves}
\end{figure}

We then performed a thermodynamical average over uncorrelated configurations sampled from BD trajectories to obtain $\langle q \rangle$ and $\langle p \rangle$. We carried out this procedure for different values of the DNA-protein interaction, represented by the strength $\varepsilon$ of the corresponding Lennard-Jones interaction and plotted the values of $\langle q \rangle$ and $\langle p \rangle$ as a function of $\varepsilon$ (\cref{fig:dense_phase_diagram_transition_curves}).

In order to to identify the transition point, we performed a fit with a sigmoid function:
\begin{equation}
  s(x)=\frac{1}{1+e^{-\lambda x}}.
\end{equation}

The inflexion point, $x^*=0$, was taken to be the transition point. Carrying out this procedure for different values of $l_p$ gave the phase diagram from \cref{fig:phase_diagrams_dense_phase:b}.

\end{subappendices}



\chapter{A side-study: computation of the structure function of a worm-like chain}
\label{ch:structure_function}

\chaptermark{Structure function of a worm-like chain}
In \cref{ch:transcription_factories}, we have seen that the structure function of a polymer is a quantity that arises in polymer field theories, and in particular in the Random Phase Approximation (RPA). Although this object has an analytical closed-form for a Gaussian chain, it is not the case for a worm-like chain (WLC). However the chromosome is a rigid biopolymer better described by the latter model. Therefore, the RPA analysis in \cref{ch:transcription_factories} has motivated additional work to compute the structure function of polymer chains with bending rigidity.

In the present chapter, we consider a discrete worm-like chain polymer model. We first introduce the pair correlation function, which is the central quantity required to compute the structure function. We will show that the pair correlation function can be expressed exactly as a power of a transfer matrix with complex entries. We then apply our result to the computation of the structure function for a worm-like chain and compare it to the values obtained with Monte-Carlo simulations, as well as with other existing methods of the literature.

\section{Relevance of the structure function in polymer field theories}
In polymer physics, the structure function is a central quantity which characterizes the density fluctuations at thermal equilibrium. More accurately, it is related to the two points correlation function for the polymer density:
\begin{equation}
  c S_N(\vec{r}) = \langle \rho\left( \vec{r}+\vec{r}') \rho(\vec{r}' \right) \rangle,
  \label{eq:structure_function_continuum}
\end{equation}
where $N$ is the length of the polymer, $\rho(\vec{r})$ is the concentration of polymer and $c=N/V$ is the mean-field concentration in a cavity with volume $V$.

The structure function has key applications in self-consistent field theories of polymer mixtures, including the RPA. Briefly, the RPA looks for Fourier modes $\vec{k}$ driving critical fluctuations (see for instance \cite{Fredrickson2005,deGennes1979}). The specific shape of the structure function can induce instabilities for non zero wave number $\vec{k}$, which is in general the signature of a microphase separation.

The analytical expression of the structure function for a Gaussian polymer is known to be the Debye function \cite{deGennes1979,Edwards1988,DesCloizeaux1990}. Yet, many polymers cannot be considered as such. The example of biopolymers like DNA, actin and microtubules is a case in point. It is then necessary to take into account a persistence length $l_p$ which characterizes the distance over which a polymer looses the memory of its orientation. This is the realm of semi-flexible polymers. In a simplified picture, such a polymer can be discretized as a sequence of monomers with inextensible bonds, free to rotate from one to the next. A common way to deal with them is to use the so-called Kratky-Porod model \cite{Kratky11061949}, or worm-like chain (WLC), which introduces a penalty proportional to a bending modulus $\kappa$ when two consecutive bonds are not aligned.

Despite the longing interest in computing the structure function for semi-flexible polymers, there is no exact analytical closed-form available. Nonetheless, several solutions have been proposed. Some of them rely on analytical approximates \cite{Kholodenko1993,Bhattacharjee1997,Yamakawa1980,DesCloizeaux1973}, while others provide numerical methods to compute the desired quantity \cite{Pedersen1996, Spakowitz2004, Zhang2014}. After presenting our method, we will discuss some of them in the sequel.

\section{Expression of the pair correlation function}
\subsection{Pair correlation and structure function}
Let us start from the discrete WLC presented in \cref{sec:model_chromosome_description}, with $N+1$ monomers. The density of monomers at position $\vec{r}$ is given by:
\begin{equation}
  \rho(\vec{r}) = \sum \limits_{n=0}^{N} \delta\left( \vec{r} - \vec{r}_n \right),
\end{equation}
where as usual, $\vec{r}_n$ is the vector of spatial coordinates for monomer $n$. If we substitute this expression in \cref{eq:structure_function_continuum}, and integrate the translational degree of freedom, then we obtain:
\begin{equation}
  S_N(\vec{r}) = \frac{1}{N} \sum \limits_{m \ne n}^{N} \langle \delta\left( \vec{r}-(\vec{r}_n - \vec{r}_m) \right) \rangle,
  \label{eq:structure_function_discrete}
\end{equation}
in which the terms such that $n=m$ have been removed when going from a continuous to a discrete chain. Let us now introduce the pair correlation function:
\begin{equation}
  g_N(\vec{r})=\left\langle \delta \left( \vec{r} -(\vec{r}_N - \vec{r}_0)\right) \right\rangle,
 \label{eq:pair_correlation}
\end{equation}
with Fourier transform:
\begin{align}
  \begin{aligned}
    g_N(\vec{k})	&= \left\langle \exp{\left( i \vec{k} \cdot (\vec{r}_N -\vec{r}_0) \right)} \right\rangle \\
    &=\dfrac{1}{Q_N} \int \prod \limits_{j=1}^{N} \ud{^2 \vec{u}_j}  \exp{\left[ - \kappa \sum \limits_{j=1}^{N-1} ( 1 - \vec{u}_{j}.\vec{u}_{j+1} ) +  i \vec{k} . \sum \limits_{j=1}^N \vec{u}_j  \right]}.
  \end{aligned}
  \label{eq:pair_correlation_ft}
\end{align}

From \cref{eq:structure_function_discrete}, we see that the Fourier transform of the structure function, $S_N(k)$, can be expressed as:
\begin{equation}
S_N(k)= \dfrac{1}{N} \sum \limits_{n \ne m}^N g_{\mid n-m \mid }(k),
 \label{eq:structure_function}
\end{equation}

Note that at $k=0$, we retrieve $S_N(0)=N+1$ which is the number of scattering units, \textit{i.e.} monomers of the chain. It is clear that the central quantity to compute is the pair correlation function, but the integral form from \cref{eq:pair_correlation_ft} is out of scope for practical use when $N$ grows to large values.

\subsection{Expression in terms of transfer matrices}
\label{sec:transfer_matrix_expression}
We extend the transfer matrix defined in \vref{eq:wlc_discrete_transfer_matrix} to Fourier modes, $\vec{k} \neq 0$:
\begin{align}
T(\vec{u} \mid \vec{u}') = \exp{\left( - \kappa (1 - \vec{u} \cdot \vec{u}') + i \vec{k} \cdot \dfrac{\vec{u}+\vec{u}'}{2} \right)}
 \label{eq:wlc_discrete_transfer_matrix_t}
\end{align}
and rewrite \cref{eq:pair_correlation_ft} as
\begin{align}
g_N(\vec{k}) = \dfrac{1}{Q_N} \int { \left[ \prod \limits_{j=1}^{N} \ud{^2 \vec{u}_j}  \,  T(\vec{u}_j \mid \vec{u}_{j-1}) \right] \exp{\left(i \vec{k} \cdot \dfrac{\vec{u}_1 + \vec{u}_N}{2} \right)}}
\label{eq:pair_correlation_ft_prod_1}
\end{align}

For $k=0$, $T$ is real and symmetric, yet for $k\neq0$, the transfer matrix is still symmetric but with complex matrix elements. Note that this kind of transfer matrix had been introduced earlier in the literature, but with $i \vec{k} = f$ real, in order to compute the relative extension of a WLC polymer when a force is applied at both ends \cite{Marko87591995, Bensimon1998}. In what follows, we will keep the same notation for $T$ but it should be kept in mind that it implicitly depends on the wave number $k$.

The integration in \cref{eq:wlc_discrete_transfer_matrix_t} is carried out over the angular variables $\varphi$ and $\theta$ in the spherical coordinates system attached to the z-axis (see \vref{eq:spherical_coordinate_system_zaxis}). However, we can formally integrate out the $\varphi$ variables. Hence we obtain a reduced transfer matrix $\hat{T}$, with matrix elements:
\begin{align}
  \hat{T}(\theta \mid \theta') = I_0(\kappa \sin\theta \sin\theta') \exp{\left( -\kappa (1 - \cos\theta \cos\theta') + ik \dfrac{\cos\theta + \cos\theta'}{2} \right)},
\end{align}
where
\begin{equation}
  I_0(z)=\int \dfrac{\mathrm{d}\varphi}{2 \pi} \, \exp{\left( z \cos \varphi \right)}
\end{equation}
is the modified Bessel function of rank $0$. The evaluation of $I_0(z)$ is performed numerically, for which several routines are available \cite{NumericalRecipes2007}. Besides, one can still use polynomial approximations to save up computational time. Using the reduced transfer matrix, the pair correlation function from \cref{eq:pair_correlation_ft_prod_1} now reads:
\begin{equation}
  g_N(\vec{k}) = \dfrac{1}{Q_N} \int \limits_0^\pi \ud{\theta} \sin{\theta} \, \Phi(\theta) \hat{T}^{N-1} \Phi(\theta),
  \label{eq:pair_correlation_final}
\end{equation}
where we defined $\Phi(\theta)=\exp{(i k \cos{\theta}/2)}$. Note that our notation for matrix-vector product implies that the correct measure, $\mathrm{d}\theta \sin{\theta}$, is taken in the integration. In particular, the product of $\hat{T}$ with the function $\Phi$ reads:
\begin{equation}
  \hat{T} \Phi(\theta) = \int \limits_0^\pi \ud{\theta'} \sin{\theta'} \, \hat{T}(\theta \mid \theta') \Phi(\theta').
\end{equation}

In conclusion, the pair correlation function has been expressed as an integral over one single angular variable $\theta$. Besides, it only depends on the norm $k$ of the Fourier mode considered. As can be expected at that point, a calculation scheme of $g_N(k)$ will consist in expanding $\Phi$ on the basis of eigenfunctions of $\hat{T}$. However it is worth pointing out that $\hat{T}$ is not hermitian in general, except when $k=0$ for which it is real and symmetric. Therefore in the general case, the eigenfunctions $Y_n(\theta)$ of $\hat{T}$ are not orthogonal, \textit{i.e.} $ \int \ud{\theta} \sin{\theta} Y_n(\theta) Y_{n'} \, (\theta) \neq \delta_{n,n'}$. We stress that the final form in \cref{eq:pair_correlation_final} is exact and can be formally rewritten as:
\begin{equation}
  g_N(\vec{k})= \frac{\left\langle \Phi \mid \hat{T}^{N-1} \mid \Phi \right\rangle_{k\phantom{=0}}}{\left\langle \Phi \mid \hat{T}^{N-1} \mid \Phi \right\rangle_{k=0}}.
  \label{eq:pair_correlation_final_brackets}
\end{equation}

\section{Application to the computation of the structure function}
In this section, we introduce our method to compute the pair correlation function $g_N(k)$ and apply it to compute the structure function of a discrete WLC with $N+1=200$ monomers. Then we compare our results to existing methods.

For computational efficiency, we have used the following expression for the structure function:
\begin{equation}
  S_N(k) = \dfrac{2}{N} \sum \limits_{n=1}^N (N-n +1) g_n(k),
\end{equation}
which is equivalent to \cref{eq:structure_function}, but reduces the computational complexity from $O(N^2)$ to $O(N)$.

\subsection{Transfer matrix method}
As announced, we devised a numerical procedure based on \cref{eq:pair_correlation_final_brackets} to compute $g_N(k)$. For convenience, we have chosen to make the change of variable $\Upsilon=-\cos \theta$, with the uniform integration measure on $\left[ -1,1 \right]$. We also chose a discretization of size $M$, through the regular subdivision:
\begin{equation}
  \Upsilon_m = -1 + m \frac{2}{M} \text{ with } m=0,\dots,M.
  \label{eq:regular_subdivision}
\end{equation}

Using this discretization, the matrix elements of the reduced transfer matrix $\hat{T}$ read:
\begin{equation}
  t_{mn}= \exp{\left( -\kappa \left(1 - \Upsilon_m \Upsilon_{n} \right) -i k \dfrac{\Upsilon_m+\Upsilon_{n}}{2}\right)} I_0\left(\kappa \sqrt{1-\Upsilon_m^2}\sqrt{1-\Upsilon_{n}^2}\right),
\end{equation}
where $I_0(z)$ is the modified Bessel function of rank $0$. Similarly, the discrete version of the function $\Phi$ is a vector with coordinates:
\begin{equation}
\phi_m=\exp\left(-ik \dfrac{\Upsilon_m}{2}\right).
\end{equation}

For each value of the wave number $k$, the discrete matrix $\hat{T}$ can be diagonalized, namely $t_{mn}=\sum_j p_{mj} \lambda_j p^{-1}_{jn}$, where $\lambda_0 > \lambda_1 > \dots \lambda_{M-1}$ are the eigenvalues and $p_{ij}$ is the matrix of coordinates for the diagonal basis. Once again, let us emphasize that both the eigenvalues $\lambda_i$ and the matrix $p_{ij}$ depend on the wave number $k$. The pair correlation in \cref{eq:pair_correlation_final_brackets} is then finally expressed as the ratio of two sums:
\begin{equation}
  g_N(k)=\frac{\left[ \phi_m p_{mj} \lambda_j^{N-1} p^{-1}_{jn} \phi_{n} \right](k)}{\left[ \phi_m p_{mj} \lambda_j^{N-1} p^{-1}_{jn} \phi_{n} \right](0)},
\end{equation}
where the summation on indices $j,m,n$ is implied.

\begin{figure}[!htbp]
  \centering
\includegraphics[width=0.5 \textwidth]{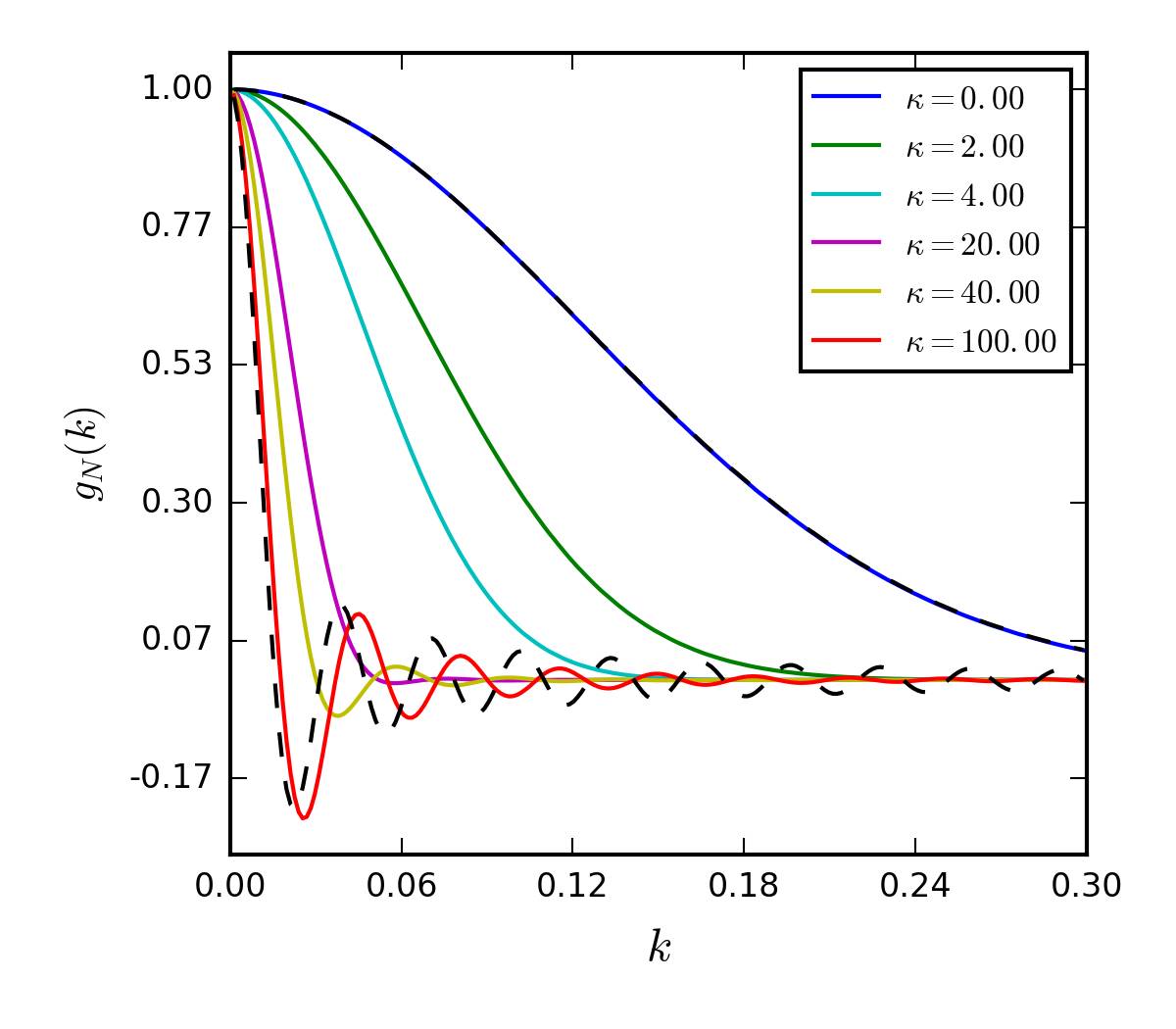}
\caption{Computation of the pair correlation function for different values $\kappa$. The dotted lines are the Gaussian, $\exp{\left( -k^2 N / 6  \right)}$, and the rod, $\sin{\left( k N \right)}/k N$ pair correlation functions. We considered a chain of length $N=200$ and used a discretization with $M=1000$.}
\label{fig:comparison_pair_functions}
\end{figure}

This procedure provides a systematic way to compute the pair correlation function of a discrete worm-like chain for any value of the rigidity parameter $\kappa$ (\cref{fig:comparison_pair_functions}). The pair correlation function obtained correctly interpolates between the Gaussian chain, for which $g_N(k) = \exp{\left( -k^2 R_g^2 \right)}$, and the rod-like chain, for which $g_N(k)=\sin{(k N)} / (k N)$.

\begin{figure}[!htbp]
  \centering
  \subfloat[]{\label{fig:effect_discretization:standard}\includegraphics[width=0.48 \textwidth]{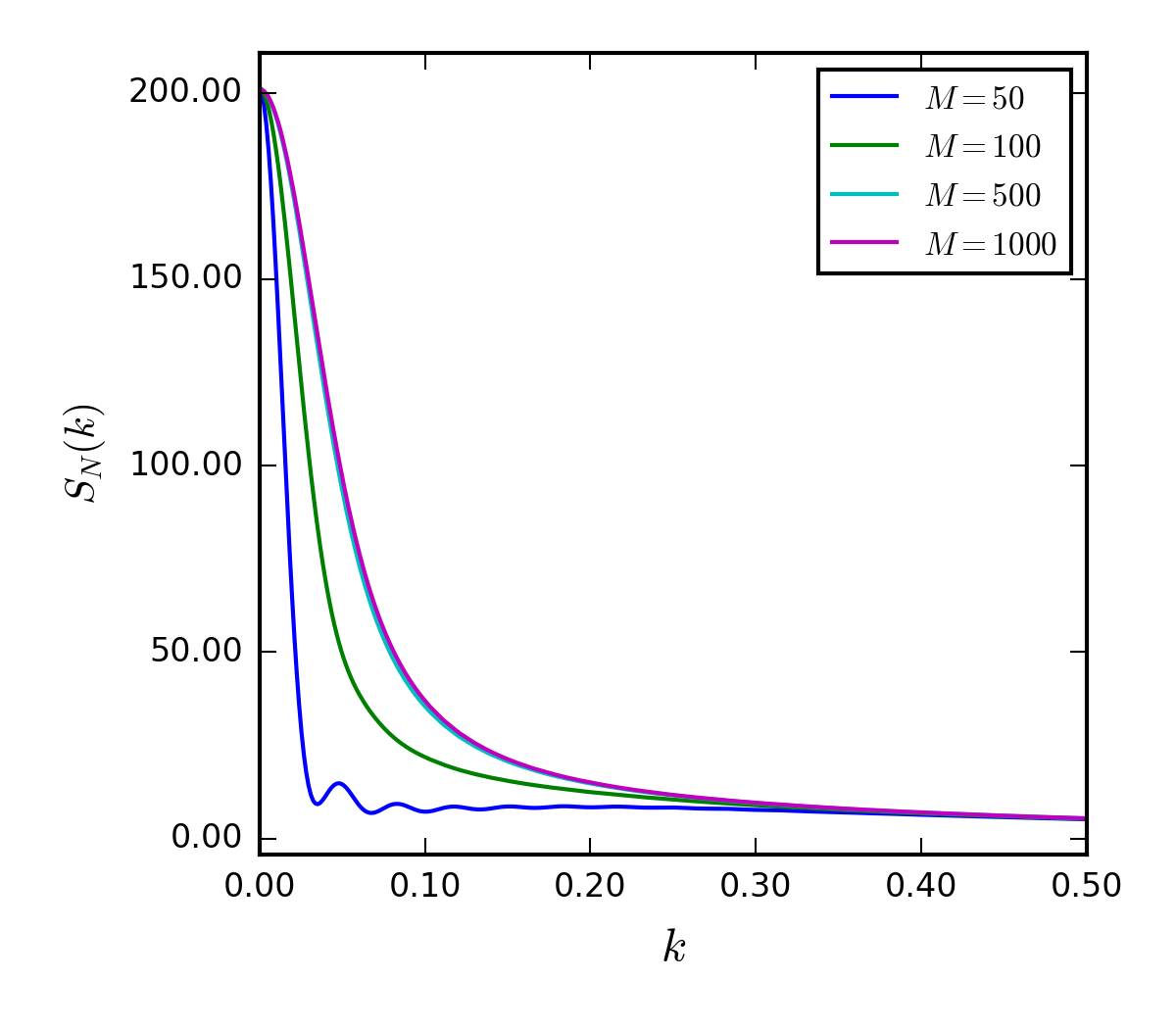}}%
  \quad
  \subfloat[]{\label{fig:effect_discretization:kratky}\includegraphics[width=0.48 \textwidth]{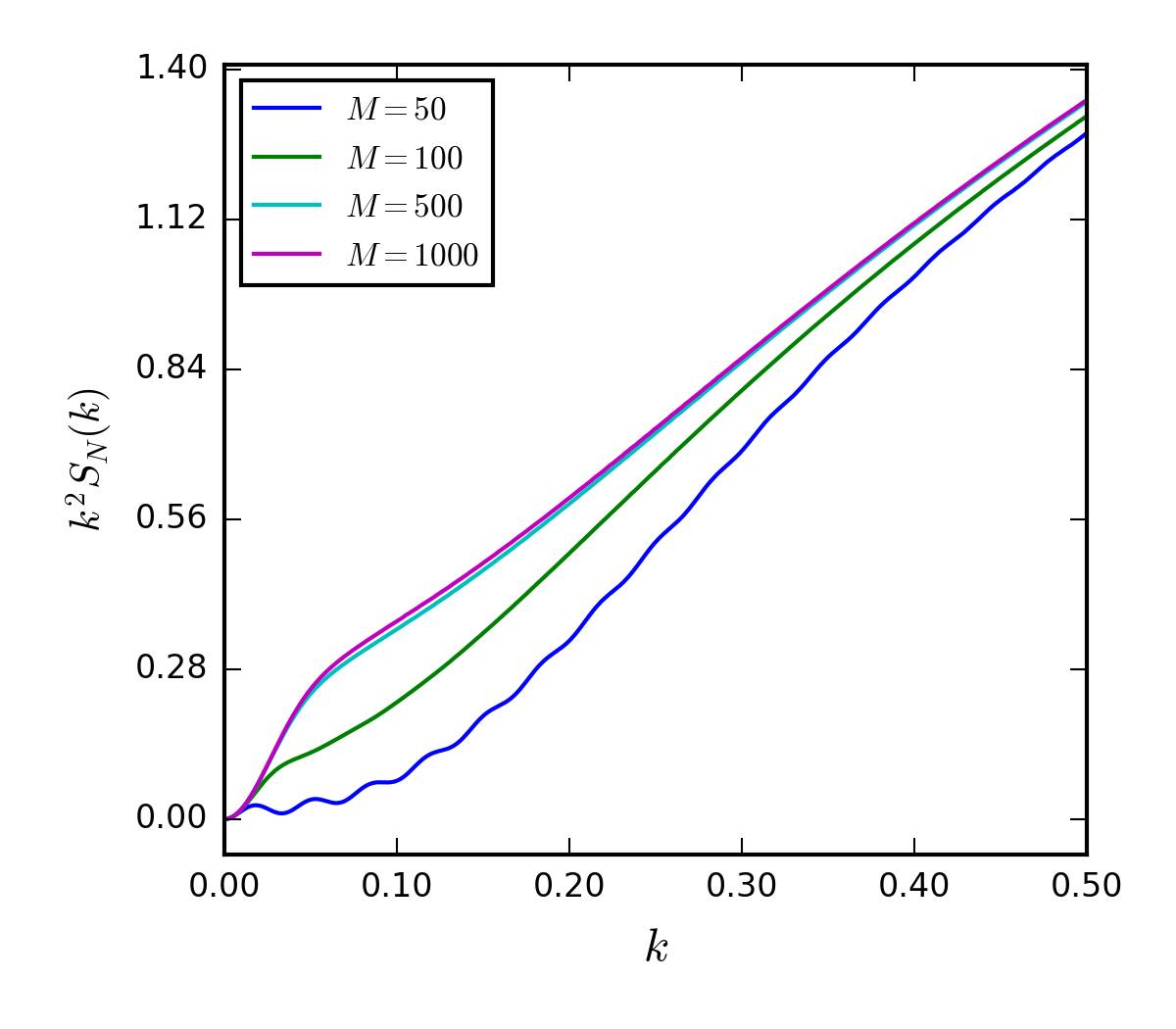}}
  \caption{Transfer matrix computation of the structure function for $M=50,100,500$ and $1000$. We considered a chain of length $N=200$ with a bending rigidity $\kappa=20$. \protect\subref{fig:effect_discretization:standard} $S_N(k)$ as a function of $k$. \protect\subref{fig:effect_discretization:kratky} $k^2 S_N(k)$ as a function of $k$ (Kratky plot).}
\label{fig:effect_discretization}
\end{figure}

In practice, the numerical complexity lies in the fact that for any wave number $k$, we need to diagonalize a $M \times M$ complex square matrix. As the rigidity of the chain increases (large $\kappa$), a discretization with a larger $M$ is required, resulting in an increased complexity. For $\kappa \gg 1$, $T(\theta \mid \theta')$ is sharply peaked around the straight bond configuration $\theta=\theta'$, with a maximum at $\theta=\theta'=0 \, (\pi)$. As $T(\theta \mid \theta')$ becomes localized near $\theta=\theta'$, replacing the integral in \cref{eq:pair_correlation_final} by a Riemann sum with a regular subdivision such as in \cref{eq:regular_subdivision} is not adapted, and $M$ needs to be increased. Therefore, for very rigid chain, the transfer matrix method is not \textit{ad hoc} since the complexity for diagonalizing a matrix of size $M$ grows like $O(M^3)$. As can be expected, the accuracy reached for the computation of the structure function $S_N(k)$ will depend on the value of $M$ (\cref{fig:effect_discretization}). In other words, for fixed $M$, the quality of our prediction falls off as the rigidity of the chain increases, especially in the small $k$ regime. On the basis of this analysis, we used a discretization $M=\num{1000}$ in further applications.

\subsection{Comparison with other methods}
\begin{figure}[!htbp]
  \centering
  \subfloat[]{\label{fig:comparison_methods:standard}\includegraphics[width=0.48 \textwidth]{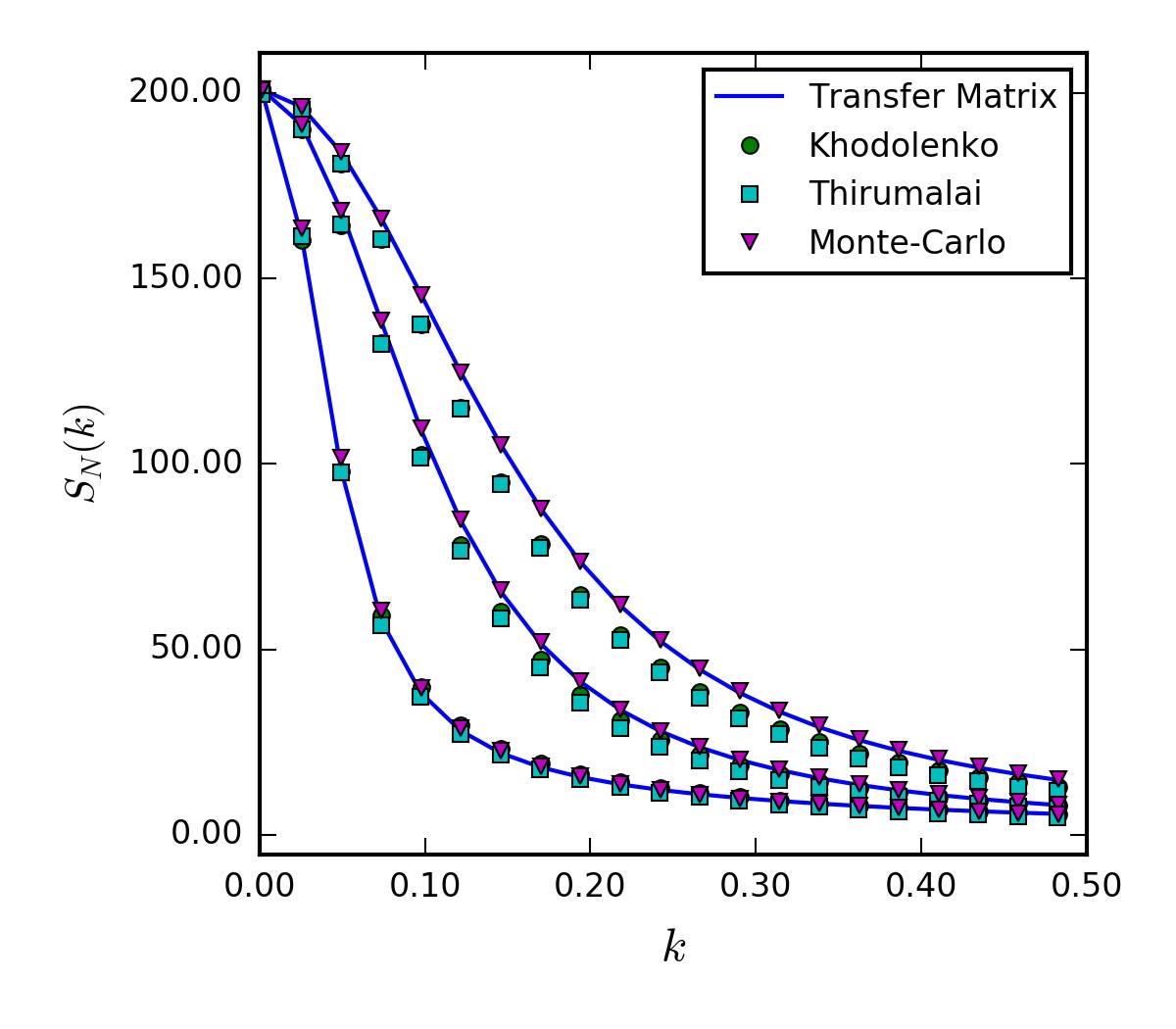}}%
  \quad
  \subfloat[]{\label{fig:comparison_methods:kratky}\includegraphics[width=0.48 \textwidth]{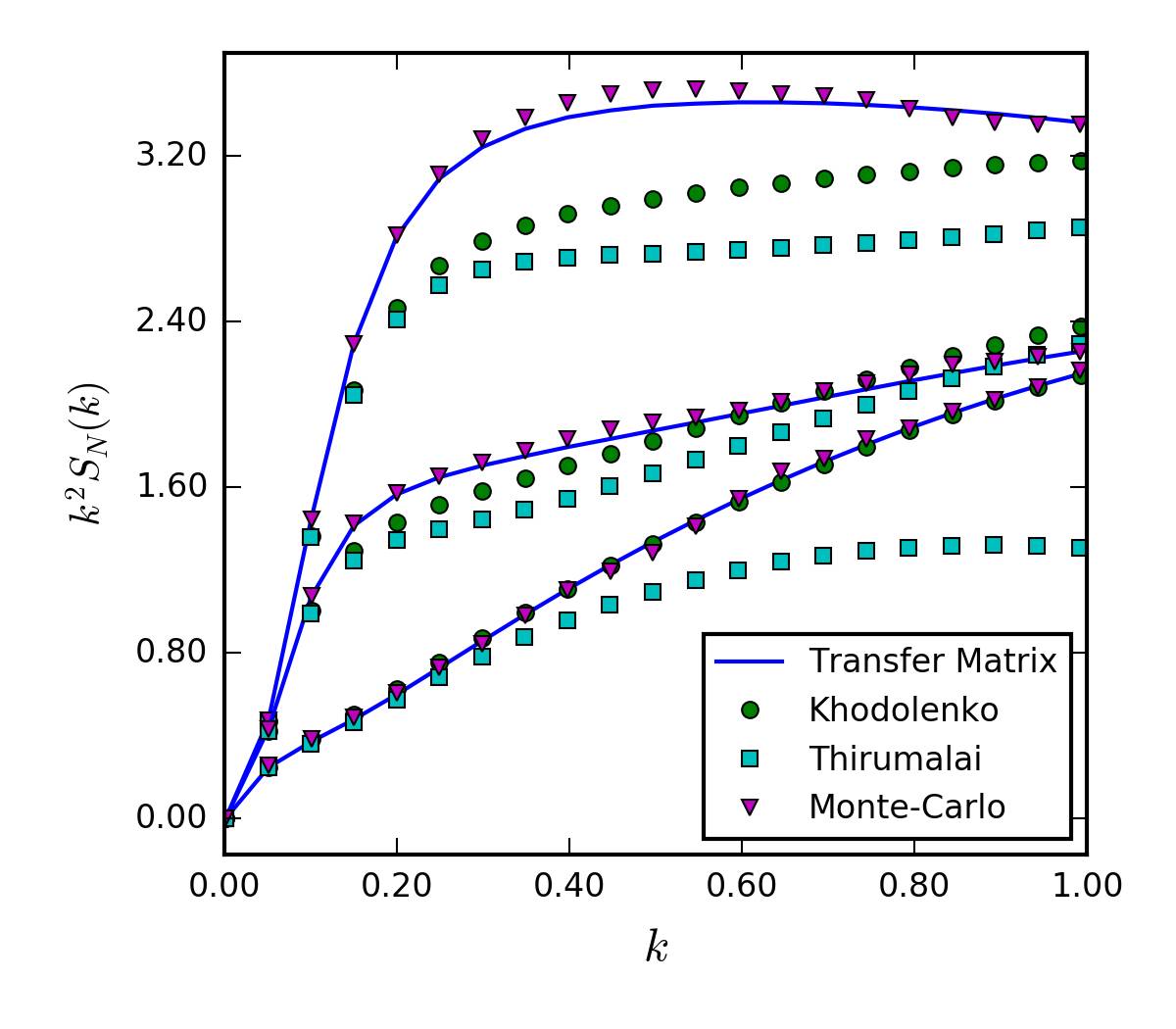}}%

\caption{Comparison of the structure function obtained with different methods, for $\kappa=2.00, 4.00,20.00$. We considered a chain of length $N=200$. \protect\subref{fig:comparison_methods:standard} $S_N(k)$ as a function of $k$. \protect\subref{fig:comparison_methods:kratky} $k^2 S_N(k)$ as a function of $k$ (Kratky plot).}
\label{fig:comparison_methods}
\end{figure}

As a reference method, we computed the structure function of a WLC using Monte-Carlo simulations. We used a standard Metropolis-Hasting Monte-Carlo algorithm to sample configurations of a discrete WLC in the Boltzmann ensemble. A configuration was defined by the $N+1$ coordinates of the monomers $\lbrace \vec{r}_i \rbrace$. At each iteration, a new configuration $\lbrace \vec{r}_i' \rbrace$ was generated from the previous one using pivot and crankshaft moves. The probability to accept the new configuration was taken as usual to be
\begin{equation}
  \proba{\lbrace \vec{r}_i \rbrace \rightarrow \lbrace \vec{r}_i' \rbrace} = \min{\left(1, \exp{\left[ - \beta U_b(\lbrace \vec{r}_i' \rbrace)-\beta U_b(\lbrace \vec{r}_i \rbrace)\right]} \right)}
\end{equation}
where $\beta U_b$ is defined in \vref{eq:wlc_discrete}. As is well known, the stationary distribution resulting from this Markov process samples the Boltzmann equilibrium. After an initial run intended to reach the Boltzmann equilibrium, we sampled \num{10000} configurations every \num{500} iterations. For $N+1=200$, the autocorrelation time appeared to be smaller than the time between two such configurations. Using the ergodicity property of Markov processes, we computed thermal averages by taking an average over the sampled configurations. In particular, the pair correlation function was computed using
\begin{align}
g_N(k) = \left\langle \cos{\left( \vec{k} \cdot (\vec{r}_N - \vec{r}_0) \right)} \right\rangle
\end{align}
which is equivalent to \cref{eq:pair_correlation_ft} because $g_N(k)$ is real. As can be seen in \cref{fig:comparison_methods}, the structure function obtained with the transfer matrix method and with Monte Carlo simulations are in good agreement.

Other methods to compute the structure function of a WLC can be found in the literature. Two good analytical expressions are available. Khodolenko \cite{Kholodenko1993} used an ansatz for the structure factor of a WLC based on a Dirac propagator equation. By design, this model smoothly interpolates between the Gaussian and rod-like chain limits. Although the formula proposed by Kholodenko seems like a good approximate for the pair correlation function, it is not the actual solution for the WLC model. Furthermore, both the physical interpretation of the parameters and the accuracy for intermediate stiffness are not clear. Bhattacharjee and co-workers \cite{Bhattacharjee1997} enforced the constraint on the bond length, $\mid \vec{u}_i^2 - 1 \mid$, through a mean-field approximation and relaxed the integration on the unit sphere to the full volume (see \vref{sec:polymer_models_bending_rigidity_gaussian}), and obtained an analytical expression for the pair correlation function. Yet, its accuracy might be called into question for moderate stiffness, due to uncertain contributions of local chain length fluctuations which are not taken into account at the mean-field level. Other methods have been proposed, giving a numerical approximate of the pair correlation function of a WLC. Spakowitz \cite{Spakowitz2004} and co-workers computed $g_N(k)$ as an infinite continued fraction, which must be truncated  for numerical evaluation. Although the numerical implementation of the continued fraction seems straightforward, further treatments are required to obtain the structure function, namely an inverse Laplace transform. Zhang and co-workers \cite{Zhang2014} used the Chapman-Kolmogorov equation satisfied by the Green function $G(\textbf{k}, \textbf{u} ; s)$ of a WLC (see \cref{app:wlc_continuous}). Physically, $G(\textbf{k}, \textbf{u} ; s)$ is the spatial Fourier transform of $G(\textbf{r}, \textbf{u} ; s)$, which is the joint probability distribution that a chain starting at the origin ends up at position $\textbf{r}$ with orientation $\textbf{u}$. The associated numerical procedure makes use first of an expansion of $G(\textbf{k}, \textbf{u} ; s)$ in terms of the spherical harmonics functions. This method shares similarities with our own approach.

We chose to compare our results (at $M=1000$) with the analytical  forms of Khodolenko and Bhattacharjee (see \cref{app:methods}), using the Monte-Carlo result as a reference. We observe that the transfer matrix method performs better than the analytical forms of Kholodenko and Bhattacharjee for moderate stiffness (\cref{fig:comparison_methods}). Conversely, for strong stiffness, the transfer matrix method would require a higher discretization $M$, and therefore it performs less well than the analytical expressions. Note that this could have been expected since both Kholodenko and Bhattacharjee forms are derived from approximations whose validity improves for stiff chains.

\section{Discussion}
In conclusion, we have presented a method to compute the structure function of a WLC, in Fourier space. The method relies on the eigenvalue decomposition of a transfer matrix with complex entries, which is performed for each value of the wave number $k$. Specifically, the pair correlation function, which is expressed as a power of the transfer matrix $T$, can be straightforwardly computed with this method.

Our method appeared to be in good agreement with a computation of the structure function obtained from Monte-Carlo simulations of a WLC. In addition, we have compared it with two of the existing analytical approximations that can be found in the literature. We stress that the structure function computed with these methods is not the WLC structure function, because the authors model the polymer bending rigidity using models which are approximation of the WLC model. This might have lead to discrepancies when comparing with our method because the bending rigidity parameter in these models is not exactly the WLC persistence length, although we followed the interpretation given by the authors of these studies. We found that our transfer matrix method performs better for moderate stiffness of the WLC. To the contrary, for large persistence length our method performs less well. This is due to a too sparse discretization of the transfer matrix $T$, resulting in a discrepancy between discrete sums and continuous integrals. A practical way to circumvent this issue would be to consider a finer discretization of the transfer matrix $T$ (larger $M$). This is of course possible, but it should be kept in mind, that the time required to compute the structure function at wave number $k$, $S_N(k)$, scales as $N$ times the time required to diagonalize $T$, which is in $O(M^{3})$.

\begin{subappendices}
\section{Other methods to compute the pair correlation function of a worm-like chain}
\label[app]{app:methods}
\paragraph{Kholodenko's method\newline}
In eq. (11) from \cite{Kholodenko1993}, the pair correlation function is computed from the expression:
\begin{align}
g_N(\vec{k}) =
\begin{cases}
  \dfrac{1}{\sqrt{1 - (k/m)^2}} \dfrac{\sinh{\left(\sqrt{1 - (k/m)^2} m N\right)}}{\sinh(m N)} & \text{ if } k < m \\
  \dfrac{1}{\sqrt{(k/m)^2-1}} \dfrac{\sin{\left(\sqrt{(k/m)^2-1} m N\right)}}{\sinh(m N)} & \text{ if } k > m
  \end{cases}
\end{align}
where $m=3/(2 l_p)$. This ansatz is obtained from the analogy of the Hamiltonians between Dirac's fermions and semi-flexible polymers.

\paragraph{Bhattacharjee's method\newline}
In eq. (15) from \cite{Bhattacharjee1997}, the pair correlation function is computed from the expression:
\begin{align}
g_N(\vec{r}) = \mathcal{N} \left[1 - (r/N)^2 \right]^{-9/2} \exp{\left( - \dfrac{3 N}{4 l_p} \dfrac{1}{1 - (r/N)^2} \right)}
\end{align}
where $\mathcal{N}$ is a normalization constant.

\end{subappendices}



\chapter{Model for the role of nucleoid-associated proteins in regulating transcription}
\chaptermark{NAPs and transcription regulation}
\label{ch:naps}

In this chapter, we propose a model providing a direct connection between regulation of the transcription and chromosome architecture.

In bacteria, an example of structuring proteins is the so-called family of nucleoid-associated proteins (NAPs). Hence we will build our analysis on what is known today about these proteins. We first start by a review of the literature on the four main NAPs in \textit{Escherichia coli} bacteria which are: H-NS, FIS, HU and IHF. We will describe the architectural changes induced on the chromosome by these proteins, and what is known of their consequences on genetic expression. In particular, H-NS leads to the formation of DNA filaments and hairpin loops which prevent RNA polymerase binding. Several studies have conjectured that small H-NS/DNA hairpin loops can be unstable or easily disrupted by perturbations, such as the binding of more dedicated transcription factors. Hence this constitutes the basis for a transcriptional switch, which motivates an investigation of the underlying physical mechanism.

Second, we will figure out what is the relevant genomic scale to model the structuring effect of NAPs on the chromosome. To serve this purpose, we will use data from \chipseq experiments. We will show that the distribution of H-NS and FIS binding sites on the \ecoli genome cannot be well modeled by a Poisson stochastic point process where the realization of stochastic events in time corresponds to the insertion of binding sites on the genome. In particular, we will show that deviations from this model occur at short genomic distances, hence giving a likely scale at which evolutionary pressure has been exerted.

Finally, we will explore in more details the formation of DNA hairpin loops under the effect of H-NS. We will show that in order to form stable hairpin loops, binding regions must have a minimum length. This result is first derived using a simple polymer model with implicit interactions, and then confirmed using Brownian dynamics simulations with explicit and divalent proteins. Then we elaborate on possible implications for a regulatory mechanism relying on the disruption of these structures by other proteins such as FIS.

\section{Introduction to nucleoid-associated proteins (NAPs)}
\subsection{What are NAPs?}
In eukaryotic cells, histones provide a first and significant level of organization of the chromosome. Since bacteria lack histones, the chromosome is not organized into nucleosomes and the relevant description of the chromosome is the naked fiber of diameter $\SI{2.5}{\nm}$ \cite{Langowski2412006}. However, proteins playing a structural role like histones exist. With no surprise, they are called histone-like proteins, or nucleoid-associated proteins (NAPs), and are well known structuring proteins. In \textit{Escherichia coli}, the NAPs family comprises 12 proteins \cite{Kahramanoglou2011}, including H-NS, FIS, IHF, HU and StpA (a close analog of H-NS). The stationary phase transcription factor DPS is to be mentioned too, although it is present in significant concentrations only during the stationary phase or in response to stress.

The presence of NAPs is a universal feature among bacteria. In particular, there are found in many \textit{Salmonella} species, like \textit{S. typhimurium} \cite{Dorman022007}. In other bacteria species, NAPs are not exactly the same as in \textit{Escherichia coli}, but often functional and structural analogs can be found. For instance in \textit{Bacillus subtilis}, FIS is present while H-NS is replaced by the protein Rok, which has a similar structure despite the absence of sequence homology. In \textit{Deinococcus radiodurans}, a radiation-resistant bacteria, H-NS and FIS are not found, but HU and DPS are present in the cell \cite{FrenkielKrispin2006}. Unless otherwise stated, we discuss in the sequel the biology of the \ecoli bacteria.

\subsection{Architectural and regulatory role}
Like histones in eukaryotes, NAPs contribute significantly to the compaction of the bacterial DNA. In \ecoli the $\SI{1.5}{\milli\meter}$ bacterial DNA is folded dramatically to fit into the \SI{1}{\micro \meter^{3}} volume of the cell. More precisely, DNA is constrained to sit in a small region near the center of the cell \cite{Dame102002}. It has been demonstrated that this is mainly due to H-NS, which induces the formation of approximately two clusters per chromosome, which constitute the bacterial nucleoid. It has been shown that Rok has a similar role in \textit{B. subtilis} \cite{Wang092011}. Consequently, H-NS is found inside the nucleoid, whereas FIS, IHF and HU are found mostly at its periphery \cite{McGovern1994,Ryter111975}.

In general, regions of the chromosome which are trapped inside the nucleoid appear to be transcriptionnally silent. Therefore, there is a connection between chromosome architecture and transcription, yet to be resolved. More generally, all NAPs were shown to have an effect on the transcription of a large number of genes. For instance, it is known that the presence of AT-rich regions upstream of gene promoters can dramatically increases transcription \cite{Bokal1997}. Incidentally, AT-tracks, which are DNA sequences consisting of the repetition of A and T nucleotides, are over-represented in the genome of \ecoli \cite{Cho062008}. Interestingly, H-NS, FIS and IHF bind preferentially to AT-rich regions \cite{Grainger092006}, suggesting that the binding of NAPs to the chromosome is correlated with the role of AT-tracks in regulating the transcription.

NAPs have a strong influence on DNA architecture in the cell and are often called structuring proteins. Yet, it was shown that removing mRNA from \ecoli bacteria had more impact on the topology and shape of the nucleoid than the removal of H-NS, FIS or IHF, suggesting that the prevalent role of NAPs is not only architectural but also regulatory \cite{Grainger092006}.

In short, the ubiquitousness of NAPs in bacteria points toward a key role maintained throughout evolution. In particular, NAPs are the most abundant transcription factors in \ecoli \cite{Ishihama2014}. Thus their high concentrations and numerous binding sites scattered across the genome suggest that even today NAPs still play a prevalent regulatory role over other less abundant transcription factors. Presumably, regulatory functions stem from architectural changes induced on the chromosome. Besides, they are commonly found in bacterial species, suggesting that they are the remnants of a long-lived evolution from a common ancestor. Altogether, these elements represent strong reasons which motivate a more profound understanding of the mechanism by which NAPs regulate gene expression.

\subsection{Characterization \textit{in vivo}}
\begin{figure}[!hbtp]
  \centering
  \includegraphics[width = 0.7 \linewidth]{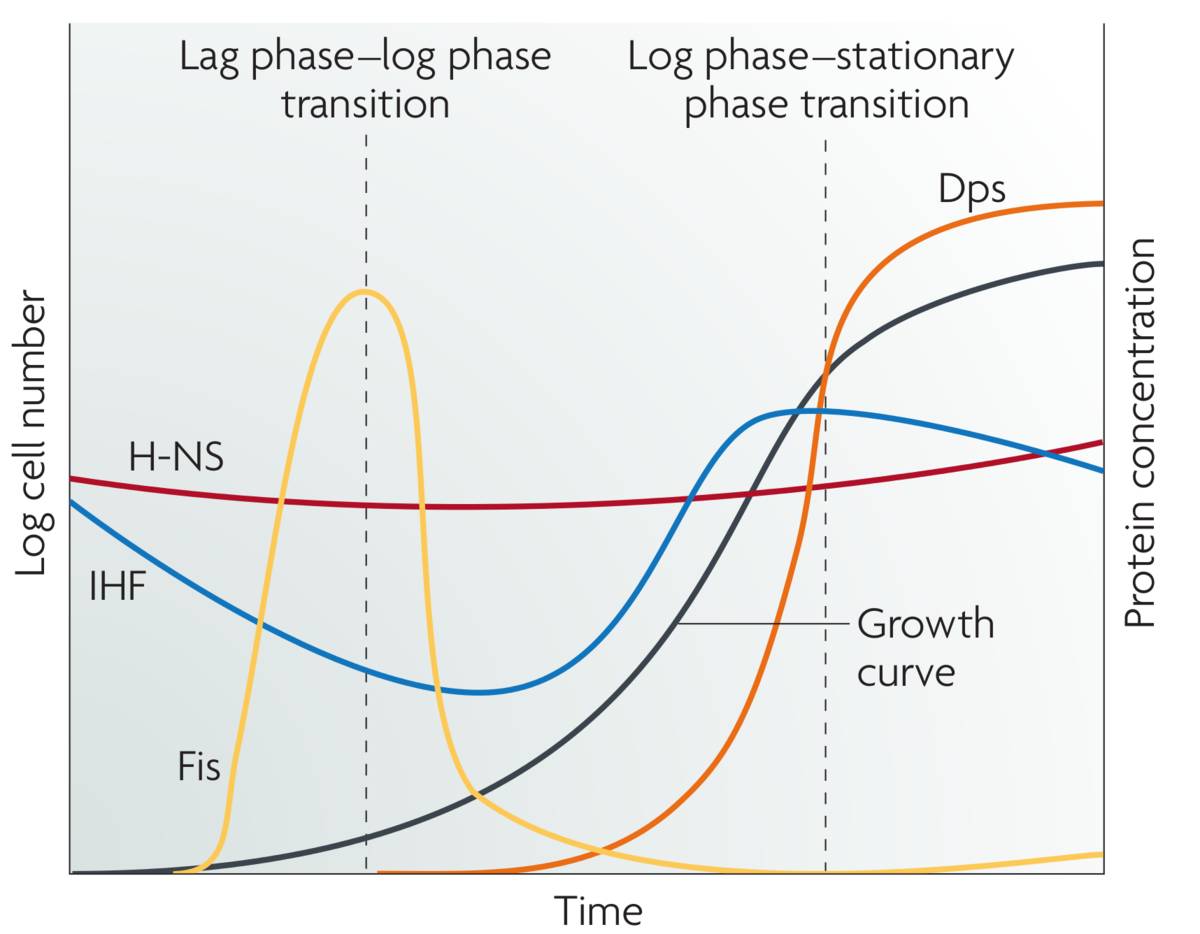}
\caption{Evolution of NAPs concentrations across the different growth phases \cite{Dillon1852010}.}
\label{fig:naps_concentration_growth}
\end{figure}

The concentration of NAPs depends on the growth phases of a cell population (\cref{fig:naps_concentration_growth}). H-NS concentration remains of the order of \SI{20000}{copies \per chromosome} \cite{Ishihama2014,Dorman042003,Navarre062007}. For this reason, it is often considered as the NAP of reference. With more than \SI{75000}{copies \per chromosome} during the exponential growth phase, FIS is the NAP with the highest concentration in the exponential growth phase. Yet its concentration plummets to less than \SI{100}{copies \per chromosome} during the stationary phase \cite{Ishihama2014,Cho062008}. Similarly, the ratio of HU to H-NS is HU:H-NS=\num{2.5} during the exponential growth phase and falls to approximately \num{1.0} during the stationary phase \cite{Dame102002}. IHF concentration is low during exponential growth phase and sharply increases at the onset of the stationary phase \cite{Dorman042003}. Eventually, NAPs can be ranked according to their concentrations in different growth phases \cite{Schneider102001,Azam63611999}:
\begin{itemize}
  \item FIS > HU > H-NS > IHF > DPS in the exponential phase;
  \item DPS > IHF > HU > H-NS > IHF in the stationary phase;
\end{itemize}
suggesting that bacterial physiology and NAPs concentrations are intimately connected.

Because of their relatively high concentrations and of the small size of bacterial cells, their observation with standard fluorescence microscopy is cumbersome \cite{Wang092011}. Indeed, let us consider a bacterial cell with volume $\SI{1}{\micro \meter^3}$, and a NAP with \num{20000} copies resulting in a number density $c = \SI{2.0e4}{\micro \meter^{-3}}$. The typical distance between two NAPs can be estimated to $d \approx c^{-1/3} \approx \SI{40}{\nm}$, which is below the visible light wavelength. This issue, also well known as the sub-diffraction limit has been addressed with modern super-resolution techniques \cite{Darzacq2013,Jin2014}.

What distinguishes NAPs from other transcription factors is not only their high copy number but also their large number of targets on the chromosome. Thus NAPs bind widely on the bacterial chromosome with a coverage of the order of one binding site for every hundred base pairs (1:\SI{100}{bp}) \cite{Wang092011}. Moreover, although less than \SI{10}{\percent} of the genome corresponds to non-coding DNA, approximately \SI{50}{\percent} of each H-NS, FIS and IHF binding sites fall in the promoter regions, suggesting a strong regulatory role for the NAPs \cite{Grainger092006}.

\section{The Nucleoid-Associated Proteins of \ecoli bacteria}
\subsection{The Histone-like Nucleoid Structuring protein (H-NS)}
\subsubsection{DNA binding}
H-NS is a small protein binding widely to DNA (\SI{17}{\percent} of the chromosome in \ecoli \cite{Kahramanoglou2011}). Consequently, it was believed for long that H-NS could bind non-specifically to DNA \cite{Dame012002,Dame092000}. In fact, state-of-the art high-throughput \chipseq experimental techniques have demonstrated that sites with a slightly increased affinity exist. Specifically, a \SI{6}{bp} long binding motif has been identified \cite{Kahramanoglou2011} (\cref{fig:naps_pwms_hns_fis}). This motif consists in an AT track, with a core of 3 consecutive nucleotides having an information content close to 2 bits of information (for the meaning of bits of information, see \cite{Stormo2013,Sheinman2012}). The average number of 3-nucleotide random draws before returning to an AT 3-nucleotide core is only 8, or equivalently \SI{24}{bp}. Therefore, despite a bias for AT-rich sequences, it turns out that the binding motif might be quite commonly found throughout the genome, which makes the non-specific binding hypothesis legitimate in a coarse-grained approach.

A H-NS monomer is able to bind DNA with its C-terminal domain. It can also can dimerize with another H-NS protein with its N-terminal domain \cite{Wang092011}. Such an H-NS dimer is a divalent protein with two DNA-binding domains. Furthermore, H-NS can gain valency by polymerizing with other dimers. Contrary to other NAPs, H-NS does not seem to induce any bending of the DNA upon binding. Thus the old claim that H-NS binds preferentially to curved regions of the DNA might find its origin in the fact that AT-tracks are indeed more flexible. Alternatively, DNA regions presenting a hairpin-like conformation might represent good candidates for subsequent binding by H-NS, since in this case H-NS can bridge the two DNA segments without any enthalpic bending penalty.

Remarkably, H-NS dimers can induce the formation of rigid DNA-H-NS-DNA filaments (or bridges) \cite{Dame092000,Dame012002}. Such filaments can nucleate from an initial binding site with higher affinity and elongate in a zipper-like fashion. At moderate H-NS concentrations ($>$1:\SI{100}{bp}), H-NS-bound DNA molecules display a characteristic structure with many filaments, usually flanked by DNA loops (\cref{fig:hns_loops_experiments:filaments}). At larger concentrations, dense structures are observed, presumably due to the existence of complexes containing H-NS oligomers with a high polymerization index.

The mechanisms for the formation of DNA/H-NS bridges has remained elusive however. For instance, it is not clear how DNA binding and polymerization of H-NS dimers result in stable filaments. On one hand, a single molecule experiment demonstrated that two double-stranded DNA molecules previously coated with H-NS fail to make filaments, suggesting that dimerization alone is not sufficient to make filaments. Instead, H-NS dimers, tetramers and other oligomers would bind several DNA sites simultaneously \cite{Dame2006}. On the other hand, the packing of DNA into the nucleoid by H-NS has been demonstrated to be highly deficient in mutants where H-NS could no longer polymerize, as evidenced by super-resolution fluorescence microscopy \cite{Wang092011}. More accurately, H-NS condenses the bacterial chromosome in approximately two clusters with diameter close to \SI{300}{\nm} in wild-type cells, but these clusters disappear in mutant cells where H-NS cannot dimerize.

\subsubsection{Thermochemical considerations}
As already mentioned, H-NS exists under the form of oligomers with different polymerization indices. \textit{In vivo} studies demonstrated that it is a dimer at low concentration and a tetramer at high concentration \cite{Dame012002}. Furthermore, single DNA molecule studies have shown that H-NS has a high off-rate from DNA, $k^{-}\approx\SI{1.5}{s^{-1}}$, suggesting that the DNA/H-NS bridges are fragile \cite{Dame2006}. This can ease the dynamical re-organization of the genome architecture. The same authors also measured the binding free energy of a H-NS dimer bound to two DNA sites and found a value close to be $ 11\, k_B T$, \textit{i.e.} a binding free energy of $\Delta^0 f \approx 5.5 \, k_B T$ per DNA-protein link.

\subsubsection{Regulatory function}
We have seen that H-NS concentration is constant in first approximation, and that it may be considered as the NAP of reference. Structural and/or regulatory changes might result from variations in the concentration ratios of other NAPs relatively to H-NS. For example, the ratio HU:H-NS is \num{2.5} in the exponential growth phase whereas it is close to $1.0$ in the stationary phase. It has been argued that HU counteracts the compaction of DNA induced by H-NS and that the balance between the action of the two NAPs have an important role in regulating the transcription \cite{Dame102002}.

H-NS over-expression in \ecoli has radical effects: it stops the cell growth, and makes the cell enter a stationary state, even when induced in the middle of the exponential growth phase. In minimal media, it was even reported to kill the cell \cite{McGovern1994,Spurio1992}. More accurately, H-NS over-expression stops the production of RNA transcripts, and therefore protein synthesis. The resulting nucleoids displayed strong morphological signatures: very dense and compact. Therefore, H-NS is generally considered as a global transcription silencer, through DNA compaction.

In physiological conditions, H-NS represses the transcription of many unrelated and non-essential genes \cite{Kahramanoglou2011,McGovern1994,Dame092000,Grainger092006,Dorman022007,Wang092011,Dame012002,Dorman042003}. We stress that H-NS is also involved in the regulation of the \textit{rrn} operon encoding rRNAs, which are extremely abundant constituents of the ribosomes (essential for the cell).

In agreement with the results obtained when H-NS is over-expressed, it was shown that genes repressed by H-NS appear to be bound by H-NS and sequestered in clusters, whereas genes which are not regulated by H-NS do not localize in such clusters \cite{Wang092011}. Consistently, a \chipseq study has shown that genes bound by H-NS are not bound by RNA polymerase (RNAP) and that their associated RNA transcripts have very low copy numbers in the cell \cite{Kahramanoglou2011}. In contrast, in mutants lacking H-NS, the same genes were significantly expressed and bound by RNAP. The study also confirmed that H-NS preferentially binds to AT-rich regions in agreement with previous claims \cite{Navarre062007,Dorman022007}. The majority of these AT-rich regions are sequences longer than \SI{1000}{bp} which are significantly enriched in genes acquired by horizontal transfer. Smaller binding regions appeared to correspond essentially to sequences in the promoters of operons or genes. In short, the mechanism of gene repression by H-NS seems to rely on the co-localization of H-NS-bound genes in dense clusters which can not be accessed by RNAP. The global regulation of transcription by H-NS might encompass a transcriptional modulation (by mild repression) of the short promoter-rich binding regions, and complete silencing of large binding regions.

\begin{figure}[htbp!]
  \centering
  \subfloat[]{\label{fig:hns_loops_experiments:filaments} \includegraphics[width = 0.8 \textwidth]{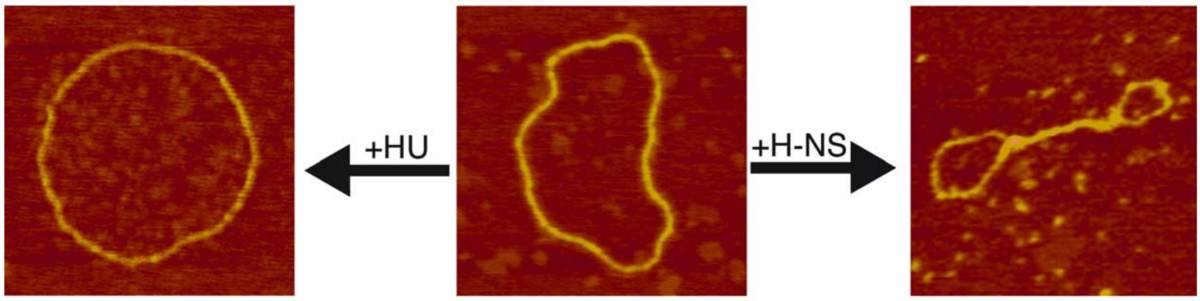}}%
  \\
  \subfloat[]{\label{fig:hns_loops_experiments:rnap_trapping}%
    \includegraphics[width = 0.4 \textwidth]{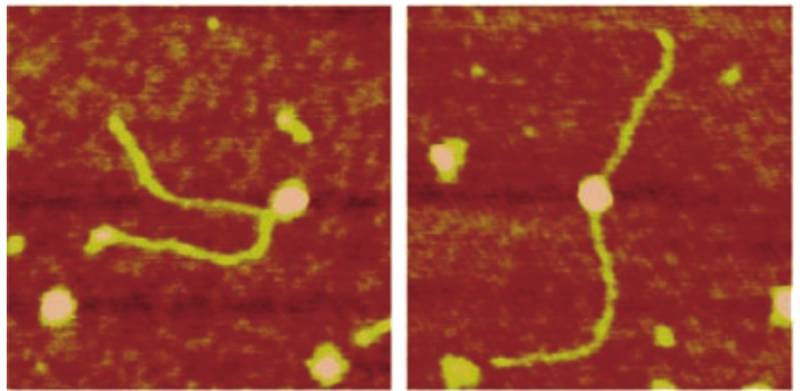}%
    \includegraphics[width = 0.4 \textwidth]{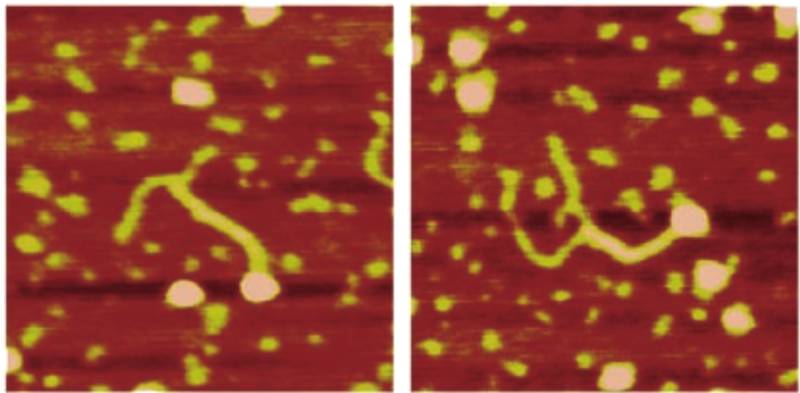}%
}%
  \\
  \subfloat[]{\label{fig:hns_loops_experiments:xenogenic_silencing}%
    \includegraphics[width = 0.3 \textwidth,valign=c]{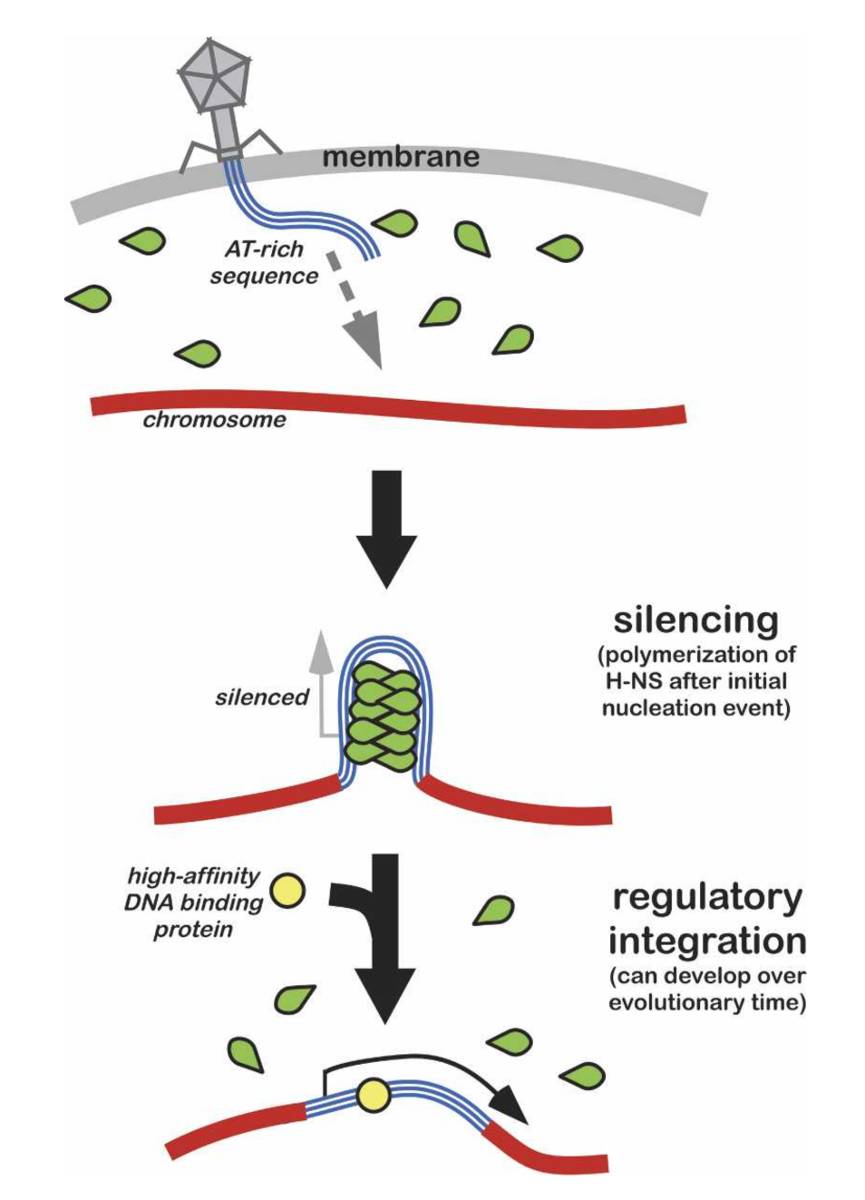}%
}%
  \quad
  \subfloat[]{\label{fig:hns_loops_experiments:chipseq}%
    \includegraphics[width = 0.5 \textwidth,valign=c]{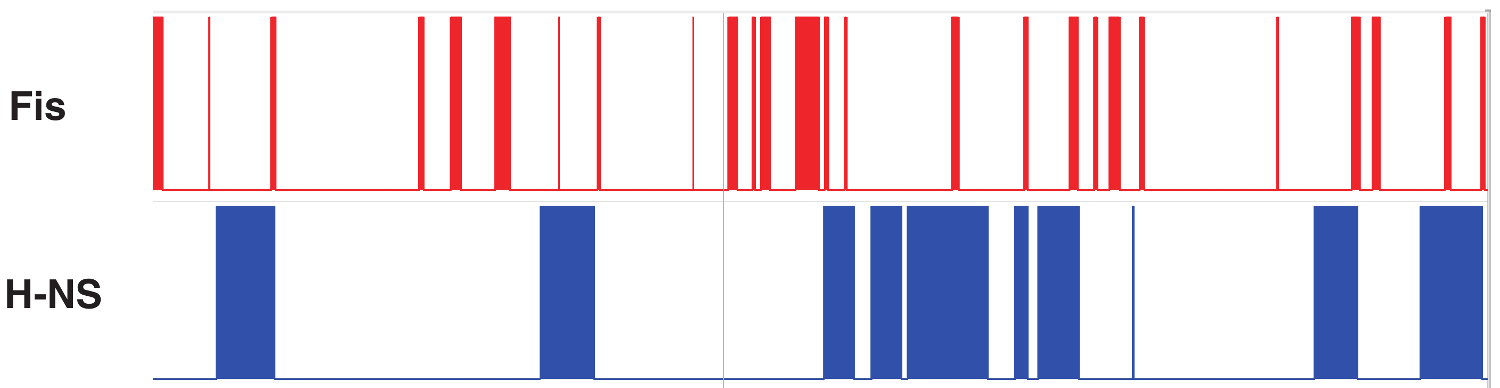}%
}

  \caption{\protect\subref{fig:hns_loops_experiments:filaments} AFM microscopy image of DNA filaments induced by H-NS \cite{Dame102002}. \protect\subref{fig:hns_loops_experiments:rnap_trapping} AFM microscopy image of a RNAP bound to a plasmid with the rrnB promoter in its center. After addition of H-NS to the solution, RNAP appears to be trapped in a DNA/H-NS hairpin complex (two right images) \cite{Dame012002}. \protect\subref{fig:hns_loops_experiments:xenogenic_silencing} Principle of xenogenic silencing by H-NS and regulatory integration \cite{Navarre062007}. \protect\subref{fig:hns_loops_experiments:chipseq} H-NS binding regions are organized into tracks, contrary to other transcription factors and NAPs such as FIS, which are the signature of filaments \cite{Kahramanoglou2011}.}
  \label{fig:hns_loops_experiments}
\end{figure}

Most AT-rich regions correspond to xenogenic DNA acquired by lateral transfer (from bacterial viruses for example). From an evolutionary perspective, it has been argued that the primary role of H-NS could have been to act as a genome sentinel by silencing the expression of xenogenic DNA \cite{Dorman022007,Navarre062007}. Subsequently, sequences and transcription factors might have evolved independently and occasionally produced mechanisms able to relieve H-NS-mediated repression. Thus, although H-NS could have acquired its prevalence in the bacterial kingdom because of its xenosilencing role, this system designed for defense might have been diverted from its original function in the course of evolution to serve other purposes, namely transcription regulation. An example is the \textit{rrnB} operon which is repressed by H-NS through the formation of a hairpin loop but activated by FIS (\cref{fig:hns_loops_experiments:rnap_trapping}).

Incidentally, an interesting candidate for a regulatory mechanism based on H-NS effect is the so-called RNAP-trapping mechanism by the \textit{rrnB} promoter \cite{Grainger092006}. AFM experiments have demonstrated that upon binding the \textit{rrnB} promoter, RNAP can be trapped in a hairpin loop at the extremity of a DNA/H-NS duplex \cite{Dame012002}. This mechanism is thought to enable a fast response to external stimulus because RNAP does not need to be recruited anymore when the H-NS repression is relieved. In the case of the \textit{rrnB} promoter, the stimuli corresponds to the binding of FIS in the promoter region. More generally, other TFs might be able to relieve the H-NS-mediated repression by disrupting the H-NS/DNA complex \cite{Dame092000,Navarre062007,Dorman022007}. On a local scale, H-NS may act in concert with other proteins resulting in specific regulatory functions. The cooperative effect of H-NS hairpin loop repression with a disrupter TF may be envisioned as a transcription ``switch''.

H-NS is also a sensor to many environmental changes. For instance, H-NS expression is increased in response to a cold shock. In \textit{Salmonella}, \SI{75}{\percent} of the genes having their expression altered by a temperature shift from \SI{25}{\degreeCelsius} to \SI{37}{\degreeCelsius} also depend on H-NS concentration \cite{Dorman042003}. Namely in \textit{S. typhimurium} it was found that an increased temperature results in a diminution of the fraction of H-NS bound to the virulence gene \textit{virF}, hence relieving its repression \cite{Dorman022007}. This suggests that the heat-shock response is mediated by a change in H-NS expression. Therefore, thermodynamical changes in the environment can alter the relative fraction of H-NS monomers and other oligomers, which induces a physiological response.

As a complementary mechanism, it has been suggested that the H-NS/DNA duplexes might prevent the supercoiling propagation along the chromosome \cite{Navarre062007}, and it was shown that H-NS over-expression correlates with a global decrease of supercoiling.

\begin{figure}[!htbp]
  \centering
  \includegraphics[width= 0.7 \textwidth]{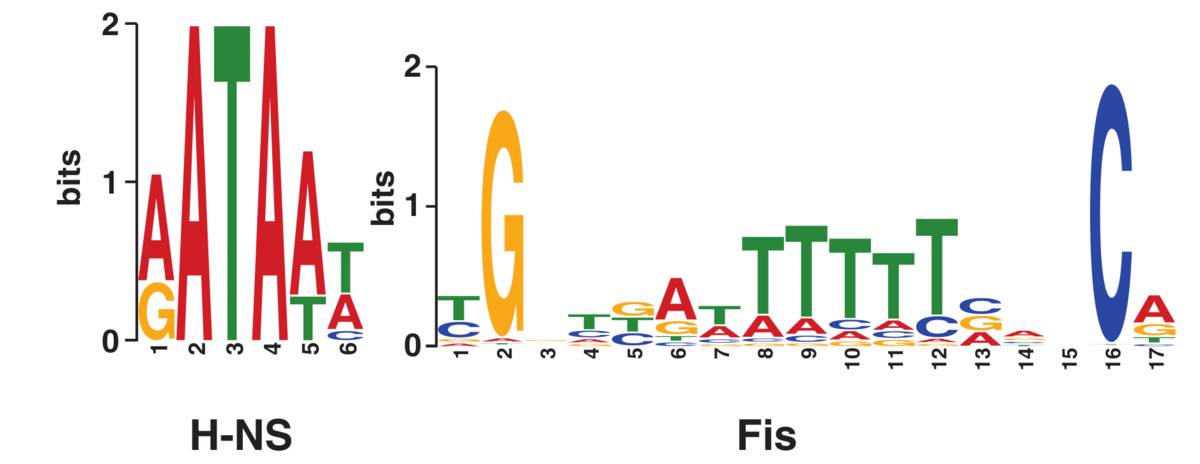}
  \caption{Position Weight Matrices for H-NS and FIS computed from high-throughput \chipseq experiment \cite{Kahramanoglou2011}.}
  \label{fig:naps_pwms_hns_fis}
\end{figure}

\subsection{The Factor of Inversion Stimulation (FIS)}
\subsubsection{DNA binding}
It was characterized early that FIS is a protein with a degenerate consensus binding sequence which can bind widely on the genome \cite{Dorman042003,Hengen121997,Skoko112005, Skoko122006, Kahramanoglou2011}. Similarly to H-NS, few sites with higher affinity can be found. In particular, \chipseq studies have identified a \SI{15}{bp} binding motif for FIS \cite{Bokal1997, Skoko122006, Kahramanoglou2011} (\cref{fig:naps_pwms_hns_fis}). Due to the presence of side chains in the protein structure, FIS effectively spans a \SI{21}{bp} region once bound to DNA \cite{Skoko112005}.

FIS is a homodimer \cite{Skoko112005, Skoko122006}, \textit{i.e.} its structure comprises two identical subunits which are assembled together. Each subunit has a helix-turn-helix structure resulting in a protein domain that can bind DNA. Hence FIS is divalent, namely it can bind two DNA sites simultaneously. Upon binding, FIS makes contacts with two consecutive major grooves on the DNA double-helix, separated by approximately \SI{11}{bp}. However, the length between the binding domains of the two subunits is too short, and consequently DNA is bent by \SIrange{40}{50}{\degree} \cite{Cho062008, Skoko122006}. Once bound, FIS can interact with flanking DNA and/or proteins with its side chains. In particular, it has been established that it can bind the $\alpha$-CTD domain of RNAP by charge complementarity. This finding supports the view that FIS can recruit RNAP to facilitate the initial binding step of RNAP to the promoter \cite{Bokal1997, Dame012002}.

The DNA structures resulting from the interaction with FIS are different from the structures induced by H-NS. In particular, FIS results in the formation of branched structures in supercoiled circular DNA molecules \cite{Schneider102001}. This effect is grounded in the formation of DNA loops. \textit{In vitro} assays have shown that DNA branching starts to appear at a protein ratio of 1:\SI{160}{bp}. For higher concentrations, starting with a ratio $<$1:\SI{80}{bp} (or a concentration $>\SI{75}{\nano M}$), DNA collapses in low mobility complexes \cite{Schneider102001}. In between, at moderate concentrations, FIS favors the formation of loops. This has also been evidenced with force-extension experiments \cite{Skoko112005, Skoko122006}, in which a stretching force is applied at both ends of a linear DNA molecule of \SIrange{50}{100}{\kilo bp}. For various FIS concentrations, a threshold force was identified, below which the DNA molecule collapses. This appeared to result from the existence of loops induced by thermal fluctuation at low forces. Such loops can be quenched by FIS proteins binding the extremities of the loop. Specifically, upon re-extension of the DNA molecule, discrete steps of DNA molecule size were observed, corresponding to the opening of such quenched loops. Their typical size was estimated to \SI{200}{bp}. From a regulatory standpoint, formation of loops in promoter regions may potentiate transcription by bringing in proximity remote regulatory sequences. On the contrary, dense aggregates occurring at high FIS concentration are expected to silence transcription in a manner similar to H-NS.

\subsubsection{Thermodynamical considerations}
The concentration of FIS peaks during the exponential growth phase, with approximately 75,000 copies/chromosome ($\sim 50 \, \mathrm{\mu M}$) \cite{Ishihama2014,Hengen121997,Bradley092007, Cho062008}. Yet, in the stationary phase, FIS concentration plummets to undetectable levels \cite{Kahramanoglou2011}. Paradoxically, despite being the most abundant NAP in this phase of growth, it remains spatially localized to the periphery of the nucleoid, whereas H-NS for instance is distributed in the whole nucleoid.

It appears that FIS binding to DNA is much stronger than that of H-NS. Indeed, starting from a DNA molecule coated with FIS, it was shown that the introduction of additional DNA molecules providing free competitor binding sites was not enough to drive the dissociation of FIS proteins from the original DNA molecule \cite{Skoko122006}. Dilution of the solution to increase the entropy of free FIS proteins was also insufficient to drive FIS dissociation. Finally, upon force re-extension of the DNA molecule and breakage of the DNA loops maintained by FIS, it appeared that FIS still remained bound to DNA. From these assays, the apparent dissociation coefficient measured was $K_d \approx 1 \, \mathrm{mM}$. Besides, \chipseq experiments revealed a very strong background signal for FIS binding in the exponential phase, confirming the high affinity of FIS with DNA genome-wide \cite{Kahramanoglou2011}.

\subsubsection{Regulatory function of FIS}
The role of FIS in regulating the transcription is less clear than with H-NS and has remained rather controversial. It has been argued that FIS is an activator for several genes, including the genes of the macrosynthesis (tRNA, rRNA) \cite{Bradley092007} and the \textit{hns} gene \cite{Dorman042003, Dorman022007}. Early \chipchip experiments also demonstrated that \SI{61}{\percent} of FIS binding occurs in promoter regions. Beside, \SI{47}{\percent} of these binding events were correlated with the binding of RNAP, supporting the role of RNAP recruiter for FIS \cite{Cho062008}. Interestingly, it is also claimed that FIS has a role in regulating DNA supercoiling since it was found to be an activator for the expression of topoisomerase I, one of the enzymes responsible for relieving supercoiling \cite{Dorman042003}.

Yet there are also many cases where FIS acts as a repressor. For instance it represses its own expression \cite{Grainger092006}. Hence there is no general tendency toward activation or repression of the transcription. Even more confusing, some equivalent genes in \textit{Escherichia coli} and \textit{S. typhimurium} appear to be contrary regulated by FIS \cite{Bradley092007}. Eventually, these results should be taken with caution since the differential expression measured for those genes is sometimes very small. A rather recent \chipseq study \cite{Kahramanoglou2011} has even concluded that despite its role as a key activator/repressor for very few genes, such as the \textit{rrnB} operon, in most cases the transcription of genes bound by FIS is not significantly affected by the \textit{fis} gene deletion.

Concerning the \textit{rrnB} operon, it has been shown that FIS binding increases the transcription of the downstream genes by 3 to 7 fold \cite{Bokal1997}. As detailed previously, this promoter is also under the repressive control of H-NS, which operates through the formation of DNA/H-NS hairpin with RNAP trapped at the apex \cite{Dame012002}. FIS has 3 binding sites upstream in the \textit{rrnB} promoter. Such bindings may interfere with H-NS binding, disrupt the DNA/H-NS filament, resulting in the de-repression of the \textit{rrnB} operon. Another evidence of this phenomenon seems to be the observation from \chipseq experiments that H-NS and FIS binding are mutually exclusive and anti-correlated \cite{Kahramanoglou2011}. Furthermore, as stated previously, H-NS has a high off-rate, suggesting that the DNA/H-NS filaments can undergo fast dissociation whereas FIS binding to DNA is strong, suggesting that it may prevent H-NS mediated repression for large times.

Several works have sought to relate the action of FIS to supercoiling in bacterial chromosome \cite{Grainger092006}. In \ecoli the typical size of a supercoiled domain is of the order of \SI{10}{\kilo bp}. \chipchip studies have shown that there is approximately two FIS proteins bound per supercoiled domain. Therefore the average spacing between consecutively bound FIS \textit{in vivo} is estimated to \SI{5}{\kilo bp}.

\subsection{The Heat-Unstable protein (HU)}
\subsubsection{DNA binding}
HU is a dimer whose monomers are coded by the \textit{hupA} and \textit{hupB} genes. Both homodimers and heterodimers are present \textit{in vivo} \cite{Swinger282004}. HU binds a \SI{9}{bp} motif on the DNA sequence \cite{Dame102002} and bends DNA by approximately 60-70\textdegree{} \cite{Wojtuszewski30962003}. The binding motif is quite degenerate and consequently, HU binding is nearly non-specific.

\subsubsection{Regulatory function}
A large number of bacteria contain proteins which are close sequence analogs to HU, pointing to an ancient and fundamental role of HU \cite{Dame102002, Swinger282004}. However, the regulatory function of HU has remained elusive. Some experimental works reported that HU increases transcription initiation. This is maybe because HU bends DNA and decreases supercoiling when binding to DNA, which may facilitate RNAP binding \cite{Dorman042003}. HU has also probably an important role in regulating DNA replication and stimulating DNA unwinding at the origin of replication (oriC) of \ecoli chromosome by regulating the assembly of the pre-replication complex \cite{Ryan13652002}.

More specifically, there are strong reasons to believe that HU can relieve the repressive action of H-NS \cite{Dorman042003, Dame102002}. Since H-NS concentration is constant during the cell-cycle, variations in HU concentration may constitute a fundamental mechanism to tune the genetic expression globally. Thus, there is an antagonism between HU and H-NS effects. The main role of HU would be to counteract the compaction of the chromosome by H-NS by opening up the chromosome in order to make it accessible for transcription \cite{Dame102002}.

\subsubsection{Thermodynamical considerations}
During the exponential growth phase, there are approximately 55,000 HU copies/chromosome, but during the stationary phase, the concentration of HU is reduced to 20,000 copies/chromosome \cite{Ishihama2014,Dorman042003, Dame102002, Swinger282004}. Furthermore, HU binding to DNA is relatively strong, with a dissociation constant $K_d \approx \num{200}-\SI{2500}{\nano M}$. As a side remark, similarly to FIS, HU tends to localize at the periphery of the nucleoid, with RNAP and ribosomes.

\subsection{The Integration Host Factor (IHF)}
\subsubsection{DNA binding}
IHF binds a \SI{13}{bp} motif on the genome \cite{regulondb_pwm}. Although its binding sequence is shorter than FIS, the binding motif is more constrained, which makes IHF the most specific of the NAPs. IHFs sharply bends DNA upon binding, by about \SI{160}{\degree}. This implies an important enthalpic cost because DNA is rigid on that scale (it has persistence length $l_p=150$ bp), which explains partially why the IHF binding is weaker than with other NAPs.

\subsubsection{Regulatory function}
As with HU, a large number of bacteria contain structural analogs to IHF, pointing to an essential role maintained throughout the evolution. It acts also probably in concert with HU to regulate DNA replication by stimulating DNA unwinding at the oriC \cite{Ryan13652002}.

It has also been conjectured that the sharp bending induced on DNA favors the nucleation of H-NS filaments. Hence, IHF might work in concert with H-NS and act as a repressor, in agreement with the finding that IHF binds mostly transcriptionnally silent regions \cite{Grainger092006, Swinger282004}. Incidentally, IHF concentration increases during the stationary phase, in which many genes are silenced.

\subsubsection{Thermodynamical considerations}
During the stationary phase, there are approximately 20,000 IHF copies/chromosome. This concentration is slightly decreased during the exponential phase but remains close to that value \cite{Ishihama2014}. IHF binding to DNA is also much weaker than that of HU for instance, with a dissociation constant $K_d \approx 20-\SI{30}{\micro M}$. This low affinity might be due to the important enthalpic cost which comes from bending DNA. It seems that the bound protein has no contact with the DNA major groove, suggesting that the binding is mainly entropic, which may be another reason for this low affinity \cite{Swinger282004}.

\subsection{Summary}
NAPs can be quite puzzling at first because they have an effect on the genetic expression of a broad class of genes and on the structure of the chromosome. As such, they illustrate the loose frontier that exists between transcription regulation and chromosome architecture, which with most likelihood implies some sort of feedback mechanism between the two processes.

Their effect on the chromosome architecture entails DNA compaction, DNA bending or the formation of specific structures. H-NS induces the formation of filaments, which makes DNA hardly accessible to other transcription factors, including RNAP. FIS is able to quench DNA loops produced by thermal fluctuations. HU tends to open the DNA double-helix by decreasing supercoiling. IHF can bend DNA in a hairpin configuration. A summary on the properties discussed in this section is given in \cref{tab:synthesis_naps}.

We have seen that NAPs are also the transcription factors with the largest concentrations in the cell. This prevalence certainly suggests that their structuring role is coupled with specific functions, probably in regulating the transcription. In this sense, it is remarkable that NAPs, or at least structural analogs, are found in different bacterial species. This fosters the view of a universal role played by NAPs in the bacterial kingdom and acquired early in the course of Evolution by giving a crucial fitness advantage. Subsequently, regulatory mechanisms based on these architectural changes induced on the chromosome may have been selected. In particular, we shall explore a model of regulation based on the formation of DNA hairpin loops (or helices) by H-NS.

\begin{table}
  \centering
  \footnotesize
  
\begin{tabular}{| L{0.13 \textwidth} | L{0.18 \textwidth} | L{0.18 \textwidth} | L{0.18 \textwidth} | L{0.18 \textwidth} |}
  \hline
  \textbf{Protein} & H-NS & FIS & HU & IHF \\
  \hline
  \textbf{Binding motif} & \SI{6}{bp} & \SI{15}{bp} & \SI{9}{bp} & \SI{13}{bp} \\
  \hline
  \textbf{Binding specificity} & nearly non-specific & nearly non-specific & nearly non-specific & specific \\
  \hline
  \textbf{Binding strength} & weak & strong & strong & weak \\
  \hline
  \textbf{DNA bending} & no & 40-\SI{50}{\degree} & 60-\SI{70}{\degree} & \SI{160}{\degree} \\
  \hline
  \textbf{Oligomerization} & yes & no & (no) & (no) \\
  \hline
  \textbf{Effect on DNA} & DNA/H-NS filaments & quenching of thermal loops with average size \SI{200}{bp} & open/rigidify the double-stranded DNA by decreasing supercoiling & DNA hairpins \\
  \hline
  \textbf{Copy number (per genome)} & \num{20000} & \num{75000} in exponential phase and $<100$ in stationary phase & \num{55000} in exponential phase and \num{20000} in stationary phase & \num{20000} \\
  \hline
\end{tabular}

  \normalsize
  \caption{Synthetic table for the properties of the four main NAPs in \textit{Escherichia coli}. We used parenthesis for properties in which some doubts remained after careful inspection of the literature.}
  \label{tab:synthesis_naps}
\end{table}

\section{Relevant scale for modeling the effect of NAPs}
\label{sec:naps_scale_modeling}
\subsection{Evolutionary selection of random insertions}
Before attempting to model the effect of NAPs on the chromosome architecture and seeking to relate this to biological processes, an important question is: what is the correct scale to investigate this effect? Indeed, performing Brownian Dynamics (BD) simulations at base pair resolution is not realistic because sampling equilibrium configurations would require too much computational power. It is therefore required to coarse-grain some molecular details. For instance, taking a unit length of about one double-helix turn, \textit{i.e.} $\SI{1}{u} \sim \SI{10}{bp} = \SI{3.4}{\nm}$ seems at first like a reasonable choice because it is close to the naked DNA diameter of \SI{2.5}{\nm} and therefore allows for a consistent modeling of the bacterial DNA using a beads-on-string polymer. It is also of the order of magnitude of a NAP binding sites (\cref{fig:naps_pwms_hns_fis}).

But even at this resolution, modeling the full \ecoli chromosome would require about \num{5e5} beads, let alone the introduction of protein beads to model NAPs which are to interact with DNA. In order to reduce the complexity and focus on the underlying physical process it seems necessary to consider shorter chunks of chromosome. But how to choose their size?

In the case of the \textit{lac} operon, repression occurs through a looping mechanism \cite{Muller-Hill1998}. It requires the promoter to contact an auxiliary site, \SI{401}{bp} downstream on the sequence, caused by the binding of the \textit{lac} repressor. Similarly, a repressor system of the coliphage $\lambda$ was evolved in \ecoli in which the simultaneous binding of a tetramer with the promoter and an auxiliary binding sites separated by \SI{3600}{bp} lead to a drastic repression of transcription \cite{Revet1999151}. In these simple examples, the natural scale for the regulation of transcription is the distance between the promoter and the auxiliary sites.

Such reasoning does not apply to NAPs because the regulatory mechanism has remained less clear, and in particular it cannot be reduced to the formation of one single loop. However, NAPs binding sites may have been acquired and maintained during Evolution by horizontal transfers \cite{Navarre062007}. Therefore, we will use this assumption to investigate at which scale was exerted evolutionary pressure.

\subsection{Model of NAPs binding sites insertion}
\label{sec:model_nap_bindingsite_insert}
Let us consider the \ecoli genome of size $N_G=\SI{4.6e6}{bp}$, with the usual genomic coordinate $s=1,\dots, N_G$. We now introduce the sequence of coordinates:
\begin{equation}
  s_0 < s_1 <s_2<\dots<s_M,
\end{equation}
which represent the starting position of M+1 binding sites on the genome. We also introduce the spacing variables $d_i=s_{i}-s_{i-1}$. Formally, this can be seen as the realization of $M$ events, starting from time $s_0$, drawn from a stochastic point process, in which the spacing distances are random variables. In the absence of evolutionary pressure, it would be reasonable to expect that random insertions of foreign DNA have been independent events. Therefore, we will consider that the spacing variables $d_i$ are drawn from independently and identically distributed (i.i.d.) random variables that we denote with capital letters, $D_i$ . Thus we may model the insertion of NAPs binding sites as a stochastic point process with independent increments:
\begin{equation}
  \mathcal{P}=\left(s_0, s_1,s_2,\dots,s_M,\dots\right) \qquad S_i - S_{i-1} = D_i \stackrel{\mathcal{L}}{\equiv} D.
  \label{eq:point_process}
\end{equation}

We may also make the more restrictive assumption on the spacing random variable $D$:
\begin{equation}
  \proba{D \geq d} = \proba{D -d_0 \geq d \mid D \geq d_0},
  \label{eq:non_aging_condition}
\end{equation}
which is a non-aging condition in standard survival stochastic point processes. As a consequence, \cref{eq:non_aging_condition} ensures that $D$ is an exponentially distributed random variable, with probability distribution function (p.d.f.)
\begin{equation}
  \mu(d)=\frac{1}{\langle d \rangle}e^{- d / \langle d \rangle},
  \label{eq:spacing_expo_law}
\end{equation}
which is equivalent to say that $\mathcal{P}$ is a Poisson stochastic point process.

\begin{figure}[!htbp]
  \centering
  \begin{tikzpicture}
\def\dz{2}
\def\rr{0.5em}
\def\tw{0.05em}
\def\th{0.8em}
\def\xa{-4}
\def\xb{-3}
\def\xc{-2.5}
\def\xd{-1.5}
\def\xe{3.50}
\def\xf{4.25}
\def\xlo{-5}
\def\xhi{5}

\tikzset{
  loci/.style = {
    shape = circle,
    align = center,
    draw  = #1,
    text=white,
    fill = black,
    solid,
    inner sep= 0pt,
    minimum size=\rr
  },
  tick/.style = {
    shape = rectangle,
    align = center,
    draw  = #1,
    text=white,
    fill = black,
    solid,
    inner sep= 0pt,
    outer sep= 0pt,
    minimum width=\tw,
    minimum height=\th
  },
}

  \node (L1) at (\xlo,\dz) {};
  \node (R1) at (\xhi,\dz) {};
  \draw[black, very thick, -{Latex[scale=1.2]}] (L1) -- (R1);
  \node[loci] (D1) at (\xd,\dz) {};

  \node (L2) at (\xlo,0) {};
  \node (R2) at (\xhi,0) {};
  \draw[black, very thick, -{Latex[scale=1.2]}] (L2) -- (R2);
  \node[loci] (A2) at (\xa,0) {};
  \node[loci] (B2) at (\xb.0,0) {};
  \node[loci] (C2) at (\xc,0) {};
  \node[loci] (D2) at (\xd,0) {};
  \node[loci] (E2) at (\xe,0) {};
  \node[loci] (F2) at (\xf,0) {};
  \draw[-{Latex[scale=0.8,sep=2pt]}] (\xa,{0.5*\dz}) -- (A2);
  \draw[-{Latex[scale=0.8,sep=2pt]}] (\xb,{0.5*\dz}) -- (B2);
  \draw[-{Latex[scale=0.8,sep=2pt]}] (\xc,{0.5*\dz}) -- (C2);
  \draw[-{Latex[scale=0.8,sep=2pt]}] (\xe,{0.5*\dz}) -- (E2);
  \draw[-{Latex[scale=0.8,sep=2pt]}] (\xf,{0.5*\dz}) -- (F2);

  \node (L3) at (\xlo,-\dz) {};
  \node (R3) at (\xhi,-\dz) {};
  \draw[black, very thick, -{Latex[scale=1.2]}] (L3) -- (R3);
  \node[loci] (A3) at (\xa,-\dz) {};
  \node[loci] (B3) at (\xb,-\dz) {};
  \node[loci] (C3) at (\xc,-\dz) {};
  \node[loci] (D3) at (\xd,-\dz) {};
  \node[loci] (E3) at (\xe,-\dz) {};
  \node[loci] (F3) at (\xf,-\dz) {};
  \draw[dotted,red,line width = 1pt, rounded corners, -{Latex[red,scale=1.0]}] (D3) -- (\xd,{-1.50*\dz}) -- (\xc, {-1.50*\dz}) -- (C3) node[anchor=north east] {$-$};
  \draw[dotted,green,line width = 1pt, rounded corners, -{Latex[green,scale=1.0]}] (D3) -- (\xd,{-0.50*\dz}) -- (\xe, {-0.50*\dz}) -- (E3) node[anchor=south east] {$+$};
  \draw[dotted,green,line width = 1pt, rounded corners, -{Latex[green,scale=1.0]}] (D3) -- (\xd,{-0.50*\dz}) -- (\xf, {-0.50*\dz}) -- (F3) node[anchor=south west] {$+$};


  \node (L4) at (\xlo,-2*\dz) {};
  \node (R4) at (\xhi,-2*\dz) {};
  \draw[black, very thick, -{Latex[scale=1.2]}] (L4) -- (R4);
  \node[tick, label={[label distance=0.5em]-90:$s_0$}] (A4) at (\xa,-2*\dz) {};
  \node[tick, label={[label distance=0.5em]-90:$s_1$}] (B4) at (\xb,-2*\dz) {};
  \node[tick, label={[label distance=0.5em]-90:$s_{i-1}$}] (D4) at (\xd,-2*\dz) {};
  \node[tick, label={[label distance=0.5em]-90:$s_{i}$}] (E4) at (\xe,-2*\dz) {};
  \node[tick, label={[label distance=0.5em]-90:$s_{i+1}$}] (F4) at (\xf,-2*\dz) {};
  \draw[-{Latex[,scale=1.0]}] (\xd,-1.75*\dz) to node[above]{$d_i$} (\xe,-1.75*\dz);

  \draw[dashed,-{Latex[scale=1.2]}] (R1) to [bend left = 45] node[right,text width=10em, outer sep=2pt] {random insertion by horizontal transfer} (R2) ;
  \draw[dashed,-{Latex[scale=1.2]}] (R2) to [bend left = 45] node[right,text width=12em, outer sep=2pt] {evolutionary selection of regulatory mechanisms} (R3) ;
  \node[label={[label distance=1.0em]0:binding sites layout}] at (R4){};
\end{tikzpicture}
  \caption{Random insertion of NAPs binding sites and evolutionary selection of sites playing an essential regulatory role.}
  \label{fig:scheme_nap_binding_selection}
\end{figure}
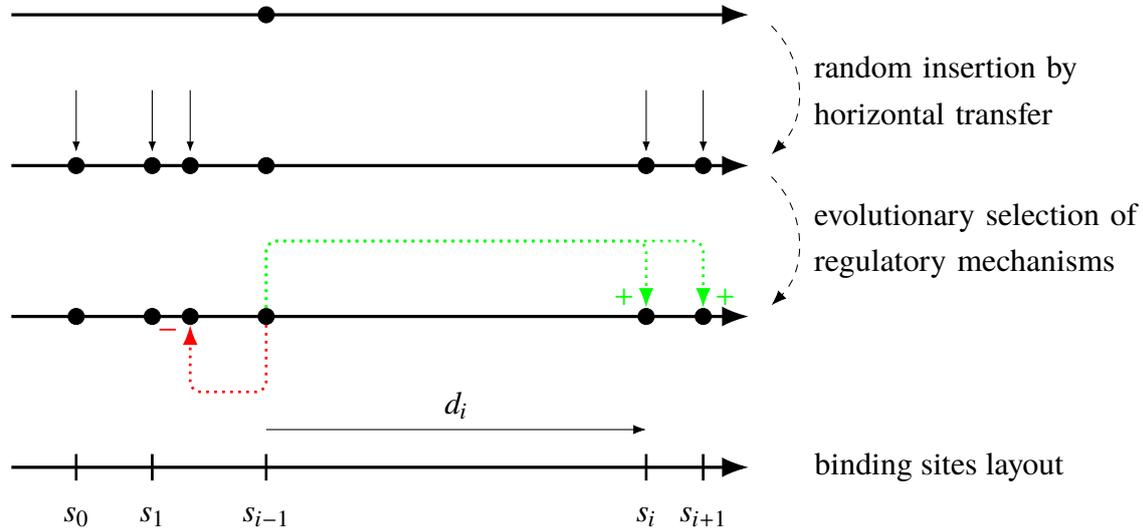

Therefore, if NAPs have a regulatory role underlying cooperative binding between distant sites along the DNA sequence, this should be visible as a bias in the insertion of binding sites throughout time (\cref{fig:scheme_nap_binding_selection}). In particular, the independence between consecutive binding sites insertions is flawed and deviation from the exponential distribution in \cref{eq:spacing_expo_law} should arise. However, it is reasonable to think that at large genomic distances, binding site insertions become uncorrelated. We may thus define a cross-over distance $d^*$ such that:
\begin{align}
  \begin{cases}
    D_{<} = \mathbbm{1}_{\{d < d^*\}} D & \text{is exponentially distributed}, \\
    D_{>} = \left(1-\mathbbm{1}_{\{d < d^*\}}\right) D & \text{is exponentially distributed}.
  \end{cases}
\end{align}

In conclusion, we propose to compute the \pdf of the spacing distance between consecutive NAPs binding sites (we will shortly present how). If it is exponentially distributed, then no evolutionary pressure has flawed the Poisson-dot-process-like insertion of the binding sites. In that case, one can doubt that any regulatory role is exerted by NAPs. Conversely, if it is non-exponentially distributed, it is the signature for the existence of a non-random layout of NAPs binding sites with a regulatory role. The cross-over between the two regimes will give us the scale that should be considered when modeling the effect of NAPs on the chromosome architecture.

\subsection{Binding sites spacing analysis from AT content}
\begin{figure}[!htbp]
  \centering
  \includegraphics[width=0.5 \textwidth]{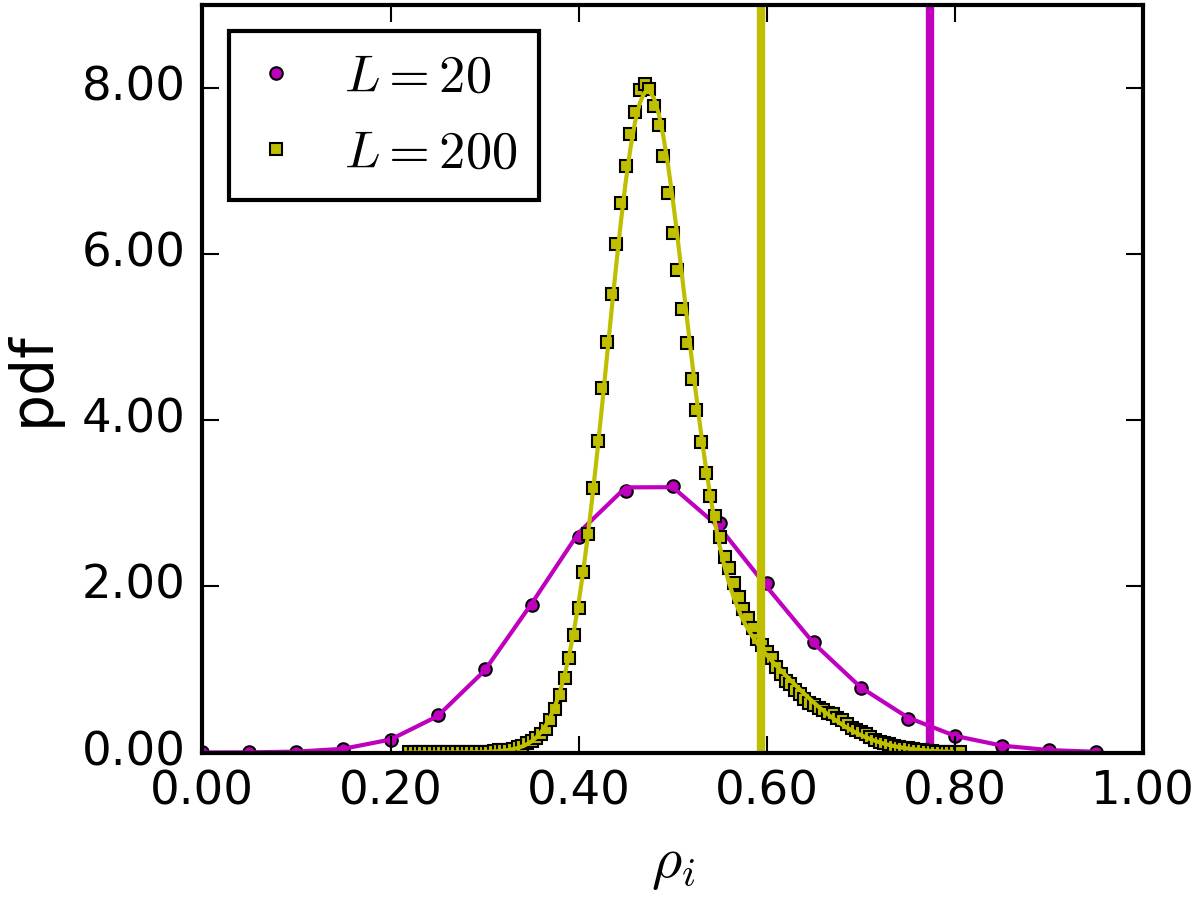}
  \caption{Probability distribution function of the AT-content $\rho_{AT}(s)$ (data points), fitted to a sum of Gaussian \pdf. We used a window size $L=20$ or $L=200$.}
  \label{fig:atcontent_density}
\end{figure}

In a first approach, we consider that NAPs binding sites correspond to AT-rich sequences. This is valid to some extent because as discussed in the last section, H-NS indeed binds preferentially to AT-rich sequences \cite{Grainger092006}. In order to perform this investigation, we use the MG1655 \ecoli genome, available from \cite{ZhouD6132013}. Then we define the density of AT nucleotides at coordinate $s$ on the genome, $\rho_{AT}(s)$, by considering a running window of size $L$:
\begin{equation}
  \rho_{AT}(s)=\frac{1}{L} \sum \limits_{k=s}^{s+L-1} \mathbbm{1}_{\{A,T\}}(b_k)
  \label{eq:rhoat_def}
\end{equation}
where $b_k \in \{A,T,G,C\}$ is the nucleotide at coordinate $k$. The corresponding \pdf can be computed and is shown in \cref{fig:atcontent_density} for different sizes of the running window.

\begin{figure}[htbp]
  \centering
  \includegraphics[width= 1 \textwidth]{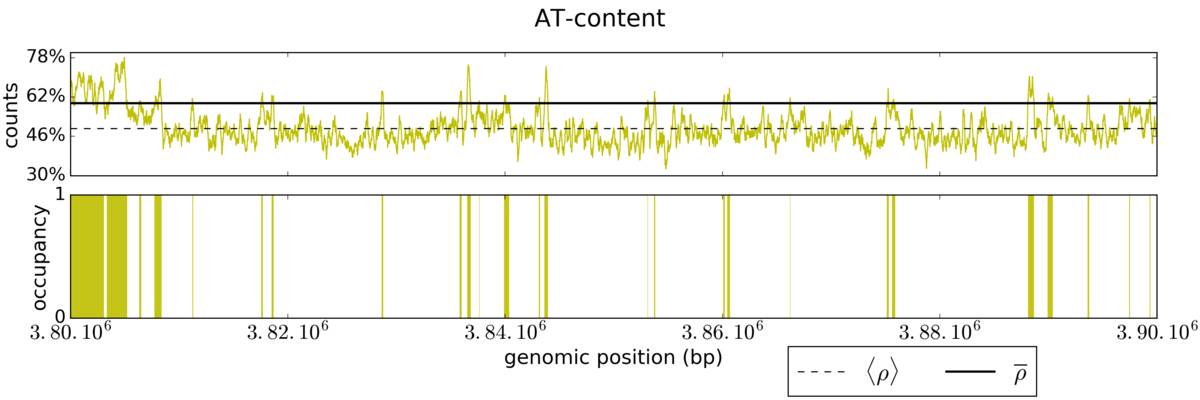}%
  \\
  \includegraphics[width= 1 \textwidth]{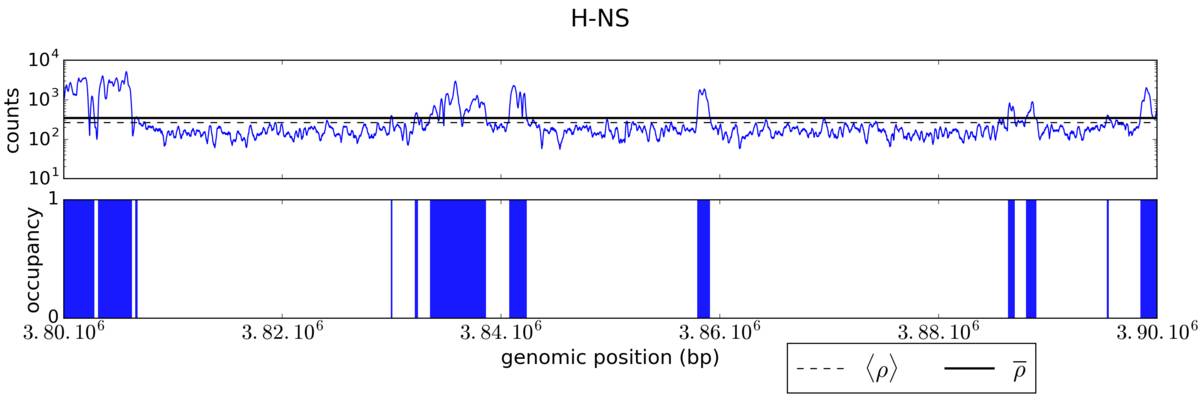}%
  \\
  \includegraphics[width= 1 \textwidth]{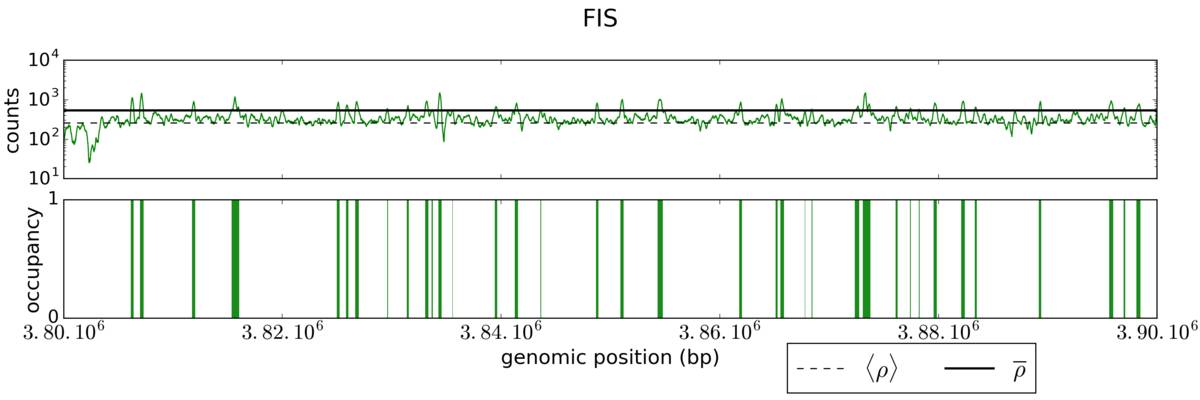}

  \caption{Binding regions in a chunk of \ecoli genome computed from AT-content ($L=200$) or from \chipseq counts with H-NS and FIS.}
  \label{fig:naps_density_window}
\end{figure}

In order to identify potential NAPs binding sites, we need to set a threshold $\overline{\rho}$ such that an occupancy variable can be defined as:
\begin{align}
  \chi_{AT}(s) =
  \begin{cases}
    1 & \text{if } \rho_{AT}(s) > \overline{\rho} \\
    0 & \text{otherwise,}
  \end{cases}
  \label{eq:occupancy_binding_atcontent}
\end{align}
and used to identify binding sites to coordinates where $\chi_{AT}(s)=1$. The threshold was set by fitting the AT-density $\rho_{AT}(s)$ with a sum of Gaussian distributions. For example, for $L=20$ the distribution of $\rho_{AT}(s)$ is well fitted by a single Gaussian \pdf whereas for $L=200$ two Gaussian \pdf were required (\cref{fig:atcontent_density}). We then set the threshold as:
\begin{equation}
  \overline{\rho} = \mu + 3 \sigma
  \label{eq:at_threshold}
\end{equation}
where $\mu$ and $\sigma^2$ are the mean and variance of the dominant Gaussian distributions. In \cref{fig:naps_density_window}, we show the binding regions obtained ($\chi_{AT}(s)=1$) for a chunk of the \ecoli genome. We have chosen genomic coordinates in the range \SIrange{3.8e6}{3.9e6}{bp} in agreement with \cite{Kahramanoglou2011}, which will be used in the next section.

\begin{figure}[!htbp]
  \centering
  \subfloat[]{\label{fig:atcontent_region_nobinding:20} \includegraphics[width= 0.40 \textwidth]{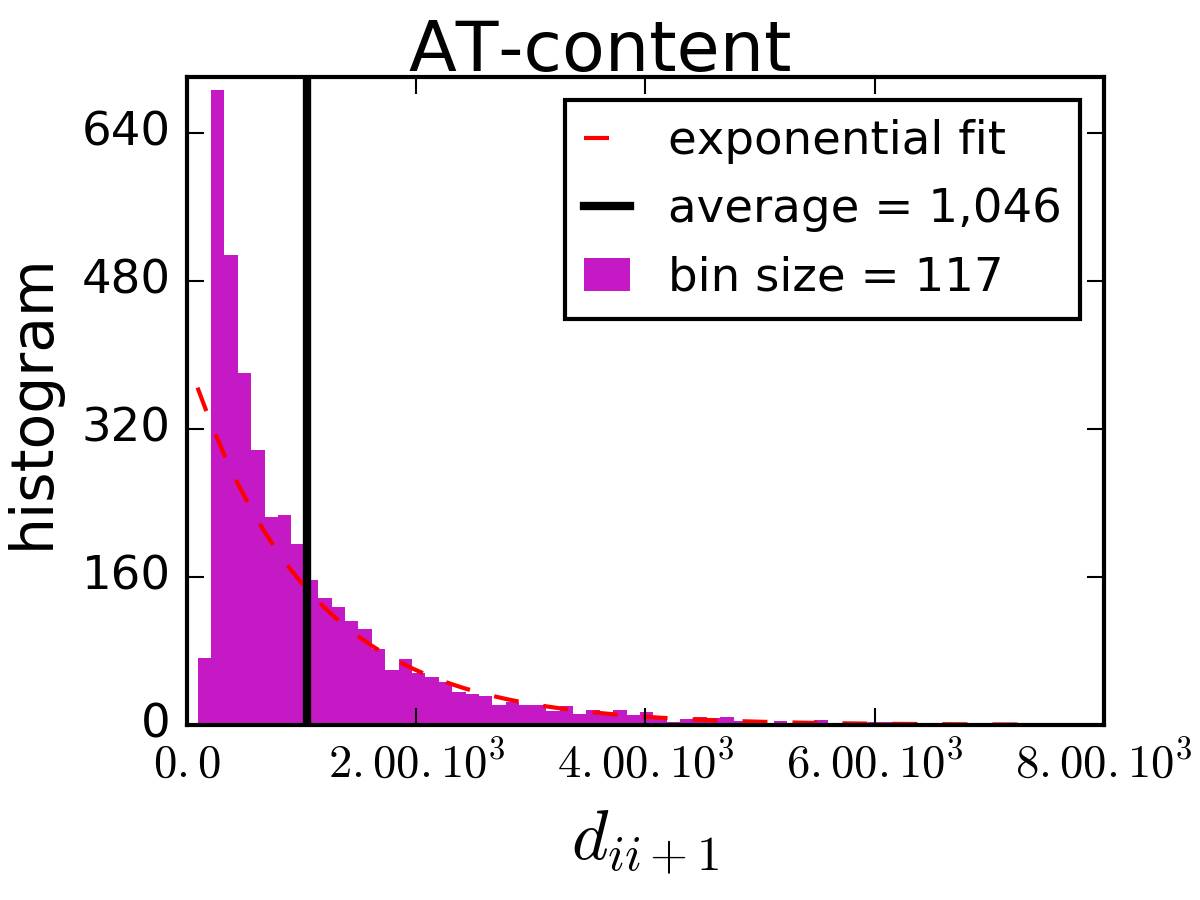}}%
  \quad
  \subfloat[]{\label{fig:atcontent_region_nobinding:200} \includegraphics[width= 0.40 \textwidth]{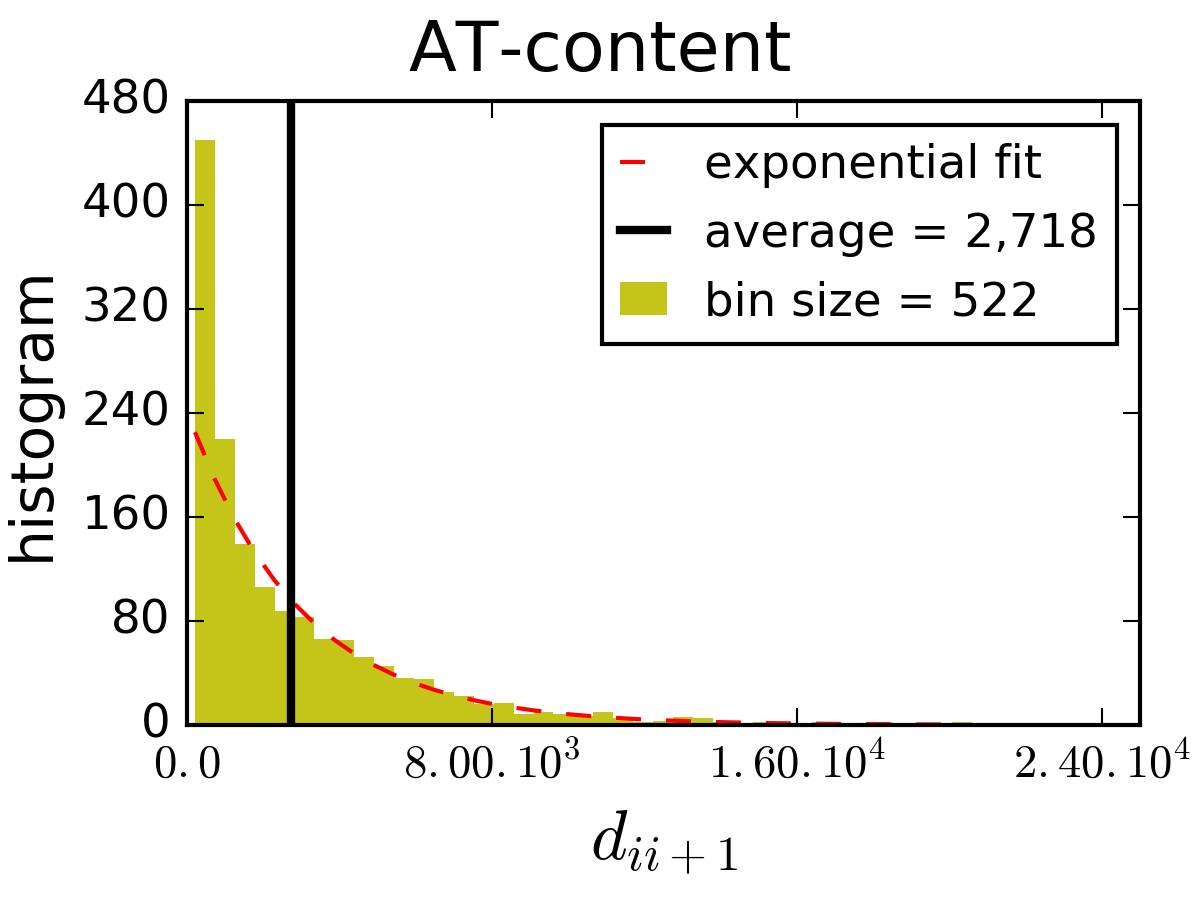}}%
  \\
  \subfloat[]{\label{fig:atcontent_region_nobinding:both} \includegraphics[width= 0.40 \textwidth]{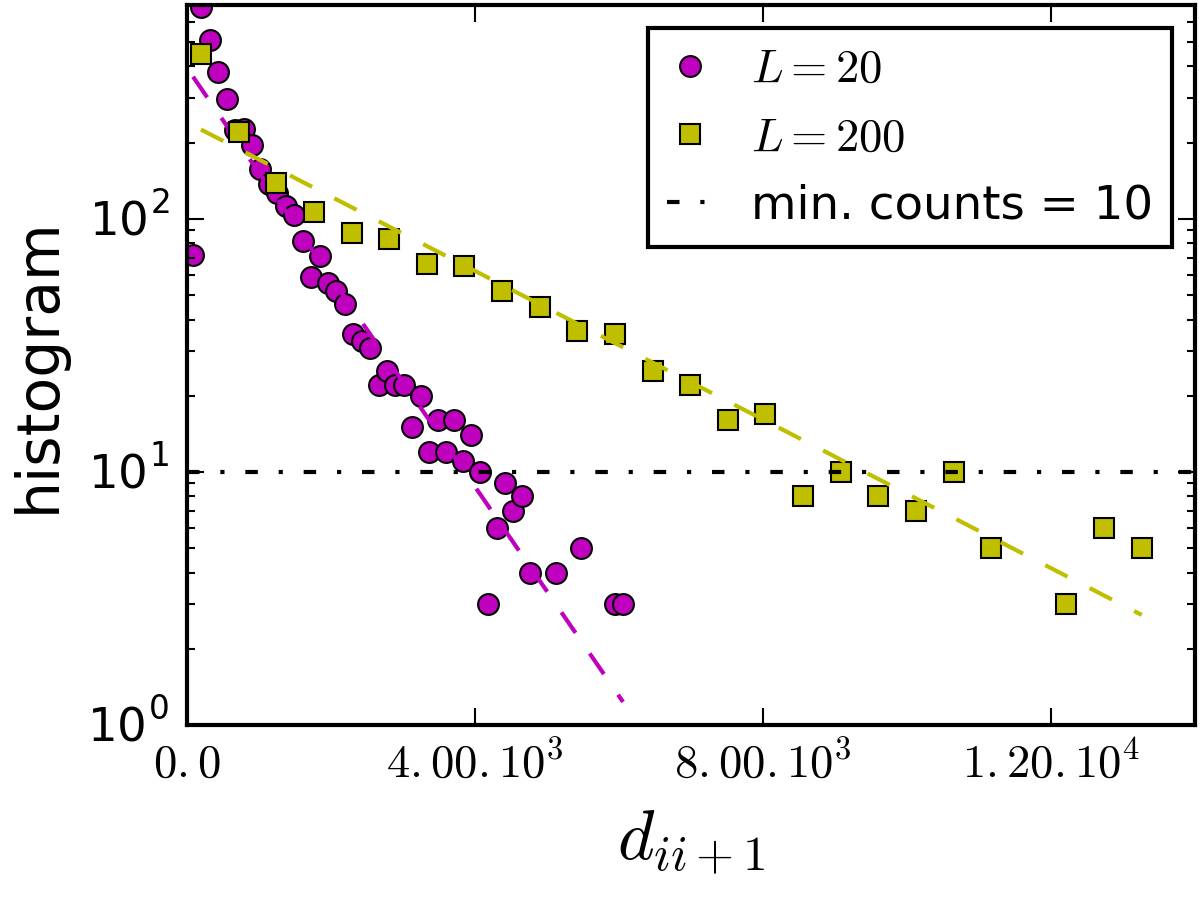}}
  \caption{Probability distribution of the distance between consecutive binding sites for \protect\subref{fig:atcontent_region_nobinding:20} $L=20$ and \protect\subref{fig:atcontent_region_nobinding:200} $L=200$. \protect\subref{fig:atcontent_region_nobinding:both} Fit with an exponential (logarithmic scale). We only considered bins of the histograms with a number of data points greater that $10$.}
  \label{fig:atcontent_region_nobinding}
\end{figure}

Eventually, we are able to analyze the presence of long-range interactions in the NAPs binding sites repartition. Following the directions given in \cref{sec:model_nap_bindingsite_insert}, we have computed the \pdf for the distance between consecutive binding regions (\cref{fig:atcontent_region_nobinding}). In other words, we computed the \pdf of the size of the empty regions. We observe that this \pdf has an exponential tail, suggesting that no evolutionary constraint is exerted at distances $d > d^*\approx \SI{2}{\kilo bp}$. On the contrary, deviations from the exponential decay are seen for $d<d^*$ and characterized by an over-represented fraction of empty regions with small sizes.

\begin{figure}[!htbp]
  \centering
  \subfloat[]{\label{fig:atcontent_autocorrelation:standard}\includegraphics[width=0.4 \textwidth]{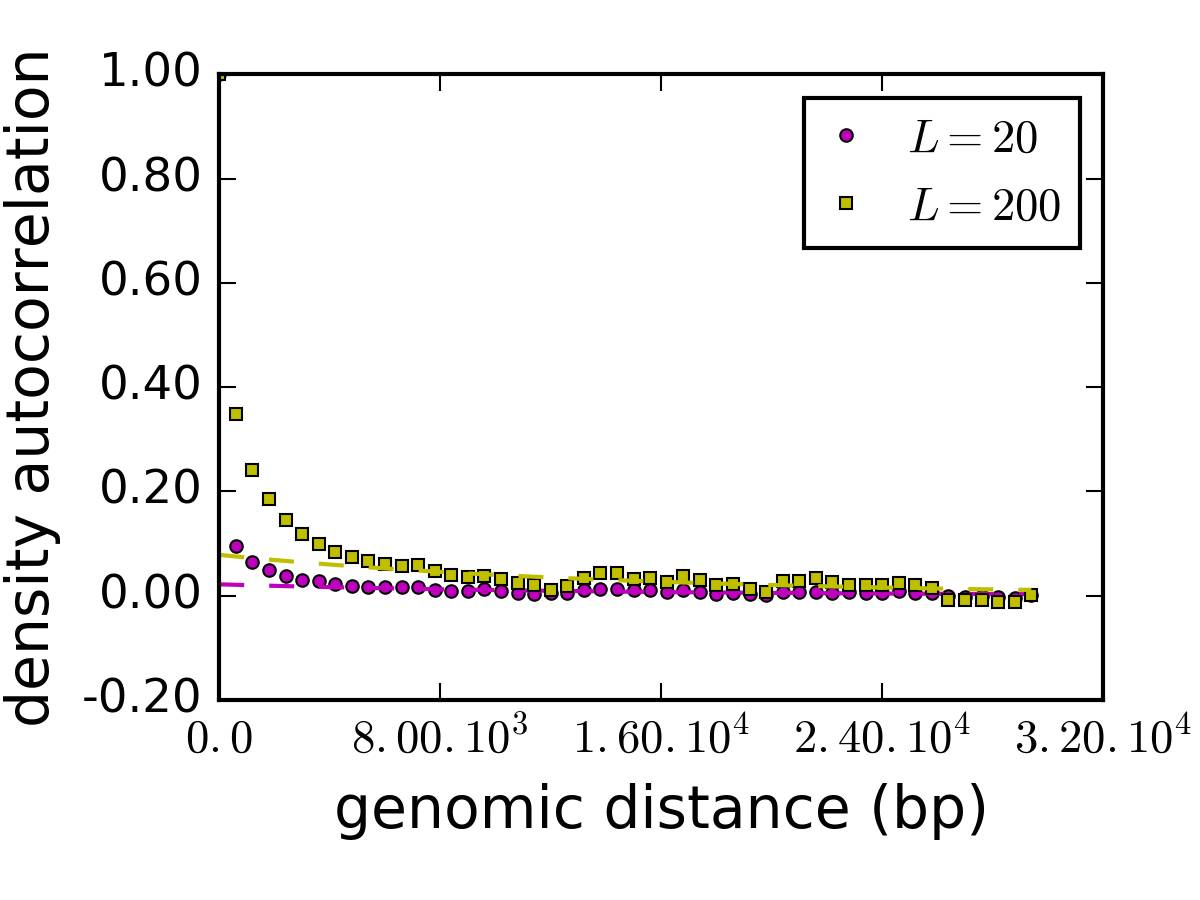}}%
  \quad
  \subfloat[]{\label{fig:atcontent_autocorrelation:log}\includegraphics[width=0.4 \textwidth]{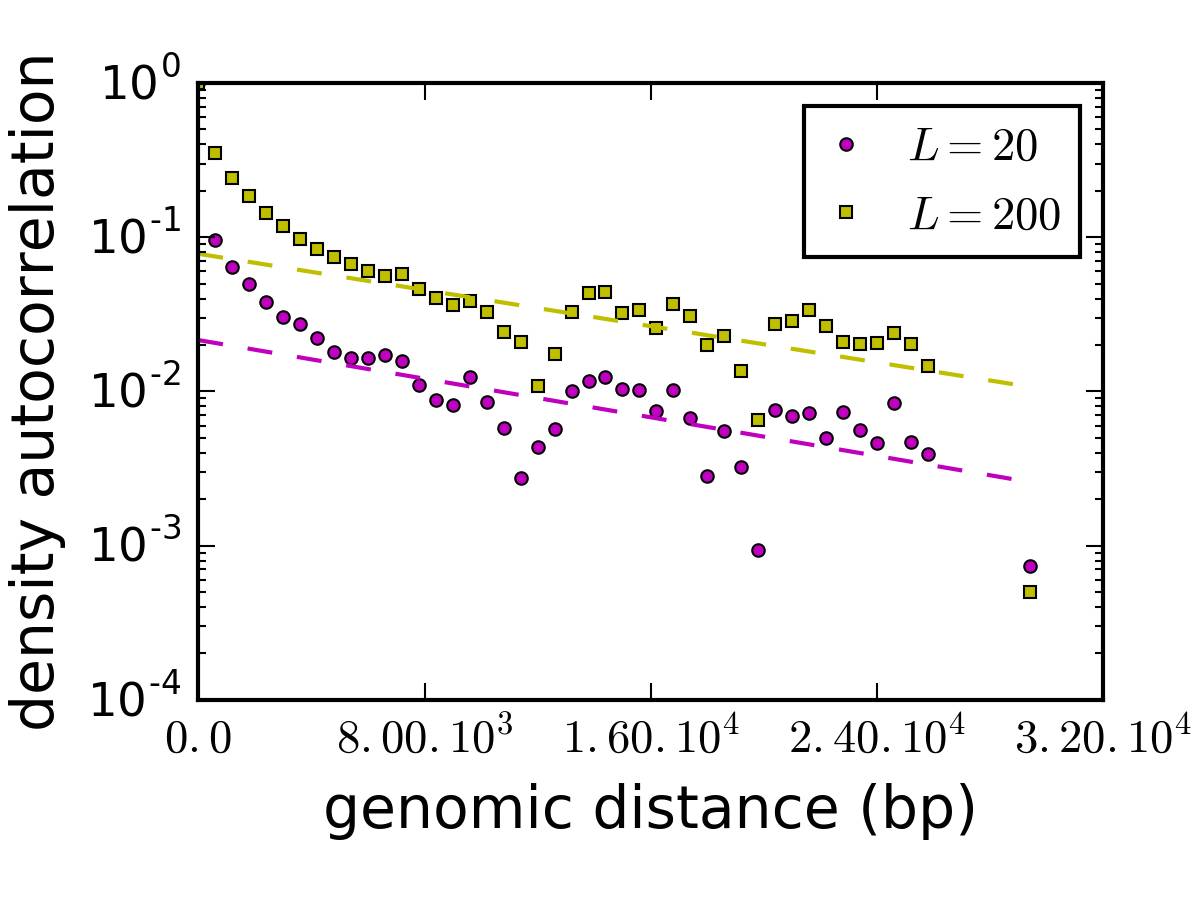}}
  \caption{Auto-correlation of $\rho_{AT}(s)$: \protect\subref{fig:atcontent_autocorrelation:standard} standard and  \protect\subref{fig:atcontent_autocorrelation:log} logarithmic scale.}
  \label{fig:atcontent_autocorrelation}
\end{figure}

For cross-validation, we have also computed the (connected) auto-correlation function of the AT-density:
\begin{equation}
  C_{AT} =  \left\langle \rho_{AT}(s+s_0)\rho_{AT}(s_0) \right\rangle - \langle \rho_{AT}(s) \rangle \langle \rho_{AT}(s_0)\rangle
  \label{eq:correlationfunc_connec_atdensity}
\end{equation}
where $s_0$ can be any coordinate on the genome. We observed a non-exponential decay at short genomic distances (\cref{fig:atcontent_autocorrelation}). Moreover, an exponential fit of the tail suggests that the cross-over indeed takes place for genomic distances of a few \SI{}{\kilo bp}.

\subsection{Binding sites spacing analysis from \chipseq data}
While the previous approach can give insight on existing correlations between NAPs binding sites on the genome, it is clear that limiting NAPs binding sites to AT-rich regions is a crude approximation of the reality. Incidentally, \chipseq experiments provide a way to measure directly the \textit{in vivo} genomic positions of NAPs binding sites. Therefore, in this section, we will consider this \chipseq experimental data and use the Poisson-point-process analogy described in \cref{sec:model_nap_bindingsite_insert}.

\chipseq experiments measure the density of binding for a protein of interest to the chromosome. Briefly, \chipseq experiments involve first a cross-linking step to fix proteins bound to DNA. Then DNA is sheared and the proteins, tagged with an anti-body, are immuno-precipitated. After purification, DNA fragments that were bound to such proteins remain and are amplified by PCR. An alignment step follows, in which the read sequences are mapped to genomic coordinates with a typical resolution of $\SI{200}{bp}$. Hence, for each bin at coordinate $s$, we obtain the number of times the protein of interest was bound to this particular location. Actually, the counts obtained represent the number of binding events up to the PCR amplification ratio. However, we shall consider that this operation only changes the normalization of the counts. In summary, a counting variable $n_{chip}(s)$ is obtained. Here we use recent high-throughput \chipseq data for H-NS and FIS \cite{Kahramanoglou2011}.

Let us now define the density of binding:
\begin{equation}
  \rho(s)=\frac{1}{\mathcal{N}} n(s), \quad \mathcal{N}=\sum \limits_{s} n(s),
  \label{eq:chipseq_density}
\end{equation}
where we have dropped the ``$chip$'' index for the sake of clarity. As before, we need to define a threshold that allows us to label each coordinate $s$ as a binding or a non-binding site.

It may seem natural to assume that the density of binding at coordinate $s$ is a Boltzmann weight \cite{Stormo2013}:
\begin{equation}
  \rho(s)=\frac{1}{Z} e^{- \beta \varepsilon(s)},
  \label{eq:chipseq_density_boltzmann}
\end{equation}
where $\varepsilon(s)$ is the binding energy of the protein to the sequence at coordinate $s$, $\beta=(k_B T)^{-1}$ is the inverse temperature and $Z$ is a normalization.

Let us also assume that there is a finite number $M$ of binding energy levels encountered throughout the genome, such that:
\begin{equation}
  \varepsilon_{M} < \varepsilon_{M-1} < \dots < \varepsilon_{1} < \varepsilon_{0}
\end{equation}
where $\varepsilon_{0}$ is the unbound state and $\varepsilon_{i}$ with $i \geq 1$ are bound states. The bound states may represent different binding modes and correspond to the existence of binding sites with different affinities, \textit{e.g.} primary and auxiliary binding sites. Then the probability for a protein to be bound to to the chromosome with energy $\varepsilon$ is expressed as a sum of delta functions:
\begin{equation}
  \proba{\varepsilon} = \sum \limits_{k=1}^{M} \alpha_k \delta(\varepsilon-\varepsilon_{k}),
  \label{eq:chipseq_density_boltzmann_energy_delta}
\end{equation}
where $\alpha_k$ is the proportion of sequences with binding energy $\varepsilon_k$ in the genome.

\begin{figure}[htbp]
  \centering
  \subfloat[]{\label{fig:naps_chipseq_counts:both}\includegraphics[width= 0.40 \textwidth]{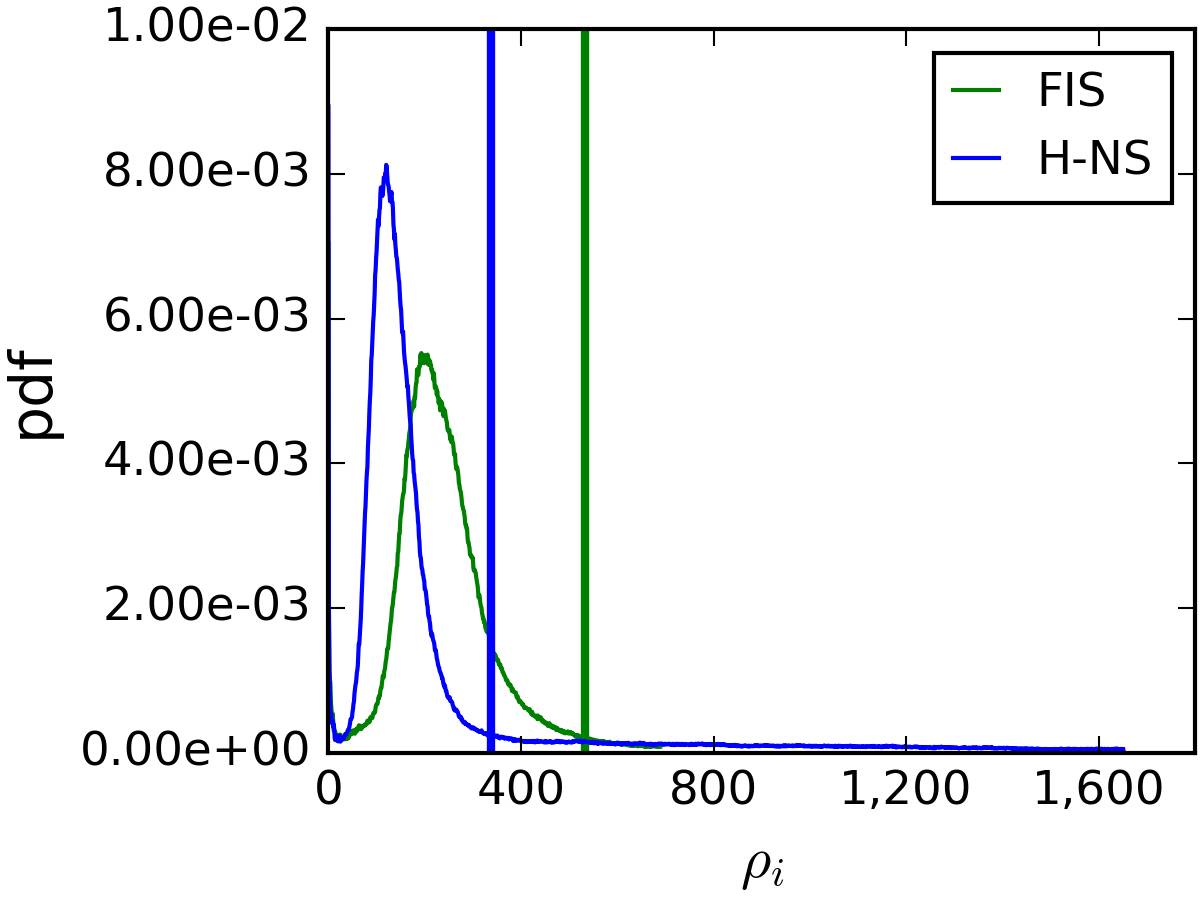}}%
  \\
  \subfloat[]{\label{fig:naps_chipseq_counts_eps:hns}\includegraphics[width= 0.40 \textwidth]{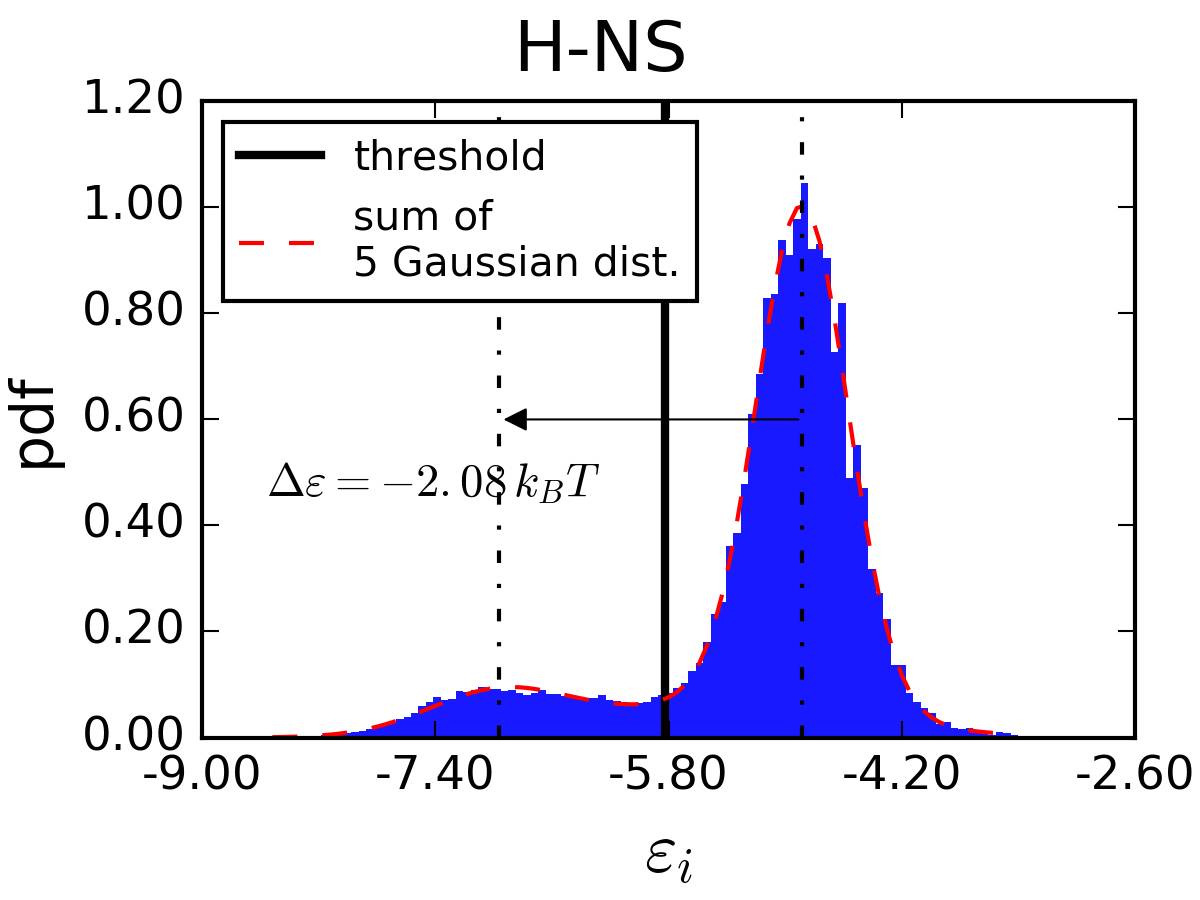}}%
  \quad
  \subfloat[]{\label{fig:naps_chipseq_counts_eps:fis}\includegraphics[width= 0.40 \textwidth]{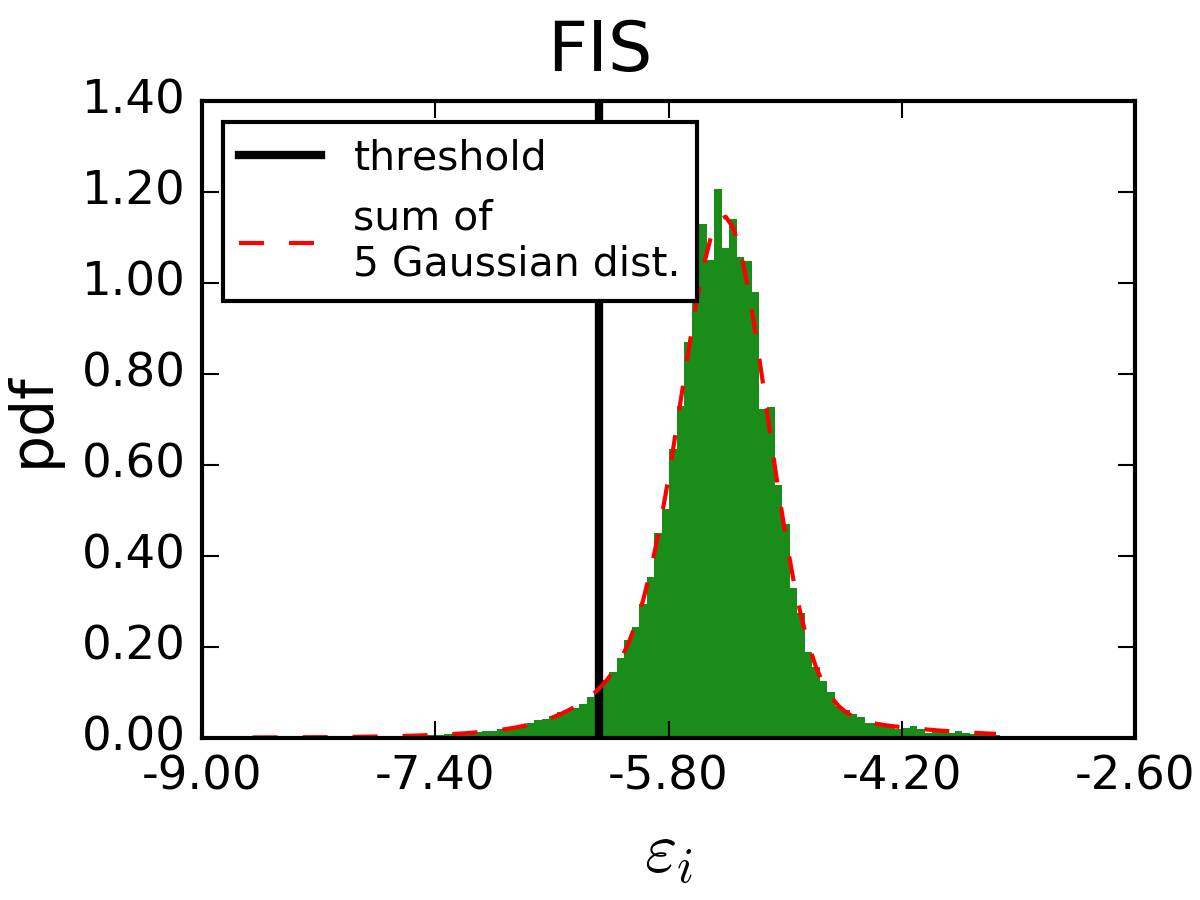}}
  \caption{\protect\subref{fig:naps_chipseq_counts:both} Distribution of the \chipseq counts $\rho(s)$ and determination of a threshold separating non-binding from binding sites. \protect\subref{fig:naps_chipseq_counts_eps:hns}-\protect\subref{fig:naps_chipseq_counts_eps:fis} Fit of $\varepsilon(s)=-\ln{\rho(s)}$ with a Gaussian multi-variate distribution from which the threshold is determined, for H-NS and FIS.}
  \label{fig:naps_chipseq_counts}
\end{figure}

In reality, the energy states might not be exactly discrete because the space of binding sequences is very large. Hence a better description might be to replace the delta-functions introduced in \cref{eq:chipseq_density_boltzmann_energy_delta} by Gaussian weights:
\begin{equation}
  \proba{\varepsilon} = \sum \limits_{k=1}^{M} \alpha_k \frac{1}{\sqrt{2 \pi \sigma_k^2}} \exp{\left( -\frac{(\varepsilon-\varepsilon_k)^2}{2 \sigma_k^2}\right)},
    \label{eq:chipseq_density_boltzmann_energy_gaussian}
\end{equation}
where $\sigma_k^2$ is the variance of the energy fluctuations of the NAP binding in mode $k$, with mean energy $\varepsilon_k$.
For both H-NS and FIS, fitting $-\ln{n(s)}$ to a sum of Gaussian distribution according to \cref{eq:chipseq_density_boltzmann_energy_gaussian} appeared to be a good approximation. For H-NS (\cref{fig:naps_chipseq_counts_eps:hns}), we can clearly distinguish two peaks in the energy levels distribution. In particular, we can assess the difference between the bound state and the unbound state to $\varepsilon_0 - \varepsilon_1 \approx 2 \, k_B T$. Note however that the energy scale in $k_B T$ depends on the PCR amplification ratio. For FIS (\cref{fig:naps_chipseq_counts_eps:fis}), the conclusion is less clear because only one mode remains in the binding energy distribution. It seems highly unlikely however that FIS be present only in the unbound state because it has a strong affinity with DNA. Instead we prefer to consider that the bulk of the proteins is actually bound to the chromosome.

In summary, we can extract the binding energies of the protein (up to a constant) from the logarithm of the \chipseq counts, and fitting the energy \pdf to a multi-variate Gaussian mixture model gives us the associated energy levels. In particular, we can use this information to define a threshold that retain only the bound states. More accurately, we considered the sum of the $M^*$ dominant Gaussian distributions such that
\begin{equation}
 \sum_{k \leq M^*} \alpha_k > \underline{\alpha},
\end{equation}
with $\underline{\alpha} = \SI{50}{\percent}$. This defines a distribution $f_b(c)$ for the bulk of the binding sequences with mean and standard deviation given by:
\begin{align}
    \varepsilon_b = \quad \frac{\sum \limits_{k=1}^{M^*} \alpha_k \varepsilon_k}{\sum \limits_{k=1}^{M^*} \alpha_k},
    \qquad
   \sigma_b^2 & = \quad \frac{\sum \limits_{k=1}^{M^*} \alpha_k (\varepsilon_k^2 + \sigma_k^2)}{\sum \limits_{k=1}^{M^*} \alpha_k} - \varepsilon_b^2,
\end{align}
which appears to be a better description of the dominant mode (unbound for H-NS) than taking the single Gaussian distribution with $k=0$. As announced, this enables the definition the thresholds:
\begin{equation}
  \underline{\varepsilon} = \varepsilon_b - 3 \sigma_b, \qquad \overline{\rho} = \exp{(-\underline{\varepsilon})},
  \label{eq:naps_chipseq_threshold}
\end{equation}
from which we defined an occupancy field, similar to \cref{eq:occupancy_binding_atcontent}:
\begin{equation}
  \chi(s) =
  \begin{cases}
    1 & \text{if } \rho(s) > \overline{\rho}, \\
    0 & \text{otherwise,}
  \end{cases}
  \label{eq:occupancy_binding_chipseq}
\end{equation}
such that coordinates where $\chi(s)=1$ are considered as binding sites.

We have applied this method to H-NS and FIS (\cref{fig:naps_chipseq_counts}) and scanned the \chipseq counts along the genome to find potential binding sites. In \cref{fig:naps_density_window}, we show the result for the same genome window as the one used in \cite{Kahramanoglou2011}. In order to attenuate inaccuracies related to the resolution of \chipseq experiments, and following the same authors, we have joined binding regions separated by less that \SI{200}{bp}. The obtained binding regions are in qualitative agreement with \cite{Kahramanoglou2011}, so we conclude that our definition for the threshold is consistent, and stick to it because it has a clearer physical interpretation in terms of binding energies.

\begin{figure}[!htbp]
  \centering
  \subfloat[]{\label{fig:naps_chipseq_region_nobinding:hns} \includegraphics[width= 0.40 \textwidth]{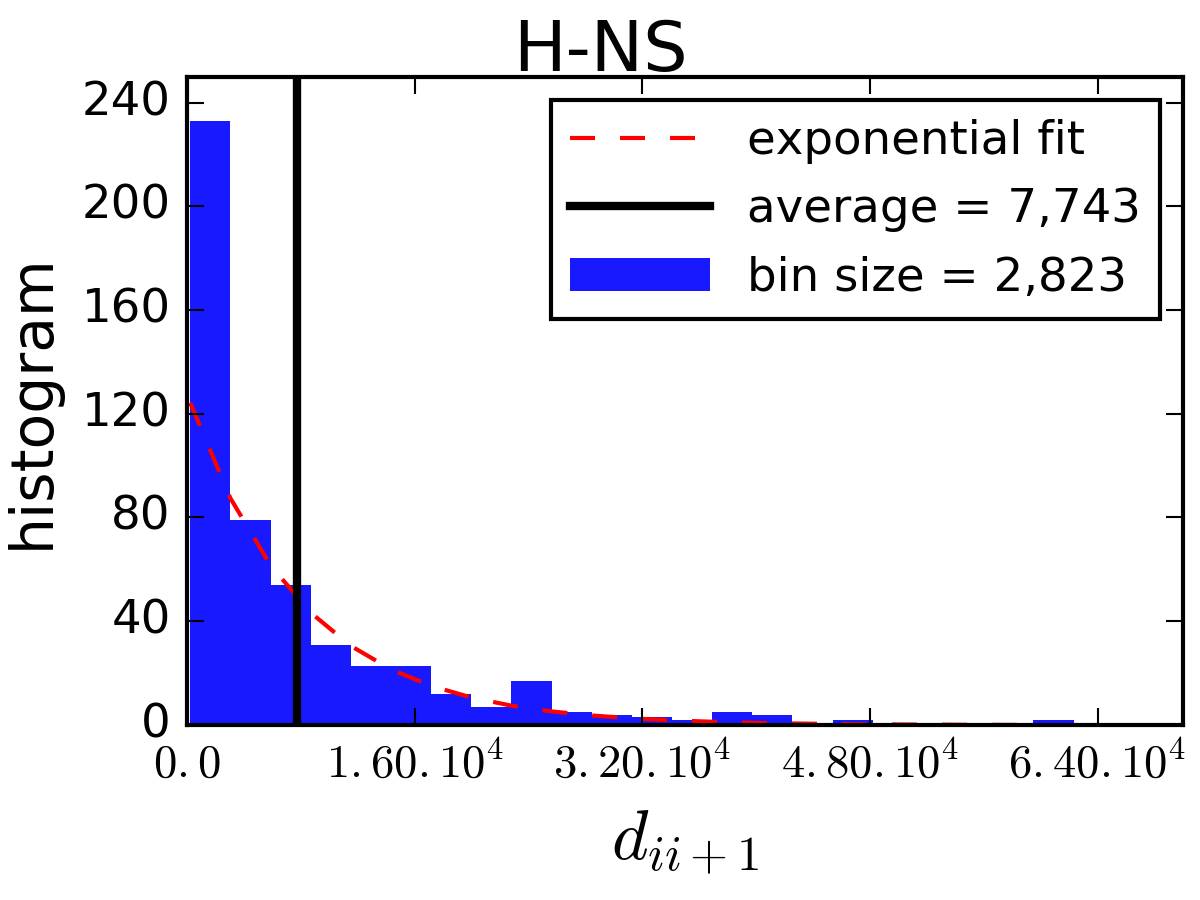}}%
  \quad
  \subfloat[]{\label{fig:naps_chipseq_region_nobinding:fis} \includegraphics[width= 0.40 \textwidth]{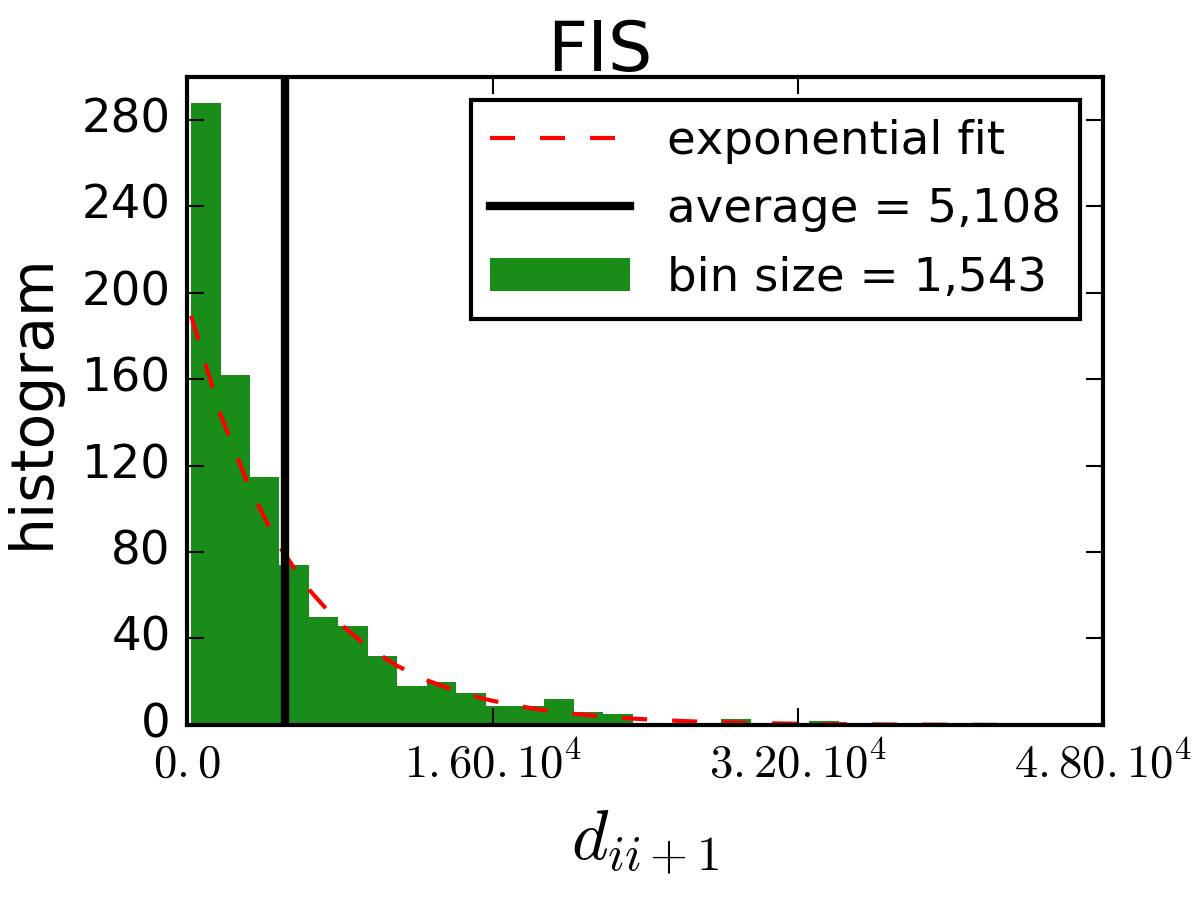}}%
  \\
  \subfloat[]{\label{fig:naps_chipseq_region_nobinding_log:both} \includegraphics[width= 0.40 \textwidth]{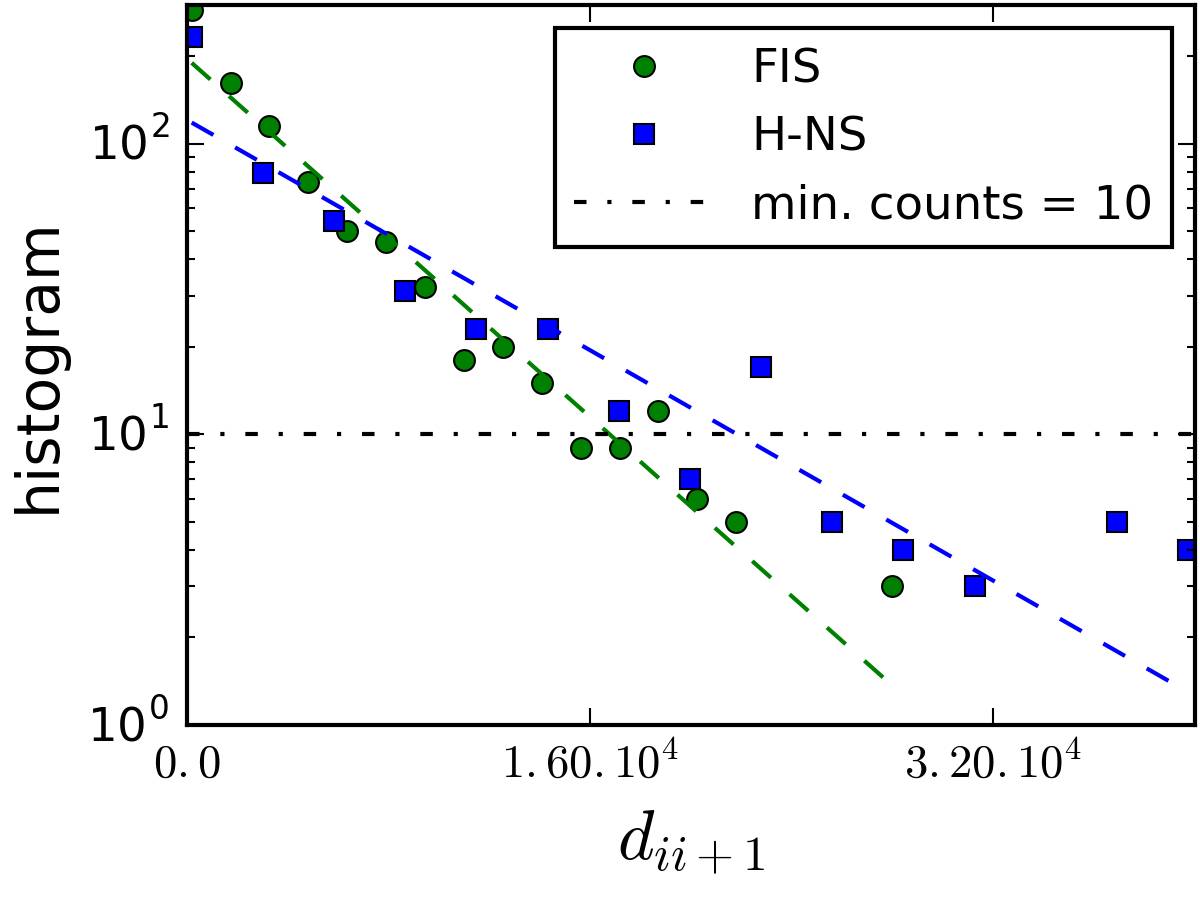}}
  \caption{Probability distribution of the distance between consecutive binding sites for \protect\subref{fig:naps_chipseq_region_nobinding:hns} H-NS and \protect\subref{fig:naps_chipseq_region_nobinding:fis} FIS. \protect\subref{fig:naps_chipseq_region_nobinding_log:both} Fit with an exponential (logarithmic scale). We only considered bins of the histograms with a number of data points greater that $10$.}
  \label{fig:naps_chipseq_region_nobinding}
\end{figure}

We are now ready to analyze the presence of long-range interactions in the NAPs binding sites repartition. Similarly to \cref{sec:model_nap_bindingsite_insert}, we have computed the \pdf for the distance between consecutive binding regions on the genome (\textit{i.e.} the \pdf for the length of the empty regions). For both H-NS and FIS, we obtain that this \pdf is well fitted by an exponential distribution, except at short genomic distances (\cref{fig:naps_chipseq_region_nobinding}). Actually, a large number of distances fall within the first bin of the histogram in our figure and result in a deviation from the exponential decay. Altogether, the distribution for the distance between consecutive binding sites can be considered as exponential for genomic distances $d > d^*$, with $d^* \approx \SI{3}{\kilo bp}$ for H-NS and FIS. Note that the exponential distribution of the distance between binding regions for FIS is consistent with a previous work which found it to be well described by an exponential \pdf with average \SI{5}{\kilo bp} \cite{Cho062008}.

\begin{figure}[!htbp]
  \centering
  \subfloat[]{\label{fig:naps_chipseq_autocorrelation:both} \includegraphics[width= 0.40 \textwidth]{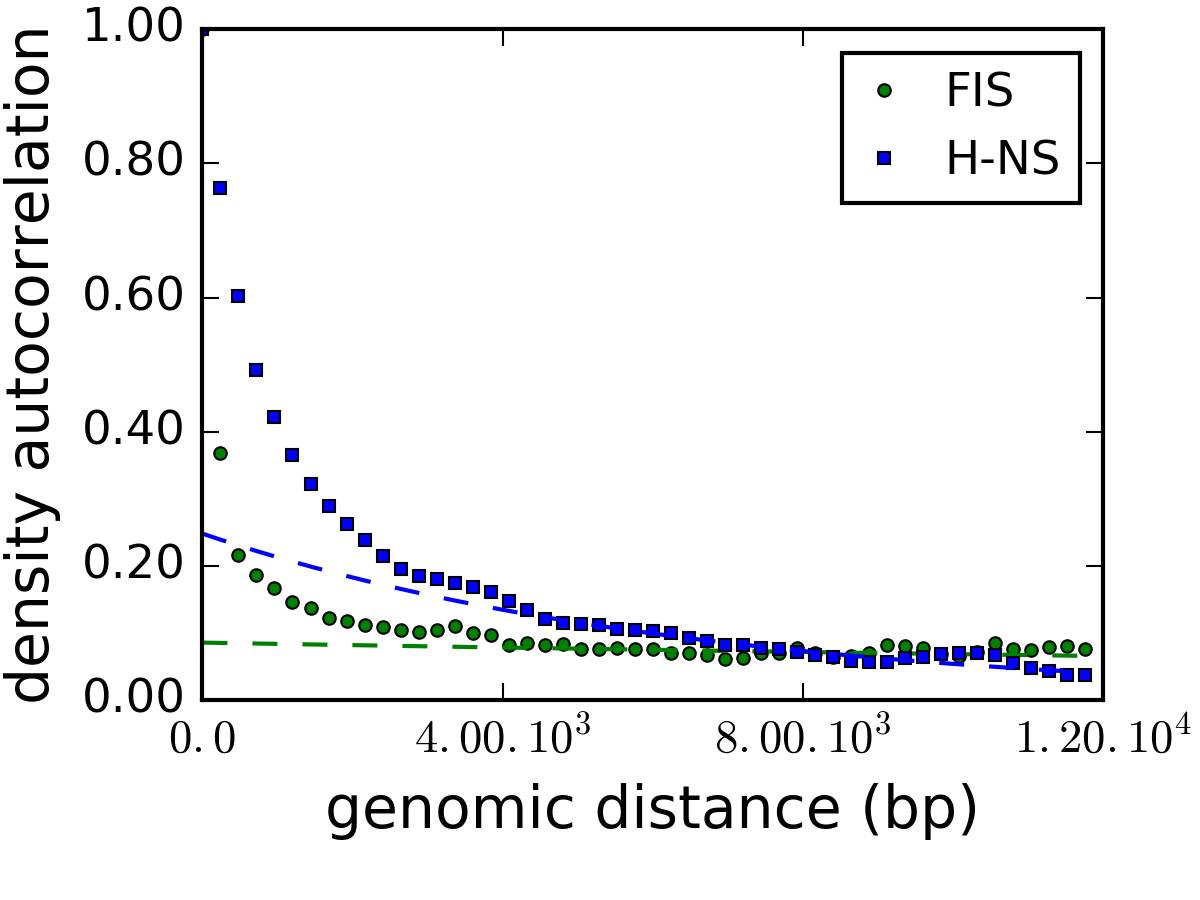}}%
  \quad
  \subfloat[]{\label{fig:naps_chipseq_autocorrelation_log:both} \includegraphics[width= 0.40 \textwidth]{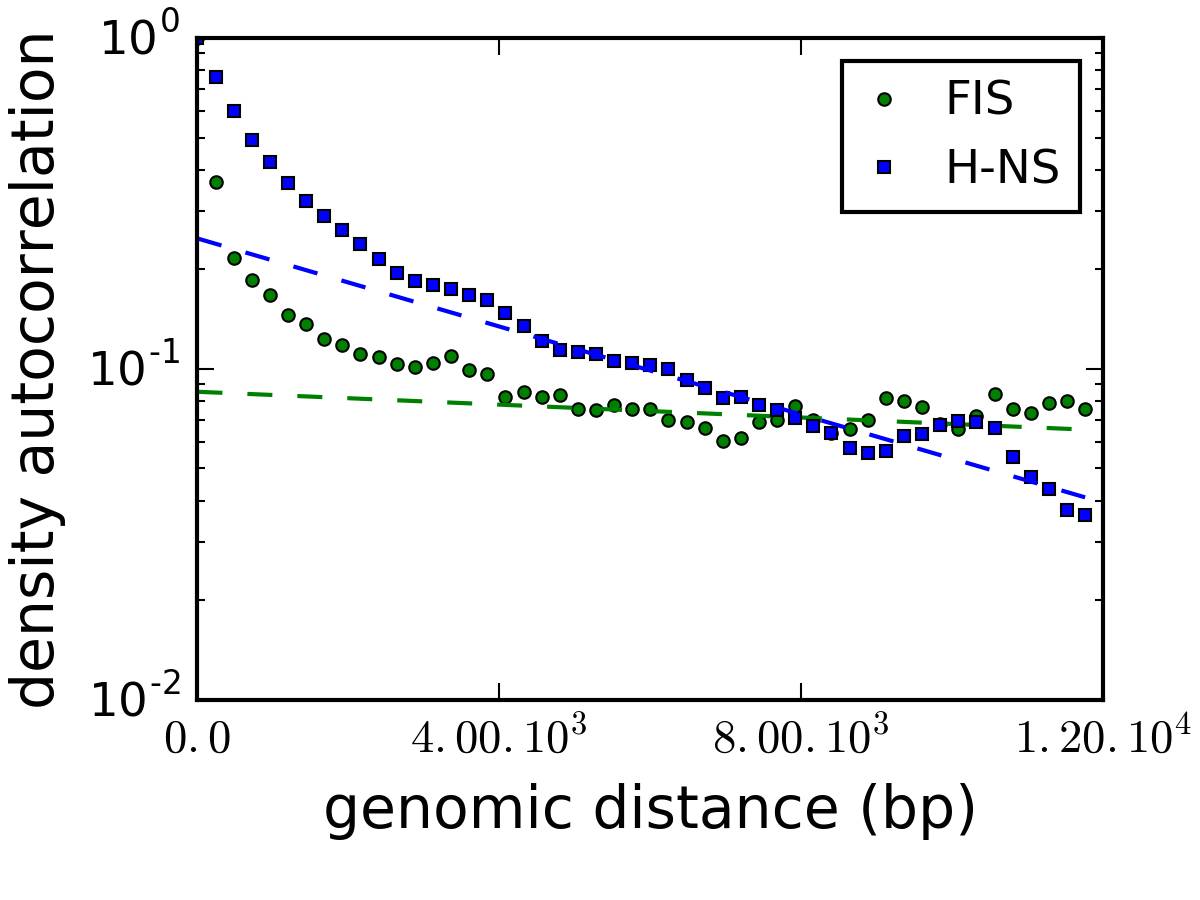}}%
  \caption{Auto-correlation of the \chipseq counts for FIS and H-NS. \protect\subref{fig:naps_chipseq_autocorrelation:both} Standard scale. \protect\subref{fig:naps_chipseq_autocorrelation_log:both} Logarithmic scale.}
  \label{fig:naps_chipseq_autocorrelation}
\end{figure}

We cross-validated our results by computing the auto-correlation function for the \chipseq counts $n(s)$:
\begin{equation}
  C(s) =  \left\langle n(s+s_0)n(s_0) \right\rangle - \left\langle n(s+s_0) \right\rangle \left\langle n(s_0) \right\rangle,
  \label{eq:naps_chipseq_autocorrelation_counts}
\end{equation}
where $s_0$ can be any coordinate on the genome. For both H-NS and FIS, we obtain that $C(s)$ decays exponentially for genomic distances larger than a few \SI{}{\kilo bp} (\cref{fig:naps_chipseq_autocorrelation}). For shorter genomic distances, it is clear that $C(s)$ does not have exponential variations, as can be seen in logarithmic scale. For genomic distances larger than \SI{5}{\kilo bp}, the auto-correlation functions seems to collapse on an exponential curve.

\subsection{Conclusion}
In this section, we have exploited an analogy between a Poisson stochastic point process and the insertion of NAPs binding sites in the genome throughout evolution. We found that the repartition of NAPs binding sites in \ecoli genome presents very few correlations at long genomic distances. This suggests that any regulatory mechanism induced by an architectural change of the chromosome following NAPs binding can be investigated for chromosome chunks of length $d<d^*$.

Using \chipseq data for H-NS and FIS, we found $d^* \approx \SI{3}{\kilo bp}$. We conclude that it is sufficient to study H-NS and FIS in regions with a size of a few $d^*$ (typically $N \approx \SI{10}{\kilo bp}$) because it is unlikely that regulatory mechanisms involving architectural changes exist at larger genomic distances. In particular, this suggests that BD simulations at a relatively low resolution (\textit{i.e.} with a monomer size $b \approx \SI{10}{bp}$) may be used to investigate the effect of NAPs on the chromosome architecture, and infer regulatory effects.

\section{Model for the regulatory effect of H-NS}
As seen in the previous section, we may assume that it is sufficient to model the effect of H-NS on the chromosome architecture on relatively short scales. In this section, we investigate a view of genetic regulation in which short H-NS binding regions of the DNA experience transitions between an open state, in which the chromatin is accessible to RNAP, and a closed (or looped) state which prevents RNAP binding and therefore represses transcription.

\subsection{Experimental evidences of H-NS loops}
Atomic-force microscopy experiments (AFM) have shown that H-NS induces the formation of DNA filaments (\vref{fig:hns_loops_experiments:filaments}) \cite{Dame102002}. These filaments result from the action of H-NS, which can bridge two neighboring DNA sequences together. Other experiments were performed with a \textit{rrnB} promoter (the main promoter of one of the seven ribosomal RNA synthesis genes in \textit{Escherichia coli}) inserted in the middle of a \SI{1200}{bp} long linear DNA fragment \cite{Dame012002}. First, when introduced into the medium, RNAP appeared to bind the promoter, and in most cases induced a local curvature of the DNA fiber. Second, the introduction of H-NS resulted in the formation of DNA filaments in the vicinity of the promoter. More accurately, the structure obtained can be compared to a hairpin loop. Remarkably in many cases, RNAP appeared to be trapped at the apex of such hairpin loops (\cref{fig:hns_loops_experiments:rnap_trapping}). Such RNAP-trapping mechanism is thought to silence transcription of the gene under the control of this promoter, but also to enable quick transcription restart once the H-NS mediated repression is relieved because RNAP does not need to be recruited from the bulk.

It was also demonstrated with high-throughput \chipseq techniques that H-NS binding sites are clustered in regions, or tracks, of varying size $L$ \cite{Kahramanoglou2011}. Regions of size $L>\SI{1000}{bp}$ appeared to correspond to genes with very low transcription levels. On the contrary, in short regions of size $L<\SI{1000}{bp}$ transcription levels were not so low. Actually, many of these short regions appeared to fall within the promoters. Following earlier discussions, we may assume that large regions are transcriptionnally silent because of the formation of DNA filaments by H-NS. In short regions however, the overall force to maintain the DNA hairpin loop is weaker, and consequently, DNA filaments may disassemble under the effect of perturbations, leading to the removal of one or several H-NS proteins from the complex. Such perturbations may result from the binding to DNA of a protein with higher affinity, or simply from the entropic fluctuations of the DNA polymer. Altogether, the disruption of DNA hairpin loops in binding regions of short lengths under the effect of thermal fluctuations or by the binding of an external protein, may lead to the derepression of the downstream genes.

In summary, we propose that H-NS regulatory functions have derived from its original sentinel role. Long binding regions are strongly repressed because they are confined in DNA filaments where RNAP cannot bind. In shorter regions, the looped state is more sensitive to perturbations and can undergo transitions to an open state, that can be used to modulate the expression of genes. This suggests that there is a characteristic length for the size of H-NS binding regions which separates the two regimes. In the sequel, we present a simplified model for the underlying physical mechanism in which such a characteristic length naturally emerges.

\subsection{Model for the formation of H-NS loops}
\subsubsection{Free energy for hairpin configurations}
As usual, we model a chunk of chromosome by a discrete polymer chain of size $N$. The chain consists of $N+1$ monomers with coordinates $\vec{r}_{0},\vec{r}_{1},\dots,\vec{r}_{N}$; $N$ bonds defined as $\vec{u}_i=\vec{r}_i - \vec{r}_{i-1}$; and $N-1$ joints characterized by an angle $\alpha_i$ such that $\cos{\alpha_i} = \vec{u}_i \cdot \vec{u}_{i+1}$. Furthermore, we use the WLC model, with internal energy given by:
\begin{equation}
  \beta U_{b}\left[ \left\{ \vec{r}_i \right\} \right] = l_p \sum \limits_{i=1}^{N-1}\left( 1 - \cos{\alpha_i} \right), \qquad \Vert \vec{u}_i \Vert = 1
  \label{eq:naps_kratky_porod}
\end{equation}
where $\beta=(k_B T)^{-1}$ is the inverse temperature and $l_p$ is the persistence length.

In first approximation, we assume that the bridging effect of H-NS can be modeled implicitly by introducing effective interactions between DNA monomers. Hence we consider a chain made of $2P$ monomers and a "sticky" sequence of $2(L+1)$ monomers in its center (\cref{fig:hairpin:noloop}). This constitutes a chain of size $N=2P+2L+1$. We are interested in the equilibrium of this system, and in particular in the probability of the configurations in which the sticky sequence is paired with itself. For simplicity, we assume that the configurations space is reduced to two configurations:
\begin{itemize}
  \item Open (o): the chain is free; in particular the sticky sequence is not paired.
  \item Closed (c): the sticky sequence is paired in a hairpin structure of length $L$.
\end{itemize}

Hence the partition function of this system reduces to a two-state model:
\begin{equation}
  Z = Z_o + Z_c.
  \label{eq:partfunc_hairpin}
\end{equation}

In the rest of this section, we shall use notations introduced in \cref{sec:model_chromosome_description} to describe a discrete worm-like chain. In particular, let us introduce again the entropy per monomer $z$ (see \vref{eq:wlc_discrete_partfunc}), and the chain propagator $q_N(\vec{u})$ (see \vref{eq:wlc_discrete_propag_reducedproba}). The first term in the right-hand side (r.h.s.) of \cref{eq:partfunc_hairpin} is simply the partition function of the free chain, \textit{i.e.}
\begin{equation}
  Z_o = 4 \pi z^{2P + 2L}.
\end{equation}

The second term is obtained as follows. We divide the polymer in three pieces (see \cref{fig:hairpin:noloop}) which are: the paired sequence of size $2L+1$ and the two dangling extremities of length $P$. The partition function for the closed configuration is then obtained by summing the Boltzmann weights of two free polymer chains of length $P$, plus the Boltzmann weight corresponding to a polymer folded in a hairpin configuration with direction $\vec{u}$. It can be expressed by using the Chapman-Kolmogorov structure of the worm-like chain propagator $q_n(\vec{u})$:
\begin{equation}
  \begin{aligned}
    Z_c &= 2 \pi \int \ud{\vec{u}_{P+1}} \ud{\vec{u}_{P+2L+1}}\, q_P(\vec{u}_{P+1}) q_P(\vec{u}_{P+2L+1}) \delta\left( \vec{u}_{P+1} + \vec{u}_{P+2L+1} \right) e^{-\beta E_L} \\
      &= 2 \pi \int \ud{\vec{u}} q_P(\vec{u}) q_P(-\vec{u}) e^{-\beta E_L},
  \end{aligned}
\end{equation}
where $E_L$ is the enthalpic contribution of the hairpin configuration. We considered this enthalpic gain to be extensive in the hairpin length and proportional to the pairing energy $-2 \varepsilon$, and the enthalpic cost comes from the chain bending rigidity. Thus we have $\beta E_L = 2l_p - 2L \varepsilon$. Note that without loss of generality, we have considered that the apex of the hairpin has a double-elbow structure with $\alpha_{P+L}=\alpha_{P+L+1}=\pi/2$. There is also a factor $2 \pi$ due to the invariance by rotation around the hairpin axis. Finally, we obtain for the Boltzmann weight for the close state:
\begin{equation}
  Z_c = \frac{1}{2} (4 \pi)^2 z^{2P} e^{-2 l_p + 2L \varepsilon}.
\end{equation}

The closed configuration will be dominant at thermal equilibrium if its free energy is lower than the open configuration free energy:
\begin{equation}
  - \ln{Z_c} < - \ln{Z_o} \quad \Leftrightarrow \quad L > L^* = \frac{l_p - \ln{\sqrt{2 \pi}}}{\varepsilon - \ln{z}}.
  \label{eq:hairpin_critical_length}
\end{equation}

Therefore, a characteristic length naturally arises that separates a regime in which the closed configuration prevails at equilibrium, for $L>L^*$, from another regime in which the open state prevails, for $L<L^*$. The precise value of $L^*$ results from a competition between the chain bending rigidity, the pairing energy and the conformational entropy per monomer, through $l_p$, $\varepsilon$ and $\ln{z}$ respectively.

The probabilities for the two states at equilibrium are then simply given by
\begin{align}
  \begin{aligned}
    &\proba{o} &=& \quad \frac{Z_o}{Z_o + Z_c} &=& \quad \frac{1}{1+\Upsilon^{1-L/L^*}} \\
    &\proba{c} &=& \quad \frac{Z_c}{Z_o + Z_c} &=& \quad \frac{\Upsilon^{1-L/L^*}}{1+\Upsilon^{1-L/L^*}} \\
  \end{aligned}
  \qquad
    \text{with} \qquad \Upsilon = 2 \pi e^{-2 l_p}
  \label{eq:hairpin_probabilities}
\end{align}

When increasing the length of the sticky sequence, the probability of the closed configuration increases progressively from $\proba{c}=0$ to $\proba{c}=1$. At $L=L^*$, the two states have the same probabilities $\proba{o}=\proba{f}=1/2$. Note that this is not a phase transition because the cross-over between the two regimes is continuous. However, this result breaks down in the limit $l_p \to \infty$. In that case, the probability of the closed configuration jumps abruptly from $\proba{c}=0$ to $\proba{c}=1$ at $L=L^*$ and it is a phase transition.

\begin{figure}[!htbp]
  \centering
  \subfloat[]{\label{fig:hairpin:noloop}
  \includegraphics[width= 0.45 \textwidth]{polymer_hairpin_strict.tex}%
  }
  \quad
  \subfloat[]{\label{fig:hairpin:loop}
  \includegraphics[width= 0.45 \textwidth]{polymer_hairpin_loop.tex}%
  }
  \caption{\protect\subref{fig:hairpin:noloop} Bridged configuration with no loop (hairpin). \protect\subref{fig:hairpin:loop} Bridged configuration with a loop.}
  \label{fig:hairpin}
\end{figure}

\subsubsection{Free energy for hairpin-plus-loop configurations}
We have just studied the case of a polymer chain containing a unique ``sticky'' region of size $2L+1$. We now consider the case in which a chain contains two sticky regions, each of size $L$, and separated by a linker of $M$ non-sticky monomers (\cref{fig:hairpin:loop}). As before, the rest of the chain is made of $2P$ monomers. The partition function of the closed configuration now reads
\begin{equation}
  Z_c = \int \ud{^2 \vec{u}} q_P(\vec{u}) q_P(\vec{-u}) e^{2 L \varepsilon} G_M(\vec{u})
\end{equation}
where $G_M(\vec{u})$ is the Boltzmann weight for the linker:
\begin{equation}
  G_M(\vec{u}) =  \int \ud{^2 \vec{u}_{\perp}} T^{M+2}\left( -\vec{u} \mid \vec{u} \right) \delta \left(  \vec{u}_{P+L+1} + \dots + \vec{u}_{P+L+M+1} - \vec{u}_{\perp} \right) \delta\left( \vec{u} \cdot \vec{u}_{\perp} \right)
  \label{eq:hairpin_model_gm}
\end{equation}
where $\vec{u}_{\perp}$ is by construction a unit vector perpendicular to $\vec{u}$, and $T$ is the transfer matrix used to describe a worm-like chain (see \vref{eq:wlc_discrete_transfer_matrix}). With the same arguments as before, we obtain a more general expression for the characteristic length in \cref{eq:hairpin_critical_length}:
\begin{equation}
  L^* = - \frac{1}{2} \frac{\ln{G_M}}{\varepsilon - \ln{z}}
  \label{eq:hairpinloop_critical_length}
\end{equation}

In particular, for a chain with no linker, $M=0$, we have $G_0=2 \pi \exp{(-2 l_p)}$ and \cref{eq:hairpinloop_critical_length} reduces to the previous expression of \cref{eq:hairpin_critical_length}.

\subsection{Investigation with Brownian Dynamics}
We have obtained the existence of a characteristic length for H-NS binding regions from a very simple polymer model. In particular we have considered implicit interactions so far. To cross validate our result, we now present a BD model with explicit proteins and compute the probabilities of the open and close states.

\subsubsection{Model for DNA and H-NS}
Following the results of \cref{sec:naps_scale_modeling}, we model the chromosome at a resolution close to the naked DNA fiber. In particular, we take monomers of size $b \approx \SI{10}{bp} = \SI{3.3}{\nm}$. It is also close to the size of one H-NS binding site. Thus we model a chunk of chromosome of size \SI{4}{\kilo bp} as a beads-on-string polymer with $N+1=400$ monomers and persistence length $l_p=15 \, b$. As usual we use a FENE potential to model the bonds elasticity, a Kratky-Porod potential to model the chain bending rigidity and a truncated Lennard-Jones potential to model excluded volume. A summary of the potentials considered for BD simulations is given in \cref{tab:naps_hnsloops_numerical_model}.

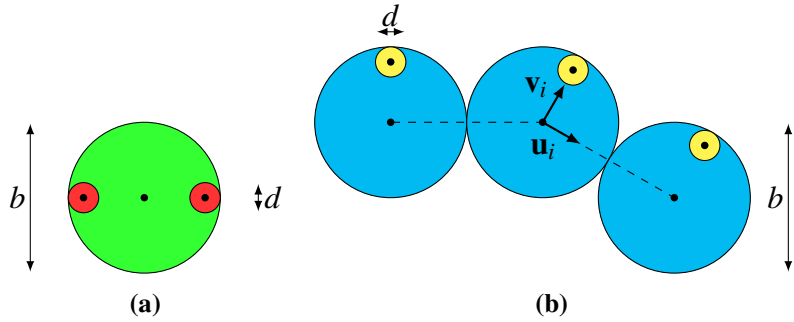
\begin{figure}[!htbp]
  \centering
  \subfloat[]{\label{fig:naps_hns_model_sketch:hns}%
  \begin{tikzpicture}[thick, nodes = {align = center}, every node/.style={inner sep=0,outer sep=0}, >=latex]
  \node[] (X') at (-1.0,0){};
  \node[] (X) at (+1.0,0){};
  \node[] (Y') at (0,-1.0){};
  \node[] (Y) at (0,+1.0){};

  \node[] (O) at (0,0){};
  \begin{scope}[on background layer]
  	\draw[fill=green!80!white] (O) circle [radius=1];
  \end{scope}
  \draw[] plot [only marks, mark size=1.0, mark=*] coordinates {(O)};

  \node[] (O1) at (-0.8,0){};
  \node[] (O2) at (+0.8,0){};
  \begin{scope}[on background layer]
    \draw[fill=red!80!white] (O1) circle [radius=0.2];
    \draw[fill=red!80!white] (O2) circle [radius=0.2];
  \end{scope}
  \draw[] plot [only marks, mark size=1.0, mark=*] coordinates {(O1)};
  \draw[] plot [only marks, mark size=1.0, mark=*] coordinates {(O2)};

  \node[] (H1) at (-1.5,-1){}; \node[] (H2) at (-1.5,1){};
  \node[] (K1) at (1.5,-0.2){}; \node[] (K2) at (1.5,0.2){};
  \draw[thin,<->] (H1) -- (H2) node[midway,left,black,outer sep=2pt] {$b$};
  \draw[thin,<->] (K1) -- (K2) node[midway,right,black,outer sep=2pt] {$d$};
\end{tikzpicture}}%
\quad
\subfloat[]{\label{fig:naps_hns_model_sketch:dna}%
  \begin{tikzpicture}[thick, nodes = {align = center}, every node/.style={inner sep=0,outer sep=0}, >=latex]
  \node[] (O1) at (0,0){};
  \node[] (O2) at (2,0){};
  \node[] (O3) at ({2+2*sqrt(3)/2},{-2*0.5}){};
  \begin{scope}[on background layer]
  	\draw[fill=cyan!80!white] (O1) circle [radius=1];
  	\draw[fill=cyan!80!white] (O2) circle [radius=1];
  	\draw[fill=cyan!80!white] (O3) circle [radius=1];
  \end{scope}
  \draw[] plot [only marks, mark size=1.0, mark=*] coordinates {(O1)};
  \draw[] plot [only marks, mark size=1.0, mark=*] coordinates {(O2)};
  \draw[] plot [only marks, mark size=1.0, mark=*] coordinates {(O3)};
  \draw[thin,dashed] (O1) -- (O2) -- (O3);

  \node[] (M1) at (0.0,0.8){};
  \node[] (M2) at ({2.0+0.8*0.5},{0.8*sqrt(3)/2}){};
  \node[] (M3) at ({2+2*sqrt(3)/2+0.8*0.5},{-2*0.5+0.8*sqrt(3)/2}){};
  \begin{scope}[on background layer]
    \draw[fill=yellow!80!white] (M1) circle [radius=0.2];
    \draw[fill=yellow!80!white] (M2) circle [radius=0.2];
    \draw[fill=yellow!80!white] (M3) circle [radius=0.2];
  \end{scope}
  \draw[] plot [only marks, mark size=1.0, mark=*] coordinates {(M1)};
  \draw[] plot [only marks, mark size=1.0, mark=*] coordinates {(M2)};
  \draw[] plot [only marks, mark size=1.0, mark=*] coordinates {(M3)};

  \node[] (H1) at ({2+2*sqrt(3)/2+1.5},{-2*0.5-1}){}; \node[] (H2) at ({2+2*sqrt(3)/2+1.5},{-2*0.5+1}){};
  \node[] (K1) at (-0.2,1.2){}; \node[] (K2) at (0.2,1.2){};
  \draw[thin,<->] (H1) -- (H2) node[midway,left,black,outer sep=2pt] {$b$};
  \draw[thin,<->] (K1) -- (K2) node[midway,above,black,outer sep=2pt] {$d$};

  \node[] (P3) at ({2+0.6*sqrt(3)/2},-0.6*0.5){};
  \node[] (Q3) at ({2+0.5*0.6},{0.6*sqrt(3)/2}){};
  \draw[thick,->] (O2) -- (P3) node[midway,below left,outer sep=2pt] {$\vec{u}_i$};
  \draw[thick,->] (O2) -- (Q3) node[midway,above left,outer sep=2pt] {$\vec{v}_i$};
\end{tikzpicture}}
\caption{Numerical model for H-NS binding to DNA. \protect\subref{fig:naps_hns_model_sketch:hns} Model for H-NS as a divalent protein. \protect\subref{fig:naps_hns_model_sketch:dna} Model for monovalent DNA monomers binding to H-NS.}
\label{fig:naps_hns_model_sketch}
\end{figure}

We also introduce the H-NS protein as a sphere with same dimensions as the DNA monomers. But in reality, H-NS is a divalent protein, with two DNA binding sites roughly making a \SI{180}{\degree} angle with the center of mass of the protein. In order to reproduce this anisotropy in the binding to DNA, we introduce two tiny spheres of diameters $d$, tangent to the protein sphere of diameter $b$ (\cref{fig:naps_hns_model_sketch:hns}). Note that such spheres are fake atoms that we have introduced only to construct a numerical model for H-NS bivalency. This design was inspired by previous work \cite{Brackley36052013}. In the sequel, we have taken $d=0.2 b$.

\begin{figure}[htbp]
  \centering
  \includegraphics[width=0.5 \textwidth]{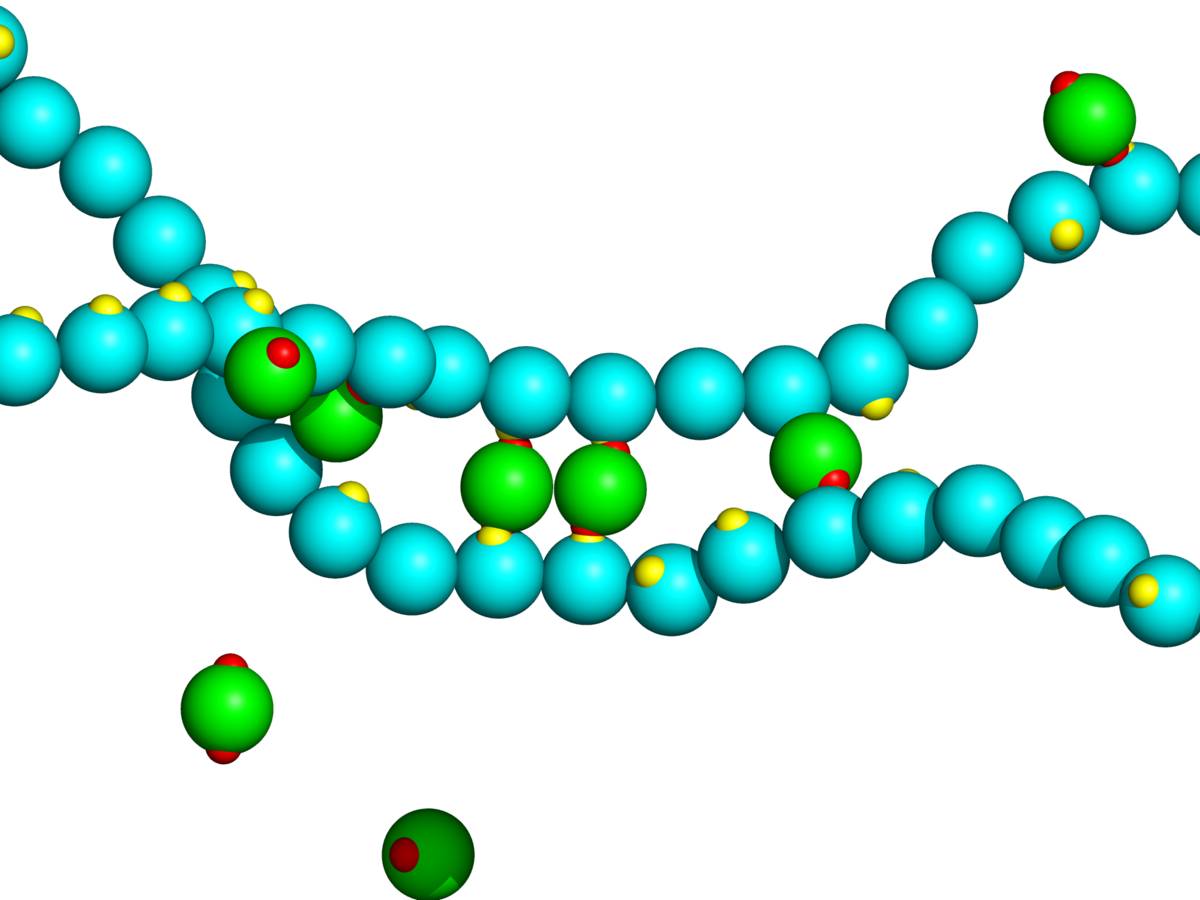}
  \caption{Snapshot of Brownian dynamics simulation implementing the model of bivalency for H-NS (green with two red binding sites) and monovalency for H-NS binding sites (cyan with one yellow binding site). Fake atoms are represented with an exaggerated diameter in order to be seen.}
  \label{fig:divalent_model_hns}
\end{figure}

Similarly, H-NS binding sites can only bind one H-NS protein at a time. Thus, for DNA monomers able to bind H-NS, we also introduce a fake atom of diameter $d$ (\cref{fig:naps_hns_model_sketch:dna}). Let us denote $\vec{v}_i$ the unit vector giving the direction to the H-NS binding site from the DNA monomer center with coordinates $\vec{r}_i$. Incidentally, the covalent bonds between H-NS and DNA are formed with bases making contacts through the major groove. However, at the scale considered, $b$ corresponds approximately to one helical turn of the major groove, which is of \SI{11}{bp}. Due to DNA torsional stiffness, H-NS binding sites on consecutive monomers should point approximately in the same direction. In order to reproduce this property in our simulations we had to introduce additional potentials. First, we introduce a bending rigidity potential to favor configurations in which the H-NS binding site points in a direction orthogonal to the bond direction. More explicitly we introduce the potential:
\begin{equation}
  U_{\perp} = k_{\perp} \sum \limits_{i=0}^{N} (1 - \sin{\gamma_i}), \qquad \sin{(\gamma_i)} = \vec{u}_i \cdot \vec{v}_i,
  \label{eq:gem_naps_model_hns_orthogonal}
\end{equation}
in which it is seen that the penalty is a minimum when $\gamma_i=\pi/2$. This ensures that the H-NS binding site remains on the surface of the tube of diameter $b$ containing the DNA monomers (\textit{i.e.} the DNA fiber). Second, we need to ensure that consecutive H-NS binding sites tend to point in the same direction. This is achieved by introducing a dihedral potential:
\begin{equation}
  U_{\sslash} = k_{\sslash} \sum \limits_{i=1}^{N-1} (1 - \cos{\varphi_i}),
  \label{eq:naps_model_hns_dihedral}
\end{equation}
where $\varphi_i$ is the azimuthal angle between H-NS binding sites for the monomers $i$ and $i-1$ in the spherical coordinate system whose zenith direction is $\vec{u}_i$. It is a dihedral potential because computing the angle $\varphi_i$ involves two DNA monomers with their respective H-NS binding sites, \textit{i.e.} four atoms. Altogether, the combination of these two potentials mimics the DNA torsional rigidity that tends to maintain consecutive H-NS binding sites aligned. In our simulations, we chose $k_{\perp} = 50 \, k_B T$ and $k_{\sslash} = 1 \, k_B T$. A snapshot of BD simulation implementing this model is shown in \cref{fig:divalent_model_hns}.

Finally, the attractive interaction between H-NS and DNA must be a Coulombic screened interaction, which can be well represented by a Yukawa potential:
\begin{equation}
  U_{coul}(r) = A \frac{e^{-r/r_{d}}}{r},
  \label{eq:naps_yukawa_potential}
\end{equation}
where $A$ is a scale in $k_B T$ measuring the strength of the interaction and $r_d$ is the range of the interaction. Since the Debye-H\"uckel length is of the order of \SI{1}{\nm} in the cytosol \cite{Kunze43892000}, the range of the interaction is small and essentially reduces to the dimensions of the objects interacting together. In our simulations, we took a cutoff $r_{d}=0.3 b \approx \SI{1}{\nm}$ (\cref{tab:naps_hnsloops_numerical_model}).

We considered a polymer of $N+1=400$ monomers and $P=100$ spheres in a cubic volume of size $80 b$ with periodic boundary conditions. A relaxation run was performed first for \num{e7} iterations in order to loose the memory of the initial configuration. We then performed a run of \num{e6} iterations with a soft pair potential to remove overlaps between atoms. Finally, we performed a run of \num{2e8} iterations with all interactions in \cref{tab:naps_hnsloops_numerical_model}, with integration time step $dt=\num{7e-4}$, from which we extracted \num{200} evenly sampled configurations.

\begin{table}[!htbp]
  \centering
  \begin{tabular}[]{l l l}
    \textbf{Property} & \textbf{Model} & \textbf{Values} \\
    \hline \\
    Bonds elasticity &
    \makecell[l]{FENE potential: \\
      $U_{fene}=-\frac{k_{e} r_0^2}{2 b^2} \sum \limits_{i=1}^{N} \ln{\left(1 - \dfrac{u_i^2}{r_0^2} \right)}$ \\
      with: \\
      $\vec{u}_{i}=\vec{r}_i-\vec{r}_{i-1}$
    } &
    \makecell[l]{
      $
      \begin{aligned}
        & k_{e} &=& \quad 30 \, k_B T \\
        & r_0 &=& \quad 1.5 \, b
      \end{aligned}
      $
    } \\
    \hline \\
    Bending rigidity &
    \makecell[l]{Worm-Like chain potential:\\
      $U_{wlc}= \beta^{-1} \sum \limits_{i=1}^{N-1} \dfrac{l_p}{b} (1-\cos{\alpha_i})$
    } &
    $
    \begin{aligned}
    & l_p &=& \quad 15 \, b
    \end{aligned}
    $ \\
    \hline \\
    \makecell[l]{Excluded volume \\ interactions} &
    \makecell[l]{ Truncated Lennard-Jones potential:\\
      $U_{ev}(r) = V_{LJ}(r) - V_{LJ}(r^{th})$, if $r<r^{th}$ \\
      with: \\
      $V_{LJ}(r)= 4 \varepsilon \left[ \left( \frac{\sigma}{r} \right)^{12} - \left( \frac{\sigma}{r} \right)^{6} \right]$
    } &
    \makecell[l]{
      $
      \begin{aligned}
        & \sigma &=& \quad b \\
        & \varepsilon &=& \quad 1 \, k_B T \\
        & r^{th} &=& \quad 2^{1/6} \sigma
      \end{aligned}
      $
    }
    \\
    \hline
    \makecell[l]{Short-range coulombic  \\ interaction} &
    \makecell[l]{
      Yukawa potential: \\
      $
      \begin{aligned}
        U_{coul} (r) = A \frac{\exp{\left( -r / r_{d} \right)}}{r}
      \end{aligned}
      $ \\
      \phantom{r}
    }
    &
    \makecell{
      $
      \begin{aligned}
        & r_{d} &=& \quad 0.3 \, b
      \end{aligned}
      $
    }
    \\
    \hline
  \end{tabular}

  \caption{Numerical model to perform Brownian Dynamics simulation of H-NS/DNA bridges}
  \label{tab:naps_hnsloops_numerical_model}
\end{table}

\subsubsection{Detection of DNA/H-NS bridges}
\begin{figure}[!htbp]
  \centering
  \subfloat[]{\label{fig:naps_helices_detection:anti1}%
    \includegraphics[width=0.4 \textwidth,valign=c]{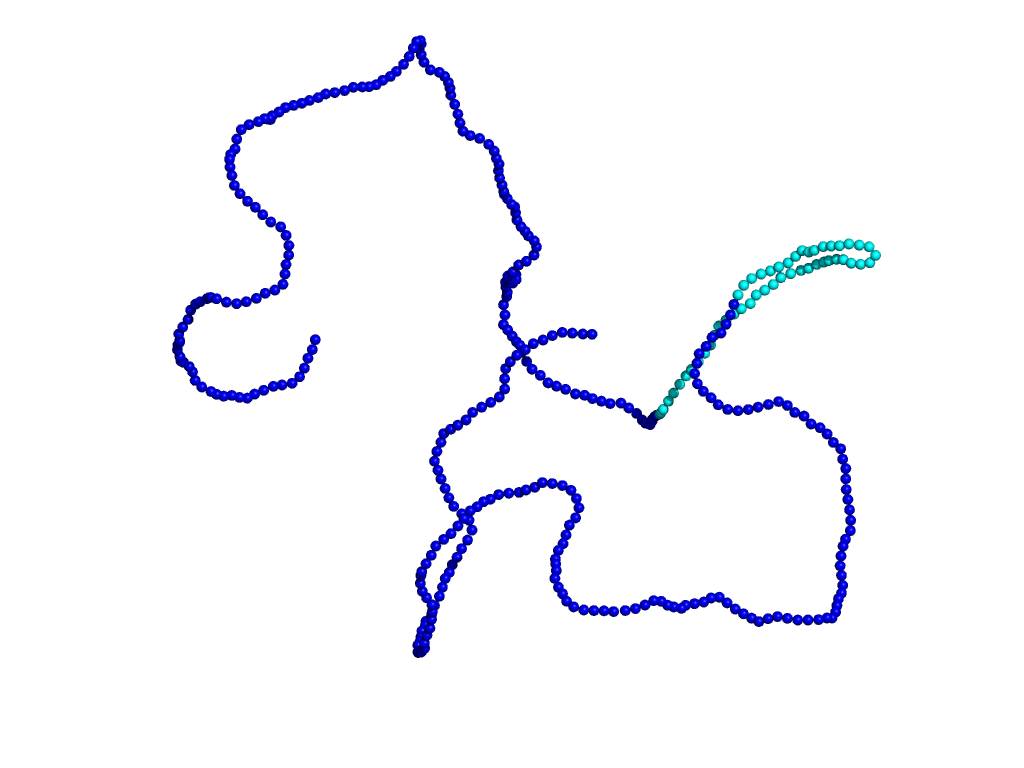}
    \quad
    \includegraphics[width=0.28 \textwidth,valign=c]{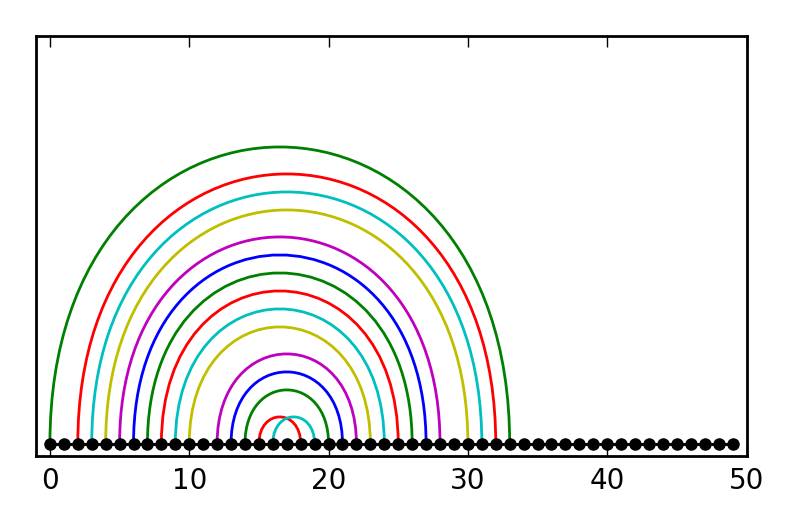}
    \quad
    \includegraphics[width=0.28 \textwidth,valign=c]{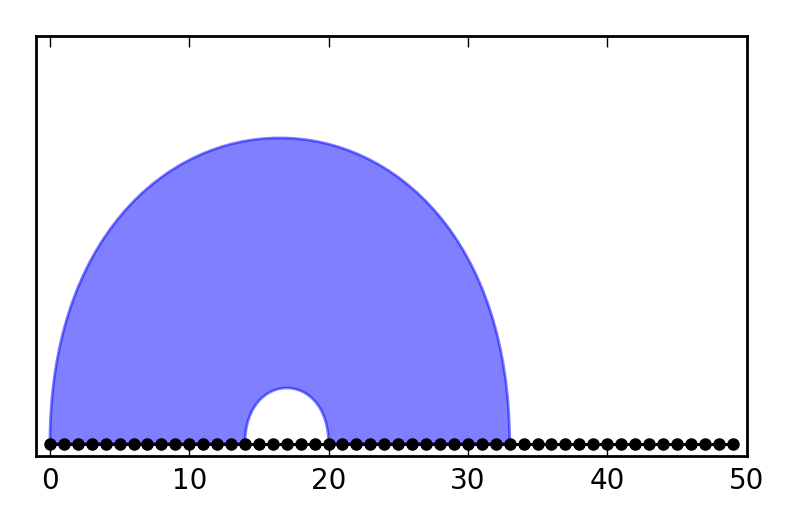}
  }
  \\
  \subfloat[]{\label{fig:naps_helices_detection:anti2}%
    \includegraphics[width=0.4 \textwidth,valign=c]{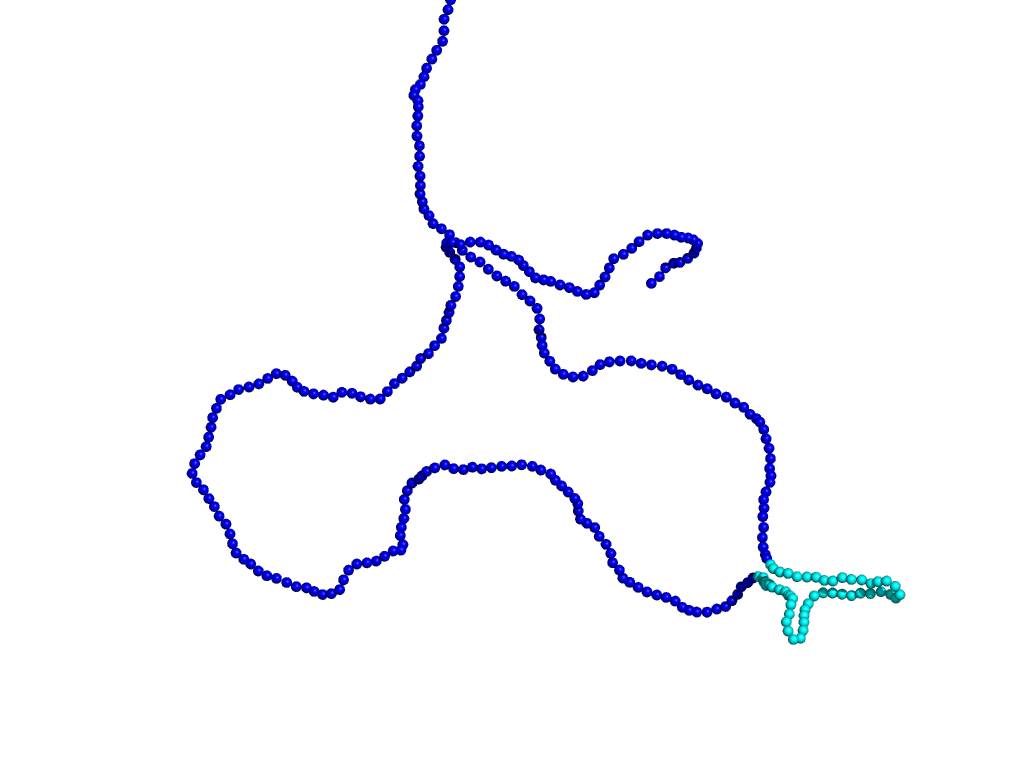}
    \quad
    \includegraphics[width=0.28 \textwidth,valign=c]{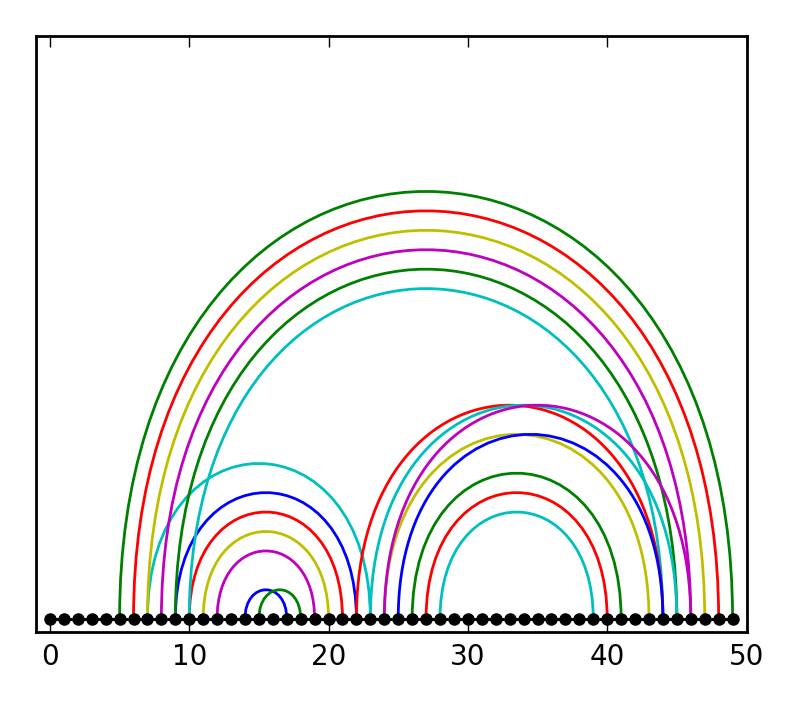}
    \quad
    \includegraphics[width=0.28 \textwidth,valign=c]{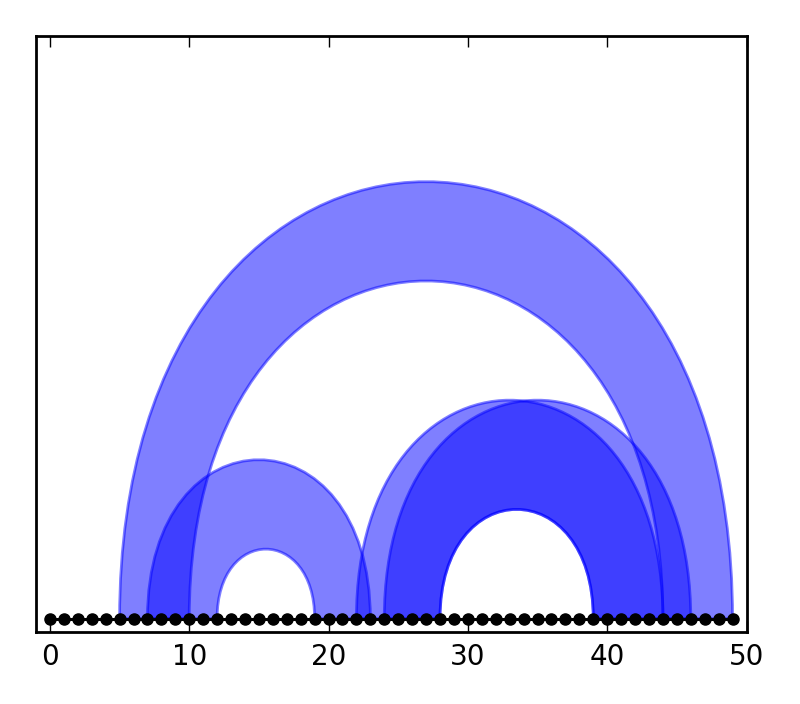}
  }
  \\
  \subfloat[]{\label{fig:naps_helices_detection:para1}%
    \includegraphics[width=0.4 \textwidth,valign=c]{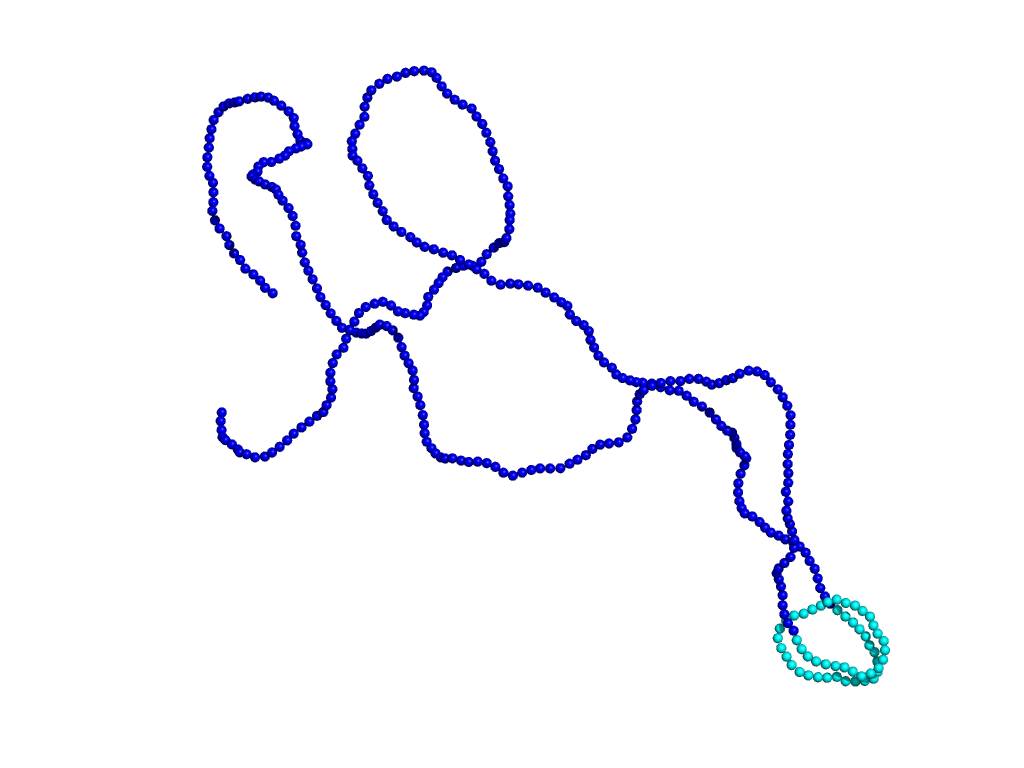}
    \quad
    \includegraphics[width=0.28 \textwidth,valign=c]{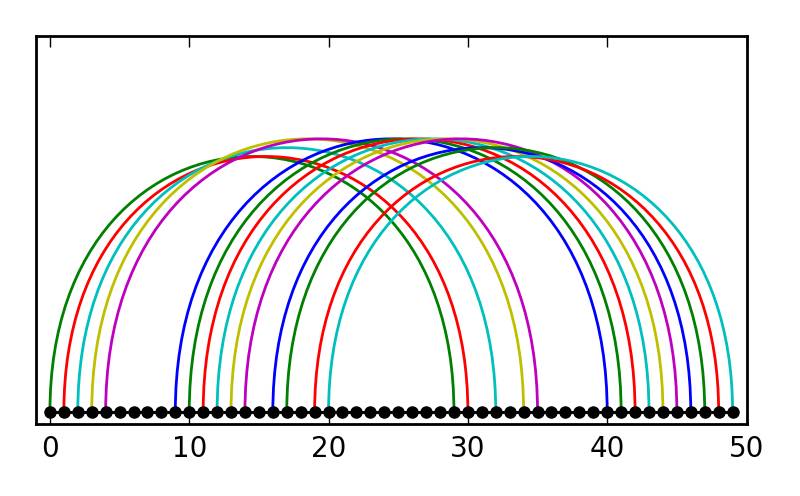}
    \quad
    \includegraphics[width=0.28 \textwidth,valign=c]{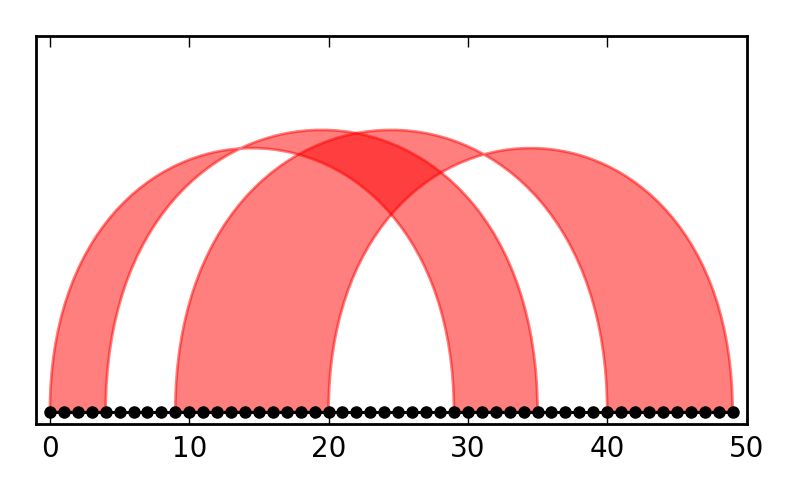}
  }
  \caption{Examples of helices detected from BD simulations. \protect\subref{fig:naps_helices_detection:anti1} One anti-parallel helix. \protect\subref{fig:naps_helices_detection:anti2} Several anti-parallel helices. \protect\subref{fig:naps_helices_detection:para1} Parallel helices arising from a toroidal shape. We show in the first column a snapshot of the conformation used. In the second column, we show the contact diagrams for the binding region in which monomers in contact are joined with an arc. In the third column, we represent the helices present in the configuration of length $H \ge 4$. All three conformations were taken from a single BD trajectory with $N=400$, $L=50$ and $A=8.0 \, k_B T$. H-NS spheres are not represented.}
  \label{fig:naps_helices_detection}
\end{figure}

The goal of our BD simulations is to assess the existence of DNA/H-NS bridges characterized in experiments and check whether they are maintained at equilibrium. Therefore, we need a strategy to detect such structures from BD trajectories. Following conventions from protein folding analysis, we define the contact diagram associated with a configuration of the binding region as a sequence of $L$ pairs $(i_e,j_e)$ with $e=1,\dots,L$ such that $j_e$ is the closest of the monomers in contact with $i_e$, and reciprocally $i_e$ is the closest of the monomers in contact with $j_e$. A contact is said to occur between monomers $i_e$ and $j_e$ when $\mid \vec{r}_{i_e} - \vec{r}_{j_e} \mid < \xi$, where $\xi$ is a threshold to be defined. In practical applications, we have taken $\xi=2.25$, and we have ignored contacts between nearest neighbors (up to third nearest neighbors). Such contact diagrams can be represented by drawing an arc for each pair of monomers in contact (\cref{fig:naps_helices_detection}).

Starting from a contact diagram, we will say that a subset of pairs $(i_e,j_e)$ with $e=1,\dots,H$ form a helix of length $H$ when there is no crossing between the arcs joining the monomers in contact. There are only two possibilities. First, when
\begin{equation}
  i_1 < i_2 < \dots < i_H < j_H < \dots < j_2 < j_1,
  \label{eq:naps_helix_anti}
\end{equation}
we will say that it is an anti-parallel (or ``-'') helix. Alternatively, when $j_H - i_H$ is sufficiently small, we may call such a helix a hairpin loop. Second, when
\begin{equation}
  i_1 < i_2 \dots < i_H < j_1 < j_2 < \dots j_H,
  \label{eq:naps_helix_parallel}
\end{equation}
we will say that it is a parallel (or ``+'') helix. Moreover, we impose that the contour distance between consecutive monomers of a helix is not too large. In other words, $\mid i_e - i_{e+1} \mid \le l_b$ and $\mid j_e - j_{e+1} \mid \le l_b$ where $l_b$ can be seen as the length of the smallest bubble allowed in a helix. In practical applications, we have taken $l_b=3$. Examples of helices detected in configurations obtained from BD simulations are shown in \cref{fig:naps_helices_detection}.

\begin{figure}[!htbp]
  \centering
  \includegraphics[width=0.5 \textwidth]{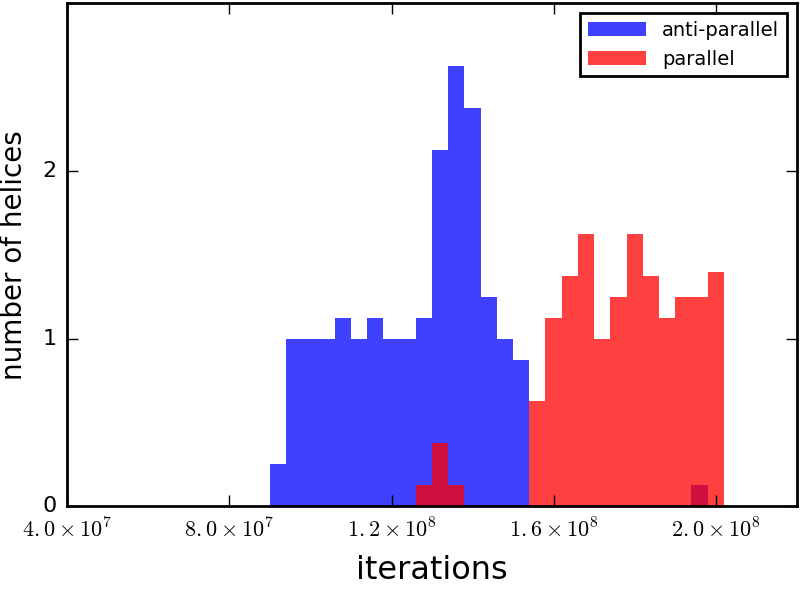}
  \caption{Time evolution of the number of helices. BD simulation performed with $N=400$, $L=50$ and $A=8.0 \, k_B T$.}
  \label{fig:naps_helices_time_evolution}
\end{figure}

For a given BD trajectory, we can compute $N_h^-(t)$ (resp. $N_h^+(t)$), which is the number of ``-'' helices (resp. ``+'' helices) present in the configuration at time $t$. These quantities display dynamical variations, as can be seen in \cref{fig:naps_helices_time_evolution}. We can compute the probability to have a ``+'' (or ``-'') helix at equilibrium as:
\begin{equation}
  \proba{\pm} = \langle \mathbbm{1}_{\mathbbm{R}_+^*}(N_h^{\pm}) \rangle,
  \label{eq:naps_helixpm_proba}
\end{equation}
where the brackets stand for a thermodynamical average performed over several configurations sampled from a BD trajectory. Finally, the probability for the existence of a helix at equilibrium is simply the sum of the probabilities of the two types of helices:
\begin{equation}
  \proba{\text{helix}} = \proba{+} + \proba{-}.
  \label{eq:naps_helix_proba}
\end{equation}

\subsubsection{Critical size of the binding region}
On the basis of these definitions, we can investigate numerically the existence of the characteristic length $L^*$ for the binding region. For several values of the binding region size $L$ we performed \num{50} independent BD simulations as detailed above. Then we computed the helix and hairpin probabilities according to \cref{eq:naps_helixpm_proba,eq:naps_helix_proba}. Note that to detect hairpin loops, we actually detected "-" helices with $j_H - i_H \le l_h=15$.

\begin{figure}[!htbp]
  \centering
  \subfloat[]{\label{fig:hns_simulation_critical:weaker} \includegraphics[width=0.45 \textwidth]{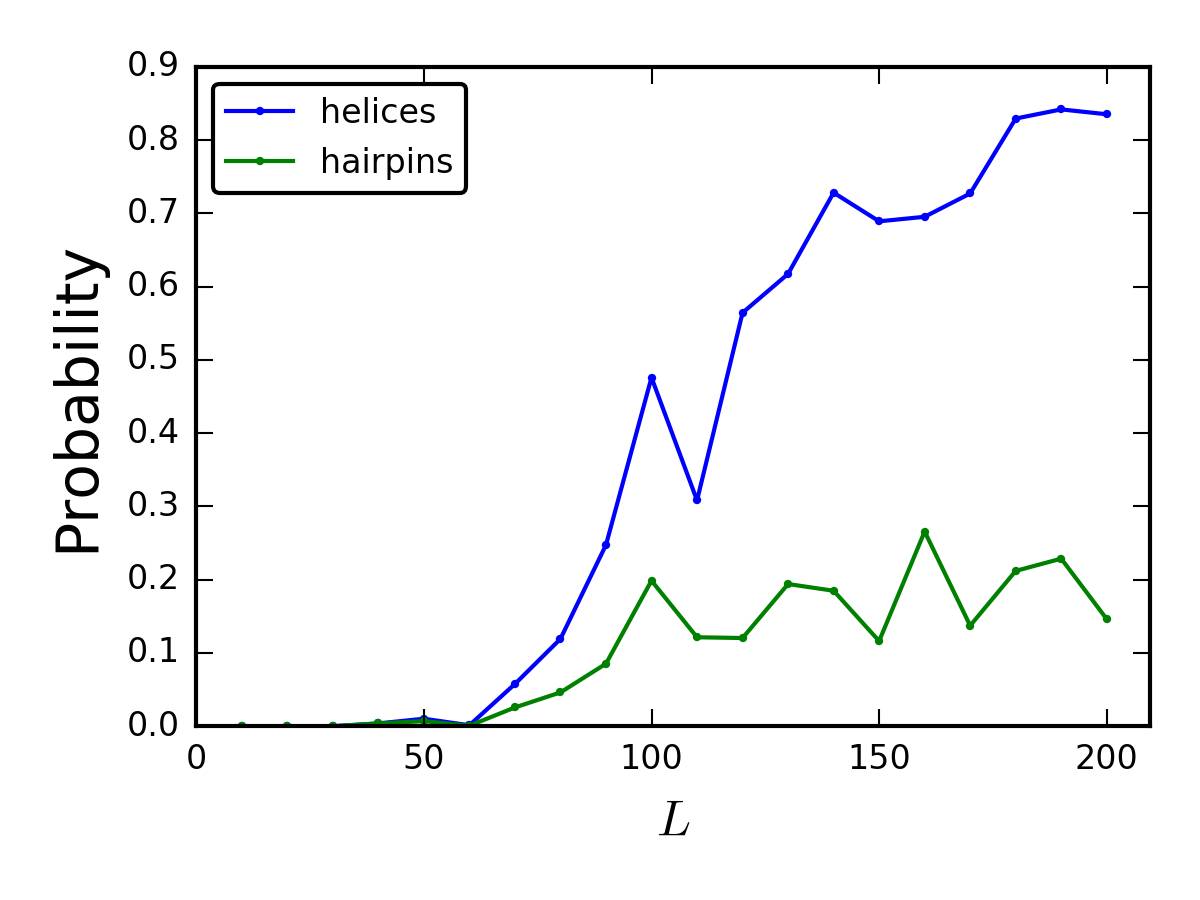}}%
  \quad
  \subfloat[]{\label{fig:hns_simulation_critical:stronger} \includegraphics[width=0.45 \textwidth]{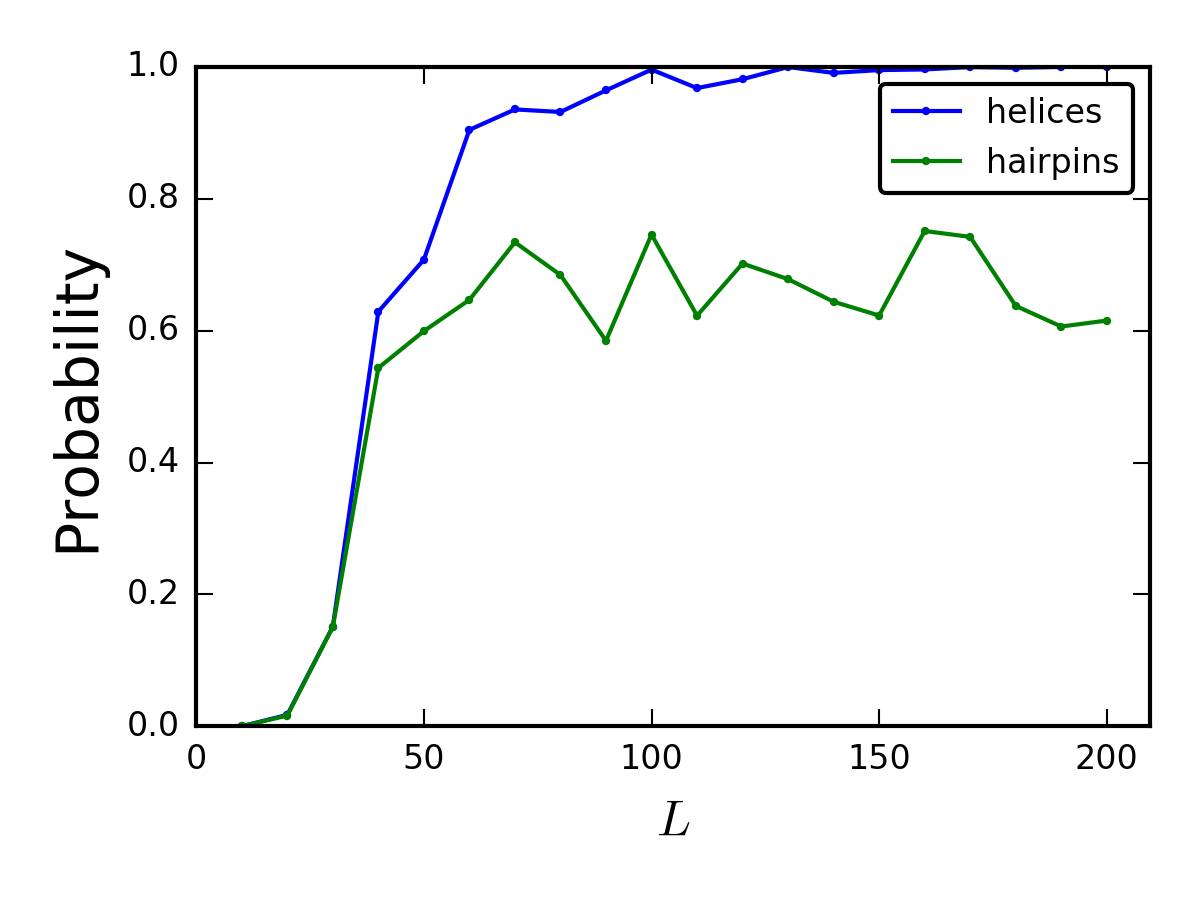}}
  \caption{Characterization of the existence of the critical length $L^*$ for the H-NS binding region. The probability for each points was obtained by performing an average as in \cref{eq:naps_helixpm_proba} over 50 BD trajectories computed independently. BD simulations were performed with $N=400$ and: \protect\subref{fig:hns_simulation_critical:weaker} $A=7.0 \, k_B T$ and \protect\subref{fig:hns_simulation_critical:stronger} $A=8.0 \, k_B T$. }
  \label{fig:hns_simulation_critical}
\end{figure}

In \cref{fig:hns_simulation_critical}, we represent these probabilities as a function of $L$ for $A=7.0 \, k_B T$ and $A= 8.0 \, k_B T$. In both case $L^*$ is clearly visible. For $L<L^*$ we have $\proba{\text{hairpin}} \approx 0$, hence such structures are not found at equilibrium. On the contrary for $L>L^*$, we have $\proba{\text{hairpin}} \lesssim 1$, hence such structures can be found at equilibrium. The fact that $\proba{\text{hairpin}}$ is close but not equal to one means however that these structures undergo dynamical fluctuations, as has already been seen in \cref{fig:naps_helices_time_evolution}. Finally, for $A=7.0 \, k_B T$ we have $L^* \approx 60$ whereas for $A=8.0 \, k_B T$ we have $L^* \approx 20 \sim l_p$. Therefore, we qualitatively recover that the critical binding region size decreases when the H-NS/DNA binding energy increases, as claimed in \cref{eq:hairpinloop_critical_length}.

\subsection{Conclusion}
In conclusion, we have investigated whether H-NS/DNA filaments characterized in AFM experiments do correspond to equilibrium structures. Our analysis was also grounded on the observation that H-NS binding regions follow a peculiar layout throughout the \ecoli genome, characterized with \chipseq assays. In particular, the fact that H-NS binding sites cluster in tracks means that binding regions can be found in the genome with various size $L$. Such a layout suggests that H-NS binding regions can fold in helices or even hairpins.

We have addressed this issue by first considering a simplified physical model in which only one binding region of size $L$ is present, and in which DNA monomers belonging to the binding region experience attractive interactions between themselves. It is an implicit model for the effect of H-NS proteins. In this framework, we have shown that a critical length for the binding region naturally emerges, separating an open regime in which the binding regions remains unfolded, from a looped regime in which the binding region folds into a hairpin conformation (possibly with a loop at the apex). We have then confirmed this prediction using BD simulations. In the latter, we have considered the H-NS proteins explicitly, and we have also designed a numerical model in order to reproduce the bivalency of the H-NS proteins and the monovalency of the H-NS binding sites.

\section{Discussion}
In this chapter, we have sought to relate directly transcription to the chromosome architecture. Naturally, we have started our investigation by reviewing the properties of NAPs, which are abundant proteins in the bacterial cell and well known for their role in structuring the chromosome. The high degree of conservation among bacteria species, combined with their prevalence over other transcription factors suggests that their role is not only architectural, but that instead they are involved in biological processes and for that reason have been selected during Evolution. In our review of the literature, we have seen that their effect on cell physiology includes some elements of transcription regulation. Therefore, NAPs seem to illustrate the connection that exists between chromosome architecture and transcription. We are convinced that this connection is dynamical and may rely on a feedback mechanism between these two components.

Much uncertainty remains however on the precise link between structures entailed by NAPs and the transcriptional response. In particular, the genomic scale of such regulatory mechanisms has remained elusive. On the basis of an analogy between a stochastic point process and the insertion of NAPs binding sites throughout Evolution, we have analyzed the presence of correlations in the NAPs binding sites insertions. Using \chipseq data available for FIS and H-NS, we have concluded that it is very unlikely that regulatory mechanisms selected by Evolution exist on genomic scales larger than \SI{10}{\kilo bp} in \ecoli bacteria. In particular, the distribution of the distances between NAPs binding sites appeared to have an exponential tail above this genomic scale.

Then we have studied in more details a typical structure induced by H-NS and well characterized in AFM experiments: DNA hairpin loops. The effect of these hairpin loops on the transcription have been thoroughly discussed in the existing literature, and appear to be at stake in the regulation of the important \textit{rrn} operon in \ecoli bacteria. In general, H-NS/DNA filaments lead to a repression of transcription by preventing RNAP binding. Yet, the result obtained from \chipseq experiments, that short H-NS binding regions are found preferentially near the promoters led us to conjecture that short hairpins may be used as dynamical switches to modulate the transcription level of downstream genes. Using a very simple polymer model of H-NS binding sites, we have found that a characteristic size for the binding region naturally emerges, resulting from the competition between the chain entropy of the DNA polymer, the bending rigidity (\textit{i.e.} the persistence length of the DNA fiber) and the bridging effect of H-NS. Binding regions with larger sizes lead to stable hairpins whereas smaller binding regions cannot form such hairpins at equilibrium. We have confirmed this finding with BD simulation, using a model in better agreement with the biological reality. Namely, we have considered divalent beads to model H-NS and monovalent beads to model H-NS binding sites on the chromosome.

The existence of a characteristic length for H-NS binding regions suggests an ambivalent role. For binding regions of large sizes, H-NS has a repressive role mediated by the formation of long DNA hairpin loops, or helices. These helices are stable and can be maintained over biological time scales. Second for regions with a size close to the characteristic length, the dynamical assembly and disassembly of DNA hairpins may modulate the genetic repression or entail fast transcriptional response, such as the so-called RNAP trapping mechanism. It also suggests that these hairpins, lying at the limit of the stability condition, can be easily disrupted by external perturbations. For instance, it could be the binding of another transcription factor with a larger affinity with DNA, such as FIS. In other words, FIS may introduce local defects to the H-NS filament. Consequently, the two remaining binding regions flanking the FIS binding site will have a size $\tilde{L} < L^*$, resulting in the disassembly of the filament. This is a possible model for transcription activation by FIS, that we call a transcriptional ``switch'', and possibly at stake at the \textit{rrn} operon. Because it is based on the local structure of the DNA, which namely depends on the presence of H-NS binding sites, it may explain why the effect of FIS on the transcription is so heterogeneous in \ecoli bacteria.

Of course, these speculations need to be refined and confirmed with further work. In particular, we plan to model a chunk of the \ecoli genome and include the real distribution of binding sites for H-NS and FIS, based on \chipseq assays. We will investigate which among the H-NS binding regions form hairpin structures. Furthermore, upon addition of FIS, we will see which ones are easily disrupted, and which ones are not. It will be interesting to see if the dynamical re-organization observed in BD simulations indeed correspond to known regulatory sites such as promoter regions. Nonetheless, we underscore that a study based solely on BD simulations has many limitations. In particular, the biological values for the binding energies are often unknown, and when experimental measures were performed, it is actually the free energy which is measured and not the enthalpic binding energy alone. Therefore, such approaches require some arbitrary choices. However, they also constitute a first step in relating more accurately architectural changes and transcription regulation.




\chapter{Reconstruction of chromosome architecture from chromosome conformation capture experiments}
\chaptermark{Reconstruction of chromosome architecture}
\label{ch:ccc}
In this chapter, we address the problem of finding a model for the chromosome architecture from contact probabilities measured in Chromosome Conformation Capture (CCC) experiments.

We start by introducing the reasons to find a better representation of the chromosome architecture. We then present in more details what are CCC techniques and how contact probability matrices can be generated. In particular we will present the methods used in this work to normalize CCC counts maps.  We conclude this introductory section by reviewing methods which have proposed models for the chromosome architecture based on CCC contact matrices.

We then move on to present our model for reconstructing chromosome architecture. It consists of a Gaussian chain polymer representation of the chromosome to which we add effective interactions between DNA monomers under the form of harmonic springs. Such effective interactions do not have any microscopic signification but instead represent a coarse-grained approach. Besides they are to be determined from an input contact probability matrix. The resulting model defines a Gaussian effective model (GEM). More formally, we may say that we address the problem of finding the connected object, as a Gaussian chain, that produces a given contact matrix.

As an important result, we will obtain an analytical closed-form relating these effective couplings to the contact probabilities at the Boltzmann equilibrium for the GEM. This method can be used to propose a physical model of the chromosome under the form of a GEM which reproduces exactly the experimental contacts. Yet it can result in a non-physical model when the correlation matrix of the GEM has negative eigenvalues. Therefore we will present an alternative method, more demanding computationally, that addresses this issue and yields a stable effective model of the chromosome.

Finally, we will apply our method to contact matrices from CCC experiments and comment on the biological significance of the architecture obtained.

\section{Introduction}
The primary function of the chromosome is to encode the genetic information of each cell individual. Yet, chromosome folding (that we call architecture) has an impact on several biological processes including replication, chromosomes segregation and transcription. On a local scale, divalent transcription factors (TFs) can bind to DNA and locally alter the structure of the chromosome, namely by forming DNA loops. In the case of the \textit{lac} operon in \textit{Escherichia coli}, the formation of a DNA loop leads to the repression of the \textit{lac} gene. On a global scale, chromosome architecture is constrained and shaped by structuring proteins, which are nucleoid-associated proteins (NAPs) in bacteria and histones in eukaryotes. In \cref{ch:naps}, we have shown how structures of the chromosome entailed by NAPs can affect transcription regulation. Therefore, a better understanding of chromosome folding seems like a keystone to unveil complex regulatory mechanisms underlying the genetic expression.

A fundamental consequence of chromosome folding on transcription is to bring co-regulated genes\index{co-regulated genes} close in space. Chromosome folding is also assumed to induce the existence of transcription factories \cite{Cook2008,Schoenfelder2010,Llopis2010,Darzacq2013}, or for instance the global silencing of genes in H-NS clusters \cite{Wang092011}. At first, such co-localization effects were called into question because the existence of molecular crowding\index{molecular crowding} together with the confinement of the chromosome in the nucleus/cell result in strong topological constraints. However, several Brownian dynamics (BD) and Monte-Carlo (MC) studies have demonstrated that co-localization of a large number of genes can be achieved despite these constraints \cite{Stefano20131,Brackley36052013}.

From a broader perspective, co-localization can be seen as a way to synchronize biological processes in the nucleus/nucleoid. For the transcription, this would be achieved by sharing higher local concentrations of RNA polymerase (RNAP) in transcription factories. In the context of epigenetics, regions on the chromosome are tagged with marks (like methylations) which affects locally genes expression. In particular, such marks can result in transcriptionnally active (euchromatin) and inactive (heterochromatin) regions. Epigenetics marks held by a region of the chromosome can also propagate to neighboring regions, and turn them for instance into actively transcribed euchromatin. Therefore chromosome architecture entails a spatial network in which biological ``signals'' can propagate to nearest neighbors. We think this constitutes a major determinant of cell physiology. Yet it is not clear whether chromosome architecture induces physiological changes by selecting genes to be transcribed, or to the contrary whether physiological changes lead to biological responses which alter chromosome architecture, or both.

Chromosome architecture can be investigated with physical models introducing binding proteins \cite{Czapla062008,Joyeux042013,2016arXiv160102822B,Pichugina2016}. However, it is a difficult problem for several reasons. First, the copy number of all TFs without distinction is huge (up to \num{e6} in \ecoli \cite{Ishihama2014}). Second, there are many different binding proteins, with different binding energies and binding sites. A common approach to address this problem is to consider a generic type of protein with average properties and representing several protein families at the same time \cite{2016arXiv160102822B,Brackley36052013}. Third, the values of the binding energies are in general not known, which leaves the investigator with a free parameter to fit (or to guess) \cite{Pichugina2016}. Simulations and theoretical studies have usually dealt with these limitations by considering simplifying assumptions which decrease the underlying complexity. For example, taking a crude model for the protein-DNA interaction, reducing the number of target types on the genome or considering several protein species as one\dots\textit{etc}. Consequently, it is hard to expect more than qualitative agreements between results of such studies and experimental data sets. Therefore, models of chromosome architecture better rooted in biological data sets and which can be used in BD studies are actively sought.

\section{Chromosome Conformation Capture experiments}
\label{sec:ccc}
\subsection{Historical context}
Chromosome Conformation Capture (CCC) techniques were developed during the years 2000s. At first, they aimed at counting the number of contacts of a particular location on the chromosome (or locus) with an other locus and were denoted by the acronym 3C. Later improvements have consisted in counting the contacts of a single locus with several other loci on the chromosome (4C), and then of many loci between themselves (5C).  Finally, the combination of CCC techniques with high throughput sequencing methods (Hi-C) brought these techniques to a larger scale by enabling the measurement of contacts between thousands of loci on the chromosome. Hence these methods yield an enormous amount of data to deal with.

During the last decade, Hi-C experiments have revolutionized experimental biology. Before them, measures of the spatial distance between different loci or genes on the chromosome were essentially performed with fluorescence techniques. Yet even with state-of-the-art techniques, like localization-based super-resolution imaging (STORM or PALM) which can be used to survey the subcellular distributions of DNA sequences tagged with a fluorophore, the resolution achieved and the amount of data generated is very humble in comparison with Hi-C methods.

For historical reasons, Hi-C techniques were first used in eukaryotic cells, like in human \cite{Lieberman-Aiden2009} or yeast \cite{Duan2010}, but they have also been used later in bacteria \cite{Jin2013,Church2011}. They have also lead us to revise our conception of chromosome architecture. In particular, contact matrices generated by these experiments generally exhibit checkerboard patterns. In eukaryotes, such patterns have been conjectured to represent a high level organization of the chromosome into Topologically Associated Domains (TADs) with a size slightly below the megabase pair. Although the biological relevance of TADs is still controversial, it has been shown in eukaryotes that TADs are highly conserved in a population of cells with the same type. Yet significant changes are visible during cell differentiation \cite{Sexton32013,Sexton4582012,Fraser2015852} and cell senescence \cite{Chandra4712015}. In bacteria, the existence of TADs is less clear given the smaller size of the bacterial chromosome. Yet changes in the physiological conditions have been shown to induce significant re-organization in the contact matrices measured \cite{Marbouty2015}. In short, Hi-C techniques have provided a novel type of biological data. In particular, it has led to studies fostering the idea that chromosome architecture and gene's expression are intimately connected.

\subsection{Method}
\begin{figure}[!htbp]
  \centering
  \subfloat[]{\label{fig:hic_experiment:procedure} \includegraphics[width = 0.7 \textwidth]{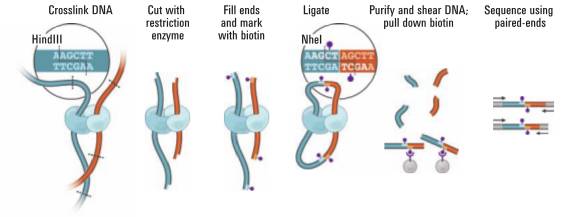}}%
  \\
  \subfloat[]{\label{fig:hic_experiment:contact_maps} \includegraphics[width = 0.7 \textwidth]{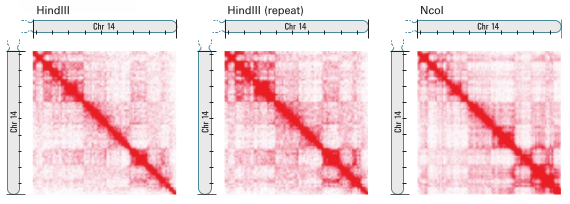}}
  \caption{Hi-C experiments (from \cite{Lieberman-Aiden2009}). \protect \subref{fig:hic_experiment:procedure} Experimental procedure to generate pairs of sequence ``reads'' corresponding to DNA segments in contacts. \protect \subref{fig:hic_experiment:contact_maps} counts maps obtained.}
  \label{fig:hic_experiment}
\end{figure}

From a practical point of view, a restriction enzyme able to cut DNA at specific restriction sites (\textit{i.e.} a nuclease) must be chosen. The DNA segments in between two restriction sites (or cuts) are called restriction fragments. A critical requirement is therefore to find an enzyme whose restriction sites are sufficiently degenerate and common in the genome to yield a subdivision of the genome as regular as possible. Typically, restriction enzymes recognize a specific DNA sequence of 6 base pairs. Hence, it cuts DNA every $4^6=\SI{4096}{bp}$ in average. Ideally, all restriction fragments should have the same size, which of course is never reached in experimental settings. The experimental procedure then relies on the following steps \cite{Lieberman-Aiden2009} (\cref{fig:hic_experiment:procedure}). First, a population of cells is cross-linked with formaldehyde and lysed. This results in the formation of covalent links between adjacent chromatin segments. Second, the restriction enzyme is introduced in order to shear the chromosome, resulting in free pairs of cross-linked restriction fragments with dangling ends. Third, a ligation step is performed in order to join the dangling ends in each restriction fragment pair, which leads to the formation of small DNA rings made of two restriction fragments. This step also removes the restriction sites and adds biotin tags in place for use in the next step. Fourth, the DNA solution is purified and all ligated fragments are obtained by immuno-precipitation of the biotin tags. Finally, the collection of fragments are amplified by PCR and sequenced, yielding a collection of ``reads''. A complex bioinformatics treatment is then required to map the reads to the original genome and identify the loci in contact. This last step has many caveats and is known to be prone to error \cite{Yaffe10592011,Imakaev9992012}. The genome is then divided into bins of equal size, longer than the restriction fragment length. The collection of mapped sequences can then be assigned to each bin and used to produce a counts map where each matrix element $n_{ij}$ is the number of contact events between bins $i$ and $j$ on the genome (\cref{fig:hic_experiment:contact_maps}). The typical size of the bins ranges from a few \SI{}{\kilo bp} to \SI{1}{\mega bp} \cite{Lieberman-Aiden2009,Marbouty2015,Belton2012268}.

\subsection{Caveats}
\label{sec:hic_caveats}
The reliability and repeatability of Hi-C experiments has been frequently called into question. Besides, processing Hi-C experiments raw measures involves a number of bioinformatics steps which are cumbersome and prone to error. Therefore, the counts maps which are a prerequisite for the computation of contact probability matrices should be considered with caution. Without pretending to exhaustivity, we review some of the experimental and methodological artifacts which should be kept in mind and that can affect the quality of the experimental data.

First, the experimental protocol involves several steps in which inaccuracies can accumulate and lead to inconsistencies. In particular, formaldehyde in aqueous solution is present in the form of methylene glycol \ce{HOCH2OH} monomers, but it also exists in the form of oligomers \ce{HO(CH2O)_nH}, where $n$ is a polymerization index. The equilibrium of the polymerization reaction depends on the formaldehyde concentration. For instance, in an aqueous solution with \SI{40}{\percent} mass fraction of formaldehyde at \SI{35}{\degreeCelsius}, the proportion of monomers in solution is only \SI{26.80}{\percent}, the rest being oligomers with $n > 1$ \cite{Gunter2000}.  This is very close to the conditions used in \cite{Lieberman-Aiden2009}, with \SI{37}{\percent} mass fraction of formaldehyde, which clearly suggests that cross-links between restriction fragments have varying size depending on the formaldehyde oligomer that made the cross-link. Furthermore, the size of the cross-link itself between formaldehyde and DNA is of varying length \cite{Jackson1251999}. For these reasons, it may be better to consider that the actual distance between a pair of cross-linked restriction fragments is a Gaussian distribution centered around a most-likely distance $\xi$, rather than always below a threshold distance $\xi$.

Another origin of inconsistency in the experimental protocol may come from the PCR amplification of the purified reads. Indeed, an important requirement is to perform as few PCR cycles as possible ($\sim 10$). This ensures a linear amplification of the reads and preserves the counts distribution up to a normalization \cite{Belton2012268}. Finally, several control experiments must be carried out to check the quality of the produced Hi-C library (the collection of read pairs). For instance, the distribution of the size of restriction fragments can be checked by gel electrophoresis. Ideally, they should all have the same size.

Another control carried out consists in re-digesting the obtained Hi-C library with the restriction enzyme to check that a complete digestion of the chromosome occurred. Note also that in the original Hi-C study \cite{Lieberman-Aiden2009}, counts map had been generated with two different restriction enzymes in order to cross-validate the obtained results. This practice has somehow been lost since all subsequent Hi-C publications have only used one restriction enzyme.

Once the Hi-C library has been obtained, it must still be processed with bioinformatics methods in order to transform the raw data of read pairs into a counts map with elements $n_{ij}$ counting the number of contacts (to a normalization) between genomic locations $i$ and $j$ \cite{Lieberman-Aiden2009,Imakaev9992012,Yaffe10592011}. In particular this implies mapping each read to a location on the genome. At a low level, a primary source of concern is to actually successfully map all reads. For instance, reads from regions with many DNA repeats (coming for instance from transposon elements) or small reads can often not be mapped uniquely to a specific genomic location. These ambiguous reads are therefore discarded. There are also cases in which a read cannot be mapped to the genome. This can originate from DNA recombinations which occur during the experimental protocol, PCR/sequencing errors, reads alignment issues... In this work, we used the results of these procedures as it is available from the literature. In particular, we worked directly with the counts maps computed in previous research works \cite{Lieberman-Aiden2009,Marbouty2015}.

\subsection{From counts to contact probabilities}
\label{sec:cmap_generation}
\subsubsection{Normalization issue}
The counts map can be used to assess the contact probability $c_{ij}$ between any pair $(i,j)$ of loci on the genome. However, this step is not straightforward because the normalization to transform counts into contact probabilities is not known.

In \cite{Lieberman-Aiden2009} the contact probabilities are computed as $c_{ij} = n_{ij} / \langle n_{ij} \rangle$, where $\langle n_{ij} \rangle$ simply means the average over genomic loci pair $(p,q)$ separated by the same contour distance, $\mid p-q\mid = \mid i-j \mid$. However there is no rigorous justification for this choice of normalization. Other studies have attempted to address the normalization problem by designing a numerical procedure that ensures that the obtained contact probabilities produce a stochastic matrix, \textit{i.e.} the line sums $\sum_j c_{ij}=1$ \cite{Imakaev9992012,Yaffe10592011,Cournac12012}. However, although this tends to smoothen the variations of the contact probabilities, we do not see clear reasons supporting the idea that the contact probability matrix should be a stochastic matrix. For instance, in the case of a Gaussian polymer, the probability that the distance between any pair $(i,j)$ of monomers vanish, $d_{ij}=0$, is:
\begin{equation}
  c_{ij} = \left( \frac{3}{2 \pi b^2} \right)^{3/2} \mid i-j \mid^{-3/2},
\end{equation}
where $b$ is the size of a monomer. It obviously does not satisfy the stochastic matrix condition. Consequently, we have chosen to consider an alternative approach (although less sophisticated than the methods just mentioned) to normalize counts maps into contact probability matrices.

In principle, the normalization factor between counts and contact probabilities should be the total number of cells in the experimental sample (possibly multiplied by the PCR amplification ratio). Then $n_{ij}$ is simply the number of cells in which a contact between loci $i$ and $j$ is observed. Assuming that the experimental sample contains $\mathcal{N}$ cells, the contact probability is then simply expressed as $c_{ij} = n_{ij} / \mathcal{N}$. In practice however, $\mathcal{N}$ is unknown. We now propose two simple approximations for this normalization.

\subsubsection{Trace normalization}
It is natural to expect that the closer restriction fragments are on the DNA sequence, the higher their contact probability is. In particular, restriction fragments falling into the same bin should always be in contact, \textit{i.e.} $c_{ii}=1$, and indeed diagonal elements $n_{ii}$ usually take the largest values. Hence, we may be tempted to assume that each diagonal element is equal to the number of cells in the sample. However, in real data sets, all diagonal elements are not equal. Thus we consider instead that the number of cells in the sample can be approximated by the average value of the diagonal elements. The contact probability is then computed as:
\begin{equation}
  c_{ij} = \frac{n_{ij}}{\mathcal{N}}, \qquad \mathcal{N} = \frac{1}{N} \sum \limits_{i=1}^N n_{ii}.
  \label{eq:contact_normalization_trace}
\end{equation}

\subsubsection{Maximum normalization}
It is possible however that diagonal counts $n_{ii}$ are abnormally high. This might be due for instance to self-ligations of isolated restriction fragments or cross-linking with sister DNA during replication. In order to circumvent this issue, we assume that there exists at least one pair of off-diagonal loci $(i_0,j_0)$ which are always in contact, \textit{i.e.} $c_{i_0 j_0}=1$. Note that this assumption is different from the stochastic matrix condition, which assumes that every monomer is always in contact with at least one other monomer. Therefore, the number of cells in the sample is estimated as the maximum of the off-diagonal counts. Actually, counts are very high not only on the diagonal, but also near the diagonal. Therefore, we may choose to discard counts such that $\mid i-j\mid < l_d$ where $l_d \ge 1$ is a length to adjust. In the end, the contact probability is computed as:
\begin{equation}
  c_{ij} = \frac{n_{ij}}{\mathcal{N}}, \qquad \mathcal{N} = \max_{\mid i-j \mid \ge l_d}{( n_{ij} )}.
  \label{eq:contact_normalization_max}
\end{equation}

\begin{figure}[!htbp]
  \centering
  \subfloat[]{\label{fig:hic_normalization:counts}\includegraphics[width=0.3 \textwidth]{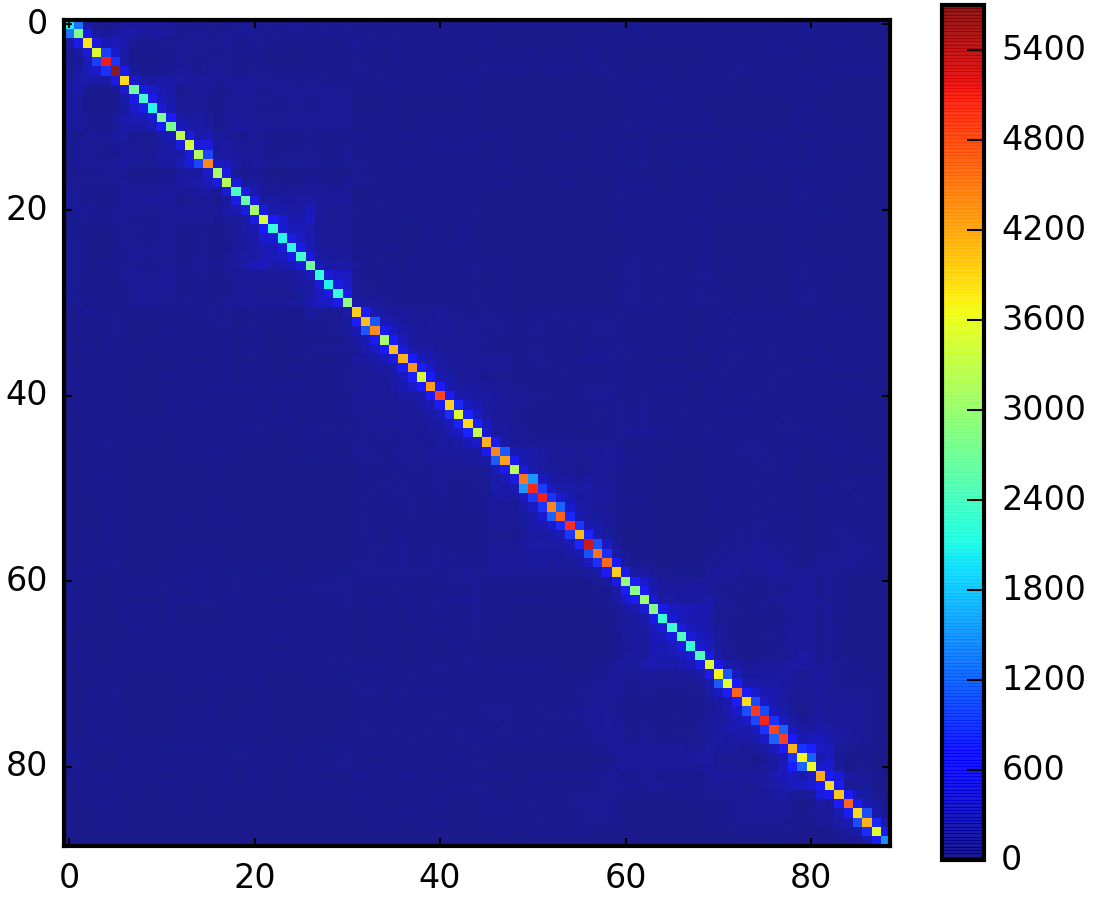}} %
  \quad
  \subfloat[]{\label{fig:hic_normalization:trace}\includegraphics[width=0.3 \textwidth]{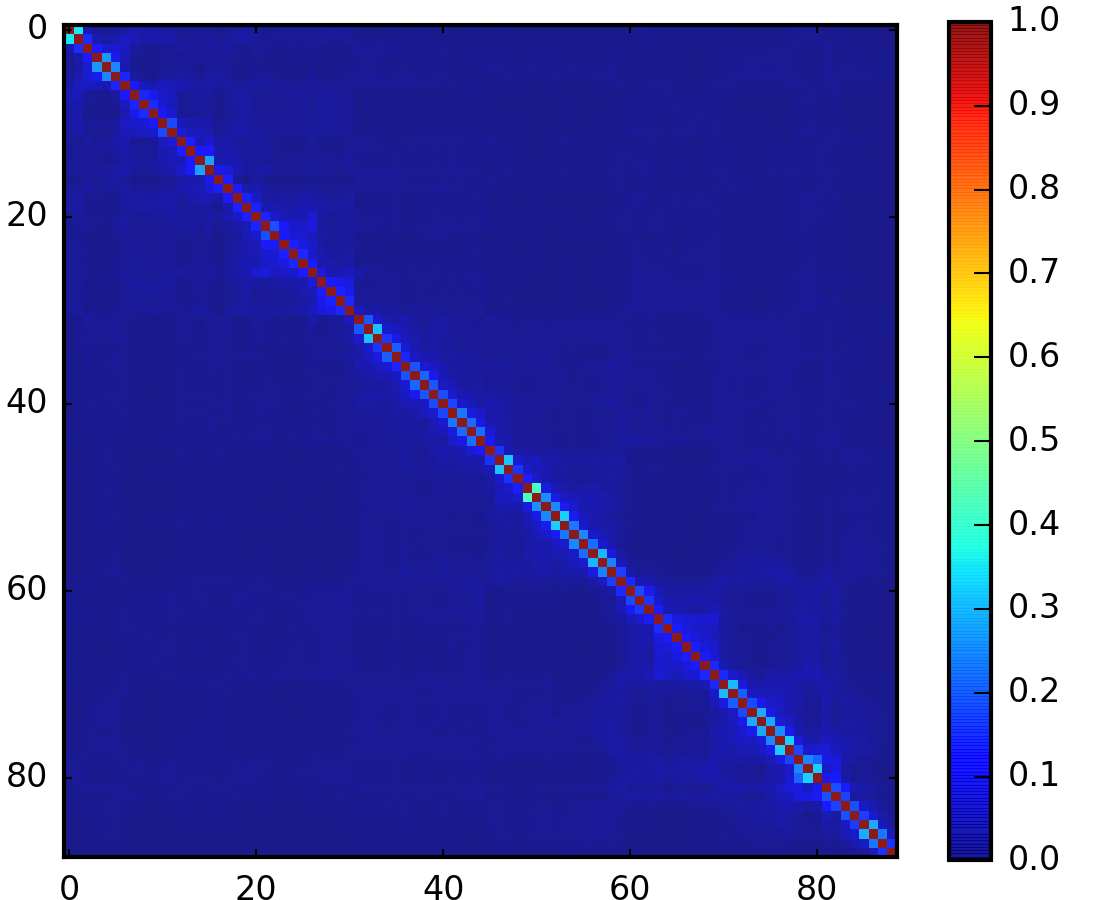}} %
  \quad
  \subfloat[]{\label{fig:hic_normalization:max}\includegraphics[width=0.3 \textwidth]{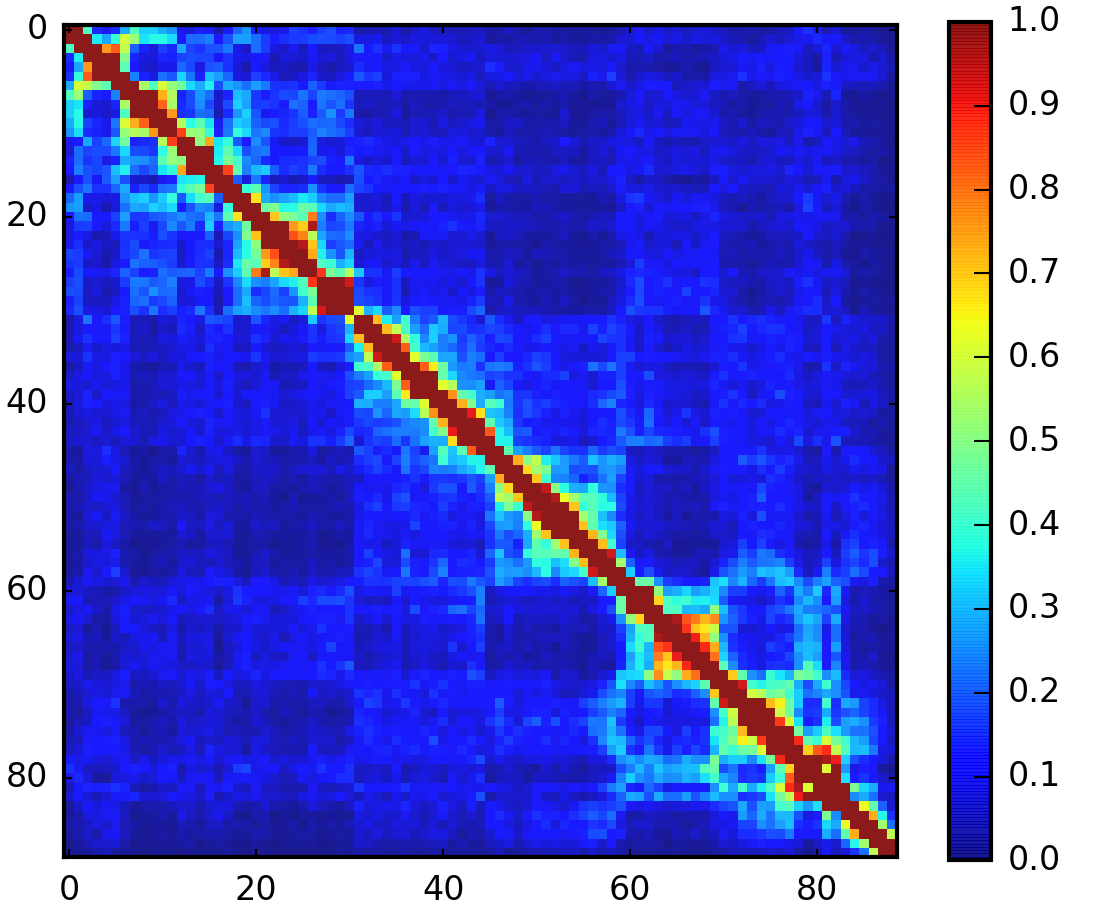}}
  \caption{Normalization of Hi-C counts into contact probabilities. \protect\subref{fig:hic_normalization:counts} Counts map of human chromosome 14 with bin size \SI{1}{\mega bp} \cite{Lieberman-Aiden2009}. \protect\subref{fig:hic_normalization:trace} ``Trace'' normalization. \protect\subref{fig:hic_normalization:max} ``Maximum'' normalization with $l_d=3$.}
  \label{fig:hic_normalization}
\end{figure}

This method with $l_d=3$ gives a contact probability matrix in qualitative agreement with \cite{Lieberman-Aiden2009} (see \cref{fig:hic_experiment,fig:hic_normalization}).

\section[Previous approaches]{Previous approaches to predict chromosome folding from Hi-C data}
We now review some of the models which have been investigated in the past to address the reconstruction of chromosome architecture from CCC data.
\subsection{Models based on an estimate of the distance matrix}
\subsubsection{Non-polymer models}
\label{subsec:hic_nonpolymer_models}
\begin{figure}[!htbp]
  \centering
  \subfloat[]{\label{fig:previous_models:umbarger}\includegraphics[width=0.45 \textwidth]{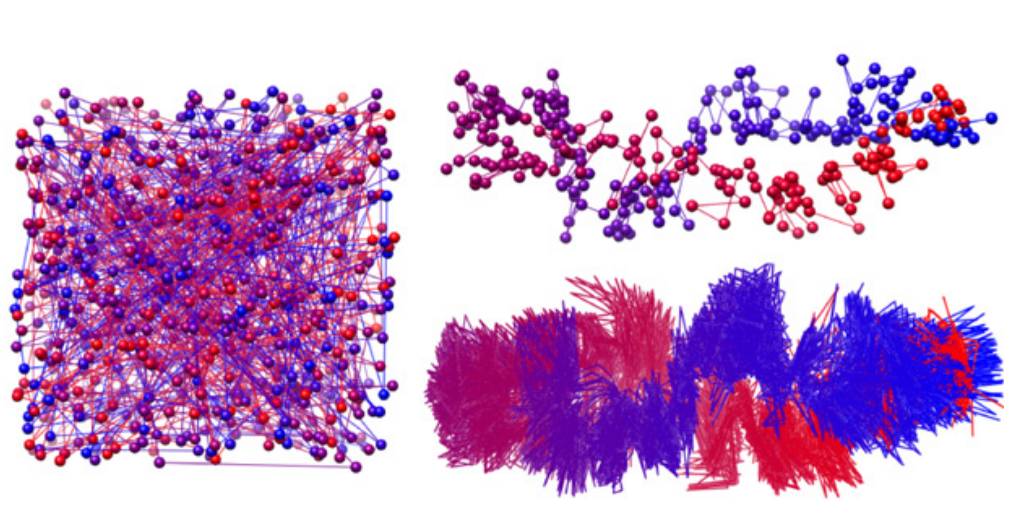}}%
  \quad
  \subfloat[]{\label{fig:previous_models:duan}\includegraphics[width=0.45 \textwidth]{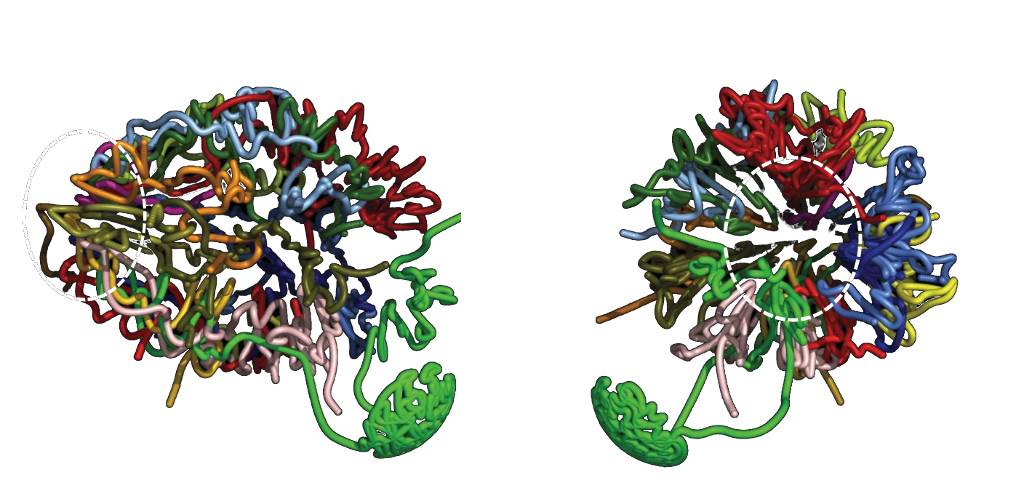}}%
  \\
  \subfloat[]{\label{fig:previous_models:jhunjhunwala}\includegraphics[width = 1 \textwidth]{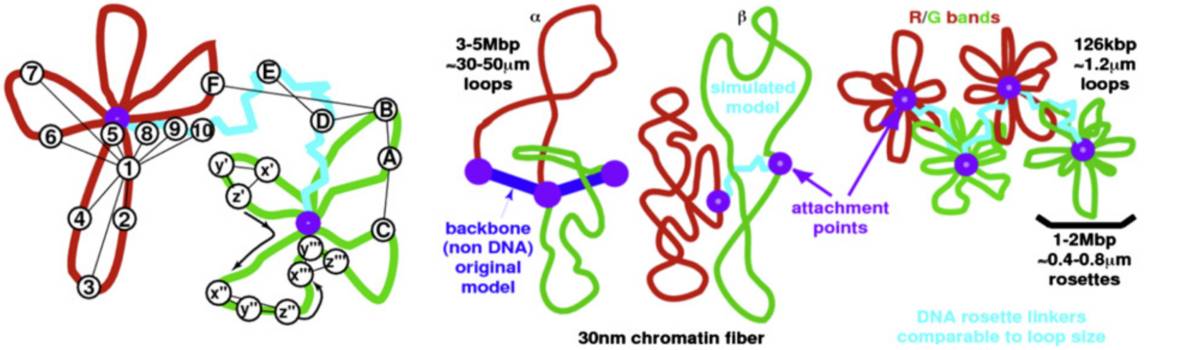}}%
  \\
  \subfloat[]{\label{fig:previous_models:mozziconacci}\includegraphics[width = 0.7 \textwidth]{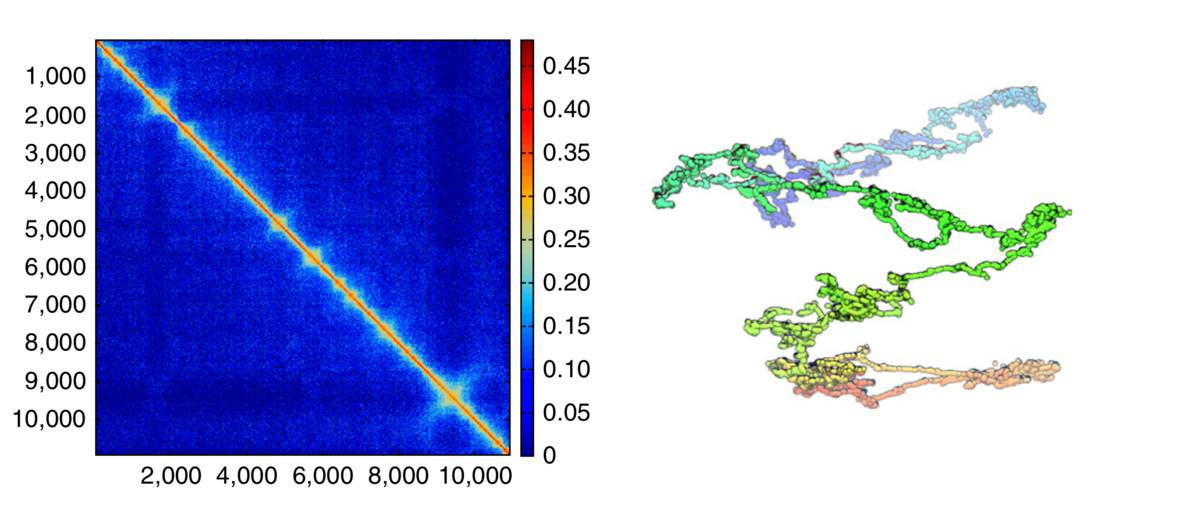}}

  \caption{Models for reconstructing chromosome architecture. \protect\subref{fig:previous_models:umbarger} Harmonic model \cite{Church2011}. \protect\subref{fig:previous_models:duan} Constraint satisfaction model \cite{Duan2010}. \protect\subref{fig:previous_models:jhunjhunwala} Random walk with tethered loops \cite{Jhunjhunwala2652008}. \protect\subref{fig:previous_models:mozziconacci} Singular value decomposition of the correlation matrix \cite{Mozziconacci2014}.}
  \label{fig:previous_models}
\end{figure}

\paragraph{Harmonic model.}
A numerical procedure relying on the introduction of harmonic potentials has been proposed to reconstruct the equilibrium configurations of the chromosome from the experimental contact probabilities \cite{Church2011}. Harmonic interactions are introduced between each restriction fragment pair $(i,j)$, such that:
\begin{equation}
  \beta U(\lbrace \vec{r}_{i} \rbrace) = \sum \limits_{i<j} \frac{k}{2} \left( r_{ij} - r_{ij}^0 \right)^2,
  \label{eq:previous_models:umbarger}
\end{equation}
in which $r_{ij} = \mid \vec{r}_{j} - \vec{r}_i \mid$ is the distance between loci $i$ and $j$, $k$ is an arbitrarily chosen elastic constant and $r_{ij}^0$ is the length of the corresponding spring. A Monte-Carlo simulation is then performed to sample equilibrium configurations of the system defined in \cref{eq:previous_models:umbarger}. These configurations are used to represent the chromosome configurations (\cref{fig:previous_models:umbarger}).

In this method, the elastic constant was assigned arbitrarily to $k= 5 \, k_B T$. The fact that this elastic constant is the same for all $(i,j)$ is a first limitation in this approach. The spring lengths are taken such that $r_{ij}^0 =  d_{ij}$, where $d_{ij}$ is the distance desired between beads $i$ and $j$. The authors assumed that the equilibrium distance between two restriction fragments is inversely proportional to the contact probability, $d_{ij} = 1 / c_{ij}$. We will come back to this assumption at the end of this section. Importantly, in a network of connected springs such as defined in \cref{eq:previous_models:umbarger}, the average distance at thermal equilibrium between a locus $(i,j)$ is in general not equal to the spring length, hence $\langle r_{ij}\rangle \neq d_{ij}$. This is an example of frustrated systems, and constitutes a fundamental limitation of this approach.

\paragraph{Constraint satisfaction.}
Another approach is to cast the problem of reconstituting chromosome architecture into a constraint satisfaction problem \cite{Duan2010}. The reformulated problem then consists in finding the coordinates $\lbrace \vec{r}_{i} \rbrace$ such that the distances between any $(i,j)$ restriction fragment pair is bounded from below and from above:
\begin{equation}
  \underline{r}_{ij} < r_{ij} < \overline{r}_{ij}.
  \label{eq:previous_model:duan}
\end{equation}

In \cref{eq:previous_model:duan} the upper bound is taken inversely proportional to the experimental contact probability, $\overline{r}_{ij} \propto 1 / c_{ij}$, and the proportionality coefficient is a parameter of the method. The lower bound $\underline{r}_{ij}$ is introduced to take into account excluded volume between restriction fragments, and to penalize contacts between adjacent fragments due to the chromosome bending rigidity. This is a constraint satisfaction problem, which can be solved with the simplex method. The obtained solution is then used to represent a chromosome configuration (\cref{fig:previous_models:duan}).

The main limitation of this approach is clearly that the choice of the lower and upper bounds must be adjusted by the user and adapted to each data set. Beside, this is not a physical model of the chromosome architecture.

\paragraph{Singular value decomposition of the spatial correlation matrix.}
Let us consider the matrix $R$ of size $d \times N$, where $d$ is the space dimension and $N$ is the number of bins in the Hi-C contact matrix. The matrix element $r_{\alpha i}$ is therefore the spatial coordinate of loci $i$ along the $\alpha$-axis ($\alpha=x,y,z$). Next we consider the Singular Value Decomposition (SVD) of $R$:
\begin{equation}
  r_{\alpha i} = \sum \limits_{\gamma =1}^d \lambda_\gamma u_{\alpha \gamma} v_{ i \gamma},
  \label{eq:previous_models:mozziconacci:svd}
\end{equation}
where $U$ and $V$ are two orthogonal matrices, and $\left\lbrace \lambda_\gamma \right\rbrace_{\gamma=1,\dots,d}$ are the singular values of $R$. Then $C=R^T R$ and $\tilde{C} = R R^T$ have the same non-zero eigenvalues, which are $\lambda_1^2$, $\lambda_2^2$ and $\lambda_3^2$ (if $d=3$). Finally we introduce the matrix of distances, $D$, with elements:
\begin{equation}
  d_{ij} = \sqrt{\sum \limits_{\alpha=1}^d \left( r_{\alpha i} - r_{\alpha j} \right)^2}.
  \label{eq:previous_models:mozziconacci:distances}
\end{equation}
It was pointed out that the correlation matrix $C$ can be obtained from the distance matrix $D$ \cite{Mozziconacci2014,Marbouty2015}. Therefore, from the knowledge of the distances, one can infer the singular values of the coordinates matrix, and obtain an approximation for $R$.

\paragraph{Relation between contact probability and average distance.}
The methods that we have presented have the inconvenient to rely on an estimate of the distances between loci on the chromosome, taken to be inversely proportional to the contact probabilities, \textit{i.e.} $d_{ij} \propto 1 / c_{ij}$. This assumption can be called into question.

\subsubsection{Polymer models}
Models presented in \cref{subsec:hic_nonpolymer_models} lack a physical model of the chromosome. In clear, the Hi-C bins define a gas of particles with coordinates $\lbrace \vec{r}_i \rbrace$ and minimizing \cref{eq:previous_models:umbarger} (resp. solving \cref{eq:previous_model:duan,eq:previous_models:mozziconacci:distances}) can result in configurations that violate topological constraints of the polymer chain representing the chromosome. Therefore, subsequent improvements have consisted in incorporating a polymer model of the chromosome when attempting to reconstruct chromosome architecture.

\paragraph{Random walk backbone with tethered loops.} Another way to look at Hi-C data is to consider that when the contact probability between loci $i$ and $j$ is high enough, it defines a DNA loop. This is the approach taken in \cite{Jhunjhunwala2652008}. In short, whenever
\begin{equation}
  c_{ij} > \underline{c},
  \label{eq:previous_models:jhunjhunwala}
\end{equation}
with an arbitrary lower bound $\underline{c}$ on the contact probability, the authors considered that the DNA subchain in the interval $[i,j]$ constitutes a loop, with $\vec{r}_i=\vec{r}_j$. The chromosome is then represented by a backbone polymer with Gaussian statistics on which are tethered polymer loops with varying sizes (\cref{fig:previous_models:jhunjhunwala}). Numerical simulations are then performed on the basis of this polymer model of the chromosome.

Although this backbone-with-loops model takes into account some sort of connectedness of the chromosome as a polymer, it is an \textit{ad hoc} model and therefore can only give rather qualitative insights.

\subsubsection{Discussion on the relation between distances and contact probabilities}
The methods introduced in \cref{subsec:hic_nonpolymer_models} assume that the distance between any restriction fragment pair $(i,j)$ can be related to their contact probability in such a way that:
\begin{equation}
  d_{ij} \propto 1 / c_{ij}.
  \label{eq:dij_cij_prop}
\end{equation}

While \cref{eq:dij_cij_prop} may appear to be a reasonable assumption, there is no fundamental reason to support it. For instance, if we model the chromosome as a polymer with scaling exponent $\nu$, we have \cite{deGennes1979}:
\begin{align}
  \begin{aligned}
    & \proba{\vec{r}_{ij}} \simeq \frac{1}{\langle r_{ij} \rangle^d} f_p \left( \frac{r_{ij}}{\langle r_{ij} \rangle} \right), \qquad  f_p(x) \underset{x\sim 0}{\sim} x^g \\
  & \langle r_{ij} \rangle \simeq b \mid i -j \mid^\nu
  \end{aligned}
\end{align}.

Let us consider that the contact probabilities are given by $c_{ij} = \proba{r_{ij}=b}$, and write $d_{ij}=\langle r_{ij} \rangle$. Then, we obtain the relation:
\begin{equation}
  d_{ij} \sim 1 / c_{ij}^{1/(d+g)}.
  \label{eq:dij_cij_polymer}
\end{equation}

For a Gaussian chain, we have $g=0$, and for a self-avoiding chain, $g=1/3$. Hence we obtain ($d=3$), $d_{ij} \sim 1 / c_{ij}^{0.33}$ and $d_{ij} \sim 1 / c_{ij}^{0.3}$, in direct contradiction with \cref{eq:dij_cij_prop}. Besides, we have seen that the contact probabilities are already an approximation obtained from the counts maps. Hence, this assumption on the relation between average distances and contact probabilities may add significant inaccuracies that one may want to avoid.

Following this line of thoughts, we emphasize that all the methods reviewed previously have in common to aim at a characterization of the 3D-folding of the chromosome. That is to say, the solution consists in a collection of coordinates $\lbrace \vec{r}_i^* \rbrace$ that represent an average conformation of the Hi-C restriction fragments. Without rejecting the quality of the research carried out, let us emphasize that reducing chromosome architecture to a mere conformation is probably unrealistic. Indeed, co-localization of loci on the chromosome results from the effect of divalent (or multivalent) proteins. Such proteins have preferred binding sites which are commonly represented with a Position Weight Matrix (PWM) \cite{regulondb_pwm}. We may estimate the strength of the binding by considering contributions of about one $k_B T$ per significant contact \cite{Sheinman2012}. For H-NS, which binds widely on the genome, we have approximately three significant contacts. For CRP, which recognizes more specific sequences, we have approximately eight significant contacts. For more specific transcription factors, we may have of the order of fifteen significant contacts. Therefore, we may consider that structuring proteins have a binding energy with DNA in the range $\varepsilon = 3-15 \, k_B T$. Consequently, as seen above, we may assess the probability to form a DNA loop between loci $i$ and $j$ as:
\begin{equation}
  \proba{i \leftrightarrow j} = \frac{1}{\mid i - j\mid^{\nu(d+g)} } e^{\beta \varepsilon},
\end{equation}
where $\nu(d+g)=2$ for a self-avoiding polymer chain with scaling exponent $\nu=3/5$. For example, considering a relatively strong transcription factor, with $\beta \varepsilon = 10 \, k_B T$, the contact probability $c_{ij} \simeq 1$ when $\mid i-j\mid = 150$ monomers and falls quickly to zero for larger contour distances. Here a monomer typically represents the scale at which a beads-on-string polymer representation of the chromosome is valid, \textit{i.e.} when the size of one monomer is of the order of the DNA fiber diameter. In bacteria for instance, the chromosome can be seen as a fiber of diameter $\SI{2.5}{\nm} \simeq \SI{7.5}{bp}$. In eukaryotes, a monomer typically represents \SI{3000}{bp}. Yet, in Hi-C contact matrices a bin typically represent \num{e3}-\SI{e6}{bp} \cite{Lieberman-Aiden2009,Marbouty2015}. This suggests that loops interactions between loci identified with Hi-C data are rather weak. In other words, thermodynamic fluctuations may provide the chromosome folding with a non negligible conformational entropy. In particular it seems a bit awkward to reduce the chromosome architecture to an average conformation.

\subsection{A polymer model reproducing experimental contact frequencies}
\label{sec:polymer_model_contacts}
Instead of finding a chromosome folding which satisfies constraints on the monomer pair distance $d_{ij}$, an alternative approach is to seek a physical model of the chromosome which reproduces the experimental contact probabilities. This has been proposed and investigated with BD simulations \cite{Jost095412014,2016arXiv160102822B}. However, as mentioned earlier, due to the complexity of chromosome interactions with proteins, this kind of studies could only be made under strong simplifying assumptions. In particular, a unique generic type of protein is included and the variety in the binding energies with different loci on the chromosome is replaced by a single binding energy (or just a few). Consequently, comparisons with experimental contact matrices have been rather qualitative.

If such simplifications are performed, we put forward the idea that chromosome architecture might be well described with an effective model in which microscopical details, such as proteins and sequence effects, are coarse-grained. In particular, the effect of structuring proteins can be taken into account implicitly by introducing an effective potential $V_{ij}(r)$ between each $(i,j)$ monomer pair. In other words, each location on the genome experiences an effective interaction with the other loci on the genome, which mimics the effect of multivalent proteins. An inspiring approach was carried out recently, in which such potentials are considered to be short-range square potentials \cite{Giorgetti9502014}:
\begin{align}
  V_{ij}(r) =
  \begin{cases}
    +\infty & \text{ if } r < \sigma \\
    - \varepsilon_{ij} & \text{ if } \sigma < r < \xi \\
    0 & \text{ otherwise,}
  \end{cases}
  \label{eq:pair_potential_square}
\end{align}
where $\sigma$ is the hard-core distance and $\xi$ is a threshold which defines at the same time the range of the potential and the distance below which monomers $i$ and $j$ are said to be in contact. By performing MC simulations on a polymer model with the pair potentials in \cref{eq:pair_potential_square}, one can obtain equilibrium configurations and use them to compute contact probabilities between monomer pairs.

Let us note $c_{ij}^{exp}$ the experimental contact probability between restriction fragments $i$ and $j$ obtained from Hi-C experiments, and $c_{ij}$ the contact probability between monomers $i$ and $j$ obtained from MC simulations of a polymer model with potentials as in \cref{eq:pair_potential_square}. We define the least-square distance between the experimental and the predicted contact matrices:
\begin{align}
  d(c_{ij},c_{ij}^{exp}) = \frac{1}{\mathcal{N}} \sum \limits_{i<j} \left( c_{ij} - c_{ij}^{exp} \right)^2,
   \label{eq:chisq_contacts}
\end{align}
where $\mathcal{N}$ is the number of monomer pairs. Finding a good model for chromosome architecture now consists in finding a collection of potentials $V_{ij}(r)$ that minimize $d(c_{ij},c_{ij}^{exp})$. The solution is achieved at the optimal values for $\sigma$, $\xi$ and the matrix of binding energy $\varepsilon_{ij}$.

In \cite{Giorgetti9502014}, a MC simulation was performed at each step of the minimization procedure, in order to re-sample equilibrium configurations of the chromosome and compute the $c_{ij}$ values. Therefore the computational burden is high.

Following these tracks, we propose in the sequel a method giving a chromosome architecture under the form of a physical model that predicts contact probabilities which match as closely as possible the experimental ones. Our approach retains some of the features introduced here, namely the representation of the chromosome with a coarse-grained polymer and effective interactions.

\section{Gaussian Effective Model}
\label{sec:gem_model}
\subsection{Model}
From now on, we model the chromosome as a polymer of length $N$, \textit{i.e.} made of $N+1$ monomers with coordinates $\lbrace \vec{r}_ i \rbrace$. Each monomer represents a Hi-C bin with size $b$. We assume that the chromosome can be well modeled by a Gaussian chain with energy:
\begin{equation}
  \beta U_{e}\left[ \left\{ \vec{r}_i \right\} \right] = \frac{3}{2 b^2} \sum \limits_{i=1}^N \left( \vec{r}_i - \vec{r}_{i-1} \right)^2,
  \label{eq:gem_gaussian_energy}
\end{equation}
where as usual $\beta = (k_B T)^{-1}$. For a more accurate theory, one may include an additional term to \cref{eq:gem_gaussian_energy} in order to model the chain bending rigidity:
\begin{equation}
  \beta U_{b}\left[ \left\{ \vec{r}_i \right\} \right] = \frac{l_p}{2} \sum \limits_{i=1}^{N-1} \left( \vec{r}_{i+1} + \vec{r}_{i-1} - 2 \vec{r}_{i} \right)^2,
  \label{eq:gem_bending_energy}
\end{equation}
where $l_p$ is the chain persistence length. In that case, the total chain energy is given by
\begin{equation}
  \beta U_0 \left[ \left\{ \vec{r}_i \right\} \right] = \beta U_{e}\left[ \left\{ \vec{r}_i \right\} \right] + \beta U_{b}\left[ \left\{ \vec{r}_i \right\} \right].
  \label{eq:gem_chain_energy}
\end{equation}

However, we have seen earlier that Hi-C experiments have a resolution which typically gives $b \approx \num{e3}-\SI{e6}{bp}$. Therefore, as a first approximation, we choose to neglect the bending rigidity of the chromosome and take $\beta U_0[\lbrace \vec{r}_i \rbrace]=\beta U_e[\lbrace \vec{r}_i \rbrace]$. We now introduce effective interactions between the monomers under the form of harmonic potentials with unknown rigidity constants. The interaction energy reads
\begin{equation}
  \beta U_I \left[ \left\{ \vec{r}_i \right\}  \right] = \frac{3}{2 b^2} \sum \limits_{i<j} k_{ij} \left( \vec{r}_i -\vec{r}_j \right)^2,
  \label{eq:gem_interaction_energy}
\end{equation}
where a coupling matrix $K$ with elements $k_{ij}$ has been introduced. Finally, the total energy (or Hamiltonian) is defined by
\begin{equation}
  \beta U\left[ \left\{ \vec{r}_i \right\} \right] = \beta U_0 \left[ \left\{ \vec{r}_i \right\} \right] + \beta U_I\left[ \left\{ \vec{r}_i \right\} \right].
  \label{eq:gem_energy}
\end{equation}

\begin{figure}[!htbp]
  \centering
  \includegraphics[scale=0.6]{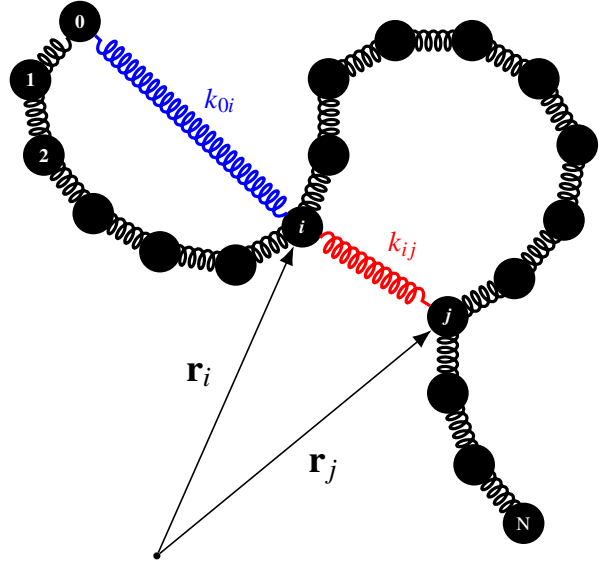}
  \caption{Gaussian Effective Model. Harmonic interactions with elastic coefficient $k_{ij}$ are added on top of the Gaussian polymer model.}
  \label{fig:gaussian_effective_model}
\end{figure}

The physical system with energy as in \cref{eq:gem_energy} defines the Gaussian Effective Model (GEM). It simply corresponds to a Gaussian chain model of the chromosome to which are added harmonic interactions between monomer pairs $(i,j)$, which are effective interactions representing the superimposition of many microscopic interactions (\cref{fig:gaussian_effective_model}). Before writing down the partition function, let us point out that this system is ill-defined at this stage. To break the translational invariance, we attach the first monomer to the origin and consider $\vec{r}_0=0$. After this preliminary remark, we can write the GEM partition function. The energy in \cref{eq:gem_energy} is quadratic, hence the partition function is computed as a Gaussian integral:
\begin{align}
  \begin{aligned}
    Z &= \int \prod \limits_{i=1}^N \ud{^3 \vec{r}_i} \exp{\left( - \beta U\left[ \left\{ \vec{r}_i \right\} \right] \right)} \\
    &= \int \prod \limits_{i=1}^N \ud{^3 \vec{r}_i} \exp{\left( - \frac{3}{2 b^2} \sum_{i,j} \vec{r}_i \cdot \vec{r}_j \sigma^{-1}_{ij} \right)} \\
    &= \left( \frac{2 \pi b^2}{3} \right)^{3N/2} \det{\Sigma}^{3/2},
  \end{aligned}
  \label{eq:gem_partition_function}
\end{align}
where we have introduced the inverse correlation matrix with elements $\sigma^{-1}_{ij}$:
\begin{equation}
  \Sigma^{-1}=T + W,
  \label{eq:gem_inverse_correlation}
\end{equation}
with:
\begin{align}
  T =
\begin{pmatrix}
  2 & -1 & \ldots & 0 & 0 \\
  -1 & 2 & \ldots & 0 & 0 \\
  \vdots & \vdots & \ddots & \vdots & \vdots \\
  0 & 0 & \ldots & 2 & -1 \\
  0 & 0 & \ldots & -1 & 1
\end{pmatrix}
,
\qquad
W =
\begin{pmatrix}
  \sum \limits_{\substack{j=0 \\ j \neq 1}} k_{1j} & -k_{12} & \ldots & -k_{1 N-1} & -k_{1N} \\
  -k_{21} &  \sum \limits_{\substack{j=0 \\ j \neq 2}} k_{2j}& \ldots & -k_{2 N-1} & -k_{2N} \\
  \vdots & \vdots & \ddots & \vdots & \vdots \\
  -k_{N-1 1} & -k_{N-1 2} & \ldots & \sum \limits_{\substack{j=0 \\ j \neq N-1}} k_{N-1 j} & -k_{N-1 N} \\
  -k_{N 1} & -k_{N 2} & \ldots & - k_{N N-1} & \sum \limits_{\substack{j=0 \\ j \neq N}} k_{N j}
\end{pmatrix}.
  \label{eq:gem_trid_redk}
\end{align}

$T$ is the tridiagonal matrix enforcing the chain structure from \cref{eq:gem_gaussian_energy} and $W$ is the matrix of reduced couplings enforcing the interactions from \cref{eq:gem_interaction_energy}. Being a Gaussian model, the system is fully determined by its correlation matrix. In particular, we have:
\begin{equation}
   \langle \vec{r}_i \cdot \vec{r}_j \rangle = \sigma_{ij} b^2,
  \label{eq:gem_sigma}
\end{equation}
where the brackets stand for a thermodynamical average with a Boltzmann weight defined from the partition function in \cref{eq:gem_partition_function}, \textit{i.e.} for any function of the monomer coordinates, $A(\lbrace \vec{r}_i \rbrace)$:
\begin{equation}
  \langle A\left( \left\{ \vec{r}_i \right\} \right) \rangle = \frac{1}{Z} \int \prod_{i=1}^N \ud{^3 \vec{r}_i} A\left( \left\{ \vec{r}_i \right\} \right) \exp{\left( - \beta U \left[ \left\{ \vec{r}_i \right\} \right] \right)}.
  \label{eq:gem_thermo_avg}
\end{equation}

Note that when $W=0$, we retrieve the standard Gaussian chain with $\langle \vec{r}_i \cdot \vec{r}_j \rangle = \min{(i,j)} b^2$. Let us emphasize that the GEM is stable only when $\Sigma$ has all its eigenvalues strictly positive. Therefore not every choice of coupling matrix $K$ leads to a physical model.

Polymer models with Gaussian interaction Hamiltonian as defined in \cref{eq:gem_interaction_energy} have received some attention in the past. They were introduced in the context of cross-linked polymers, in order to predict the size of a collapsed polymer of size $N$ with $M$ cross-links \cite{Solf66551995,Kantor52631996,Bryngelson5421996}. However, an essential difference with the GEM presented here is that in those studies, the effective interactions were uniform, \textit{i.e.} $k_{ij} = k$. More recently, such model has been re-introduced to account for the particular scaling of the gyration radius of the chromosome in the interphase nucleus \cite{Bohn0518052007}. More accurately, the radius of gyration is reaching a plateau for genomic distances larger than a few \SI{}{\mega bp}, $\langle R_g^2 \rangle \sim O(1)$. This scaling was recovered by considering the above GEM model, in which the $k_{ij}$ are Bernoulli random variables with probability distribution function (p.d.f.) such that $ \proba{k_{ij}=k} = p \delta(k_{ij}-k) + (1-p) \delta(k_{ij})$. Under this assumption, each non-zero value $k_{ij}$ defines a loop between monomer $i$ and $j$ with harmonic spring constant $k$. For this reason, this model was named Random Loop Model. The theoretical results obtained on the radius of gyration scaling were confirmed later with BD simulations \cite{Mateos-Langerak38122009}.

\subsection{Naive approach}
\label{sec:gem_model_naive_approach}
\subsubsection{Rationale}
As argued in \cref{sec:polymer_model_contacts}, a reasonable strategy seems at first to seek the coupling matrix $K$ which minimizes $d(c_{ij},c_{ij}^{exp})$, as defined in \cref{eq:chisq_contacts}, between the experimental contact probabilities and the ones predicted by the GEM. The optimal coupling matrix, $K^{opt}$ can then be used as a model for chromosome architecture. Intuitively, one may expect the contact probabilities to be proportional to the values of the couplings:
\begin{equation}
  k_{ij} = \Lambda c_{ij},
  \label{eq:kij_cij_prop}
\end{equation}
where $\Lambda$ is a scaling coefficient which needs to be adjusted. For each value of $\Lambda$, we can run BD simulations in order to sample equilibrium configurations of the corresponding GEM. Then, for any $(i,j)$ pair of monomers, we can compute the contact probabilities as:
\begin{equation}
  c_{ij} = \left\langle \theta \left(\xi - r_{ij} \right) \right\rangle,
\end{equation}
where the brackets here mean that we perform an average over the system configurations, $\theta$ is the theta function (\textit{i.e.} the indicator function of $\mathbb{R}_+$), and $\xi$ is a threshold distance below which a contact is said to occur. Therefore the computed contact probabilities depend on the coupling scale $\Lambda$ and on the threshold $\xi$. However, the threshold is clearly arbitrary and should be chosen in order to best fit the experimental contacts. Therefore, we may define the optimal threshold $\xi^{opt}$ that minimizes the contacts least-square distance:
\begin{equation}
  \xi^{opt}(\Lambda) =  \underset{\xi}{\mathrm{argmin}}\left[ d\left( c_{ij},c_{ij}^{exp}\right) \right], \qquad c_{ij}^{opt}(\Lambda) = c_{ij}(\Lambda, \xi^{opt}).
  \label{eq:naive_thres_opt}
\end{equation}
and similarly, we may define the optimal scale, $\Lambda^{opt}$ as:
\begin{equation}
  \Lambda^{opt} = \underset{\Lambda}{\mathrm{argmin}}\left[ d \left( c_{ij}^{opt},c_{ij}^{exp} \right) \right].
  \label{eq:naive_scale_opt}
\end{equation}

Since the method to compute the contact probabilities $c_{ij}$ relies on BD simulations, we may consider other polymer models than the Gaussian chain defined in \cref{eq:gem_gaussian_energy}. In particular, we can add excluded volume interactions between monomers.

\subsubsection{Application}
\begin{figure}[!htbp]
  \centering
  \subfloat[]{\label{fig:naive_approach_max2:emat} \includegraphics[width = 0.45 \textwidth]{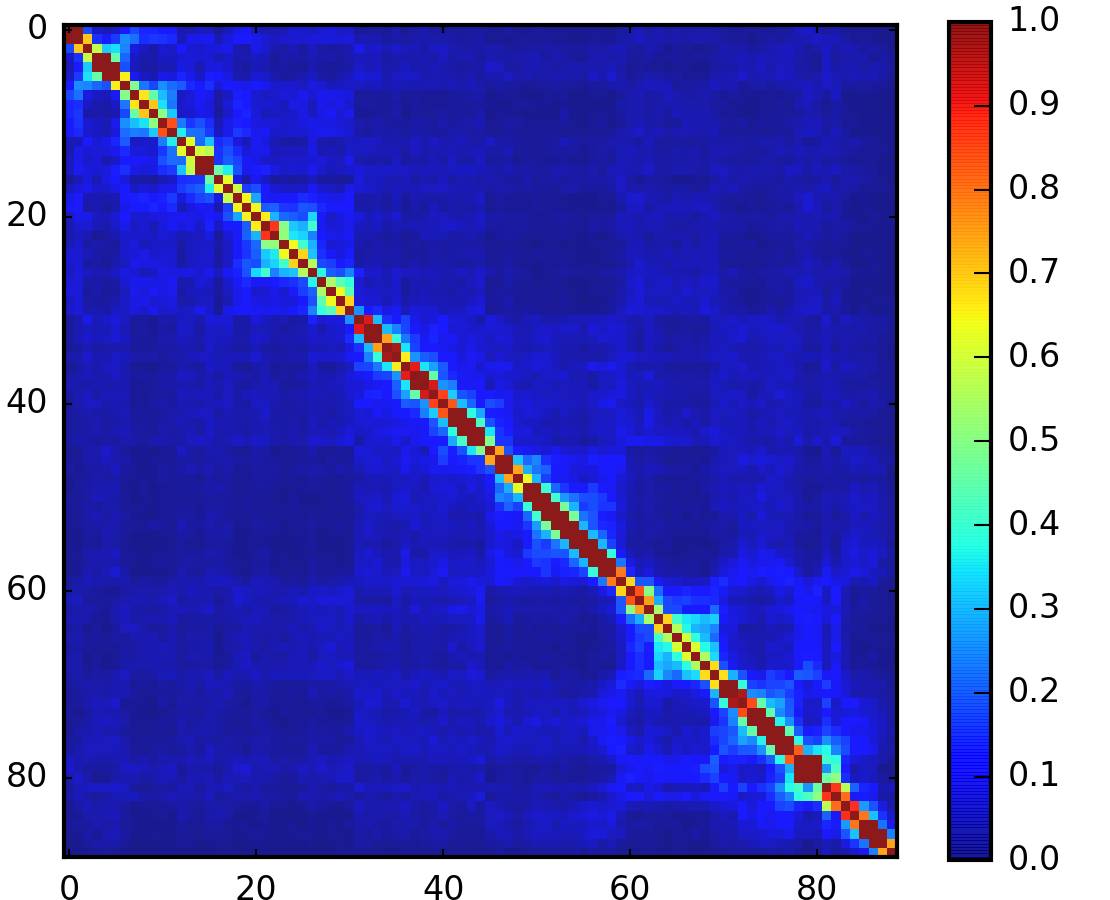}}%
  \quad
  \subfloat[]{\label{fig:naive_approach_max2:emat_hist} \includegraphics[width = 0.45 \textwidth]{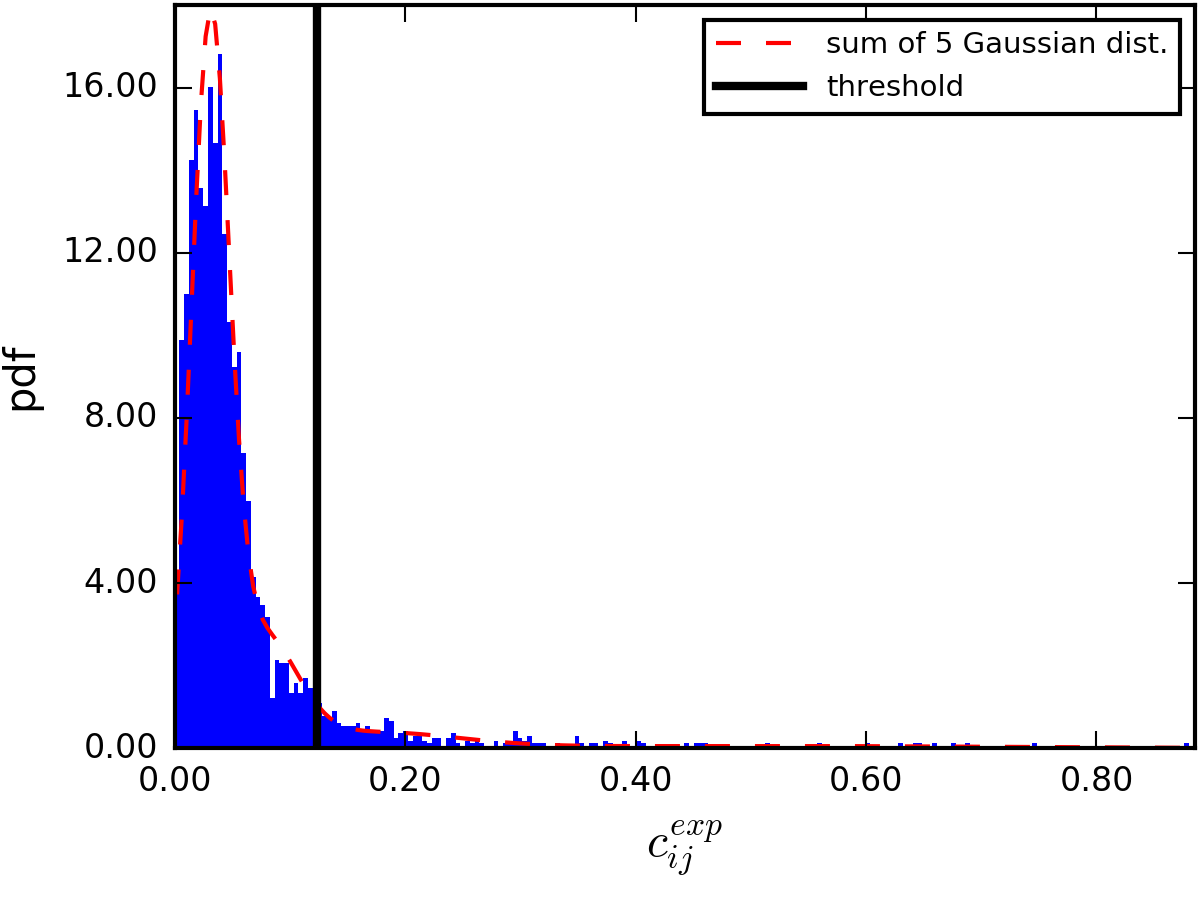}}%
  \\
  \subfloat[]{\label{fig:naive_approach_max2:cmat} \includegraphics[width = 0.45 \textwidth]{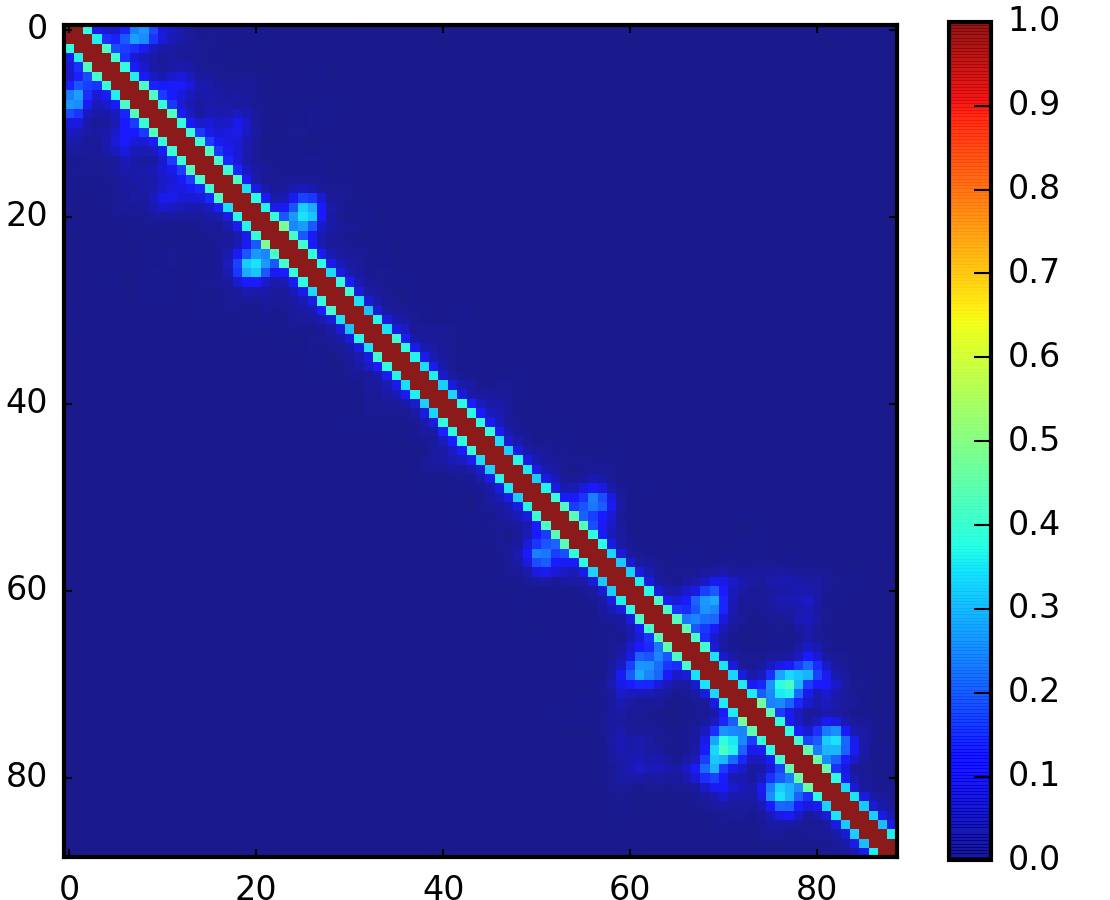}}%
  \quad
  \subfloat[]{\label{fig:naive_approach_max2:scale_optimal}\includegraphics[width=0.45 \textwidth]{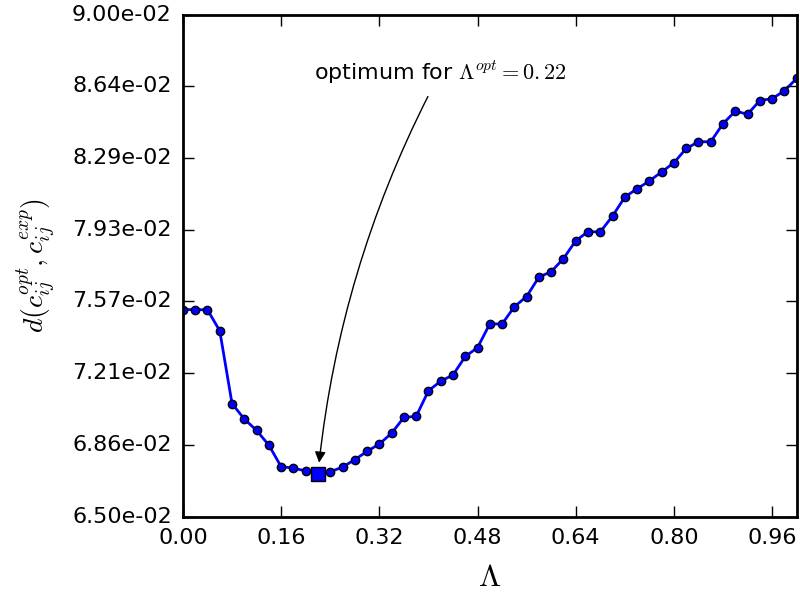}}%
   \\
  \subfloat[]{\label{fig:naive_approach_max2:kmat} \includegraphics[width = 0.45 \textwidth]{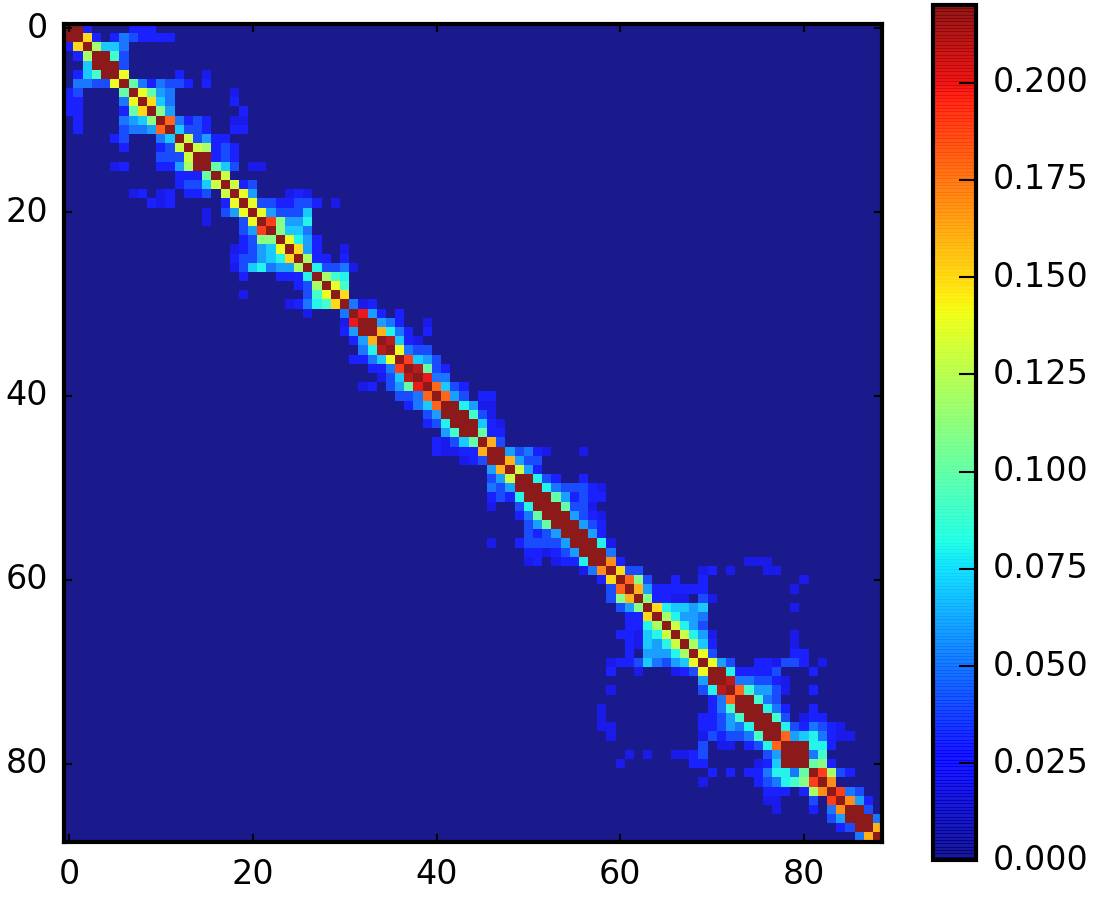}}%
   \quad
  \subfloat[]{\label{fig:naive_approach_max2:conf} \includegraphics[width = 0.45 \textwidth]{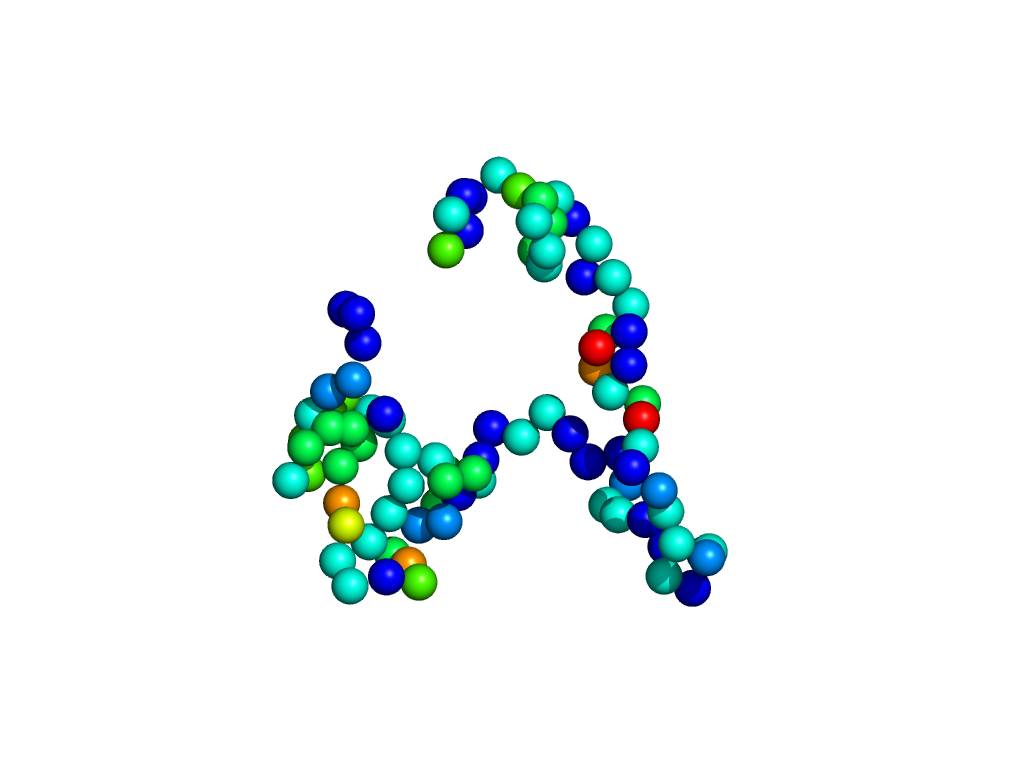}}%
  \caption{Application of a naive approach to identify an optimal GEM model matching the Hi-C contact matrix generated from ref. \cite{Lieberman-Aiden2009} with bin size $\SI{1}{Mbp}$. \protect\subref{fig:naive_approach_max2:emat} Experimental contact matrix constructed from the counts map using a ``Maximum'' normalization with $l_{d}=2$ (see \cref{sec:cmap_generation}). \protect\subref{fig:naive_approach_max2:emat_hist} Probability distribution of the experimental contacts. A threshold is defined from a fit with a sum of Gaussian distributions. \protect\subref{fig:naive_approach_max2:cmat} Final contact matrix obtained from a Brownian dynamics trajectory with $\Lambda^{opt}=0.06$. The threshold used to compute the contact probabilities is $\xi^{opt}=1.82$. \protect\subref{fig:naive_approach_max2:scale_optimal} Plot of the least-square distance $d(c_{ij}^{opt},c_{ij}^{exp})$. Each point is obtained by performing a BD simulation and finding the optimal threshold $\xi^{opt}$ that minimizes $d(c_{ij},c_{ij}^{exp})$. \protect\subref{fig:naive_approach_max2:kmat} Coupling matrix $k_{ij}^{opt}$ corresponding to the final GEM obtained by applying \cref{eq:kij_cij_prop} with $\Lambda=\Lambda^{opt}$. \protect\subref{fig:naive_approach_max2:conf} Snapshot of a Brownian dynamics configuration for the optimal GEM with $\Lambda=\Lambda^{opt}$. The color represents the intensity of the bond rigidity between monomers pairs.}
  \label{fig:naive_approach_max2}
\end{figure}

We have applied this method to Hi-C data from the human chromosome 14 \cite{Lieberman-Aiden2009}. The experimental contact matrices were computed by applying either the ``Trace'' or ``Maximum'' method on the available counts map, as described in \cref{sec:cmap_generation}. In \cref{fig:naive_approach_max2:emat}, we show the experimental contact matrix obtained by using the ``Maximum'' normalization method with $l_d=2$. The \pdf of the corresponding contact probabilities, $c_{ij}^{exp}$, is shown in \cref{fig:naive_approach_max2:emat_hist}.

In order to have a reasonable amount of non-zero $k_{ij}$, we fitted the $c_{ij}^{exp}$ distribution with a sum of $M$ Gaussian distributions, \textit{i.e.}
\begin{equation}
  \proba{c_{ij}^{exp}=c} = \sum \limits_{k=1}^{M} \alpha_k \frac{1}{\sqrt{2 \pi \sigma_k^2}} \exp{\left( - \frac{1}{2} \frac{(c-\mu_k)^2}{\sigma_k^2} \right)}.
\end{equation}

We then considered the sum of the $M^*$ dominant Gaussian distributions such that
\begin{equation}
 \sum_{k \leq M^*} \alpha_k > \underline{\alpha},
\end{equation}
with $\underline{\alpha} = 75 \%$. This defines a distribution $f_b(c)$ for the bulk of the $k_{ij}$, with mean and standard deviation given by:
\begin{align}
  \begin{aligned}
    & \mu_b & = & \quad \frac{\sum \limits_{k=1}^{M^*} \alpha_k \mu_k}{\sum \limits_{k=1}^{M^*} \alpha_k} \\
    & \sigma_b^2 & = & \quad \frac{\sum \limits_{k=1}^{M^*} \alpha_k (\mu_k^2 + \sigma_k^2)}{\sum \limits_{k=1}^{M^*} \alpha_k} - \mu_b^2,
  \end{aligned}
\end{align}
which we used to define the threshold:
\begin{equation}
  \underline{c} = \mu_b + 3 \sigma_b.
  \label{eq:naive_approach_threshold_eij}
\end{equation}

Therefore, the coupling matrix $K$ was set using \cref{eq:kij_cij_prop}, but we discarded all values such that $c_{ij}^{exp} < \underline{c}$. We applied this procedure for $0 \le \Lambda \le 1$, using a discretization such that $\mathrm{d}\Lambda = \num{2e-2}$.

For the BD simulations, we modeled the chromosome as a flexible polymer with FENE bonds. We also introduced Lennard-Jones interactions to model excluded volume interactions between monomers. In order to sample equilibrium configurations, we first performed a relaxation run of \num{e7} iterations with integration time step $dt=\num{e-2}$ and without excluded volume interactions. The value of $\Lambda$ was progressively increased to reach its final value. This relaxation run is meant to loose the memory of the initial condition and to sample many configurations without topological constraints. We then performed an intermediate run of \num{e6} iterations with integration time step $dt=\num{e-3}$ in which overlaps between monomers are removed. Finally, the main run consists of \num{e8} iterations of Langevin dynamics with integration time step $dt=\num{e-3}$, in which the excluded volume interactions are modeled with a Lennard-Jones potential. From this final trajectory, we extracted \num{e3} evenly sampled configurations that we used to compute the model contact probability matrix.

After sampling a BD trajectory for each value of $\Lambda$, we computed the model contact matrix for $0 \le \xi \le 2$ with $\mathrm{d}\xi=\num{2e-2}$, and selected the threshold minimizing the distance between experimental and predicted contacts. In \cref{fig:naive_approach_max2:scale_optimal} we showed the distance $d(c_{ij}^{opt},c_{ij}^{exp})$ as a function of the couplings scale $\Lambda$. This distance reaches a minimum at $\Lambda=\Lambda^{opt}$, which corresponds to the optimal GEM given the experimental contacts. The coupling matrix at the optimal scale, $K=K^{opt}$, is shown in \cref{fig:naive_approach_max2:kmat} and the corresponding predicted contacts obtained with the optimal threshold applied to the BD trajectory is shown in \cref{fig:naive_approach_max2:cmat}.

\subsubsection{Conclusion}
We have presented here a simple method to model chromosome architecture. If we assume that a proportionality relation holds between couplings and contact probabilities, the value obtained for $\Lambda^{opt}$ determines the closest GEM reproducing the experimental contacts. The proportionality hypothesis from \cref{eq:kij_cij_prop} requires solely to adjust the scale $\Lambda$, and the numerical procedure is therefore computationally less demanding than adjusting all $k_{ij}$ independently as in \cite{Giorgetti9502014}. However, it is clear that this proportionality assumption has no rigorous justification. \textit{A fortiori} there is no better reason to choose this one rather than the $d_{ij} \sim 1 / c_{ij}$ assumption made in the literature. Therefore, in the sequel we present another approach rooted in an analytical expression of the contact probabilities $c_{ij}$ of a GEM.

\begin{figure}[!htbp]
  \centering
  \subfloat[]{\label{fig:naive_approach_max3:emat} \includegraphics[width = 0.45 \textwidth]{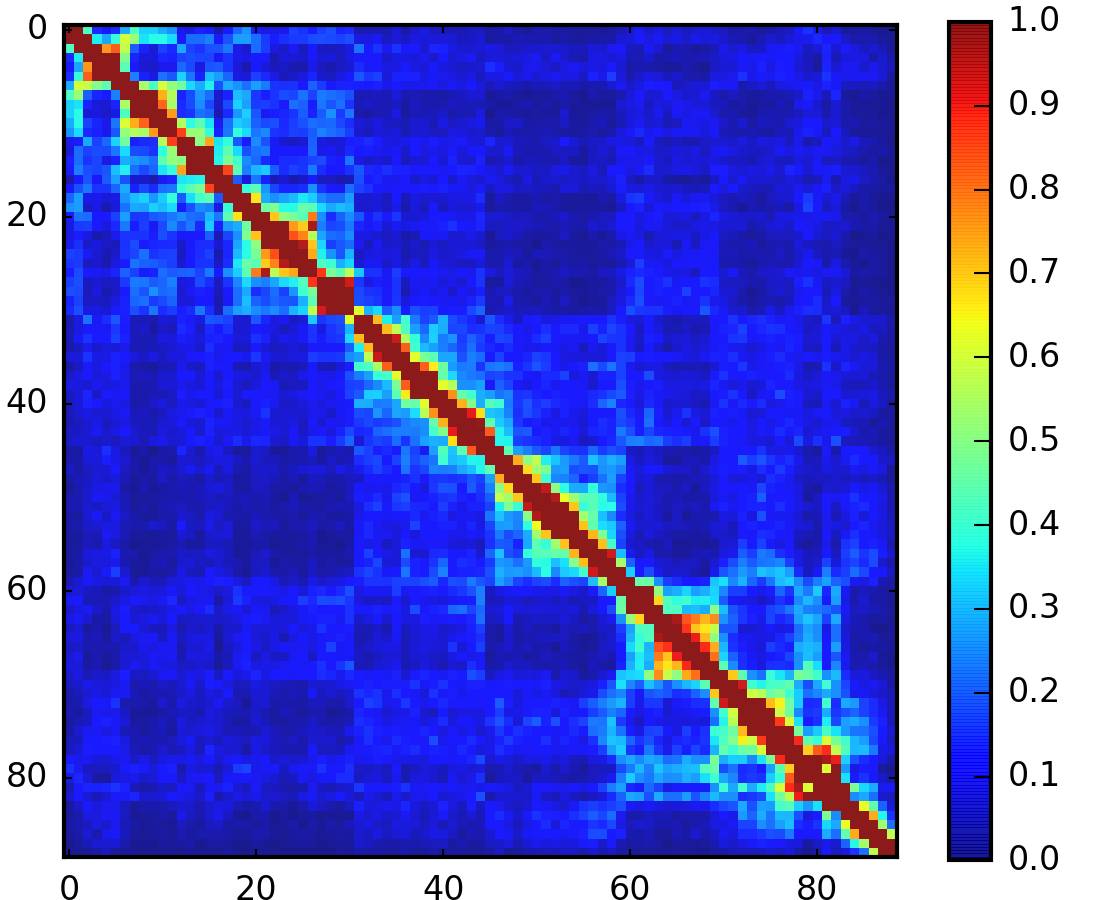}}%
  \quad
  \subfloat[]{\label{fig:naive_approach_max3:emat_hist} \includegraphics[width = 0.45 \textwidth]{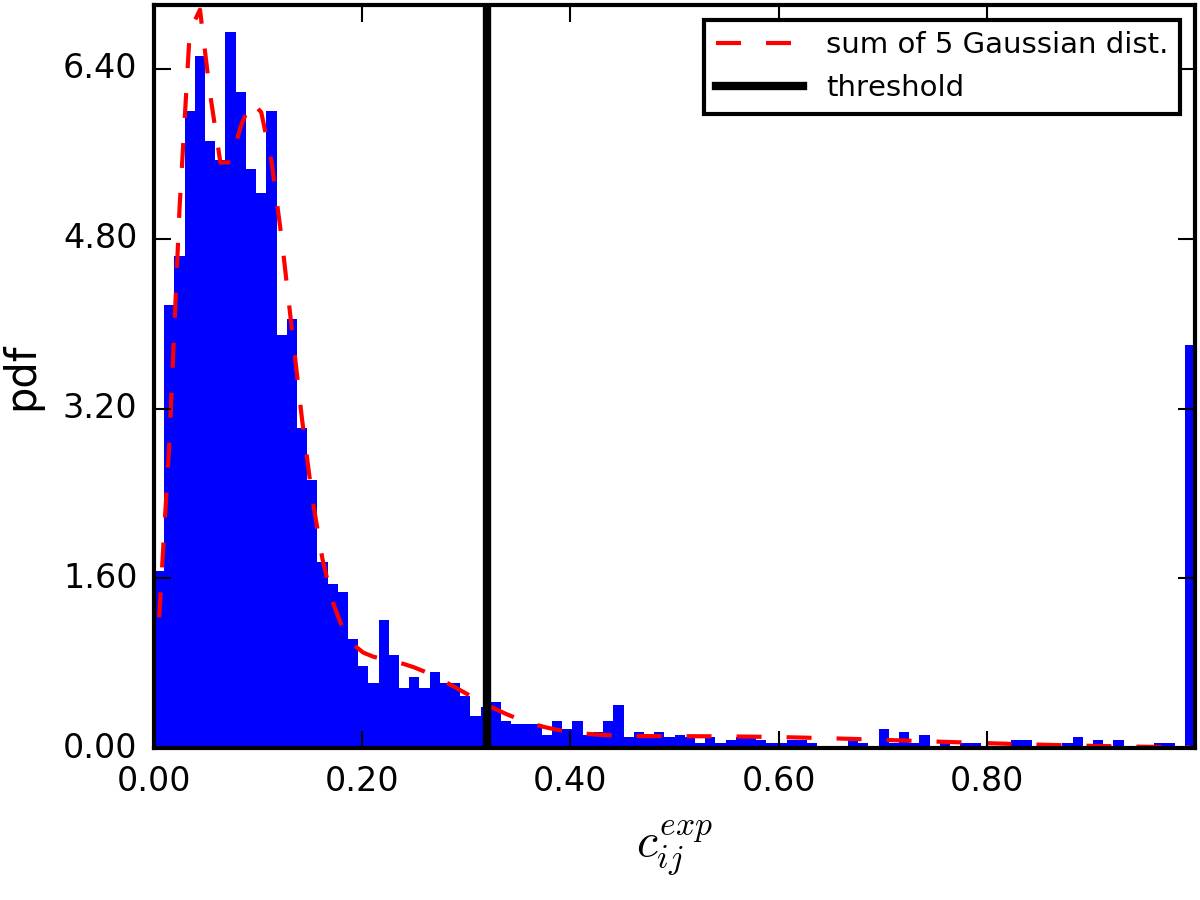}}%
  \\
  \subfloat[]{\label{fig:naive_approach_max3:cmat} \includegraphics[width = 0.45 \textwidth]{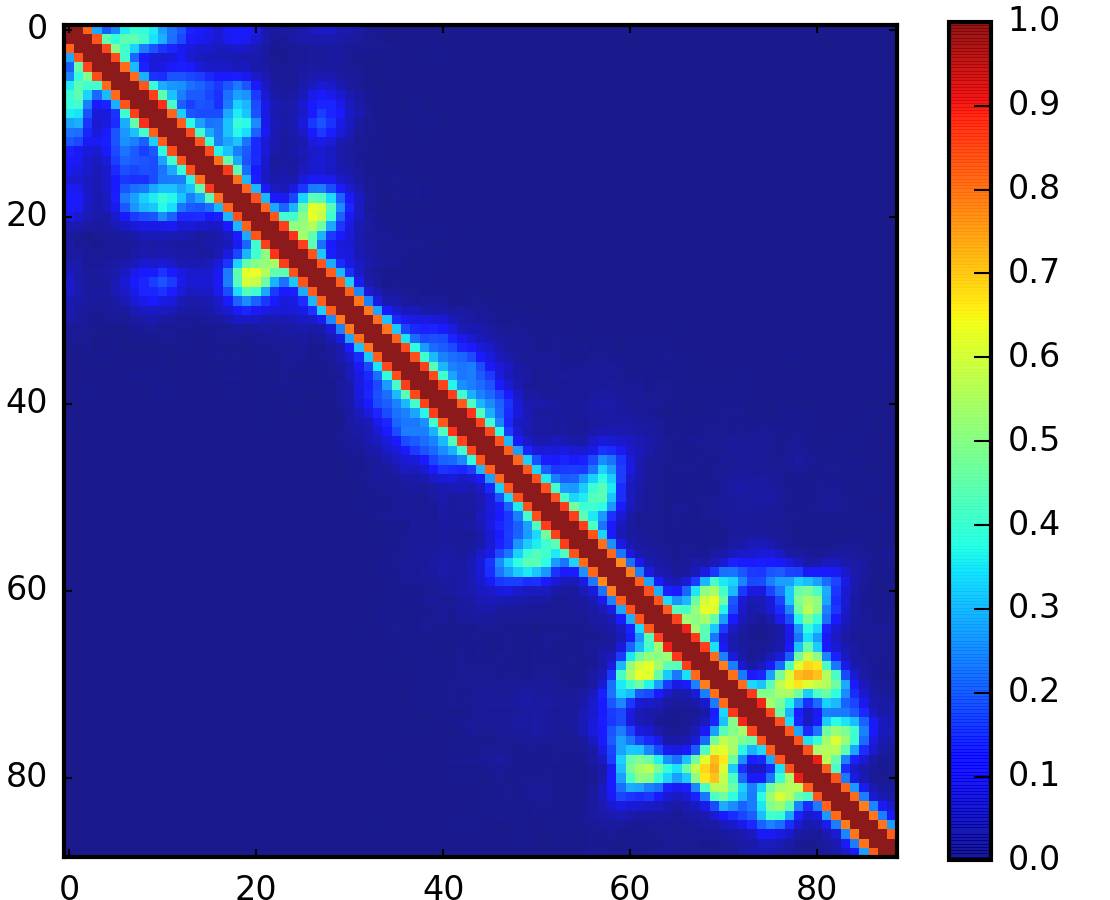}}%
  \quad
  \subfloat[]{\label{fig:naive_approach_max3:scale_optimal}\includegraphics[width=0.45 \textwidth]{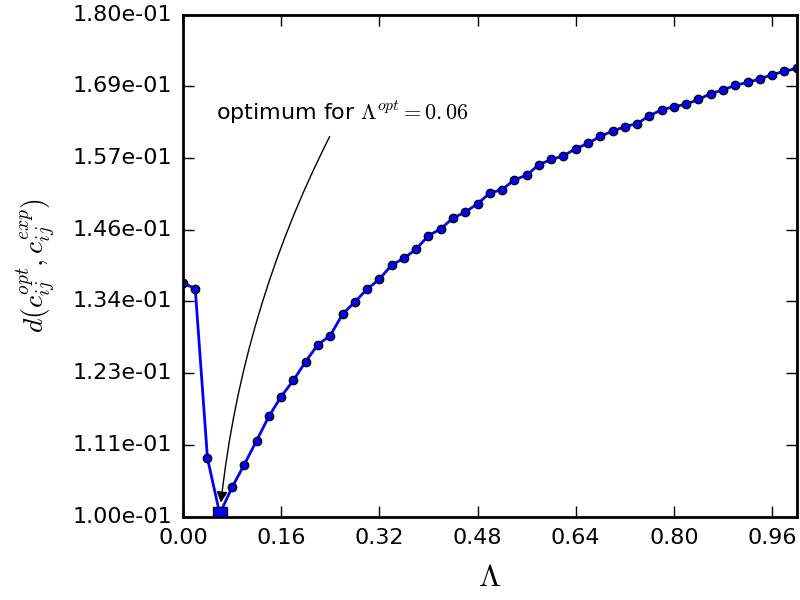}}%
   \\
  \subfloat[]{\label{fig:naive_approach_max3:kmat} \includegraphics[width = 0.45 \textwidth]{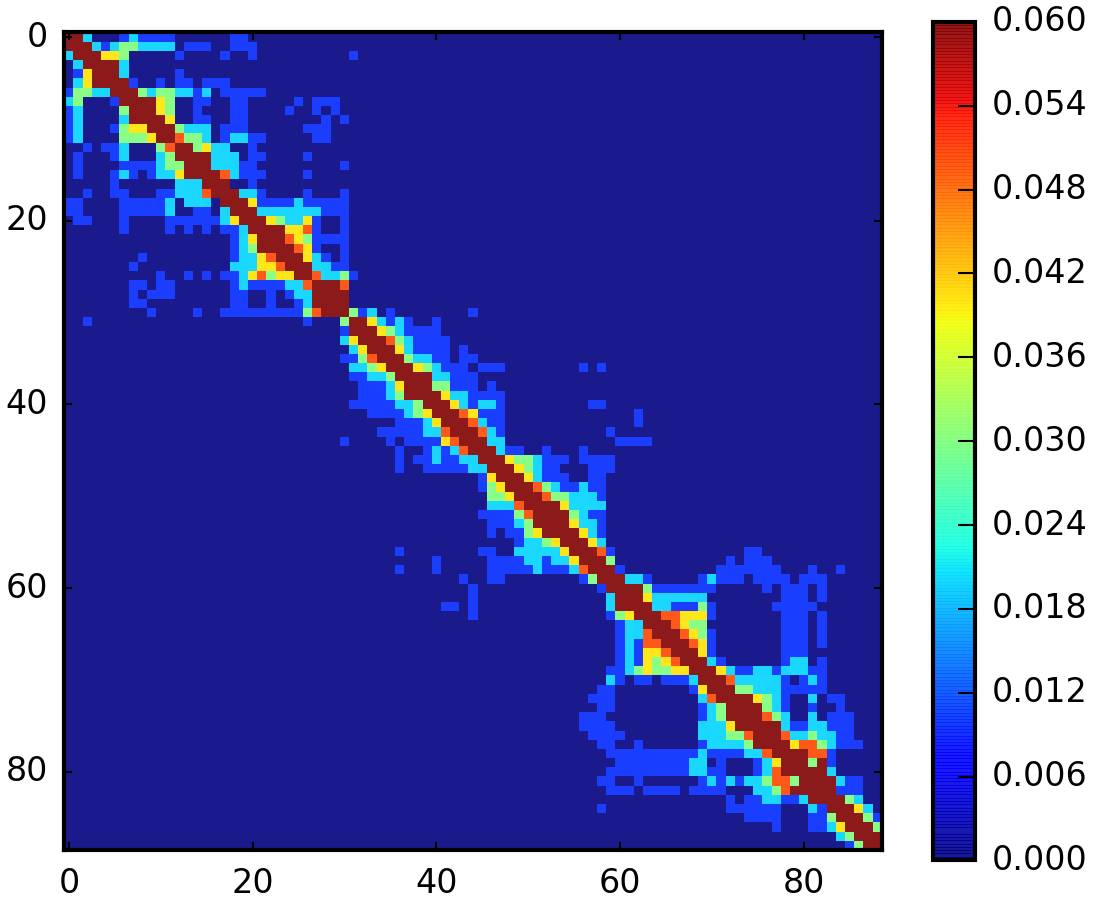}}%
   \quad
  \subfloat[]{\label{fig:naive_approach_max3:conf} \includegraphics[width = 0.45 \textwidth]{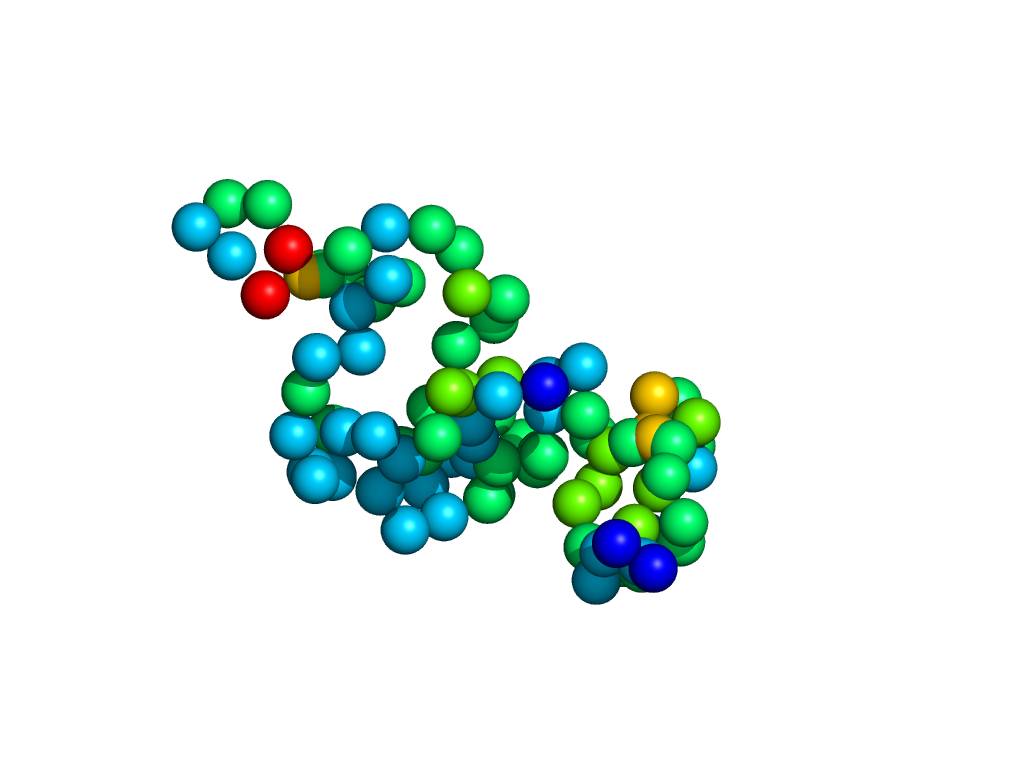}}%
  \caption{Same as \cref{fig:naive_approach_max2} but with a ``Maximum'' normalization with $l_d=3$ for the Hi-C contact probability matrix.}
  \label{fig:naive_approach_max3}
\end{figure}

\subsection{One-to-one correspondence between couplings and contact probabilities}
We now investigate a more rigorous choice for the couplings $k_{ij}$ of the GEM. Let us formally express the contact probability between monomers $i$ and $j$ as:
\begin{align}
  \begin{aligned}
  c_{ij} & = \langle \mu(r_{ij}) \rangle  \\
  &= \int \ud{^3 \vec{r}} \mu(r) \langle \delta(\vec{r}_{ij} - \vec{r}) \rangle,
  \end{aligned}
  \label{eq:gem_contact_proba_formal}
\end{align}
where $\mu(r)=\theta(\xi-r)$. As before, $\theta$ stands for the theta function and $\xi$ is the threshold distance below which a contact is said to occur. In order to make progress, we need to express the \pdf of the pair distance, $\langle \delta(\vec{r}_{ij} - \vec{r})\rangle$. This quantity can be evaluated by standard Gaussian calculus using the weight in \cref{eq:gem_thermo_avg}. We obtain:
\begin{equation}
  \langle \delta(\vec{r}_{ij} -\vec{r}) \rangle = \left( \frac{2 \pi b^2 \gamma_{ij}}{3} \right)^{-3/2} \exp{\left( - \frac{3}{2 b^2} \frac{r^2}{\gamma_{ij}} \right)},
  \label{eq:gem_pair_distance}
\end{equation}
where we have introduced the matrix $\Gamma$ of the average square distances whose matrix elements are:
\begin{align}
  \begin{aligned}
    & \gamma_{ij} &=& \quad \sigma_{ii} + \sigma_{jj} - 2 \sigma_{ij} &=& \quad \frac{1}{b^2} \langle r_{ij}^2 \rangle && \text{for } \quad 0 < i < j \le N, \\
    & \gamma_{0j} &=& \quad \sigma_{jj} &=& \quad \frac{1}{b^2} \langle r_{j}^2 \rangle && \text{for } \quad 0 < j \le N.
  \end{aligned}
  \label{eq:gem_gammaij}
\end{align}

The pair distance is a Gaussian distribution, hence the integral in \cref{eq:gem_contact_proba_formal} can be calculated and yields:
\begin{align}
  \begin{aligned}
    c_{ij} &= F_T(\gamma_{ij}) \\
    &= \mathrm{erf}{\left( \frac{X}{\sqrt{2}} \right)}  - \sqrt{\frac{2}{\pi}} X \exp{\left( -\frac{X^2}{2} \right)}, \qquad X=\frac{\xi}{a}\sqrt{\frac{3}{\gamma_{ij}}},
  \end{aligned}
  \label{eq:gem_contact_proba_theta}
\end{align}
where we have introduced the standard error function:
\begin{equation}
  \mathrm{erf}{(x)} = \frac{2}{\sqrt{\pi}} \int \limits_{0}^x \ud{t} e^{-t^2}.
  \label{eq:error_function}
\end{equation}

The function $F_T(\gamma_{ij})$ is a bijection. Hence, to any contact probability $c_{ij}$, \cref{eq:gem_contact_proba_theta} associates a unique average square distance $\gamma_{ij}$. The correlation matrix $\Sigma$ of the GEM can then be determined using \cref{eq:gem_gammaij}, and finally the coupling matrix $K$ can be obtained by inverting $\Sigma$ and using \cref{eq:gem_inverse_correlation,eq:gem_trid_redk}. Thus, we have found a unique mapping between the couplings and the contact probabilities of a GEM. A strategy to infer chromosome architecture then consists in using this mapping to obtain the GEM that reproduces the experimental contact probabilities. Let us emphasize that this is an exact result.

\subsection{Form factors}
So far, we have considered that the measure in \cref{eq:gem_contact_proba_formal} was a theta function, \textit{i.e.} $\mu(r) = \mu_T(r)$ with
\begin{equation}
  \mu_T(r) = \theta(\xi -r).
  \label{eq:gem_mu_theta}
\end{equation}

In the context of Hi-C experiments, this is equivalent to considering that every restriction fragment pair separated by a distance $r < \xi$ is cross-linked. Or in other words, the probability that restriction fragments separated by a distance $r$ cross-link is
\begin{equation}
  \proba{\text{cross-link between i and j} \mid r_{ij}=r} =
  \begin{cases}
    1 & \text{ if } r < \xi \\
    0 & \text{ otherwise }.
  \end{cases}
\end{equation}

However, there are many experimental artefacts that make this assumption quite unrealistic. In particular as already pointed out in \cref{sec:hic_caveats}, the chemical compound used to cross-link DNA, which is formaldehyde, is known to polymerize in aqueous solution. Thus formaldehyde oligomers with different polymerization indices are present in solution, resulting in cross-links with varying lengths. For that reason, the cross-linking probability may be more accurately represented by a measure which ensures that most of the cross-links occur for distances $r<\xi$, but also allow for few cross-links to occur when $r > \xi$. To serve this purpose, we introduce a Gaussian measure:
\begin{equation}
  \mu_G(r) = \exp{\left( - \frac{3}{2} \frac{r^2}{\xi^2} \right)},
  \label{eq:gem_mu_gaussian}
\end{equation}
and an exponential measure:
\begin{equation}
  \mu_E(r)= \exp{\left( - \frac{r}{\xi} \right)}.
  \label{eq:gem_mu_exponential}
\end{equation}

Let us emphasize that the form factor $\mu(r)$ is not a \pdf so it does not need to be normalized. It should rather be considered as a probability for a Bernoulli random variable. For a restriction fragment pair separated by a distance $r$, the probability to cross-link is $\mu(r)$ and the probability not to cross-link is $1 - \mu(r)$. Note that $\mu(0)=1$.

The contact probability in \cref{eq:gem_contact_proba_formal} can be re-computed for each form factor to obtain a mapping between contact probabilities and couplings, similarly to \cref{eq:gem_contact_proba_theta}. We obtain for the Gaussian form factor:
\begin{align}
  \begin{aligned}
    c_{ij} &= F_G(\gamma_{ij}) \\
           &= \left( 1 + \frac{b^2 \gamma_{ij}}{\xi^2} \right)^{-3/2},
  \end{aligned}
  \label{eq:gem_contact_proba_gaussian}
\end{align}
and for the exponential form factor:
\begin{align}
  \begin{aligned}
    c_{ij} &= F_E(\gamma_{ij}) \\
    &= (1 + Y^2) \left( 1 - \mathrm{erf}{\left( \frac{Y^2}{2} \right)} \right) \exp{\left( \frac{Y^2}{2} \right)} - Y \sqrt{\frac{2}{\pi}}, \qquad Y = X^{-1} = \left( \frac{\xi}{a} \sqrt{\frac{3}{\gamma_{ij}}} \right)^{-1}.
  \end{aligned}
  \label{eq:gem_contact_proba_exponential}
\end{align}

\begin{figure}[!htbp]
  \centering
  \includegraphics[width = 0.6 \textwidth]{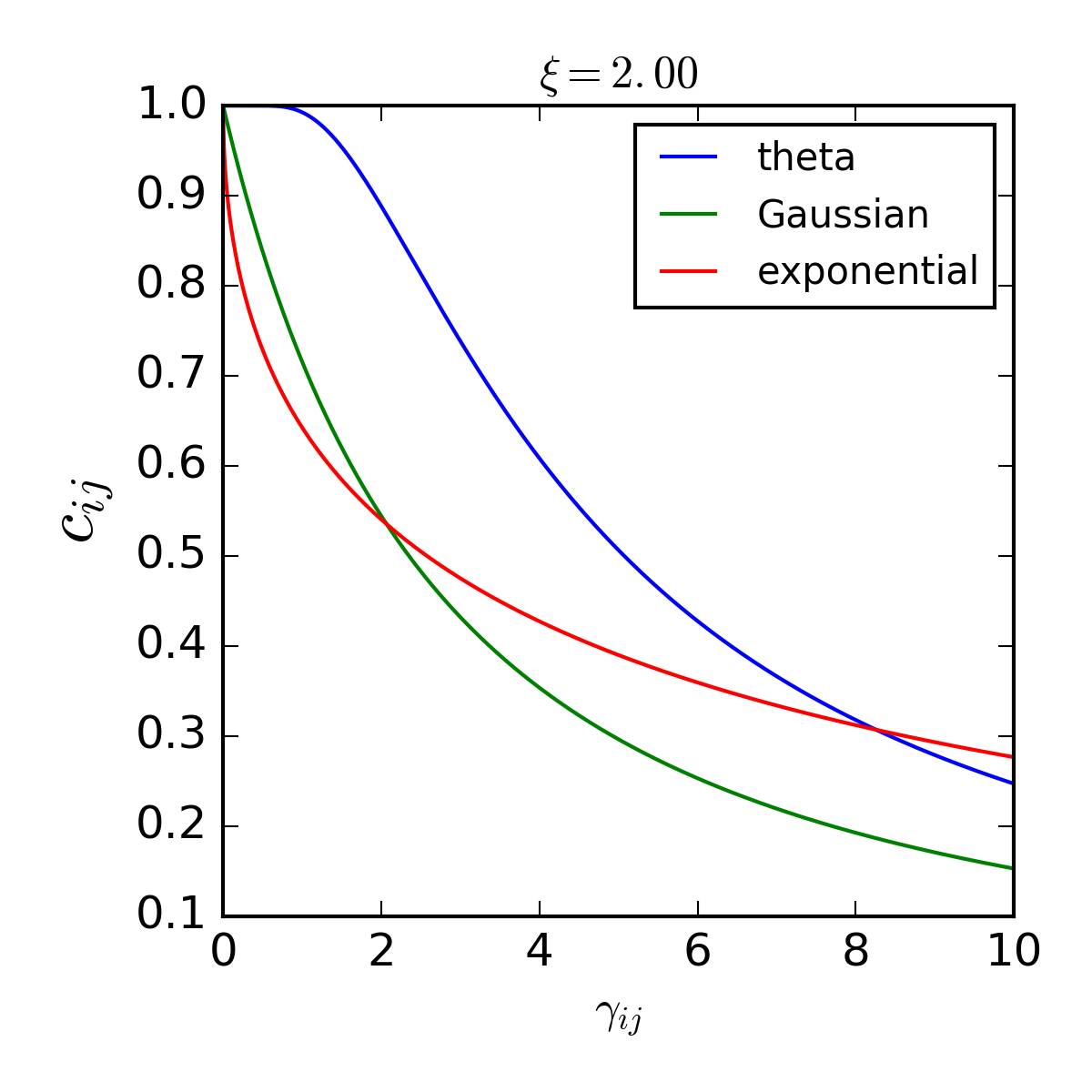}
  \caption{Comparison of mappings between contact probabilities and average square distances for the theta, Gaussian and exponential form factors. The functions are defined in \cref{eq:gem_contact_proba_theta,eq:gem_contact_proba_gaussian,eq:gem_contact_proba_exponential}. We used $b=1$.}
  \label{fig:gem_form_factors}
\end{figure}

In addition to representing more faithfully the experimental conditions, the Gaussian and exponential form factors can be seen as regularization parameters for the contact probabilities. Namely, the saturation of $ c_{ij} \to 1$ as $\gamma_{ij} \to 0$ is less pronounced than with the theta form factor (\cref{fig:gem_form_factors}). In that respect, the Gaussian form factor appears to be the best because $F_G(\gamma_{ij})$ tends to have a greater slope for $c_{ij}$ in the range \numrange{0.1}{1.0}. Thus it is less sensitive to inaccuracies in the measured contact probabilities.

\subsection{Conclusion}
In conclusion, we have introduced a physical model for chromosome architecture. We called this model a Gaussian Effective Model because all interactions between loci on the chromosome have been replaced by effective Gaussian potentials with rigidity coefficients $k_{ij}$. Within this simplified framework, we have been able to compute an analytical expression for the contact probability $c_{ij}$ between monomers $i$ and $j$. It turns out that the contact probability matrix is uniquely determined by the couplings, and reciprocally. We will sometimes refer to this property as the ``GEM mapping'' in the sequel. Importantly, this mapping relies on the choice of a threshold $\xi$ and on a form factor $\mu$. We have found that a Gaussian form factor has the advantage of decreasing the mapping sensitivity to the inaccuracies in the $c_{ij}$ values, and account for formaldehyde polymerization which results in Hi-C contacts to be detected for distances of varying lengths.

\section[Artificial contact probability matrices]{Reconstruction from artificial contact probability matrices}
\subsection{Artificial contact probability matrices}
We plan to use the GEM mapping as a method to give a prediction of chromosome architecture from Hi-C contact probabilities under the form of a GEM. In order to validate the method, we first apply it to artificial contact matrices generated with BD simulations from GEM whose couplings $k_{ij}^{th}$ are known. We have carried out this validation for various sizes ranging from $N+1=20$ to $N+1=1000$. However, in this section, we present the results for $N+1=200$ because it is a reasonable compromise between a not-too-small contact matrix, and not-too-large computational time for BD simulations.

\begin{figure}[!htbp]
  \centering
  \includegraphics[width = 0.7 \textwidth]{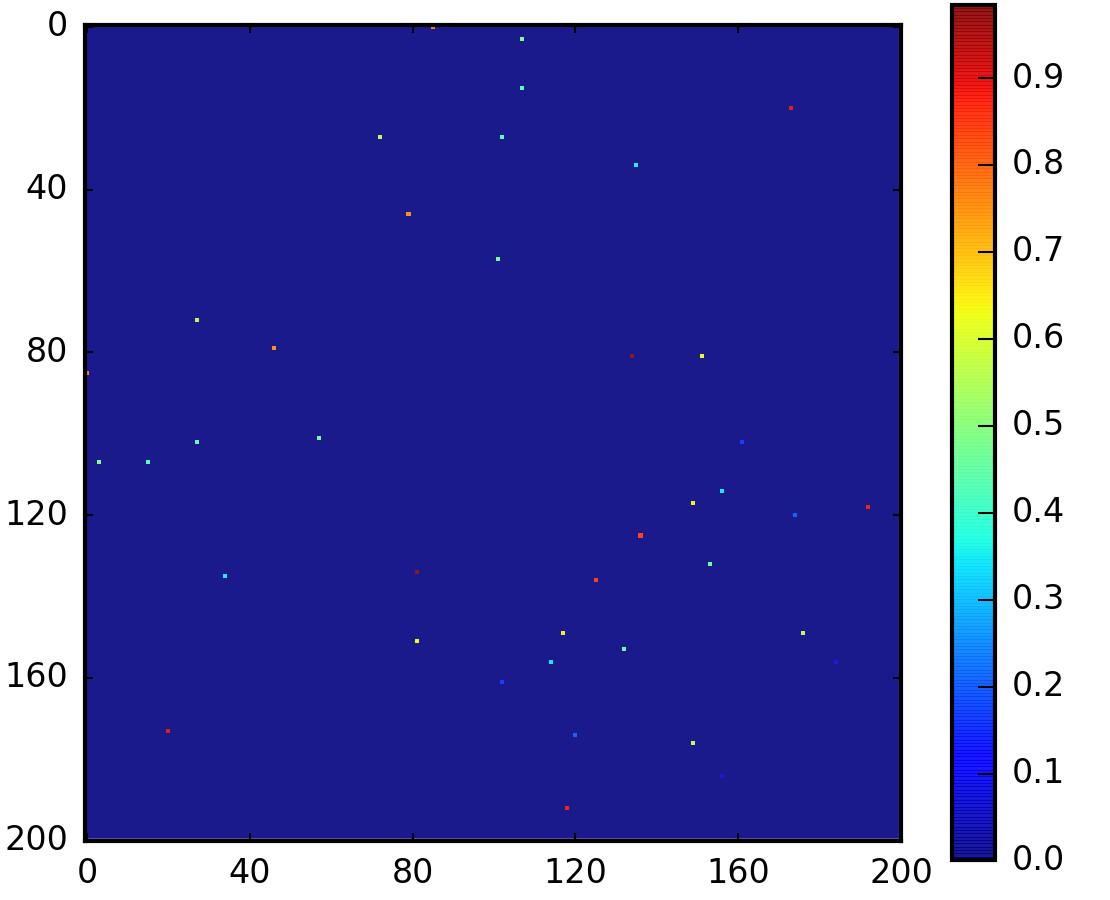}
  \caption{Theoretical coupling matrix $k_{ij}^{th}$ for $N=200$. It is made of $N_c=20$ uniform random variable with scale $\Lambda=1$.}
  \label{fig:gem_kij_th_random}
\end{figure}

In order to construct arbitrary coupling matrices $k_{ij}^{th}$, we randomly choose $N_c$ elements and assign to them a value such that:
\begin{equation}
  k_{ij}^{th}=\Lambda U,
  \label{eq:gem_kij_th_random}
\end{equation}
in which $U$ is a uniform random variable between 0 and 1, and $\Lambda$ is a scale parameter. We therefore obtain a coupling matrix with $N_c$ non-zero elements, which represent the number of constraints of the GEM. An example of such a matrix is shown in \cref{fig:gem_kij_th_random}.

\begin{figure}[!htbp]
  \centering
  \subfloat[]{\label{fig:gem_cij_comparisons:cijth_gauss} \includegraphics[width = 0.30 \textwidth]{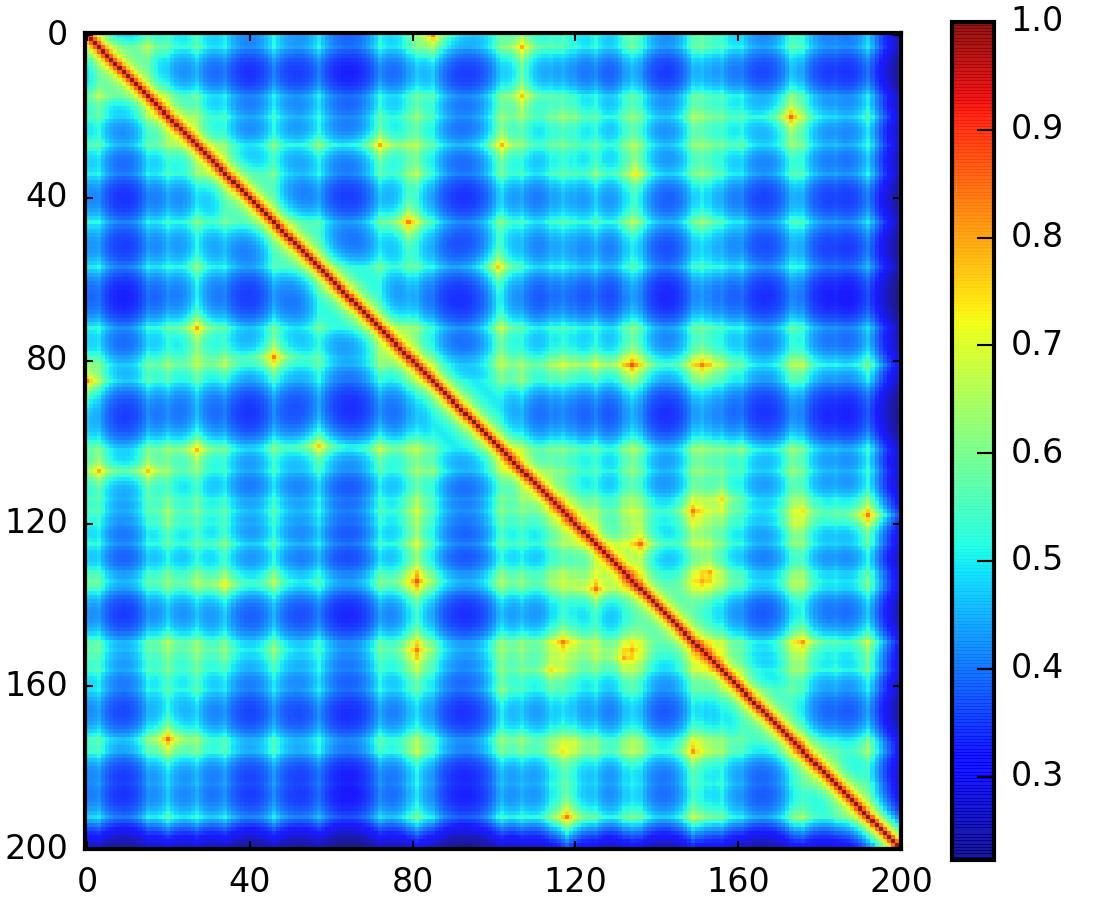} }%
  \\
  \subfloat[]{\label{fig:gem_cij_comparisons:cijexp_gauss_n10} \includegraphics[width = 0.25 \textwidth]{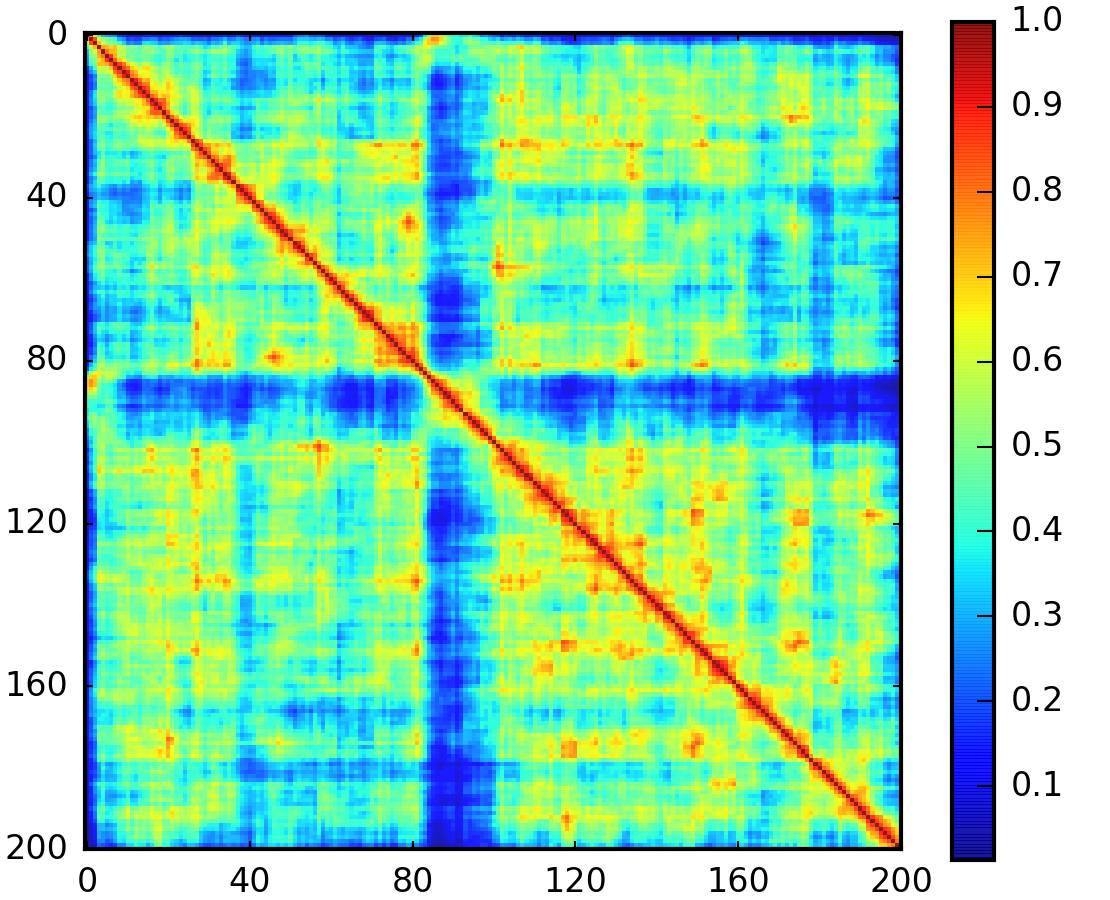} }%
  \quad
  \subfloat[]{\label{fig:gem_cij_comparisons:cijexp_gauss_n100} \includegraphics[width = 0.25 \textwidth]{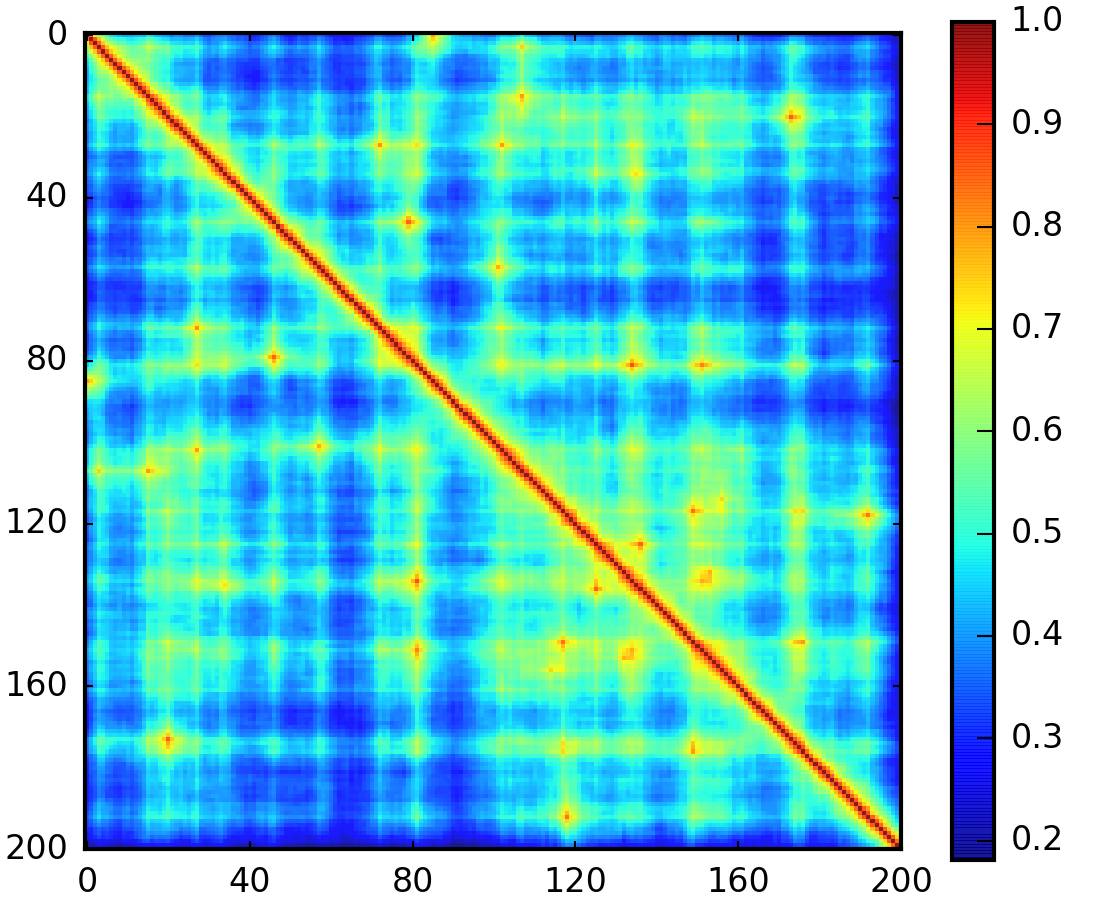} }%
  \quad
  \subfloat[]{\label{fig:gem_cij_comparisons:cijexp_gauss_n1000} \includegraphics[width = 0.25 \textwidth]{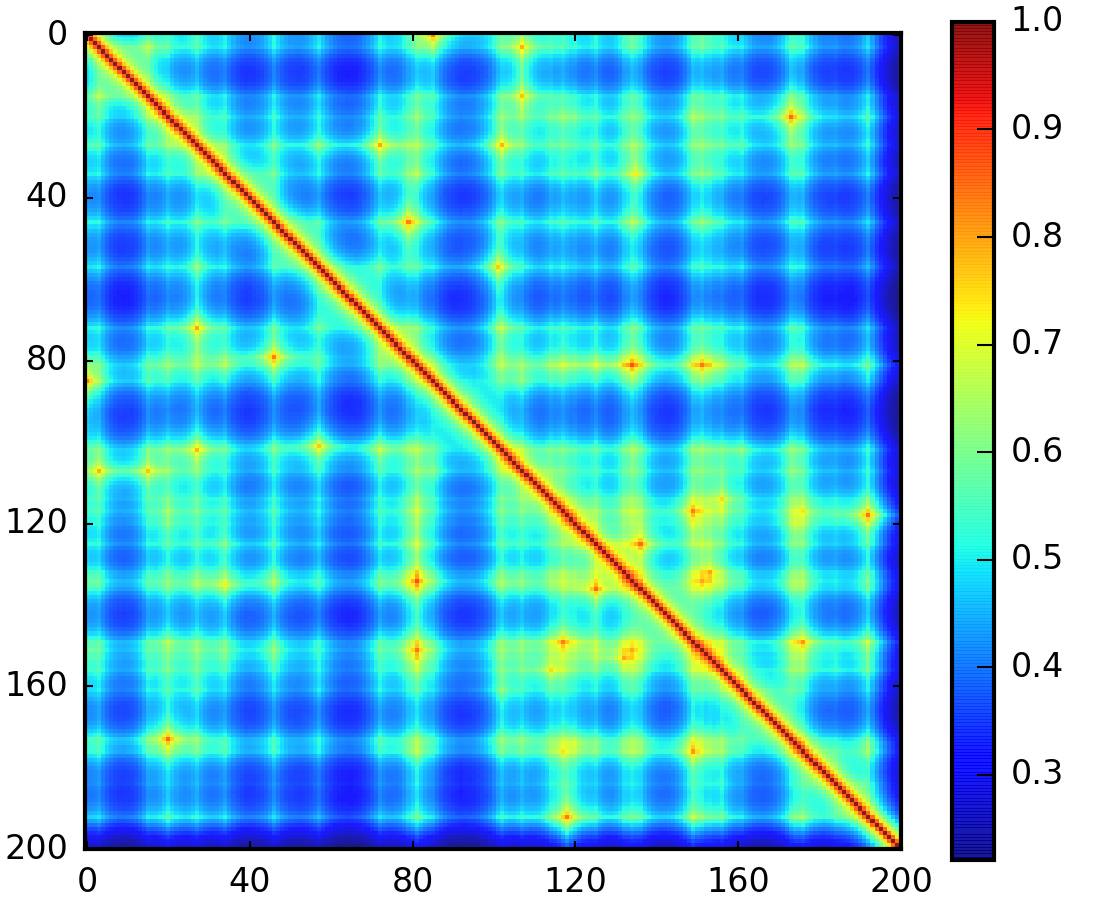} }
  \caption{\protect\subref{fig:gem_cij_comparisons:cijth_gauss} Theoretical contact probability matrix $c_{ij}^{th}$ computed from the GEM with couplings $k_{ij}^{th}$ ($N_c=20$ and $\Lambda=1$). \protect\subref{fig:gem_cij_comparisons:cijexp_gauss_n10}, \protect\subref{fig:gem_cij_comparisons:cijexp_gauss_n100} and \protect\subref{fig:gem_cij_comparisons:cijexp_gauss_n1000} Experimental contact probability matrix $c_{ij}^{exp}$ obtained from BD simulations of the GEM using $10$, $100$ or $1000$ sampled configurations. We used $\xi^{th}=\xi^{exp}=3.00$ and $\mu^{th}=\mu^{exp}=\mu_G$.}
  \label{fig:gem_cij_comparisons}
\end{figure}

Using the one-to-one mapping between the coupling matrix and the contact probability matrix of a GEM, we can compute the theoretical contact probabilities, $c_{ij}^{th}$, associated to the theoretical couplings $k_{ij}^{th}$ of the model. In order to check the validity of this mapping, we run BD simulations of a GEM with the aforementioned couplings. The chain internal energy was Gaussian, as defined in \cref{eq:gem_gaussian_energy}. Simulations were run for \num{e8} iterations with integration time step $dt=\num{e-3}$, from which $1000$ configurations evenly sampled were extracted in order to compute the experimental contact matrix $c_{ij}^{exp}$, with threshold $\xi^{exp}$ and form factor $\mu^{exp}$. Note the distinction that we have introduced for the form factors. Indeed, we need to use a theoretical form factors $\mu^{th}$ when computing the theoretical contact probabilities $c_{ij}^{th}$ from the theoretical couplings; and we also need to specify a form factor $\mu^{exp}$ which is used to compute the contact probabilities $c_{ij}^{exp}=\langle \mu^{exp}(r_{ij}) \rangle$ from configurations sampled with BD simulations. In the sequel, unless stated otherwise, we used a Gaussian form factor to compute both the theoretical and experimental contacts, \textit{i.e.} $\mu^{th}=\mu^{exp}=\mu_G$. As shown in \cref{fig:gem_cij_comparisons}, the theoretical and experimental contact probability matrices are in very good agreement. The difference can be attributed to thermal fluctuations and the finite number of configurations used to compute the experimental contacts.

In conclusion, the correspondence found in \cref{sec:gem_model} between couplings and contact probabilities of a GEM has been checked with BD simulations. We now move on to use this relation in order to infer the couplings from a given experimental contact probability matrix. In the rest of this section, we will call experimental contact probabilities the probabilities computed from BD simulations.

\subsection{Direct method for reconstructing a Gaussian effective model}
\label{sec:gem_results_direct_inversion}
Here we provide a first method to derive chromosome architecture from experimental contacts. We shall use the GEM mapping to express the model average square distances from the experimental contacts:
\begin{equation}
  \hat{\gamma}_{ij} = F_G^{-1}(c_{ij}^{exp})
  \label{eq:gem_direct_mapping}
\end{equation}
where we used hats to emphasize that this is a prediction of GEM matching the experimental contacts. Actually, because the previous relation is exact, the predicted  and experimental contacts are the same and $\hat{c}_{ij} = c_{ij}^{exp}$. The predicted couplings, $\hat{k}_{ij}$, can then be simply computed using \cref{eq:gem_gammaij,eq:gem_inverse_correlation}. Thus this method is quite straightforward.

However, what should be the threshold $\xi$ used in \cref{eq:gem_direct_mapping}? In \cref{sec:gem_model_naive_approach}, free parameters were adjusted in order to minimize the distance between predicted and experimental contacts. But precisely because $\hat{c}_{ij} = c_{ij}^{exp}$, we cannot hope to use this method to find the optimal threshold $\xi^{opt}$. In the present case, because we know the couplings of the underlying GEM model used to generate the experimental contacts, it is pretty obvious that the optimal threshold should minimize $d(k_{ij},k_{ij}^{th})$. In other words, the predicted couplings should be as close as possible to the theoretical ones. Consequently, we define:
\begin{equation}
  \xi^* = \underset{\xi}{\mathrm{argmin}}(d(k_{ij},k_{ij}^{th})),
  \label{eq:gem_directinv_thres_star}
\end{equation}
which can be seen as a hidden optimal threshold.

In practical applications however, experimental contact matrices are not generated from an underlying GEM, therefore we have to find another criterion to choose the threshold in the reconstruction procedure. Intuitively, one may expect that a good GEM candidate should not alter the rigidity of the underlying polymer chain. In other words, couplings near the diagonal should be close to zero, so that the sum in \vref{eq:gem_inverse_correlation} leaves the $t_{ij}$ elements unchanged. In order to do this, we monitored the norm of the matrix $\Delta_l$ obtained by taking only $k_{ij}$ values such that $\mid i - j \mid \le l$, and assigning other values to zero. Thus we define:
\begin{equation}
  \xi^{opt} = \underset{\xi}{\mathrm{argmin} \Vert \Delta_l \Vert}.
  \label{eq:gem_directinv_thres_opt}
\end{equation}

\begin{figure}[!htbp]
  \centering
  \subfloat[]{\label{fig:gem_directinv_covariations:gauss} \includegraphics[width = 0.45 \textwidth]{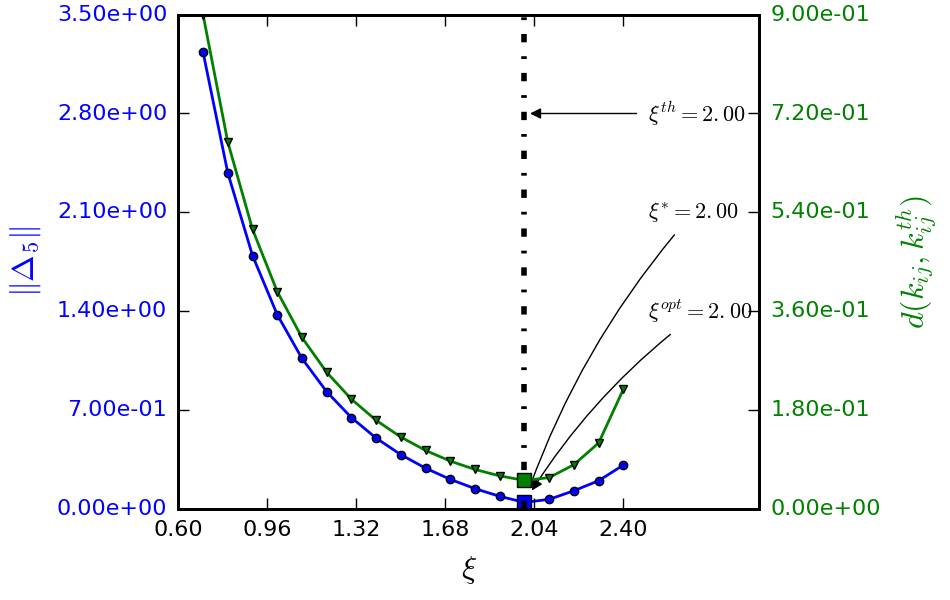}}%
  \quad
  \subfloat[]{\label{fig:gem_directinv_covariations:exp} \includegraphics[width = 0.45 \textwidth]{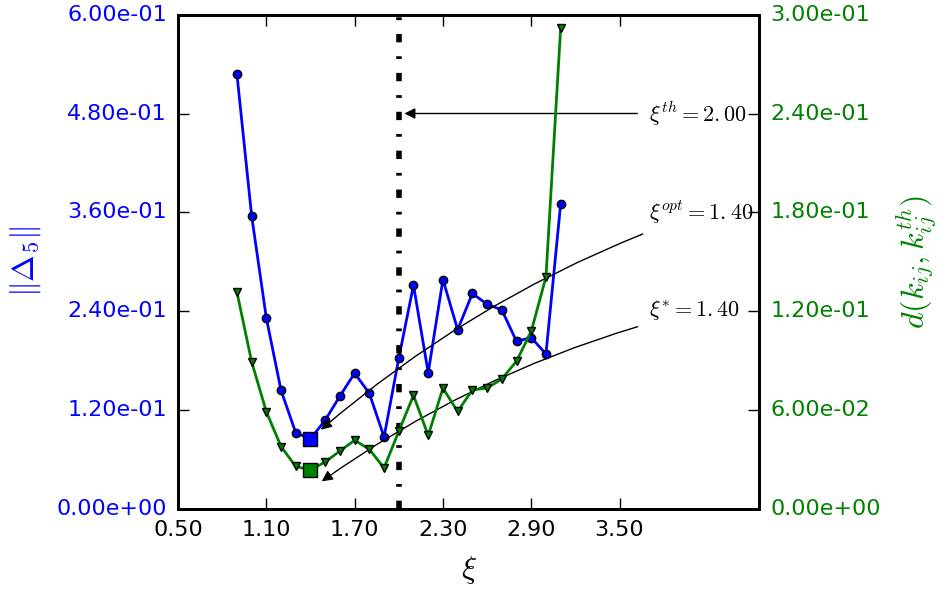}}
  \caption{Comparison of $d(\hat{k}_{ij},k_{ij}^{th})$ and $\Vert \Delta_5 \Vert$ for different thresholds in the direct reconstruction procedure with $\mu=\mu_G$. Experimental contact matrix $c_{ij}^{exp}$ were computed from a BD trajectory using a threshold $\xi^{exp}=2.00$ and: \protect\subref{fig:gem_directinv_covariations:gauss} a Gaussian form factor $\mu^{exp}=\mu_G$; \protect\subref{fig:gem_directinv_covariations:exp} an exponential form factor $\mu^{exp}=\mu_E$.}
  \label{fig:gem_directinv_covariations}
\end{figure}

In \cref{fig:gem_directinv_covariations}, we show that $d(k_{ij},k_{ij}^{exp})$ and $\Vert \Delta_l \Vert$ have approximately the same variations, and in most cases, their minimum is achieved for the same threshold, \textit{i.e.} $\xi^*=\xi^{opt}$. For the experimental contact matrices computed from the same BD trajectory, we used either a Gaussian form factor ($\mu^{exp}=\mu_G$) or an exponential form factor ($\mu^{exp}=\mu_E$). We then applied \cref{eq:gem_direct_mapping} with a Gaussian form factor in both case to obtain a candidate GEM. Note that we did not show the result for an experimental contact matrix using a theta form factor ($\mu^{exp}=\mu_T$) because the inversion procedure gave an unstable GEM (the correlation matrix is not positive definite). When applying the retrieval method to the Gaussian contact matrix we retrieve that the optimal threshold is $\xi^{opt}=\xi^{exp}$. On the contrary, when we apply the retrieval method to the exponential contact matrix, we have $\xi^{opt} \neq \xi^{exp}$. This is due to the discrepancy between the exponential form factor used to compute the experimental contacts and the Gaussian form factor of the retrieval method ($\mu \neq \mu^{exp}$). Furthermore, we see that when there is such a discrepancy, the variations of both criteria become jagged as the threshold $\xi$ used in the retrieval method increases (\cref{fig:gem_directinv_covariations:exp}). This leads to the existence of several local minima that makes the definition of the optimal threshold in \cref{eq:gem_directinv_thres_opt} ambiguous. To solve this ambiguity, we took for $\xi^{opt}$ the first local minimum found from the left, \textit{i.e.} when increasing progressively the threshold from small values.

\begin{figure}[!htpb]
  \centering
  \includegraphics[width=0.7 \textwidth]{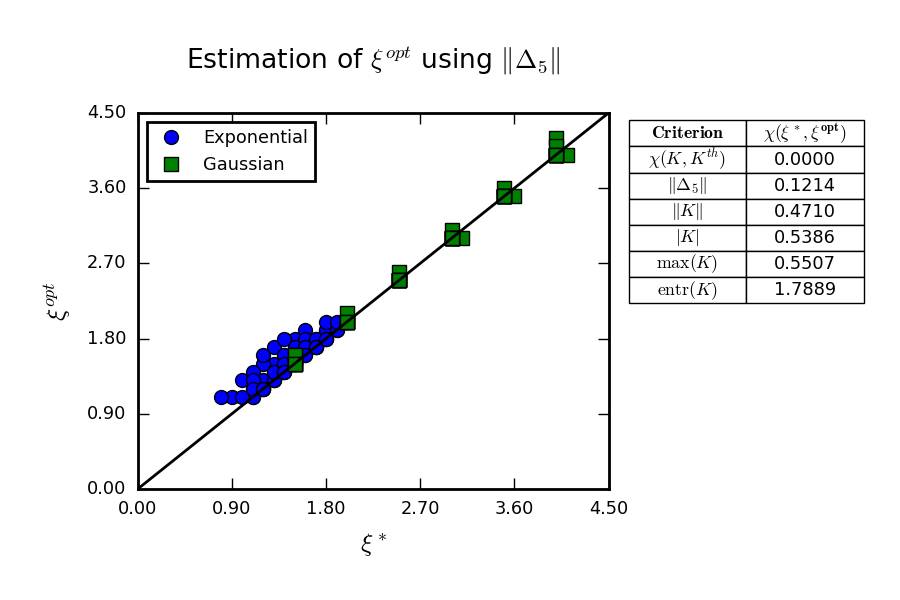}
  \caption{Least-square difference between $\xi^{opt}$ and $\xi^*$. The points correspond to the optimal threshold obtained using the $\Vert \Delta_5 \Vert$ criterion. We used the direct reconstruction procedure applied to experimental contact matrices computed from BD simulations of a GEM with $N=200$, $N_c=5,\, 20,\, 50, \, 100$ and $\Lambda=1,\, 5, \, 10$, and using a Gaussian or an exponential form factor $\mu^{exp}$. The form factor used in the retrieval procedure was Gaussian, $\mu=\mu_G$ .}
  \label{fig:gem_directinv_threshold_chisq_full}
\end{figure}

\begin{figure}[!htpb]
  \centering
  \includegraphics[width=0.7 \textwidth]{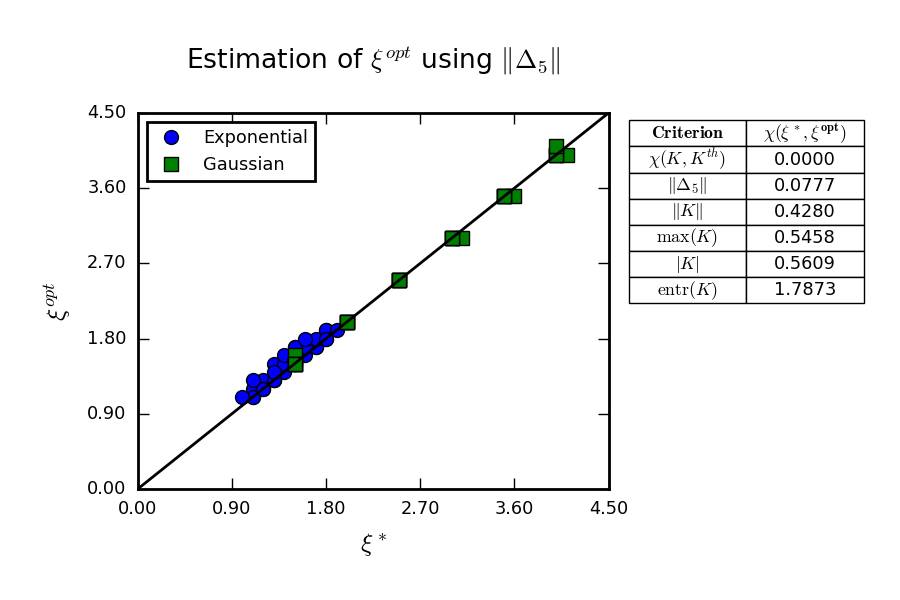}
  \caption{Same as \cref{fig:gem_directinv_threshold_chisq_full} but with $N_c=20,\, 50, \, 100$ and $\Lambda=1$ only.}
  \label{fig:gem_directinv_threshold_chisq_limited}
\end{figure}

Nonetheless, we do not always find $\xi^{opt} = \xi^*$. This is expected because the criterion used to define $\xi^{opt}$, that is to say $\Vert \Delta_l \Vert$, is rather phenomenological. Therefore, we investigated to which extent this criterion can be trusted in order to find the best GEM matching the experimental contacts. In order to do this, we computed the least-square distance between $\xi^{opt}$ and $\xi^*$ obtained for a large number of experimental contacts:
\begin{equation}
  \chi^2({\xi_a^{opt}},{\xi_a^*}) = \frac{1}{2} \sum \limits_{a} \mid \xi_a^{opt} - \xi_a^* \mid^2,
  \label{eq:gem_chisq_thres_star_opt}
\end{equation}
where the index $a$ runs over different experimental contact matrices. More accurately, we sampled BD trajectories for a number of constraints $N_c=5,\, 20,\, 50,\, 100$ and scaling coefficient $\Lambda=1,\, 5,\, 10$, from which we computed experimental contact matrices using either a Gaussian ($\mu^{exp}=\mu_G$) or an exponential form factor ($\mu^{exp}=\mu_E$). We then applied the retrieval procedure and computed $\xi^{opt}$ and $\xi^*$ as explained above. The results obtained suggest that $\xi^{opt}$ is a good approximation of $\xi^*$ (\cref{fig:gem_directinv_threshold_chisq_full}). Actually, we also carried out this analysis for other criteria. Namely we monitored: $\Vert K \Vert$, the Froebenius norm of the coupling matrix; $\vert K \vert=\sum \mid k_{ij} \mid$; $\max{(K)}=\max{(k_{ij})}$ and $\mathrm{entr}(K)$, the entropy of the \pdf of the couplings $k_{ij}$. Yet, $\Vert \Delta_l \Vert$ appeared to be the best criterion in the sense of \cref{eq:gem_chisq_thres_star_opt}. We also found that important deviations of $\xi^{opt}$ from $\xi^*$ occurred mostly for experimental contact matrices obtained from GEM with very few constraints or with a large couplings scale $\Lambda$. Indeed, the same analysis carried out by discarding GEMs with $N_c=5$ and $\Lambda > 1$ significantly improved the performances of the $\Vert \Delta_l \Vert$ criterion (\cref{fig:gem_directinv_threshold_chisq_limited}). Finally, most deviations of $\xi^{opt}$ from $\xi^*$ occurred when $\mu^{exp} \neq \mu$. In that case, $\xi^{opt}$ has a tendency to slightly overestimate $\xi^*$. Altogether, the definition taken for $\xi^{opt}$ gave consistent results.

In conclusion, the bijective relation that exists between the couplings and the contact probabilities of a GEM can be used to propose a chromosome architecture under the form of a GEM with couplings $\hat{k}_{ij}$. The GEM obtained has the property to exactly reproduce the experimental contacts. However, the computation of the GEM couplings from the contact probabilities requires to choose a threshold, which is a parameter in the form factor (see \vref{eq:gem_mu_gaussian}). In this section, we have shown that choosing the threshold that minimizes the norm $\Vert \Delta_l \Vert$, where $\Delta_l$ is the matrix of couplings in which only the diagonal band of length $l$ has been retained, appeared to be a good estimate of the optimal threshold. In particular, we used home-made GEMs together with their contact matrices computed from BD simulations to ensure that the distance $d(\hat{k}_{ij}, k_{ij}^{th})$ between predicted and theoretical couplings is minimum at $\xi=\xi^{opt}$. From a computational standpoint, this method is particularly efficient since it only requires to invert the correlation matrix $\Sigma$ in order to obtain the coupling matrix $\hat{k}_{ij}$.

\subsection{Stability analysis}
In the last section, we have presented a method to compute the GEM reproducing a given experimental contact probability matrix. However, nothing ensures that the GEM obtained is stable, \textit{i.e.} that the correlation matrix $\Sigma$ has only positive eigenvalues. Incidentally, we found that applying the direct reconstruction method to contact matrices generated using a theta form factor (which is maybe the simplest definition of a contact matrix) resulted in unstable GEMs. This is a fundamental weakness of the direct reconstruction method, and it is therefore desirable to better understand under which conditions such instabilities occur. In particular, we may expect that Hi-C contact matrices contain some noise due to inaccuracies in the measures or biases inherent to the experimental procedure. Thus, before presenting an alternative method in the next section, we analyze here the effect of corrupting contact probability matrices with noise on the performance of the direct reconstruction method.

\begin{figure}[!htbp]
  \centering
  \includegraphics[width = 0.6 \textwidth]{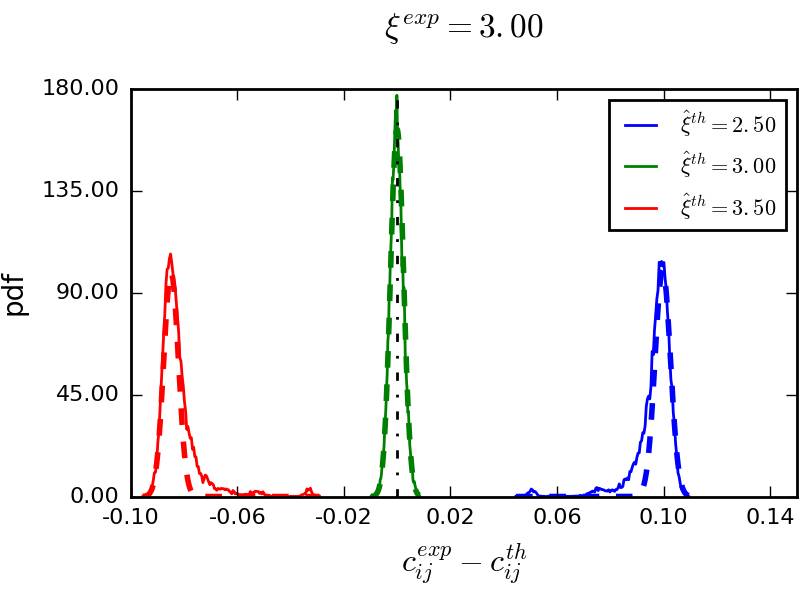}
  \caption{Distribution of the noise $\eta_{ij}=c_{ij}^{exp}-c_{ij}^{th}$, fitted to a Gaussian distribution. We used a Gaussian form factor to compute both contact matrices. $N=200$, $N_c=20$ and $\Lambda=1$.}
  \label{fig:gem_noise_gaussianity}
\end{figure}

Let us start again from our artificial GEM with couplings $k_{ij}^{th}$. We compute the associated contact matrix $c_{ij}^{th}$, using a threshold $\xi^{th}$ and a form factor $\mu^{th}$. When we perform BD simulations of this system, we obtain configurations from which we compute the experimental contact matrix $c_{ij}^{exp}$, using a threshold $\xi^{exp}$ and a form factor $\mu^{exp}$. We assume $\mu^{th}=\mu^{exp}$. Thermal fluctuations, together with the finite number of such configurations results in $c_{ij}^{exp} \neq c_{ij}^{th}$. We may therefore write the experimental contact probabilities as:
\begin{equation}
  c_{ij}^{exp} = c_{ij}^{th} + \eta_{ij},
  \label{eq:gem_noise_etaij}
\end{equation}
where $\eta_{ij}$ can be considered as a noise with unknown distribution, corrupting the ``true'' contact matrix. For $N=200$, $N_c=20$ and $\Lambda=1$, we computed the \pdf of the difference $c_{ij}^{th} - c_{ij}^{exp}$. We used a Gaussian form factor for both the experimental and the theoretical contact matrices, $\mu^{th}=\mu^{exp}=\mu_G$, and we took $\xi^{exp}=2.50$ and different values for $\xi^{th}$ (\cref{fig:gem_noise_gaussianity}). We obtained that when $\xi^{th}=\xi^{exp}$ the \pdf of $\eta_{ij}$ fits well a centered Gaussian distribution. Actually, we also obtained this result when computing the noise with $\mu^{exp}=\mu^{th}=\mu_E$ or $\mu^{exp}=\mu^{th}=\mu_T$.

Consequently, instead of running BD simulations in order to compute experimental contact matrices $c_{ij}^{exp}$, we may construct pseudo-experimental contact matrices by adding a Gaussian noise with mean and variance given by
\begin{equation}
  \langle \eta_{ij} \rangle = 0, \qquad \langle \eta_{ij}^2 \rangle = \varepsilon^2,
  \label{eq:gem_noise_etaij_mean_var}
\end{equation}
to the theoretical contact matrix $c_{ij}^{th}$. This trick allows us to investigate the stability of the direct reconstruction method as a function of the noise amplitude $\varepsilon$. Furthermore, it also allows us to explore more values for $N_c$ than if we had to run systematically a BD simulation.

\begin{figure}[!htpb]
  \centering
  \subfloat[]{\label{fig:gem_noise_nc_eta_ij:large}\includegraphics[width = 0.45 \textwidth]{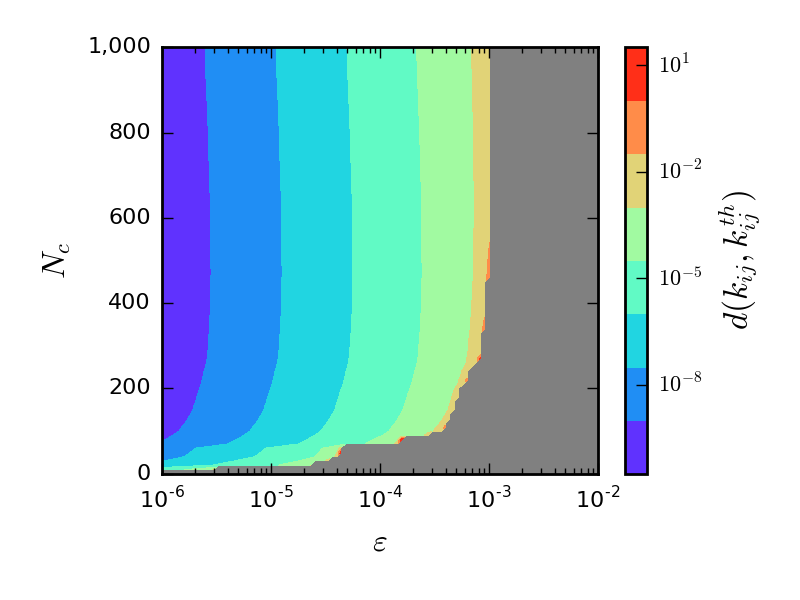}}%
  \quad
  \subfloat[]{\label{fig:gem_noise_nc_eta_ij:zoom}\includegraphics[width = 0.45 \textwidth]{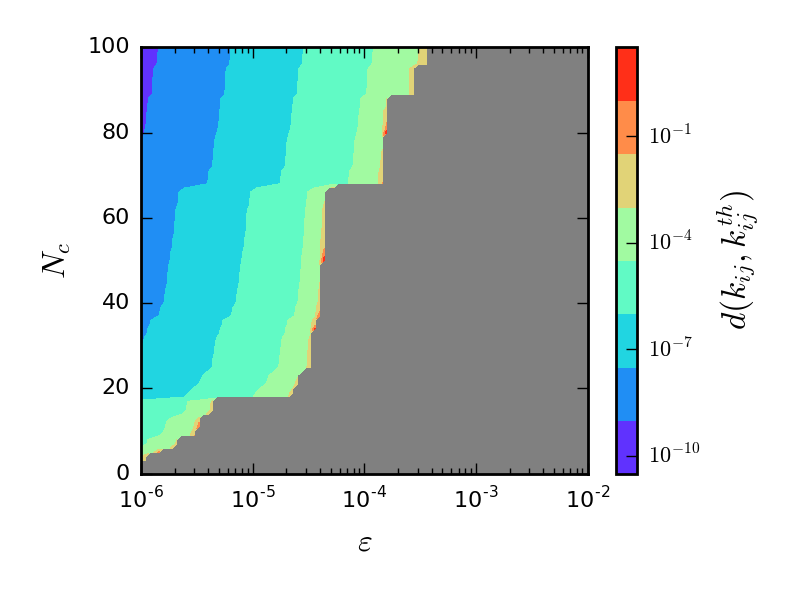}}
  \caption{Performance of the direct reconstruction method when the theoretical contact probabilities $c_{ij}^{th}$ are blurred with a Gaussian noise such that $\langle \eta_{ij} \rangle = 0$ and $\langle \eta_{ij}^2 \rangle = \varepsilon^2$. We used $N=1000$. The region in which the predicted couplings $\hat{k}_{ij}$ define an unstable GEM was shaded in grey. \protect\subref{fig:gem_noise_nc_eta_ij:large} $N_c=0,\dots,1000$. \protect\subref{fig:gem_noise_nc_eta_ij:zoom} Zoom for $N_c=0,\dots,100$.}
  \label{fig:gem_noise_nc_eta_ij}
\end{figure}

Following this observation, we explored the stability of the direct reconstruction method in the $(\varepsilon,N_c)$ plane. Note that for this study only, we used a larger size of polymer and considered $N=1000$. For each value of $N_c$, we generated a random coupling matrix $k_{ij}^{th}$ with scale $\Lambda=1$, and the associated theoretical contact probabilities $c_{ij}^{th}$. We used $\xi^{th}=3.00$ and $\mu^{th}=\mu_G$. Then we computed a pseudo-experimental contact probability matrix $c_{ij}^{exp}$ by adding to the theoretical contact probabilities a centered Gaussian noise with standard deviation $\varepsilon$. Following our previous observation, we assume that the contact probabilities obtained are a good approximation of the experimental contact probabilities that would be obtained by performing a BD simulation of the GEM and computing the contact probabilities with $\xi^{exp}=\xi^{th}$ and $\mu^{exp}=\mu^{th}$. Then we applied the direct reconstruction procedure to $c_{ij}^{exp}$ using $\mu=\mu^{exp}$ and $\xi=\xi^{exp}$, which is the optimal threshold. We therefore obtained a predicted GEM with couplings $\hat{k}_{ij}$ that we compared to the theoretical couplings by computing $d(\hat{k}_{ij}, k_{ij}^{th})$.
The result of this analysis is shown in \cref{fig:gem_noise_nc_eta_ij}, in which we shaded in grey the region where the predicted couplings $\hat{k}_{ij}$ result in an unstable GEM with a correlation matrix $\Sigma$ having negative eigenvalues. We obtain that for each value of the number of constraints, $N_c$, there is an upper bound $\overline{\varepsilon}$ on the noise amplitude such that for $\varepsilon > \overline{\varepsilon}$, the direct reconstruction method fails, in the sense that the predicted GEM is unstable. It is remarkable that for $\varepsilon < \overline{\varepsilon}$ the direct reconstruction methods perform very well, with $d(\hat{k}_{ij}, k_{ij}^{th}) \lesssim \num{e-2}$ in the worse cases. Therefore, the reconstruction appears to be robust to noise until some critical value of the noise amplitude is reached. Then the method suddenly starts to fail. We also note that the value of $\overline{\varepsilon}$ seems to depend on the number of constraints of the underlying GEM. In particular, it is clear that the performances of the direct reconstruction method get worse when $N_c \to 0$. Specifically, for $N_c=0$, we observe that even blurring the theoretical contacts with a noise of amplitude as small as $\varepsilon=\num{e-6}$ is sufficient to make the retrieval fail. To the contrary, the value of $ \overline{\varepsilon}$ seems to be maximum in a range of constraints between $N_c=\SI{10}{\percent} N$ and $N_c = \SI{100}{\percent} N$.

In conclusion, we have shown that the primary reason causing the direct reconstruction method to fail is when the predicted couplings produce an unstable GEM. By definition of the Gaussian effective model, this means that the correlation matrix $\Sigma$ of the Gaussian model has negative eigenvalues. This occurs suddenly when the experimental contacts are corrupted with a noise whose amplitude is above a critical value. However, when the noise's amplitude (or experimental precision error) is below this threshold, the predicted couplings appeared to be close to the theoretical ones.
Hence whenever the direct reconstruction method gives a stable GEM, we may consider that this is the theoretical GEM from which the experimental contact matrix was generated. Therefore, when applying this method to an experimental matrix from a Hi-C experiment, we may consider similarly that whenever the reconstructed GEM is stable, it constitutes a reliable model for the chromosome architecture.

\subsection{Reconstruction of a stable Gaussian effective model}
\subsubsection{How to ensure the stability of the reconstructed Gaussian effective model?}
When the input experimental contacts are very noisy, we have seen that the direct reconstruction procedure fails because the associated GEM is unstable. However, in the space of contact matrices, there may exist a nearby contact matrix which can be mapped to a stable GEM. This remark motivates the design of an alternative method which aims at reconstructing the closest stable GEM. In particular, the predicted contact probabilities may not exactly reproduce the experimental ones. This suggests an approach in which one seeks to minimize the distance $d(c_{ij},c_{ij}^{exp})$ between the contact matrix predicted by a GEM and the experimental contact matrix, under the constraint that the GEM is stable.

A rigorous enforcement of this principle would be to ensure that the correlation matrix of the candidate GEM has strictly positive eigenvalues. Yet, this constraint seems difficult to implement in practice. Instead we turn our attention to the more restrictive condition:
\begin{equation}
  k_{ij} > 0,
  \label{eq:gem_minimization_couplings_positivity}
\end{equation}
which ensures the positivity of the couplings. It is clear that \cref{eq:gem_minimization_couplings_positivity} is a sufficient although not necessary condition for $\Sigma$ to be a positive definite matrix. Indeed, if it is so, then the sum in \vref{eq:gem_interaction_energy} is always positive.

\subsubsection{Minimization procedure}
Finding the best stable GEM matching an input experimental contact matrix can therefore be recast in a minimization problem on the $k_{ij}$ variables, with Lagrangian:
\begin{equation}
  \mathcal{L} = \mathcal{A} + \mathcal{B}.
  \label{eq:gem_minimization_lagrangian}
\end{equation}

The two functionals in the right-hand side (r.h.s.) of \cref{eq:gem_minimization_lagrangian} have the expressions:
\begin{equation}
  \mathcal{A} = \frac{1}{2} \Vert \Sigma^{exp} \cdot \Sigma^{-1} - I  \Vert^2,
  \label{eq:gem_minimization_A}
\end{equation}
where $I$ is the identity matrix, and:
\begin{equation}
  \mathcal{B} = \sum \limits_{i<j} \theta(-k_{ij}) \left( \frac{\mid k_{ij} \mid}{\underline{k}} \right)^p.
  \label{eq:gem_minimization_B}
\end{equation}

For $\mathcal{A}$, we have preferred the expression in \cref{eq:gem_minimization_A} to $d(c_{ij},c^{exp})$ because it is quadratic in the $k_{ij}$, which is desirable for a function to minimize. In particular, in the absence of $\mathcal{B}$ the minimization would reduce to the minimization of a quadratic function whose regularity and convexity ensure straight convergence to the global minimum with standard minimization techniques. We can hope that the addition of the $\mathcal{B}$ functional will not alter too much this property. From a computational standpoint, computing $\mathcal{A}$ and its derivatives is very straightforward while computing $d(c_{ij},c_{ij}^{th})$ requires first to to invert $\Sigma^{-1}$ in order to compute the average square distances $\gamma_{ij}$ from which can be computed the contact probability matrix $c_{ij}$. The latter option involves therefore an additional computational burden that we want to avoid. However, due to the GEM mapping between the inverse correlation matrix with elements $\sigma_{ij}^{-1}$ and the $c_{ij}$, both criteria are equivalent. We give the derivatives of $\mathcal{A}$:
\begin{align}
  \frac{\partial \mathcal{A}}{\partial k_{ij}} =
  \left\lbrace
  \begin{aligned}
    & \frac{\partial \mathcal{A}}{\partial w_{jj}} & \text{ if } 0 = i < j \\
    & \frac{\partial \mathcal{A}}{\partial w_{ii}} + \frac{\partial \mathcal{A}}{\partial w_{jj}} - \frac{\partial \mathcal{A}}{\partial w_{ij}} & \text{ if } 0 < i < j,
  \end{aligned}
  \right.
  \label{eq:gem_minimization_A_derivatives1}
\end{align}
with:
\begin{align}
  \begin{aligned}
    & \frac{\partial \mathcal{A}}{\partial w_{ii}} &=& \quad \left[ S^T (S \cdot \Sigma^{-1} - I ) \right]_{ii} \\
    & \frac{\partial \mathcal{A}}{\partial w_{ij}} &=& \quad \left[ S^T (S \cdot \Sigma^{-1} - I ) \right]_{ij} + \left[ S^T (S \Sigma^{-1} - I ) \right]_{ji},
  \end{aligned}
  \label{eq:gem_minimization_A_derivatives2}
\end{align}
where we have used $S=\Sigma^{exp}$ to alleviate notations, and $w_{ij}$ is a matrix element of the reduced coupling matrix. Note the particular shape for the derivative of the off-diagonal elements in the second line of \cref{eq:gem_minimization_A_derivatives2} which appears when we enforce that the coupling matrix $k_{ij}$ is symmetric.

The $\mathcal{B}$ functional has been chosen arbitrarily to enforce the positivity of the couplings, as required from \cref{eq:gem_minimization_couplings_positivity}. The theta function ensures that the penalty is applied only when some couplings become negative. Besides, $\underline{k}$ can be seen as the modulus of the smallest negative coupling allowed. Indeed, due to the power law in \cref{eq:gem_minimization_B}, the penalty increases abruptly when $k_{ij} < - \underline{k}$. The values of $p$ and $\underline{k}$ have been adjusted from our particular experience. Actually, decreasing $\underline{k}$ results in a more stringent constraint and tends to increase the number of iterations required for the minimization to converge. To a lesser extent increasing $p$ also resulted in a more stringent constraint, but the consequences on the convergence speed were less visible. In practice, we typically used $\underline{k}=0.1$ and $p=8$.

Last but not least, we enforced $k_{ii+1}=0$ and removed these variables from the minimization. We have made this choice to ensure that the bonds rigidity of the Gaussian chain in \cref{eq:gem_gaussian_energy} is not modified.

For the practical implementation of the minimization, we used a standard steepest descent method. In other terms, $k_{ij}$ values were updated according to the equation of motion:
\begin{equation}
  \frac{\partial k_{ij}}{\partial t}(t) = - \frac{\partial \mathcal{L}}{\partial k_{ij}}(\lbrace k_{ij}(t)\rbrace),
  \label{eq:gem_minimization_motion}
\end{equation}
or more precisely its discretized version:
\begin{equation}
  k_{ij}^{(n+1)} = k_{ij}^{(n)} - h \frac{\partial \mathcal{L}}{\partial k_{ij}}^{(n)},
  \label{eq:gem_minimization_steepdesc}
\end{equation}
where $n$ is the time (or iteration) and $h$ represents the time step. Actually, we have also tried more sophisticated methods such as the conjugate-gradient method. However, although the number of iterations required to converge is significantly decreased, each step then requires to perform a line minimization, with several evaluations of $\mathcal{A}$ per iteration. Yet, evaluating $\mathcal{A}$ requires of the order of $O(N^2)$ operations. Therefore we have found that using a simple steepest descent method resulted in a faster convergence to the minimum. Following ideas developed in \cref{sec:gem_model_naive_approach}, we chose to initialize the couplings to $k_{ij}^{(0)} = c_{ij}^{exp}$.

\subsubsection{Speeding up convergence}
While the computation of $\mathcal{A}$ has a complexity in $O(N^2)$, the computation of the gradient of $\mathcal{A}$ requires of the order of $O(N^3)$ operations. This scaling seems at first particularly unadapted to deal with contact matrices with size $N \sim \num{e2}$ or $\num{e3}$ like in Hi-C experimental data sets. However this issue can be circumvented.

The key is to reduce the complexity of the gradient evaluation. It turns out that during the minimization, only a few $k_{ij}$ tend to non-zero values and represent significant constraints. The bulk of the $k_{ij}$ actually decreases quickly to near zero values, which then fluctuate in the vicinity of zero, with a magnitude well below the relevant couplings scale chosen for the procedure: $\mid k_{ij} \mid \ll \underline{k}$. Computational time spent to perform the dynamics on these couplings may be regarded as wasted because they do not correspond in the end to significant constraints and it is of little interest to know whether these couplings have a magnitude near zero or exactly equal to zero.

Therefore, every $n_t$ iterations, we performed a ``trim'' operation. We set all couplings such that $\mid k_{ij} \mid < \underline{k}$ to $k_{ij}=0$ and removed them from the minimization. As a consequence, computing the gradient of $\mathcal{A}$ becomes of complexity $O(M N^2)$ where $M$ is the number of $k_{ij} \neq 0$. In general, $M$ quickly decreases to $M \sim N_c < N $. Hence this trick enables us to save a significant amount of time in the minimization. For practical implementations, we typically used $n_t=100$.

\subsubsection{Results}
In order to validate the minimization method, we applied it to experimental contact matrices obtained from BD simulations performed on our artificial GEMs. We used a form factor $\mu^{exp}$ and a threshold $\xi^{exp}$ but as before, we assumed that this information is hidden in the reconstruction procedure.

Starting from the experimental contact probabilities $c_{ij}^{exp}$, we perform a minimization on the $k_{ij}$ as described above. The couplings $\hat{k}_{ij}$ where $\mathcal{L}$ is minimum define the best stable GEM matching the experimental contacts. Nonetheless, to compute $\Sigma^{exp}$ we had to choose a form factor $\mu$ and a threshold $\xi$. The goal is to find the GEM whose associated contact probability matrix $\hat{c}_{ij}$ is as close as possible to the experimental one. Therefore we define the optimal threshold as the one minimizing the distance to the experimental contacts:
\begin{equation}
  \xi^{opt} = \underset{\xi}{\mathrm{argmin}}(d(\hat{c}_{ij},c_{ij}^{exp})).
  \label{eq:gem_minimization_optimal_threshold}
\end{equation}

\begin{figure}[!htbp]
  \centering
  \subfloat[]{\label{fig:minimization_N200_gauss:Nc20} %
    \includegraphics[width=0.4 \textwidth, valign=c]{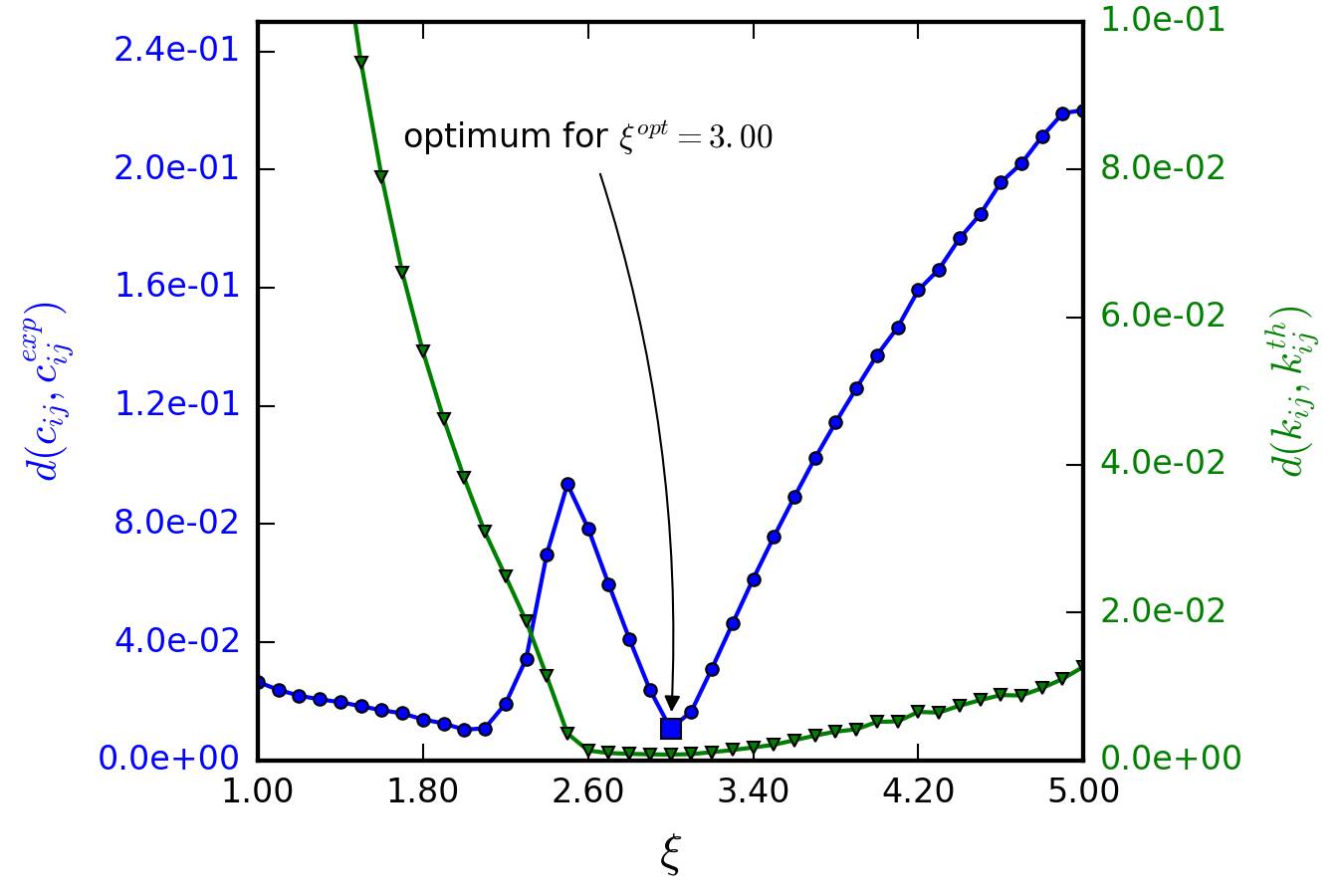}%
    \includegraphics[width=0.28 \textwidth, valign=c]{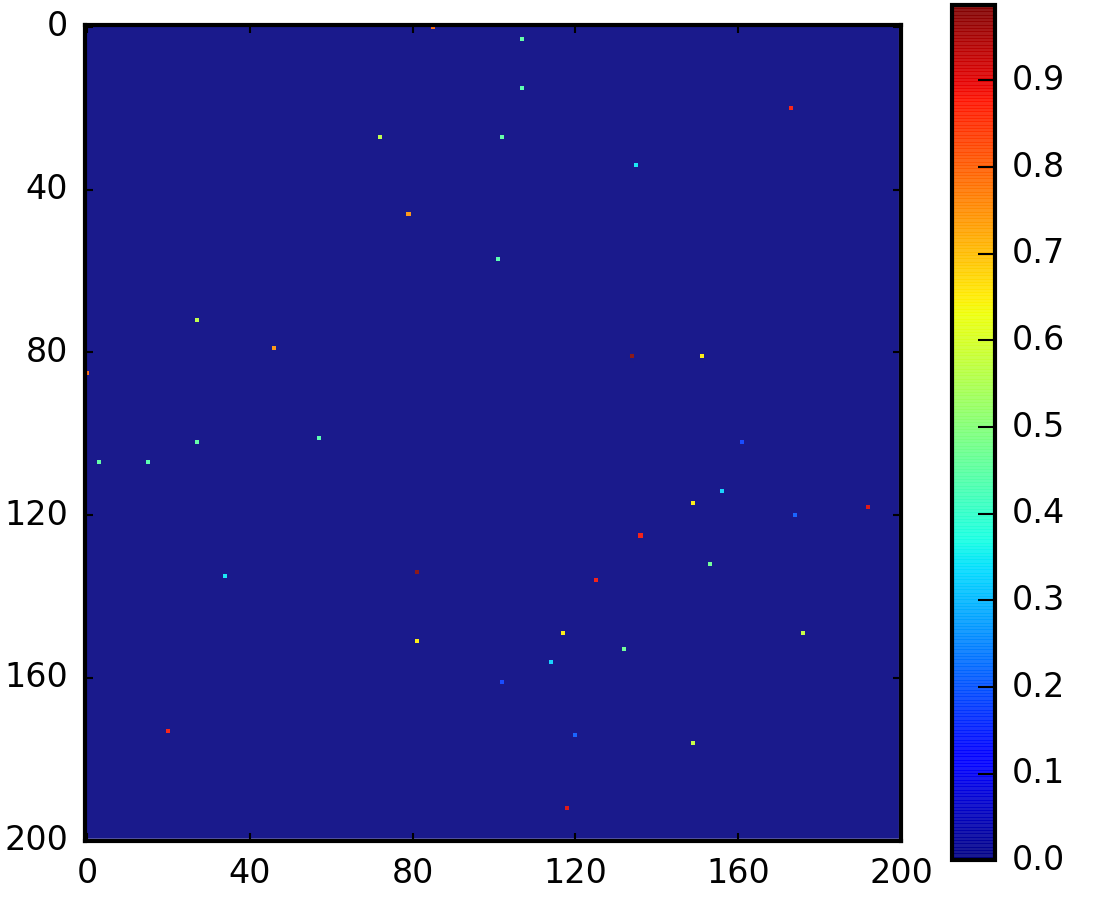}%
    \includegraphics[width=0.28 \textwidth, valign=c]{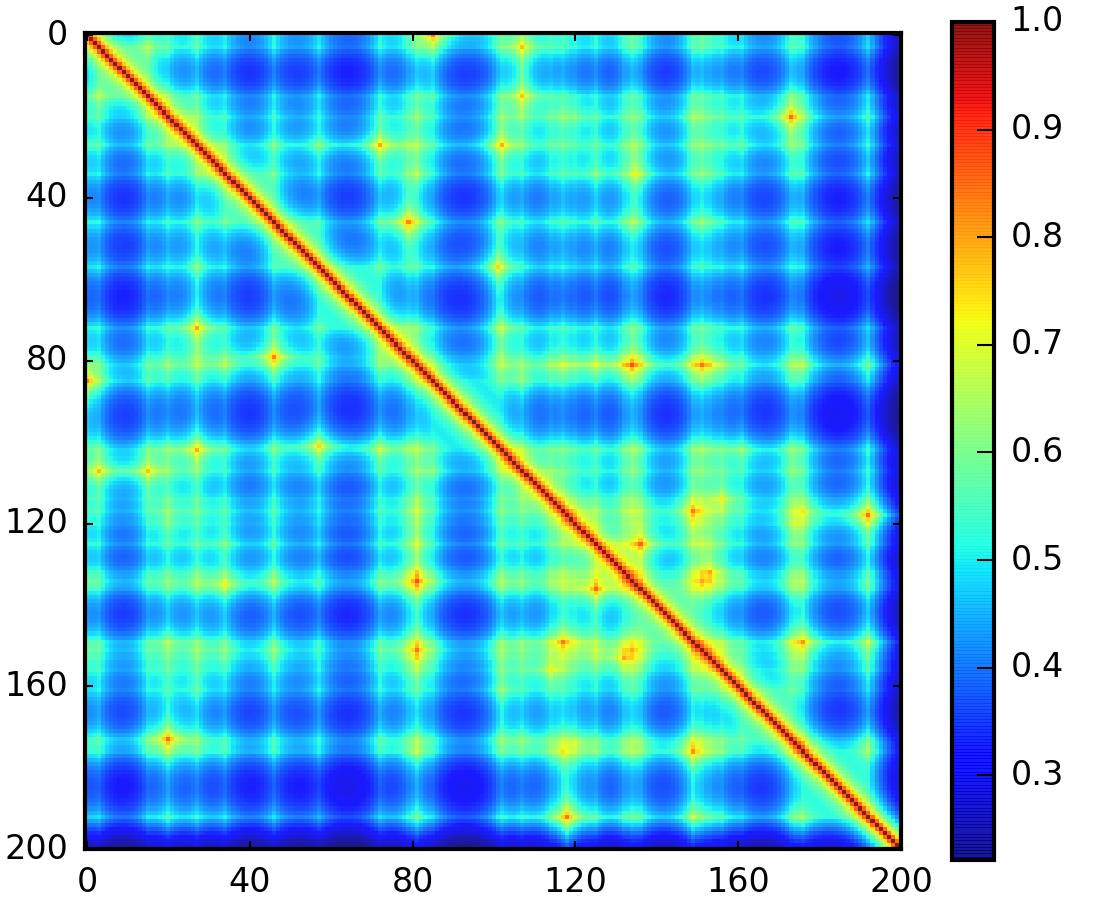}}
    \\
  \subfloat[]{\label{fig:minimization_N200_gauss:Nc50} %
    \includegraphics[width=0.4 \textwidth, valign=c]{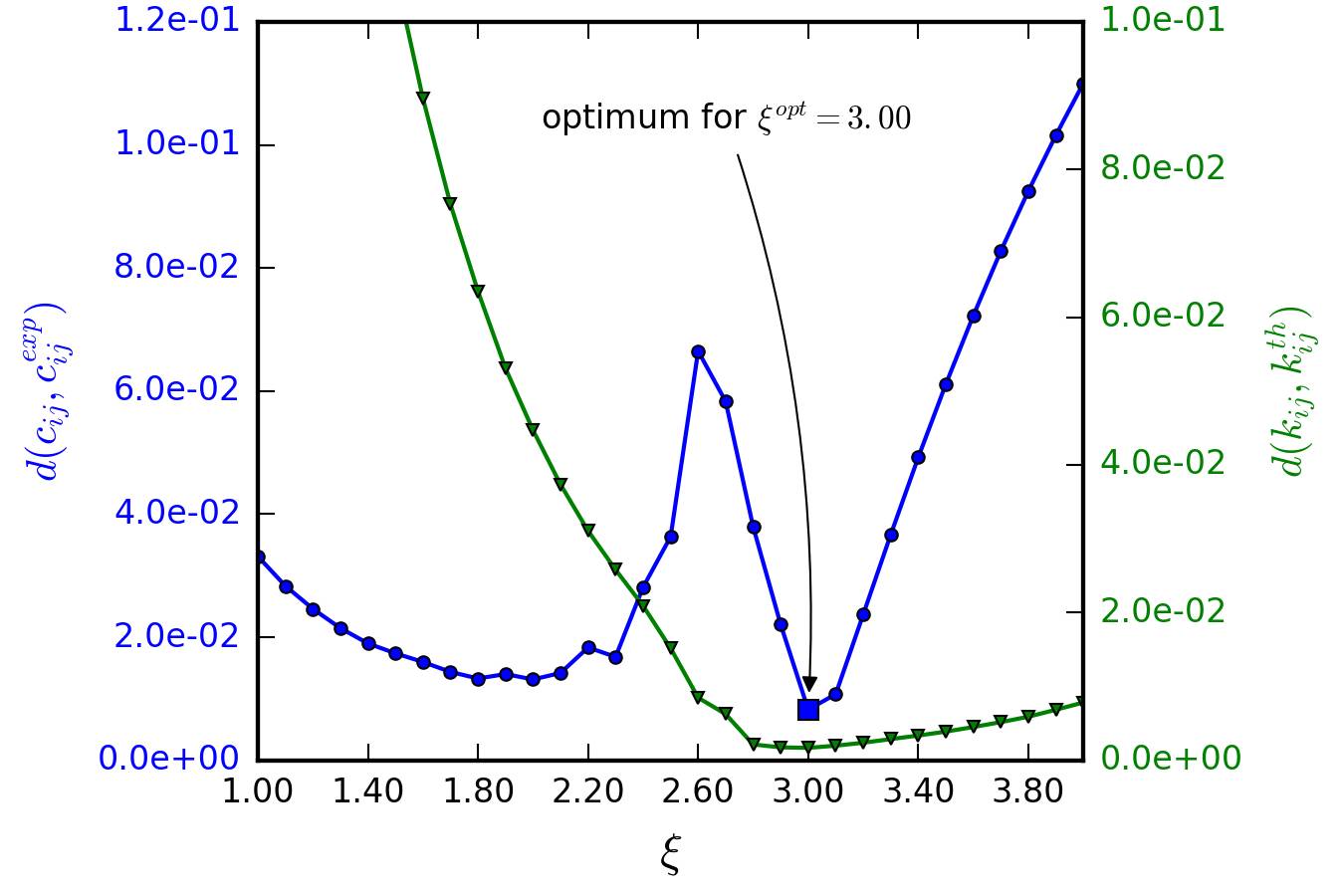}%
    \includegraphics[width=0.28 \textwidth, valign=c]{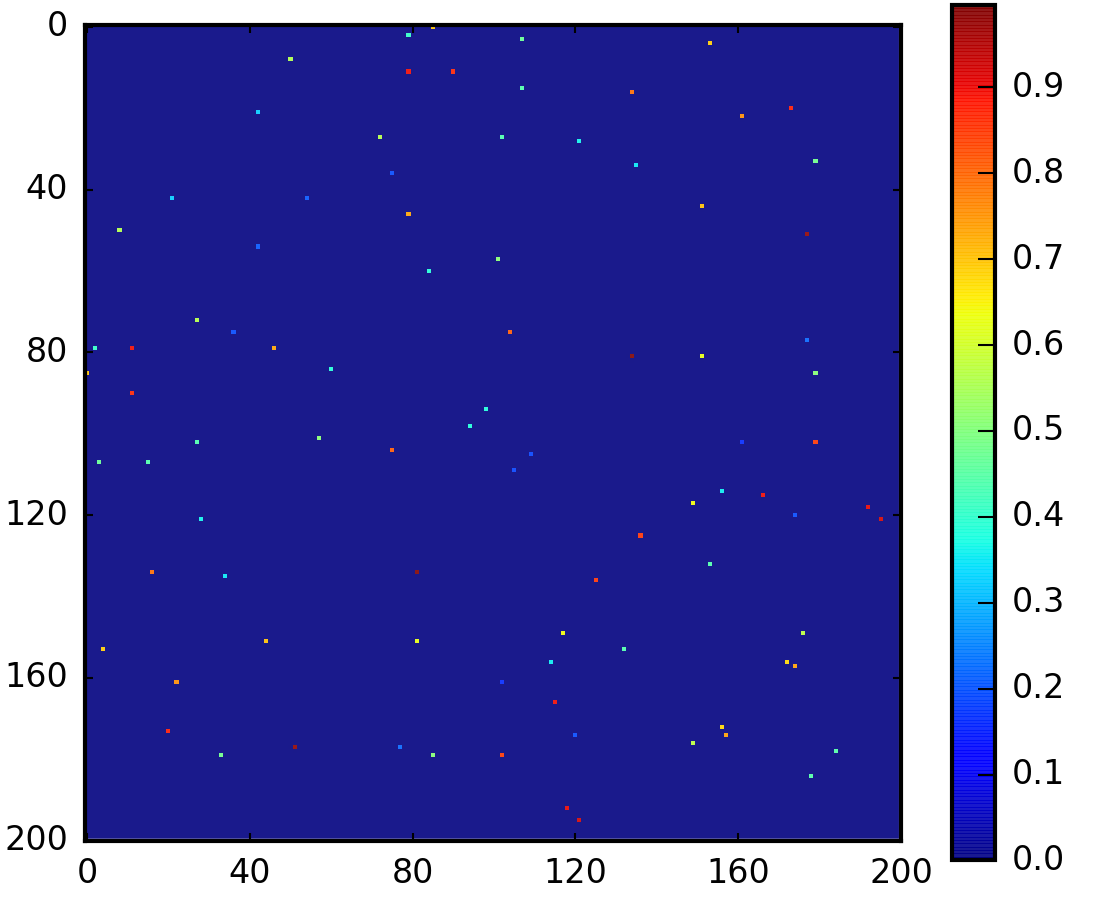}%
    \includegraphics[width=0.28 \textwidth, valign=c]{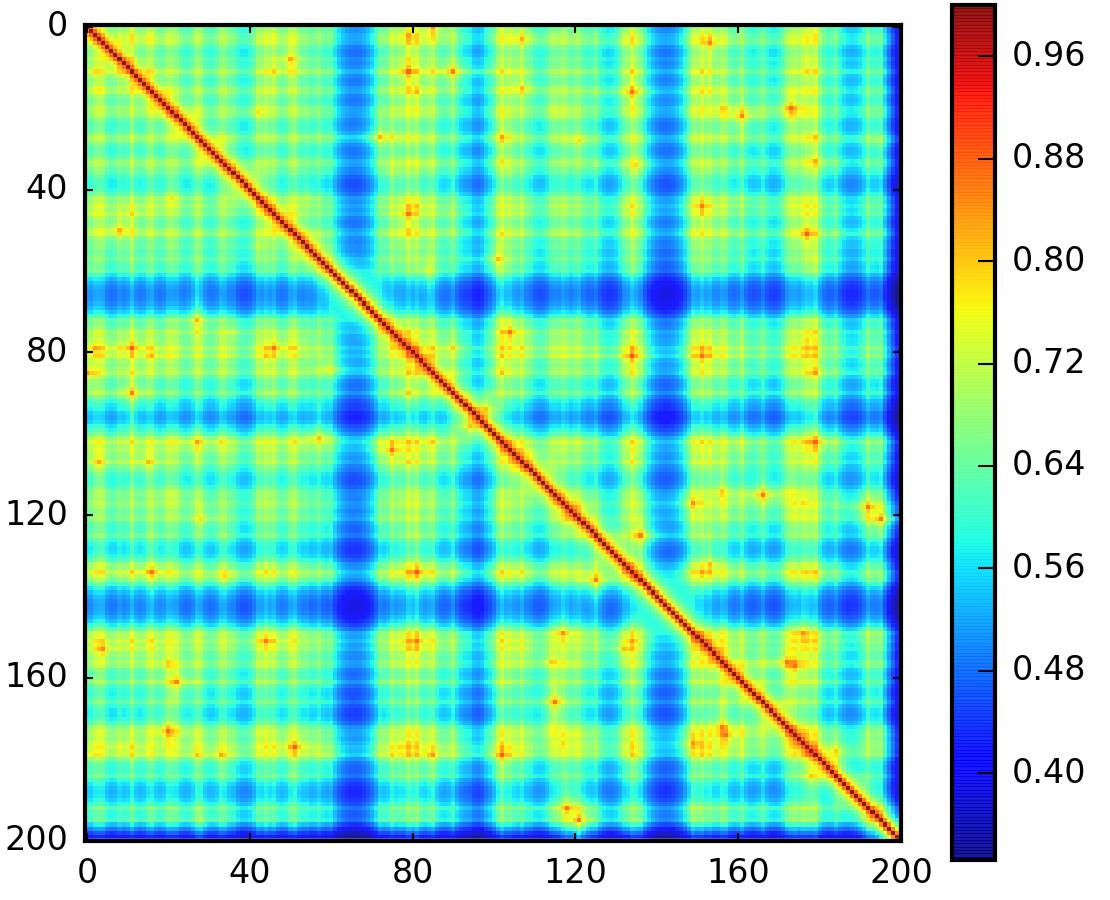}}
    \\
  \subfloat[]{\label{fig:minimization_N200_gauss:Nc100} %
    \includegraphics[width=0.4 \textwidth, valign=c]{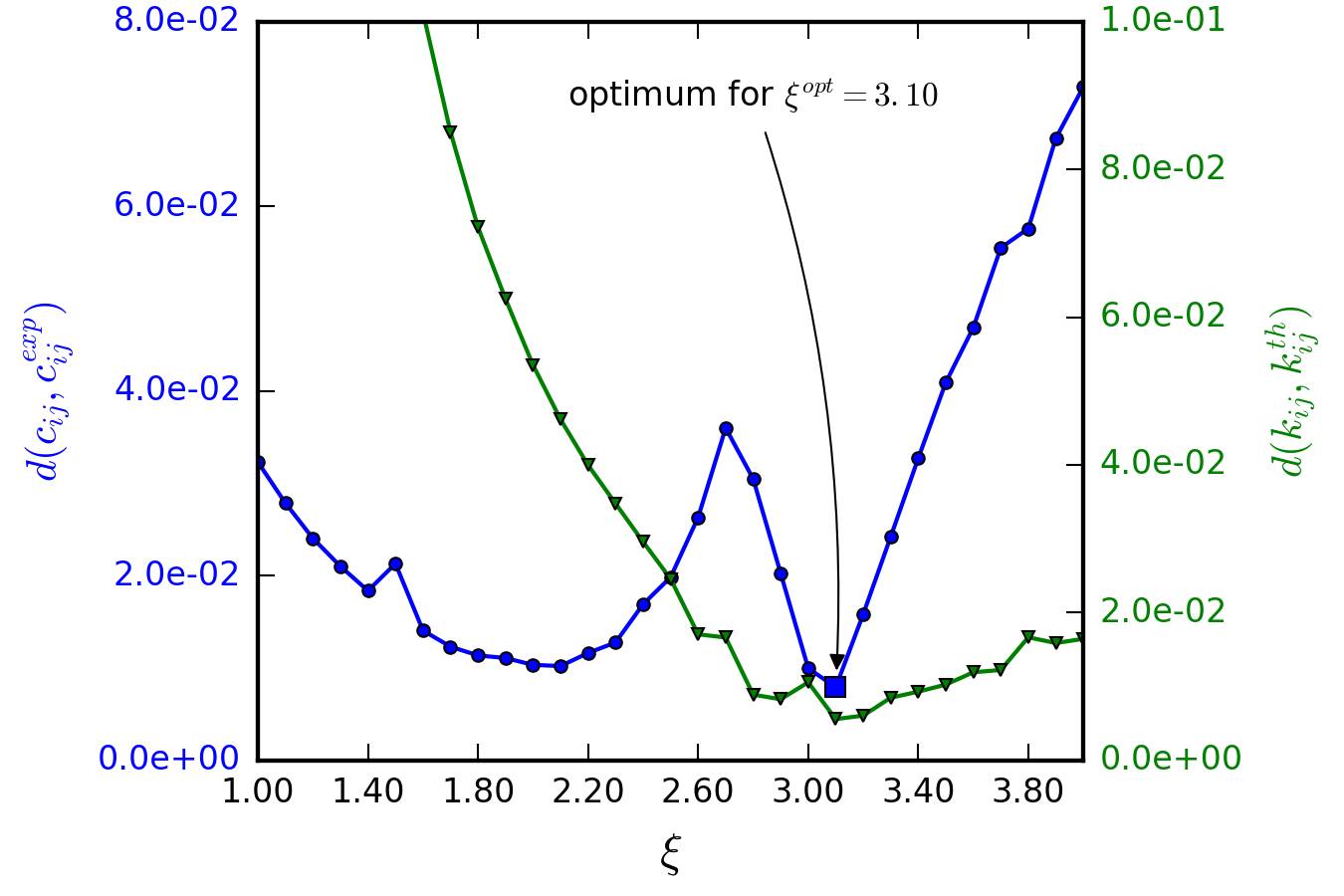}%
    \includegraphics[width=0.28 \textwidth, valign=c]{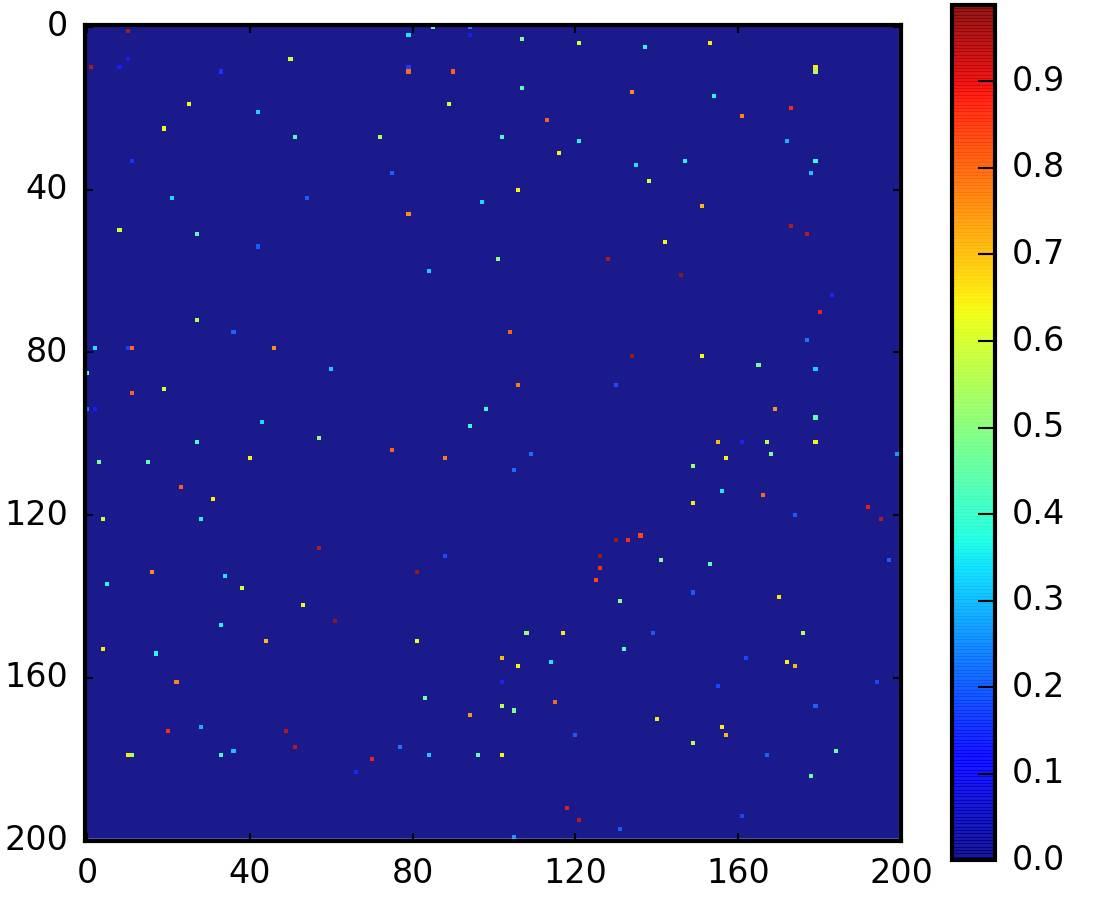}%
    \includegraphics[width=0.28 \textwidth, valign=c]{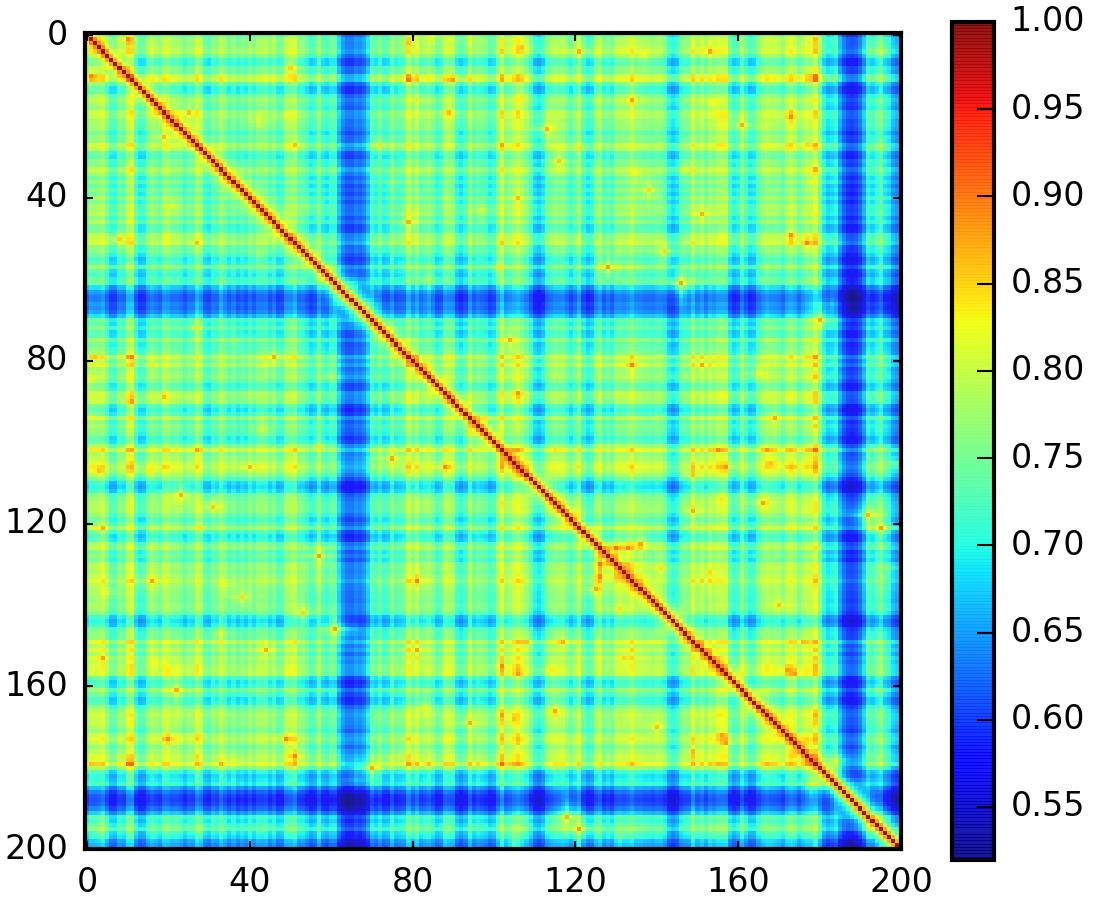}}
    \caption{Application of the minimization method to experimental contact matrices generated from a BD trajectory of a GEM ($N=200$ and $\Lambda=1$). We used $\xi^{exp}=3.00$ and $\mu^{exp}=\mu_G$ (Gaussian form factor). We show in the first column $d(\hat{c}_{ij},c_{ij}^{exp})$ and $d(\hat{k}_{ij},k_{ij}^{th})$ as a function of the threshold $\xi$ used in the minimization. The optimal threshold $\xi^{opt}$ minimizing $d(\hat{c}_{ij},c_{ij}^{exp})$ is shown. The optimal coupling matrix $k_{ij}^{opt}$ and the associated contact matrix $c_{ij}^{opt}$ are shown in the second and third column. \protect\subref{fig:minimization_N200_gauss:Nc20} $N_c=20$. \protect\subref{fig:minimization_N200_gauss:Nc50} $N_c=50$. \protect\subref{fig:minimization_N200_gauss:Nc100} $N_c=100$.}
  \label{fig:minimization_N200_gauss}
\end{figure}

\begin{figure}[!htbp]
  \centering
  \subfloat[]{\label{fig:minimization_N200_theta:Nc20} %
    \includegraphics[width=0.4 \textwidth, valign=c]{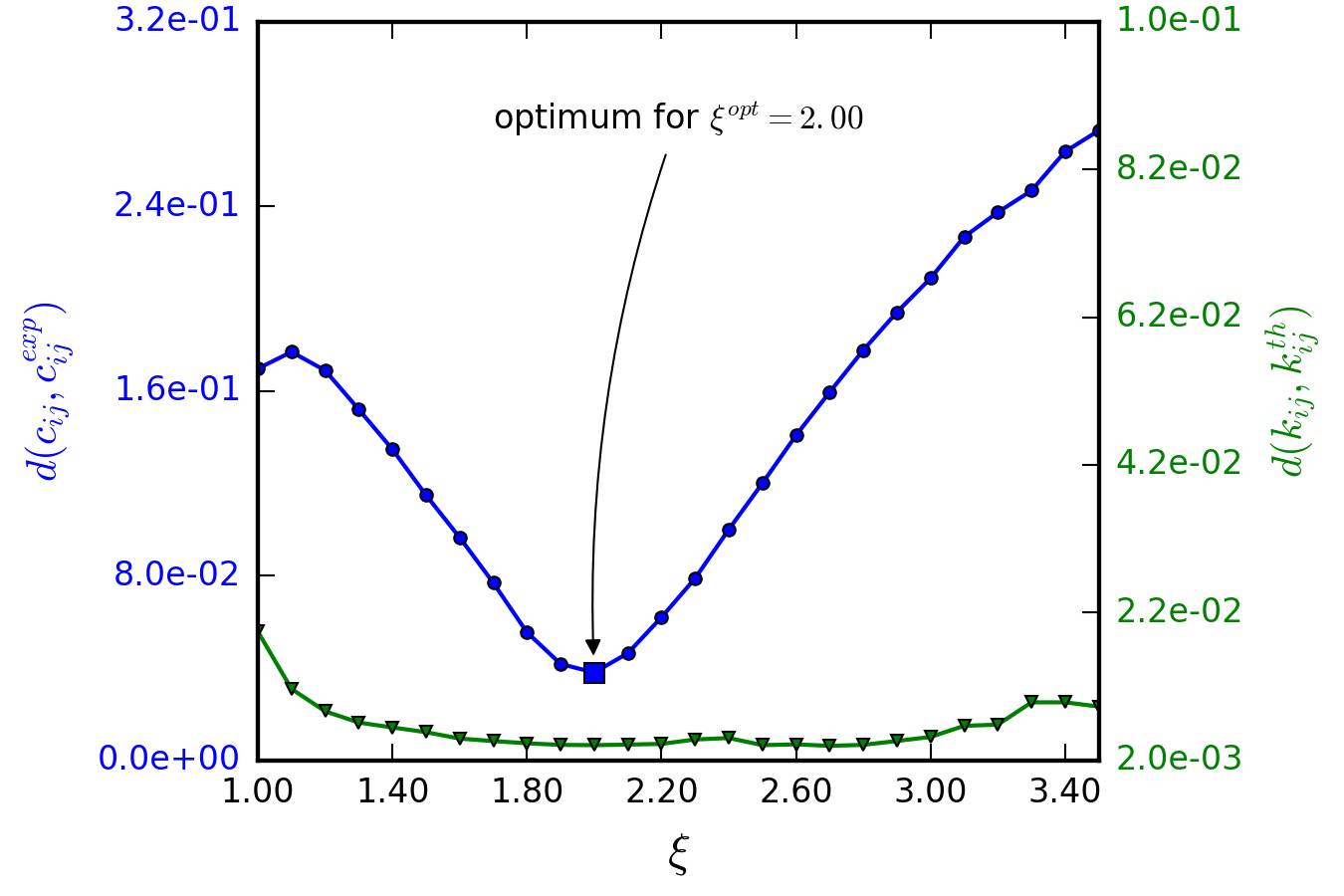}%
    \includegraphics[width=0.28 \textwidth, valign=c]{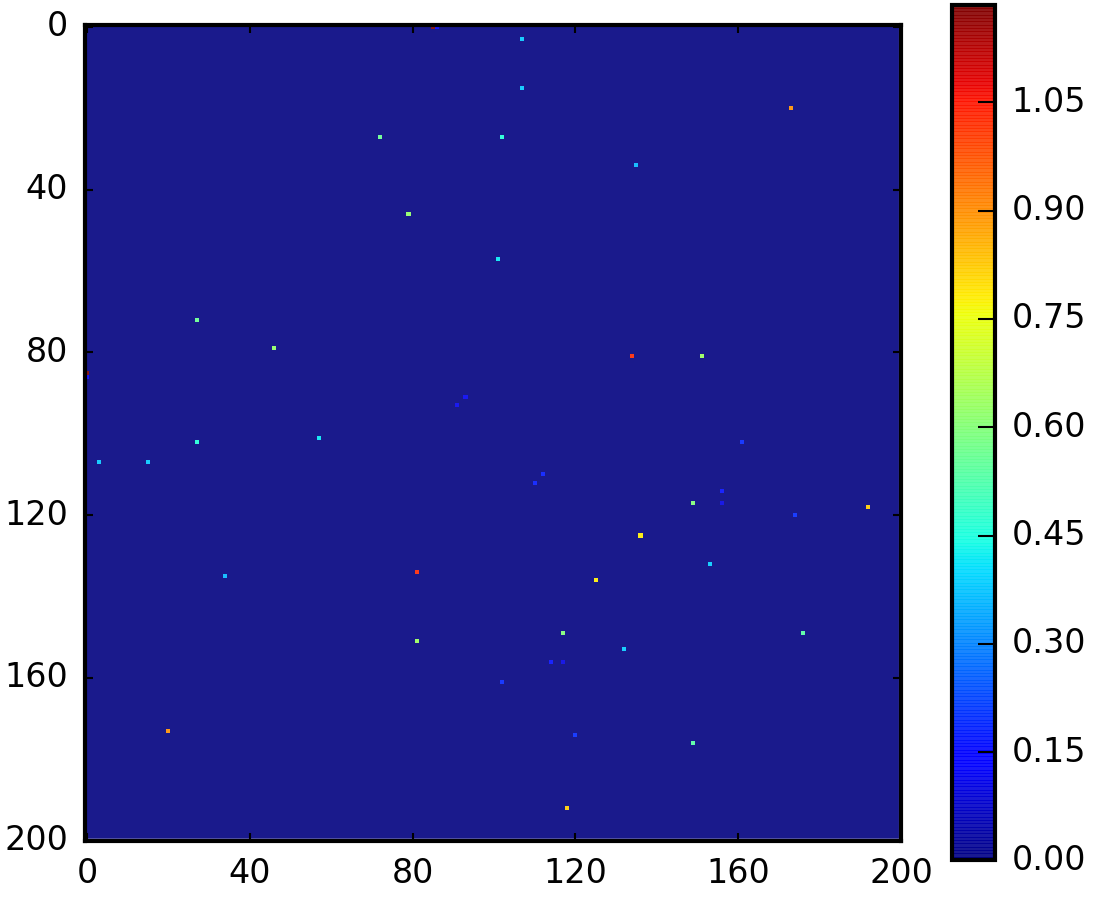}%
    \includegraphics[width=0.28 \textwidth, valign=c]{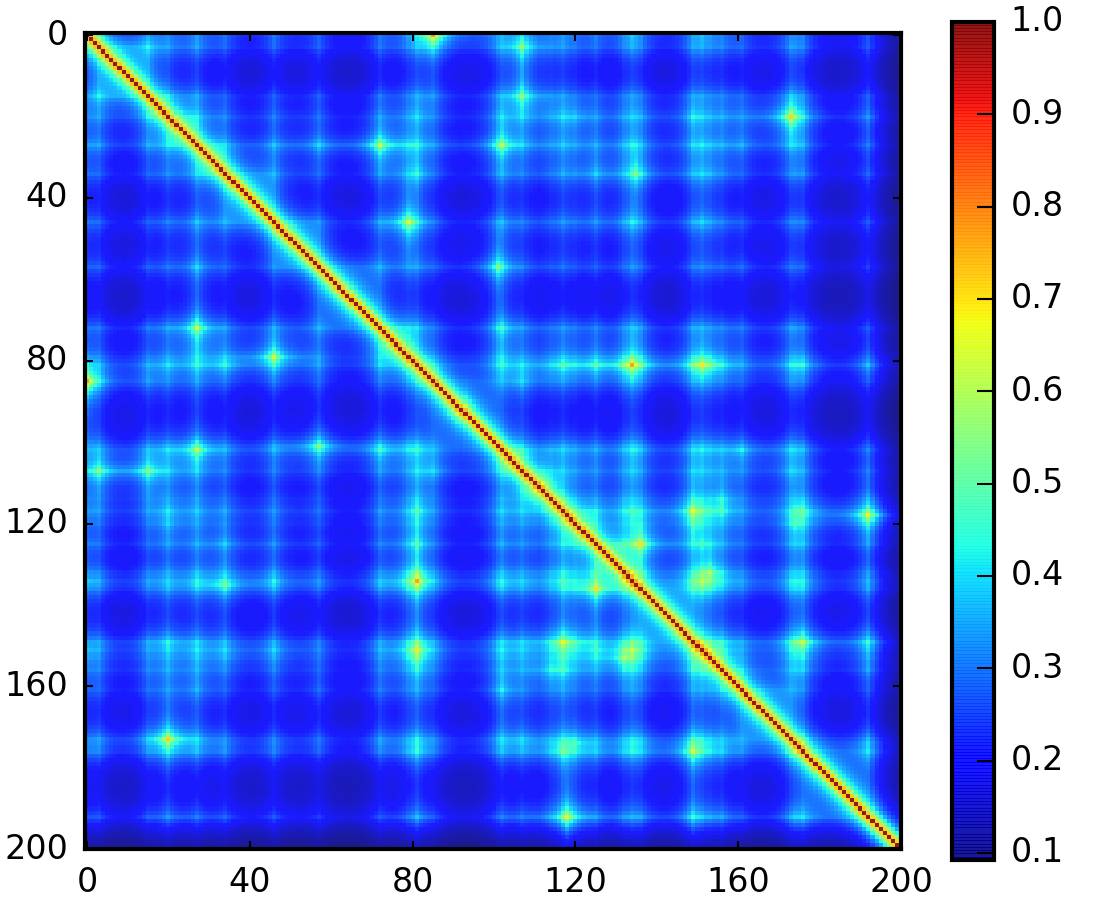}}
    \\
  \subfloat[]{\label{fig:minimization_N200_theta:Nc50} %
    \includegraphics[width=0.4 \textwidth, valign=c]{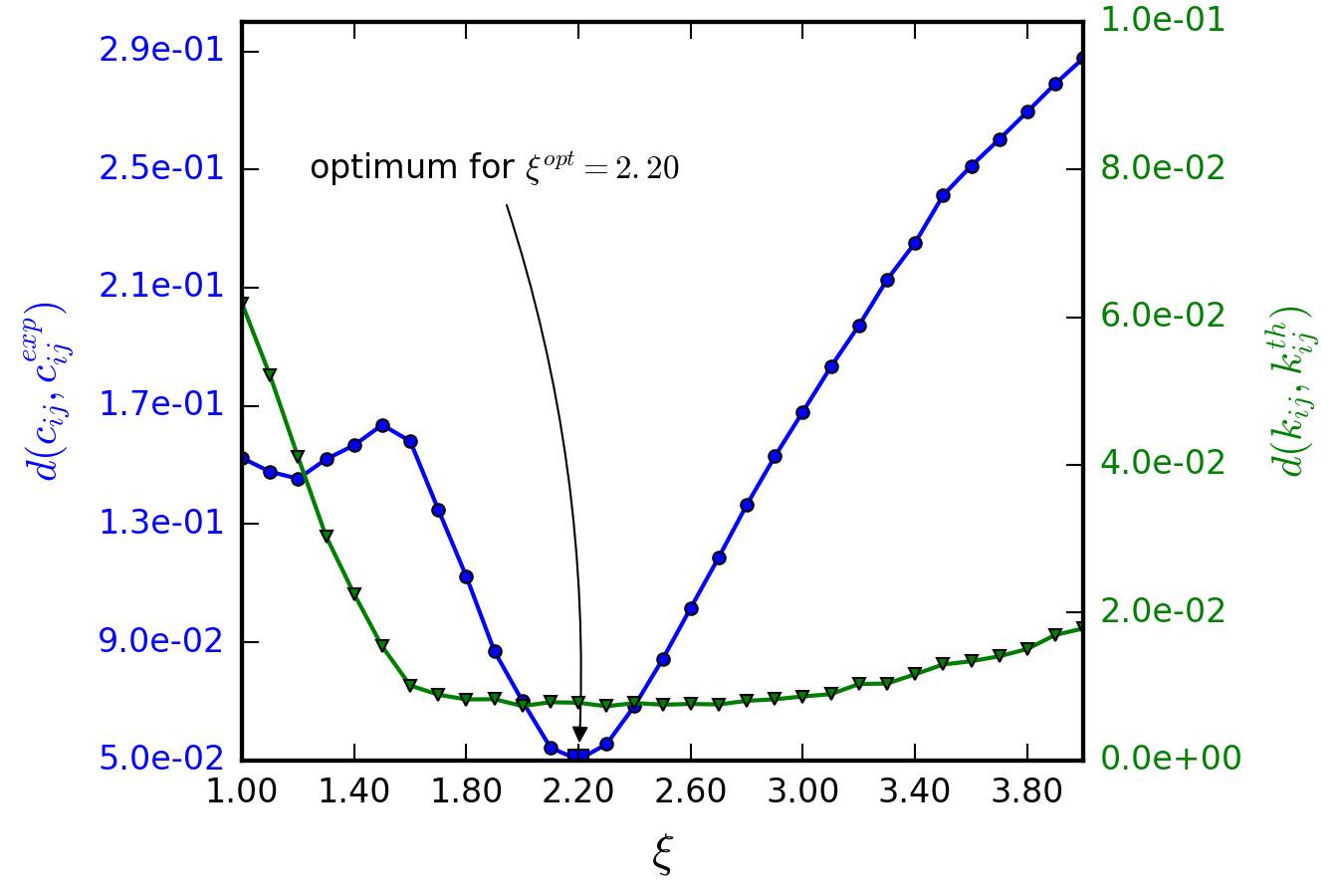}%
    \includegraphics[width=0.28 \textwidth, valign=c]{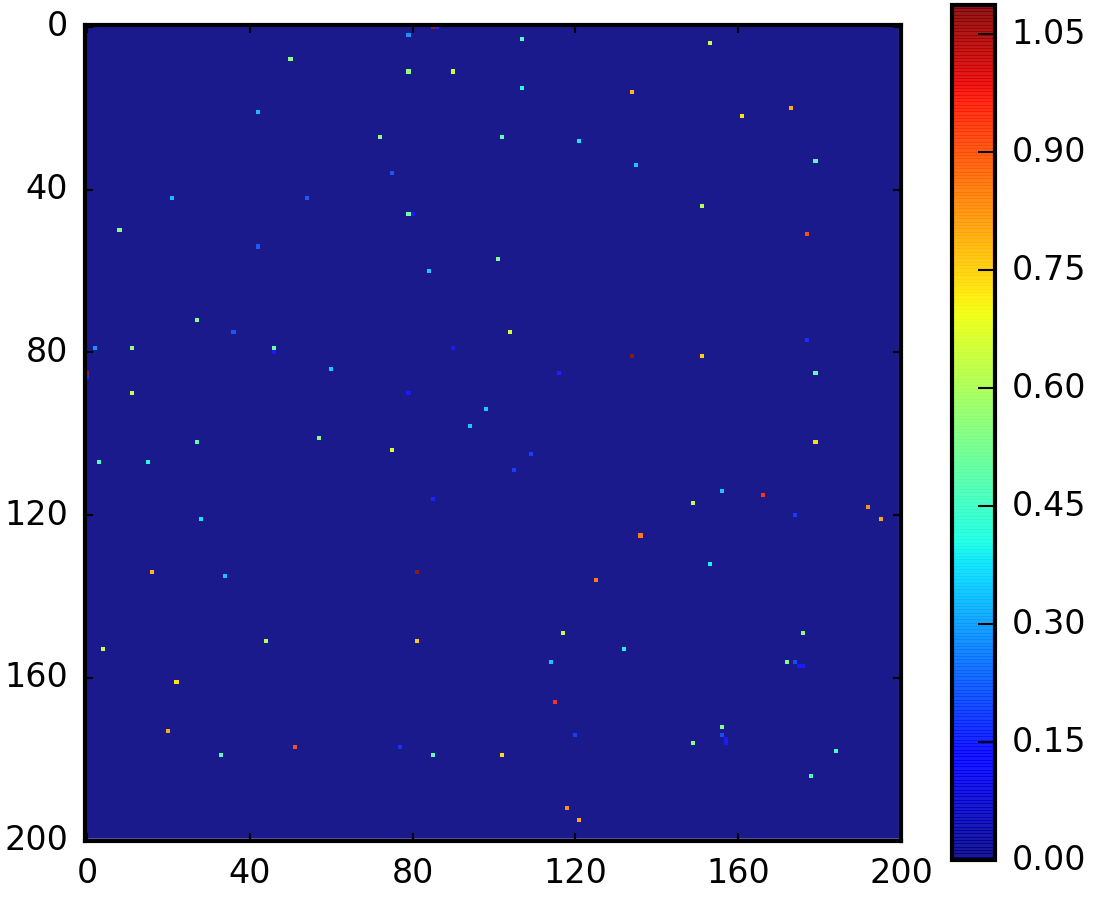}%
    \includegraphics[width=0.28 \textwidth, valign=c]{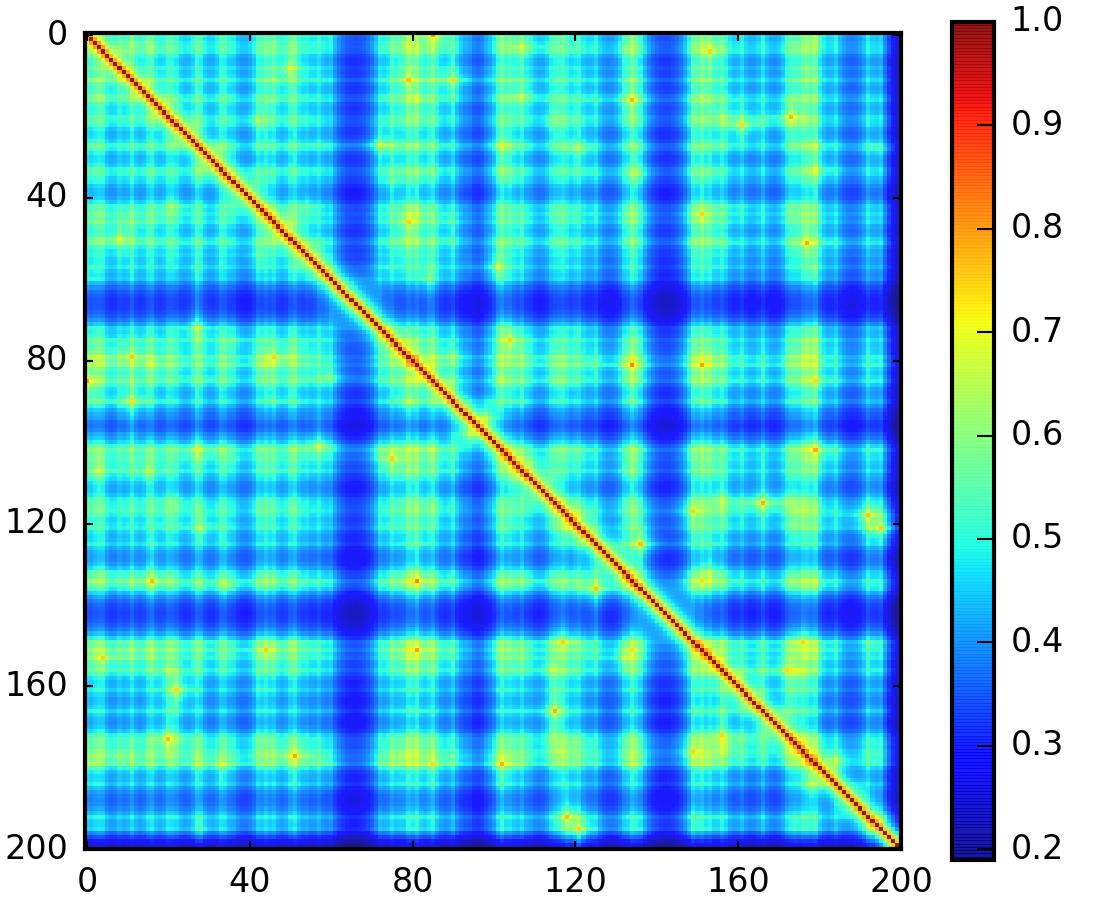}}
    \\
  \subfloat[]{\label{fig:minimization_N200_theta:Nc100} %
    \includegraphics[width=0.4 \textwidth, valign=c]{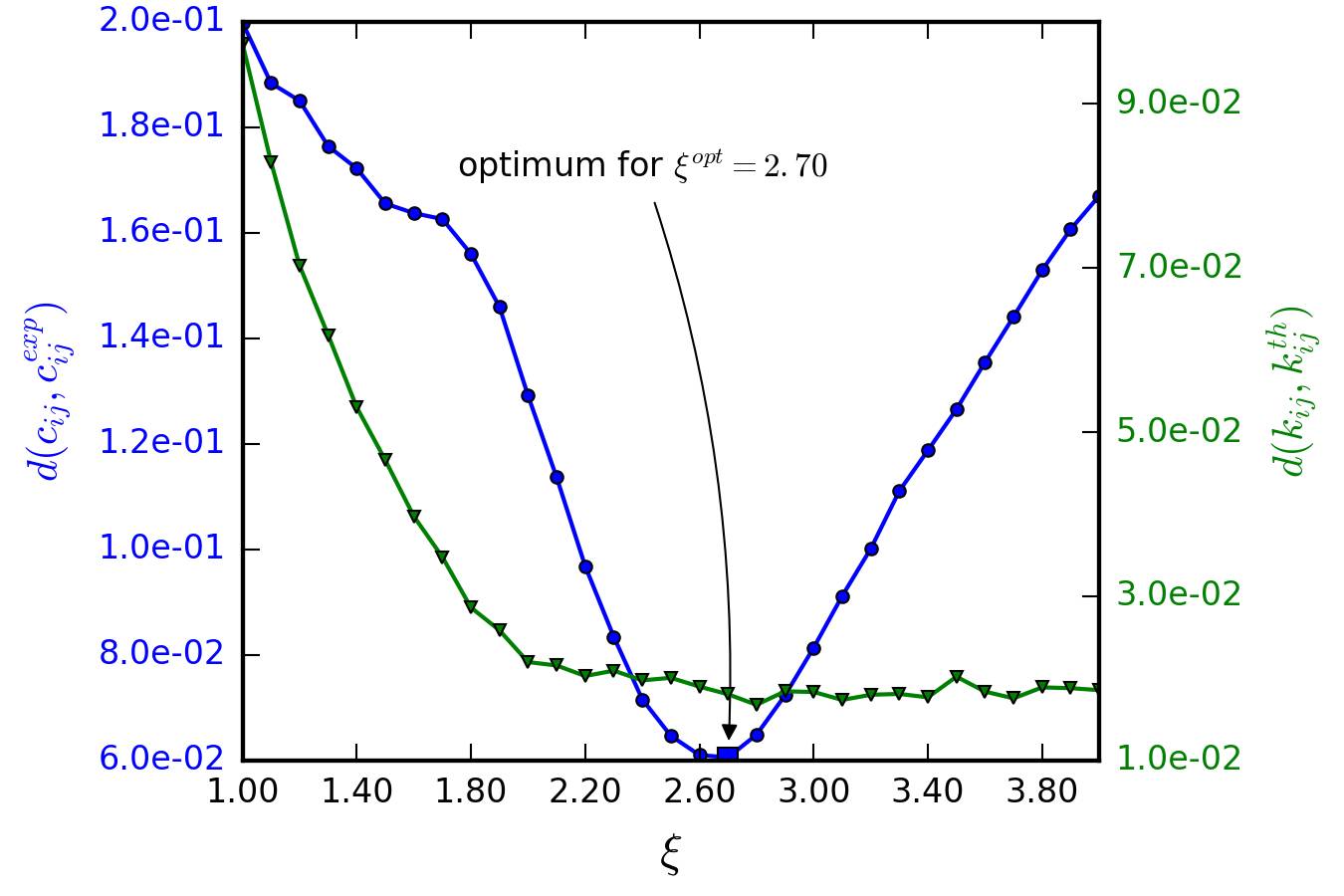}%
    \includegraphics[width=0.28 \textwidth, valign=c]{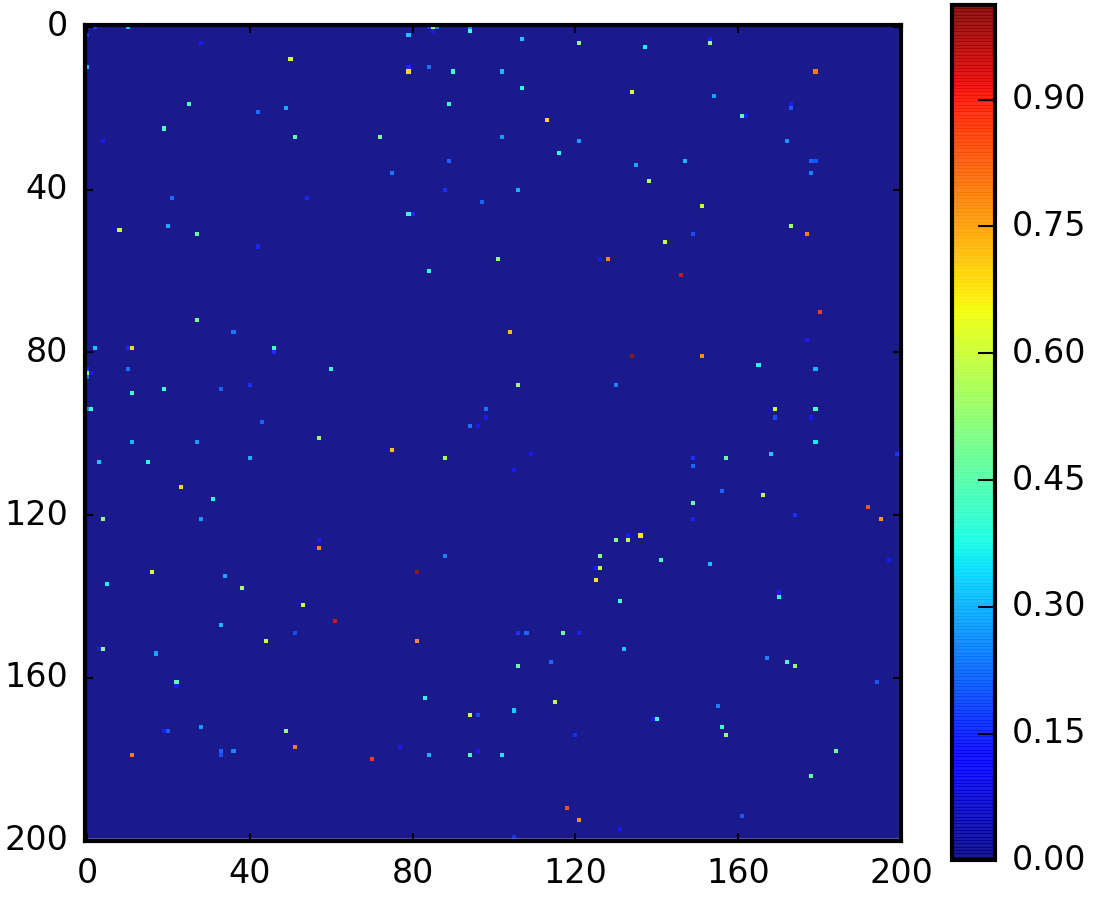}%
    \includegraphics[width=0.28 \textwidth, valign=c]{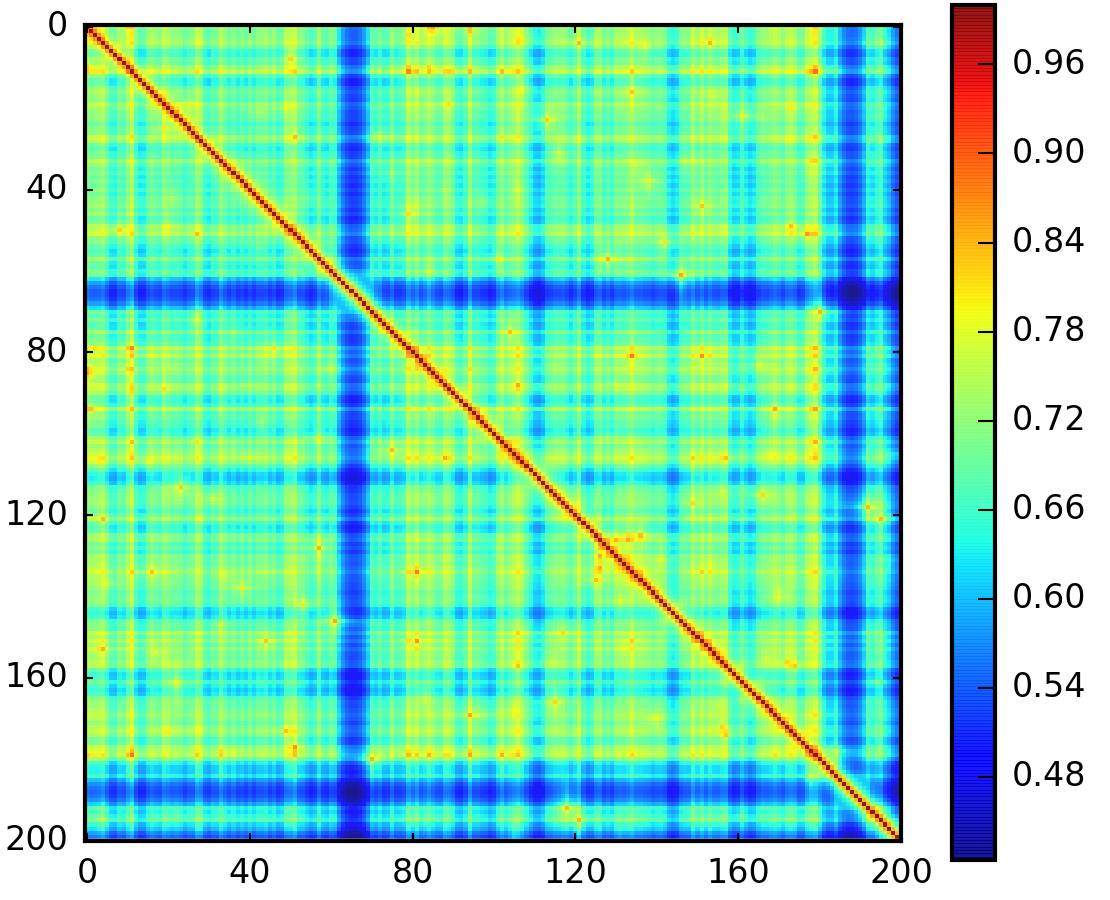}}
    \caption{Same as \cref{fig:minimization_N200_theta} but the experimental contact matrices were generated from the BD trajectories with a theta form factor instead: $\mu^{exp}=\mu_T$ and $\xi^{exp}=1.50$. \protect\subref{fig:minimization_N200_theta:Nc20} $N_c=20$. \protect\subref{fig:minimization_N200_theta:Nc50} $N_c=50$. \protect\subref{fig:minimization_N200_theta:Nc100} $N_c=100$.}
  \label{fig:minimization_N200_theta}
\end{figure}

We considered BD trajectories of GEMs with $N=200$, $\Lambda=1$ and $N_c$ constraints. We then applied the minimization procedure to the experimental contact matrices computed using a Gaussian form factor $\mu^{exp}=\mu_G$ and a threshold $\xi^{exp}=3.0$. In \cref{fig:minimization_N200_gauss}, we report the results for $N_c=20,\,50,\,100$. In the first column we represented $d(\hat{c}_{ij},c_{ij}^{exp})$ as a function of the threshold $\xi$. Since we know the true couplings of the GEM used to produce the experimental contacts, we also represented the distance between the retrieved couplings $\hat{k}_{ij}$ and the theoretical ones, \textit{i.e.} $d(\hat{k}_{ij},k_{ij}^{th})$. We observe that $d(\hat{c}_{ij},c_{ij}^{exp})$ display one narrow local minimum and another ``fat'' local minimum for smaller values of $\xi$, which is rather unexpected. However, $d(\hat{k}_{ij},k_{ij}^{th})$ diverges near the ``fat'' minimum so we conclude that this is an unphysical minimum. We do not have clear explanation for the existence of this secondary minimum, but we suspect that it is due to the positivity constraint on the $k_{ij}$. That being said, if we admit that the first minimum encountered from the right (\textit{i.e.} when decreasing progressively $\xi$ from large values) corresponds to the optimal threshold $\xi^{opt}$ as defined in \cref{eq:gem_minimization_optimal_threshold}, then the performances of the method are very good. In particular we recover that $\xi^{opt} = \xi^{exp}$. In the second and third columns, we show the optimal coupling matrix $k_{ij}^{opt}$ and contact matrix $c_{ij}^{opt}$ when $\xi=\xi^{opt}$. Note that $k_{ij}^{opt}$ reproduces $k_{ij}^{th}$ to a very good precision. We also carried out the same procedure on contact maps such that $\mu^{exp}=\mu_T \neq \mu$ and $\xi^{exp}=1.5$ (\cref{fig:minimization_N200_theta}). Due to the discrepancy between the form factors, we now have $\xi^{opt} \neq \xi^{exp}$. Although a little bit less accurate than with $\mu^{exp}=\mu_G$, we still obtain satisfactory results and the optimal couplings are very close to the theoretical ones. Interestingly, when $\mu^{exp} \neq \mu$, the ``fat'' minimum of $d(\hat{c}_{ij},c_{ij}^{exp})$ almost vanishes and we are left with a pronounced and dominant global minimum at $\xi=\xi^{opt}$.

Let us emphasize that despite its apparent computational burden, this method is still more efficient than the method proposed in \cite{Giorgetti9502014}. Indeed, the latter one also uses a minimization scheme such as \cref{eq:gem_minimization_motion}, yet at each step $n$, evaluating $\mathcal{L}$ requires to perform a full Monte-Carlo simulation to sample configurations of the system in the canonical ensemble and use them to compute the contact matrix associated to the values of the couplings at time $n+1$. Besides, the free parameters in this same approach ($\sigma$ and $\xi$ in \cref{eq:pair_potential_square}) are adjusted by hand while here we adjust the free parameter $\xi$ in order to find the optimal GEM matching the experimental contacts. That being said, when the system size is not too large (say $N<100$), their approach allows to virtually consider any polymer model, and any type of monomer-monomer interaction, while our approach is valid only when the system's Hamiltonian is Gaussian.

In conclusion, we have presented here a method to reconstruct the true couplings of an underlying GEM from an input experimental contact matrix. In contrast to the direct reconstruction method, it ensures that the obtained GEM is stable. It is therefore safer to apply to noisy experimental contact matrices, coming for instance from Hi-C experiments. As a drawback, the minimization involves a heavier computational burden that we somehow attenuated by trimming small couplings values during the minimization. The method has $\xi$ as a free parameter, which is chosen \textit{a posteriori} to minimize the distance between the experimental and the reconstructed contact probabilities. Applying this method to BD trajectories of our artificial GEMs has proven quite successful. Hence we now attempt to apply it to real contact matrices coming from Hi-C experiments.

\section[Hi-C contact probability matrices]{Reconstruction from Hi-C contact probability matrices}
In this section, we shall use Hi-C data from the human chromosome 14 \cite{Lieberman-Aiden2009}. Because we did not have clear arguments to prefer one normalization over the others, we have normalized the counts map in a contact probability matrix according to each of the methods presented in \cref{sec:cmap_generation} and applied the reconstruction procedure to obtain a Gaussian effective model of the chromosome.

\begin{figure}[!htbp]
  \centering
  \subfloat[]{\label{fig:gem_minimization_lieberman_max2:cexp} \includegraphics[width=0.45 \textwidth]{{HIC_gm06690_chr14_chr14_1000000_obs_raw_data_cmap_max2}.jpeg}}%
  \quad
  \subfloat[]{\label{fig:gem_minimization_lieberman_max2:copt} \includegraphics[width=0.45 \textwidth]{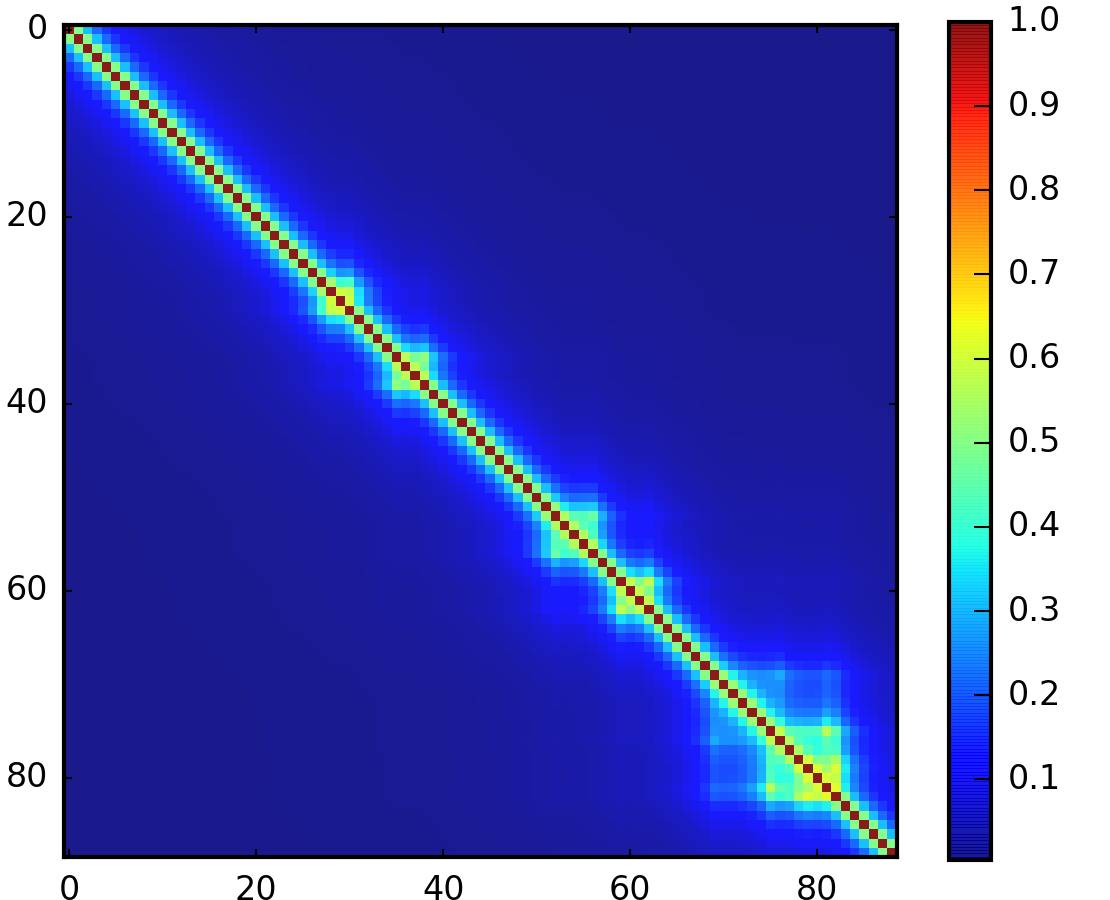}}%
  \\
  \subfloat[]{\label{fig:gem_minimization_lieberman_max2:conf} \includegraphics[width=0.45 \textwidth]{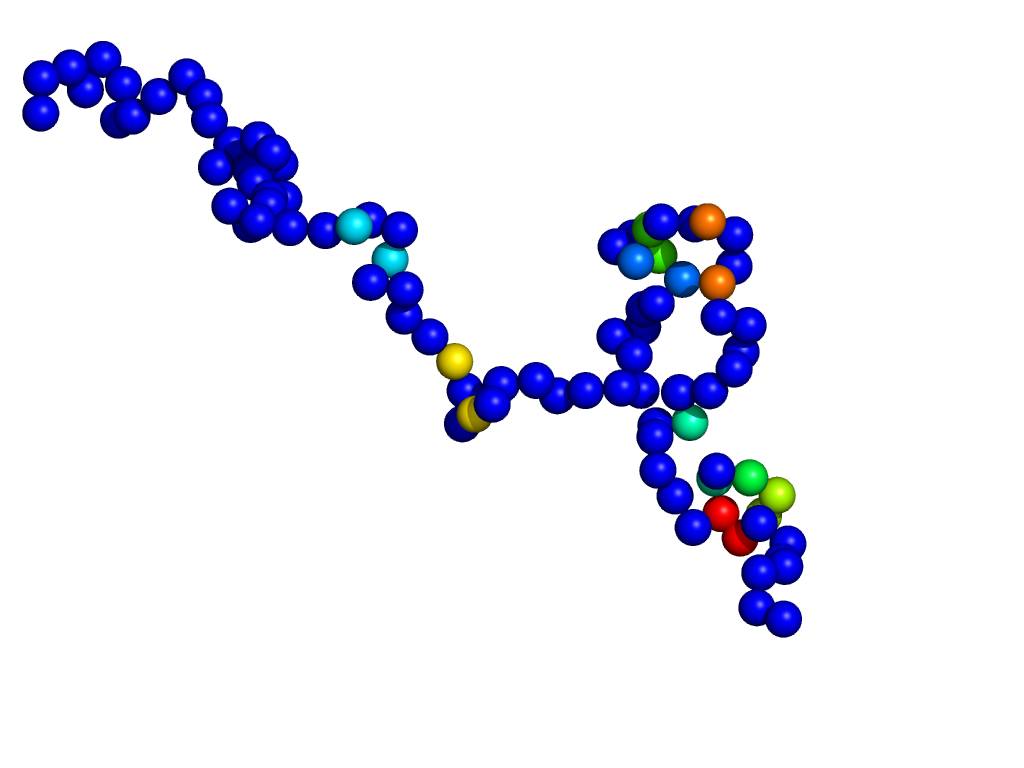}}%
  \quad
  \subfloat[]{\label{fig:gem_minimization_lieberman_max2:kopt} \includegraphics[width=0.45 \textwidth]{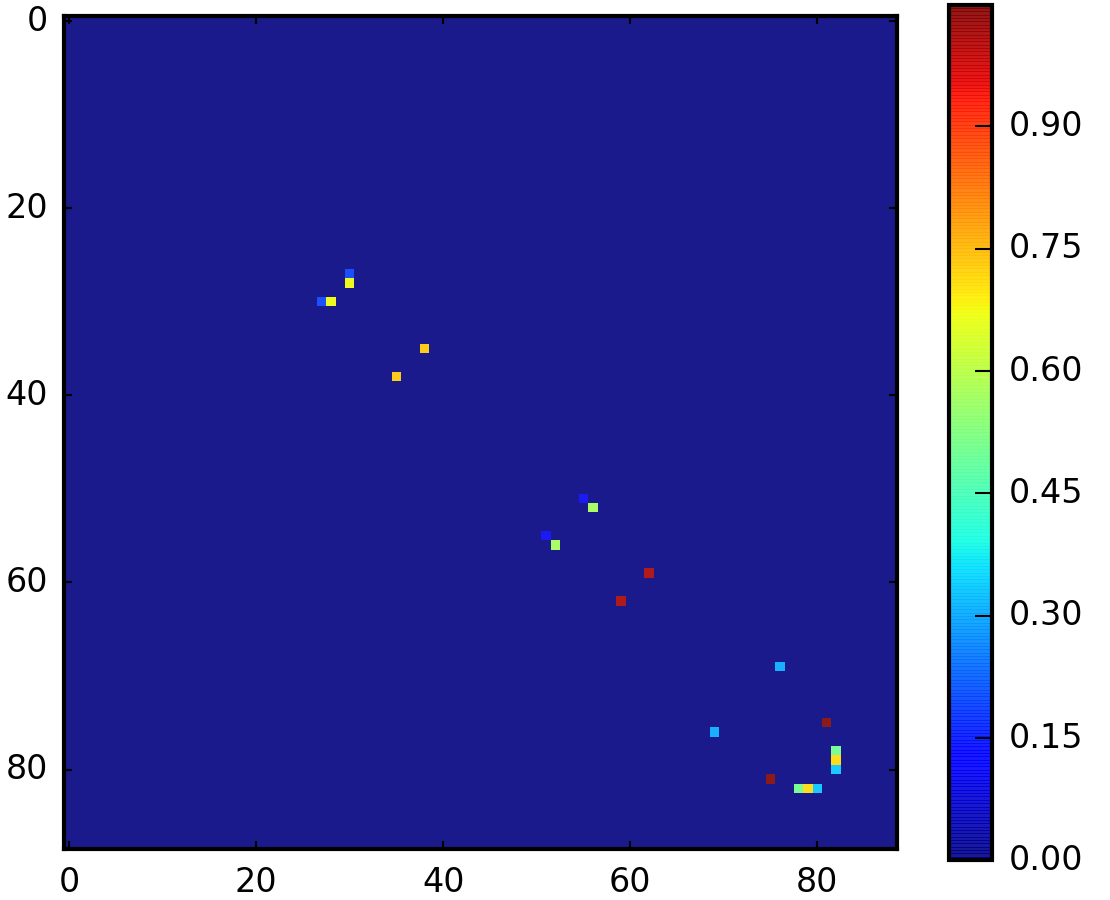}}%
  \\
  \subfloat[]{\label{fig:gem_minimization_lieberman_max2:dist} \includegraphics[width=0.45 \textwidth]{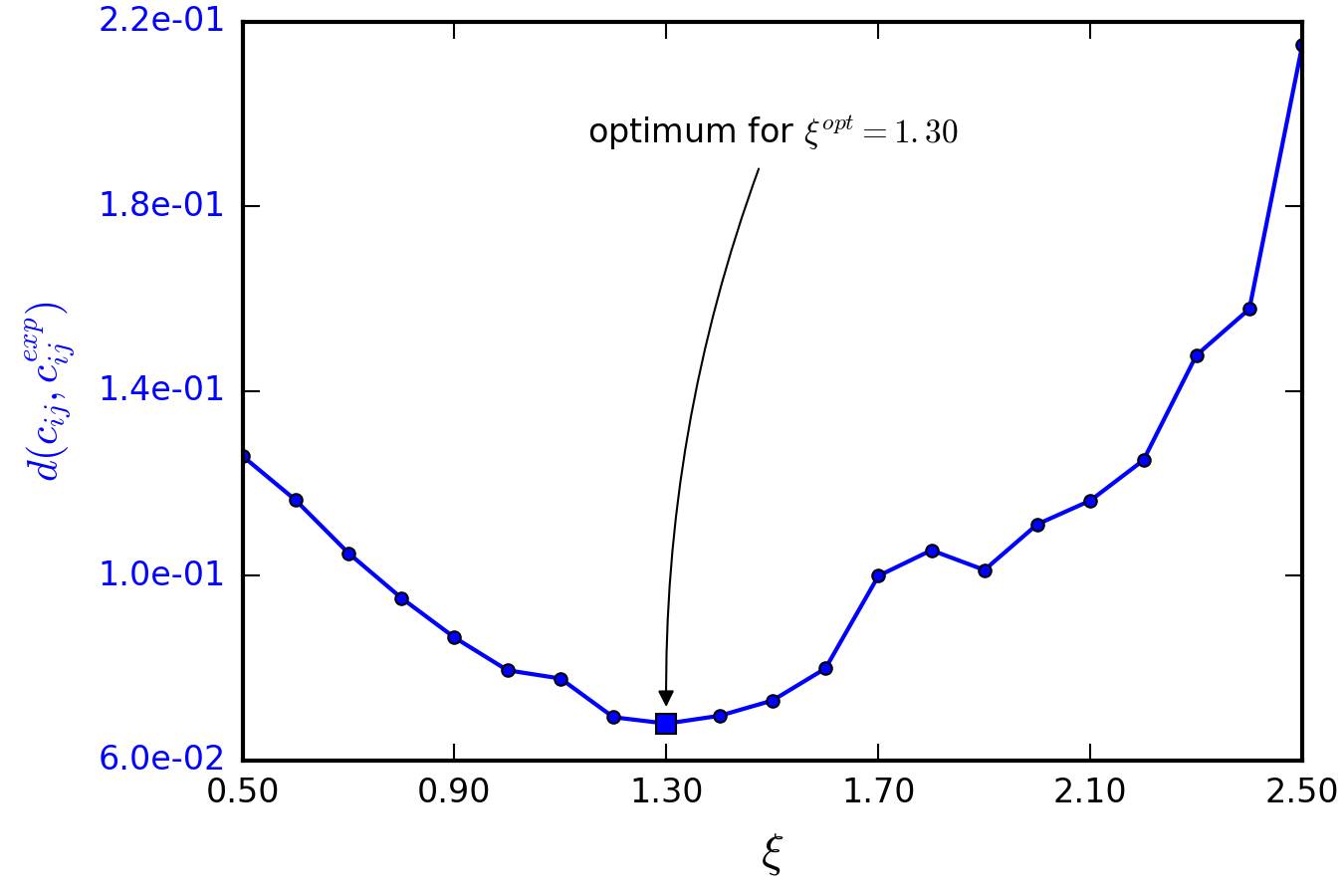}}

  \caption{Application of the minimization procedure to Hi-C data from the human chromosome 14 with bin resolution \SI{1}{\mega bp} \cite{Lieberman-Aiden2009}. \protect\subref{fig:gem_minimization_lieberman_max2:cexp} Experimental contact probability matrix obtained from the counts map with the ``Maximum'' normalization and $l_d=2$. \protect\subref{fig:gem_minimization_lieberman_max2:copt} Contact probability matrix of the optimal GEM. \protect\subref{fig:gem_minimization_lieberman_max2:conf} Snapshot of a BD simulation of the optimal GEM. \protect\subref{fig:gem_minimization_lieberman_max2:kopt} coupling matrix of the optimal GEM. \protect\subref{fig:gem_minimization_lieberman_max2:dist} Result of the minimization method.}
  \label{fig:gem_minimization_lieberman_max2}
\end{figure}

\begin{figure}[!htbp]
  \centering
  \subfloat[]{\label{fig:gem_minimization_lieberman_max3:cexp} \includegraphics[width=0.45 \textwidth]{{HIC_gm06690_chr14_chr14_1000000_obs_raw_data_cmap_max3}.jpeg}}%
  \quad
  \subfloat[]{\label{fig:gem_minimization_lieberman_max3:copt} \includegraphics[width=0.45 \textwidth]{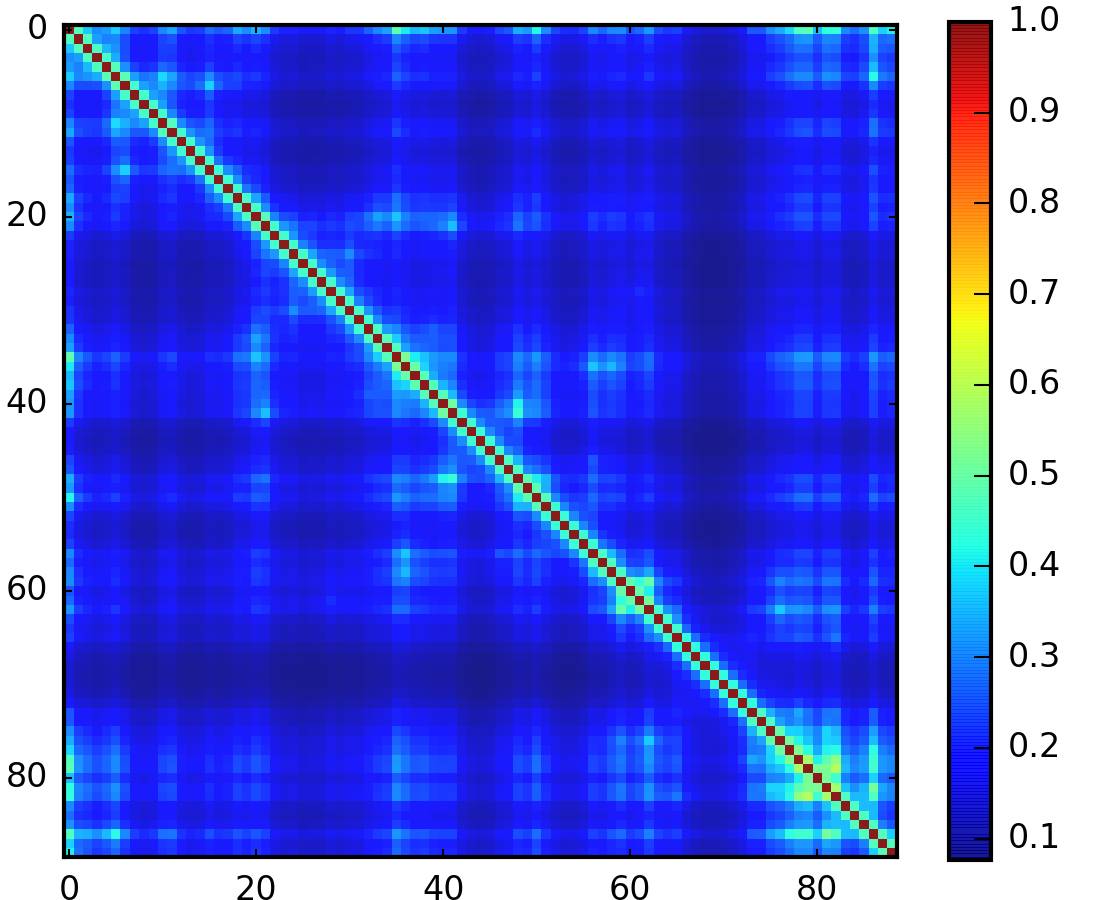}}%
  \\
  \subfloat[]{\label{fig:gem_minimization_lieberman_max3:conf} \includegraphics[width=0.45 \textwidth]{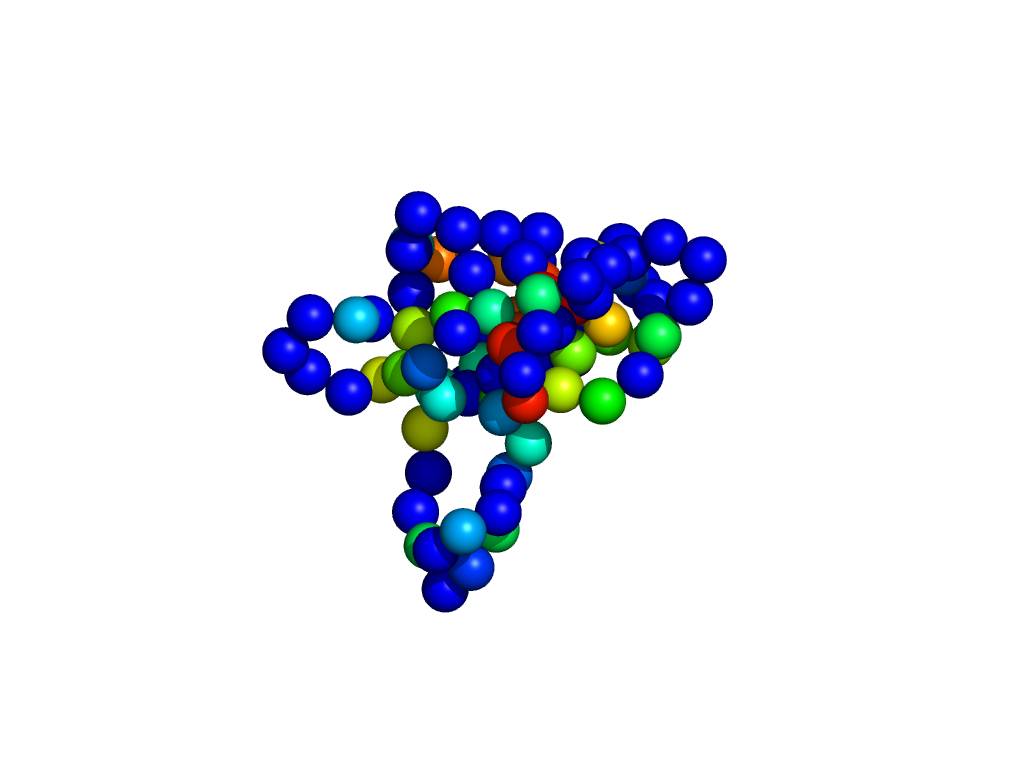}}%
  \quad
  \subfloat[]{\label{fig:gem_minimization_lieberman_max3:kopt} \includegraphics[width=0.45 \textwidth]{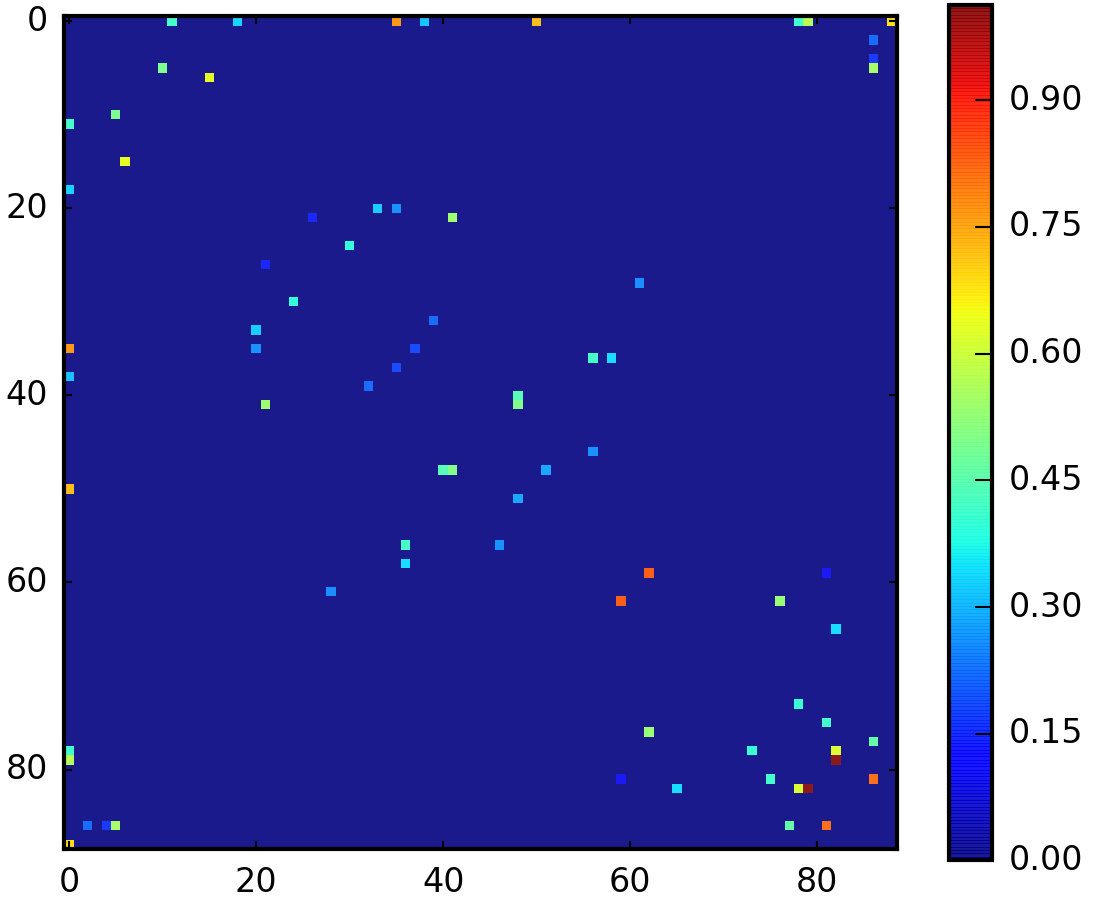}}%
  \\
  \subfloat[]{\label{fig:gem_minimization_lieberman_max3:dist} \includegraphics[width=0.45 \textwidth]{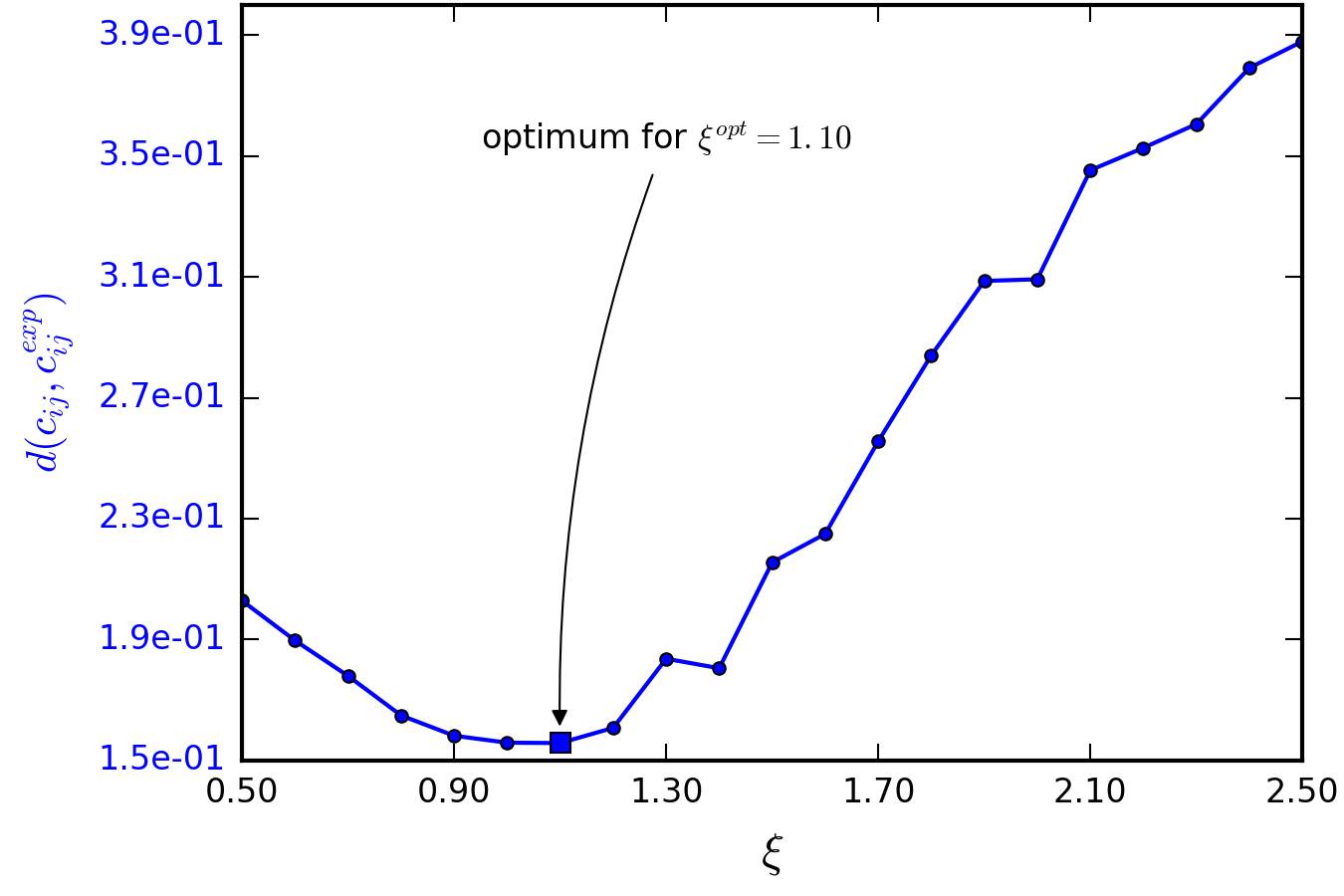}}

  \caption{Same as \cref{fig:gem_minimization_lieberman_max2} but with $l_d=3$ rather than $2$.}
  \label{fig:gem_minimization_lieberman_max3}
\end{figure}

First, we have tried to use the direct reconstruction method. However, the method has failed because the GEMs obtained were unstable. Therefore we have implemented the minimization method. We show the results for contact matrices normalized with the ``Maximum'' method and $l_d=2$ or $l_d=3$. For $l_d=2$ (\cref{fig:gem_minimization_lieberman_max2}), the optimal GEM has a contact probability matrix very close to the experimental one, with $d(c_{ij}^{opt},c_{ij}^{exp}) < 0.1$. Yet, with this choice of normalization, only short-range interactions emerge from the optimal coupling matrix. For $l_d=3$ (\cref{fig:gem_minimization_lieberman_max3}), we obtain a more complex architecture, with the presence of long-range interactions in the optimal coupling matrix. Besides, there are compartments visible in the contact probability matrix associated to the reconstructed GEM, in good global agreement with the experimental contact probability matrix. However, we now have $d(c_{ij}^{opt},c_{ij}^{exp}) > 0.1$. The configurations obtained with BD simulations of this GEM are typical of a collapsed polymer and may model ``rosette'' structures conjectured for chromosome architecture \cite{Cook12010}. Noting that the experimental contact probability matrix generated with $l_d=3$ has globally entries with larger values and less smooth variations, we may interpret this increased distance, $d(c_{ij}^{opt},c_{ij}^{exp})$, by saying that it is harder to fit the experimental contacts with a GEM than when $l_d=2$.

\section{Discussion}
\paragraph{Representation of the chromosome architecture with a Gaussian effective model\newline}
On the basis of an inspiring study \cite{Giorgetti9502014}, in this chapter we have sought to propose a polymer model of the chromosome that reproduces the contact probabilities measured in Hi-C experiments. In order to address this issue, we have investigated a physical model that we called Gaussian effective model (GEM). Specifically, we started from a Gaussian chain model of the chromosome, and added effective interactions between monomers under the form of harmonic springs with rigidity coefficients $k_{ij}$. The problem of reconstructing the chromosome architecture is then equivalent to find the $k_{ij}$ values such that the GEM reproduces, at Boltzmann equilibrium, the experimental contact probabilities, $c_{ij}^{exp}$. As a central result of our investigation, we found that within this theoretical framework, the contact probabilities of the GEM, $c_{ij}$, are uniquely related to the matrix of couplings, $c_{ij} \Leftrightarrow k_{ij}$. In particular, an analytical closed-form has been obtained. This mapping depends on two parameters which are: a threshold $\xi$ and a form factor $\mu$, which specify the probability that two monomers separated by a distance $r$ are found in contact. These parameters depend on the particular experimental setup and need to be adjusted \textit{a posteriori}. Besides, in our investigation we have chosen to use a Gaussian form factor, $\mu_G(r) = \left( 1 + b^2 \gamma_{ij} / \xi^2 \right)^{-3/2}$, in order to account for the dispersion in the cross-linking distances due to formaldehyde polymerization in Hi-C experiments.

\paragraph{Validation of the analytical relation between couplings and contact probabilities\newline}
In order to validate the analytical closed-form obtained, we have generated Brownian dynamics (BD) trajectories of artificial GEM with known couplings, $k_{ij}^{th}$, from which we extracted $M$ configurations to compute virtual contact matrices. We compared these matrices to the matrix of contacts predicted by the GEM mapping, $c_{ij}^{th}$. We found that the distance between the two was very small. For example, with $M=1000$ configurations, we had $d(c_{ij}^{th},c_{ij}^{exp}) \sim \num{5e-3}$, where the distance actually represents the average difference between matrix elements. Hence we concluded that the analytical relation obtained was reliable.

\paragraph{Direct reconstruction method\newline}
This led us to propose a first method to reconstruct chromosome architecture. Starting from an experimental contact matrix, $c_{ij}^{exp}$, this method uses the aforementioned mapping to give the couplings which define a GEM with the same contact probability matrix, $\hat{c}_{ij}=c_{ij}^{exp}$. This method performed very well on virtual contact matrices generated from BD simulations. Namely, the original couplings $k_{ij}^{th}$ were retrieved within a very good accuracy. Yet, we were faced with an unexpected problem when applying this method to experimental contact matrices from Hi-C experiments. It turns out that the GEM mapping does not ensure that the model obtained is physical. That is to say, the reconstructed couplings may result in a Gaussian correlation matrix with negative eigenvalues, corresponding to an unstable GEM. We characterized this phenomenon by showing that the method is sensitive to the presence of noise in the experimental contact probabilities. More accurately, when the noise amplitude increases above a certain value, which depends on the number of non-zero couplings in the GEM (\textit{i.e.} the number of constraints $N_c$), the reconstruction can fail even though the underlying contact matrix was generated from a stable GEM. The effect can be particularly devastating on the input data. For instance, for contact matrices generated from a free polymer ($N_c=0$), a noise of amplitude of \num{e-6} was sufficient to result in an unstable GEM.

\paragraph{Finding a stable Gaussian effective model\newline}
To address the stability issue, we had to consider an alternative approach. Instead of considering that an input contact probability matrix should be directly associated to a GEM, we rather decided to find the closest element, in the space of contact matrices, which is associated to a stable GEM. To implement this approach, we considered the minimization of a function of the $k_{ij}$ variables, and consisting of two terms. The first term was the distance between the experimental contacts and the contact matrix associated to the current $k_{ij}$ vector through the GEM mapping. The second term corresponded to a constraint on the $k_{ij}$ to ensure that the GEM is stable. This minimization is carried out for several values of the threshold used in the mapping, and the GEM having the closest contact matrix, $\hat{c}_{ij}$, to the experimental one is retained.

We demonstrated that this method gave consistent results by applying it as before to contact probability matrices generated from BD trajectories. We found that the minimum of the distance $d(\hat{c}_{ij},c_{ij}^{exp})$ indeed corresponded to a satisfactory retrieval of the hidden couplings. The method however is less accurate than the direct reconstruction discussed above. First, by construction the reconstructed contact matrix is no longer equal to the experimental one, $\hat{c}_{ij} \ne c_{ij}^{exp}$. In particular, at the optimum, the distance between the two matrices was of the order $d(\hat{c}_{ij},c_{ij}^{exp}) \sim \numrange{5e-2}{e-1}$. Furthermore, the distance as a function of the threshold displayed a secondary and unexpected minimum for small values of the threshold. We suspect that it is due to the stringent constraint imposed on the $k_{ij}$ to enforce the GEM stability. Indeed, the condition that we impose is $k_{ij} > 0$. Although having positive couplings is sufficient to obtain a positive definite Gaussian correlation matrix, it is not a necessary condition. Therefore, this condition may be too strong and prevent proper relaxation of the couplings to a GEM with an associated contact matrix closer to the experimental one.

\paragraph{Application to matrices from Hi-C experiments\newline}
Finally, we applied the reconstruction by minimization method to Hi-C experimental data sets \cite{Lieberman-Aiden2009}. It is not clear in our view how to properly normalize the experimental counts in contact probabilities. Therefore we used the ``Maximum'' normalization method presented in \cref{sec:ccc} with $l_d=2$ or $l_d=3$. In the first case, we obtained a predicted GEM where all the non-zero couplings correspond to short contour distances between monomer pairs. The distance between predicted and experimental contacts was quite small, with $d(\hat{c}_{ij},c_{ij}^{exp}) \simeq \num{5e-2}$. In contrast, with $l_d=3$ the predicted GEM presented several long-range interactions. The distance between predicted and experimental contacts was less satisfactory, namely $d(\hat{c}_{ij},c_{ij}^{exp}) > 0.1$. An interpretation is that the GEM obtained is overfitting the variations of the bulk of the $c_{ij}$, namely the checkerboard patterns in the Hi-C experimental contact probability matrix. Hence a future improvement may consist in filtering first the experimental contacts $c_{ij}^{exp}$ in order to smoothen the background variations. Incidentally, methods that normalize the Hi-C counts to produce stochastic contact matrices precisely achieve this effect. Therefore it may be interesting to resort to these methods \cite{Imakaev9992012,Yaffe10592011}.

Furthermore, we performed BD simulations of the reconstructed GEM. For $l_d=2$ the polymer configurations sampled corresponded to an open coil whereas for $l_d=3$ they were rather those of a globule. Although at first, open coil configurations may seem a more reasonable model of the chromosome, the globule configurations have the advantage to reproduce the effect of cell wall confinement. Hence, maybe an appropriate normalization of the Hi-C counts should precisely correspond to a transition from an open coil to a globule for the associated GEM. Incidentally, studies on cross-linked polymers have shown that for ideal Gaussian chain as it is the case here, this transition should occur when the number of non-zero $k_{ij}$ is of the order of the number of monomers, $N_c \sim N$ \cite{Kantor52631996}. This gives a criterion to assess the relevance of the predicted GEMs.

\paragraph{Computational efficiency\newline}
Our investigation has led us to use three reconstruction methods to obtain a GEM of the chromosome. Let us now briefly review their computational advantages and drawbacks.

Our first attempt has been to try a naive approach relying on BD simulations. In this approach, we did not use the GEM mapping to relate the couplings to the contact probabilities. Instead, we postulated that they are proportionally related with a scaling coefficient $\Lambda$, namely $k_{ij}=\Lambda c_{ij}$. We then performed BD simulations for several values of $\Lambda$ and chose the value that minimizes the distance between the Hi-C and the virtual contact probability matrices from BD simulations. This approach is obviously subject to the same criticism that we made for studies using Monte-Carlo (MC) simulations \cite{Giorgetti9502014}. Yet an important difference is that in our approach the minimization is performed as a function of just one variable, $\Lambda$, and it is therefore sufficient to run BD simulations for several values of this scaling coefficient. In contrast, in \cite{Giorgetti9502014} the minimization runs over all the $k_{ij}$, which is much more complex. Hence the naive approach presented in this chapter is more scalable and can be used on contact matrices of larger size.

Our second attempt has been to perform the GEM mapping directly on the experimental contacts. From a computational standpoint, this approach only requires to invert a $N \times N$ matrix where $N+1$ is the size of the probability contact matrix. Hence it is a particularly appealing method, in which we placed much hope at first. However we have seen that it can result in an unstable GEM. Altogether, it may be a good practice to systematically try this method before one of the other two methods.

Our third and last attempt has consisted in the minimization of the distance between an experimental contact matrix and another one corresponding to a stable GEM. Similarly to \cite{Giorgetti9502014}, this method assumes a minimization as a function of all the $k_{ij}$. Yet, in our case each iteration only requires to evaluate a cost function $\mathcal{L}$ together with its gradient, in comparison with a full BD or MC simulation. At first we though that this would provide us with a significant advantage. Yet evaluating the gradient of $\mathcal{L}$ is an $O(N^3)$ operation, therefore involving a significant computational burden. Presently, with a non-optimized code it takes more than two days on a multi-threaded CPU with twelve cores to minimize $\mathcal{L}$ for $N=1000$ (and a few minutes for $N=100$). It is not clear at the moment whether this procedure can be significantly improved to yield more reasonable time. Besides, within our current setup, this method must be repeated for each values of the threshold $\xi$ in the GEM mapping. Unfortunately, most Hi-C contact matrices have a size such that $N \sim 1000$.

\paragraph{Conclusion\newline}
Despite some caveats just discussed, we find that the methods presented in this chapter to reconstruct chromosome architecture are rather novel and pave the way for interesting applications. It is clear that a Gaussian effective model cannot help us to better describe the biological processes at the molecular level. However it can be used to propose a mesoscopic model for the chromosome. Namely the GEM obtained can be used as a basis to perform BD simulations. In comparison with BD models where interactions between DNA and proteins are chosen arbitrarily it has the advantage to be by construction better rooted in biological experiments.

Last but not least, from a theoretical standpoint, the correspondence found between the couplings of a GEM and the Boltzmann contact probabilities constitute a (humble) contribution to the problem of finding if a contact matrix can be produced by a connected physical object such as a polymer. As far as we know, advances on this subject in the literature are rather rare.

\chapter{Concluding remarks}
The binding of many divalent proteins to the chromosome entails the formation of DNA loops or compact structures, which result in a chromosome architecture (or folding) that we do not fully understand. Biological assays have demonstrated that this organization is intimately related to the transcription. Namely, the observation that co-regulated genes tend to be close in space and the characterization of transcription factories have been milestones in this new thinking. In this thesis, I have investigated the physical origin of such structures and proposed models that underlie their existence. To serve this purpose, I have combined analytical approaches from statistical physics with Brownian dynamics simulations. A major challenge in this endeavor has been two reconcile the microscopic scale of the molecular biology with the mesoscopic scale of chromosome folding.

In \cref{ch:transcription_factories,ch:naps}, I have considered first principles models to identify the physical mechanisms responsible for features characterized in experiments. Namely, the approaches undertaken explicitly considered the effect of proteins on the chromosome. In \cref{ch:transcription_factories}, I have proposed a model for the existence of transcription factories. I concluded that such clusters can indeed occur at equilibrium under the effect of a generic type of binding protein. I also proposed that at small scales, binding proteins can induce the collapse of the chromosome in a crystalline phase. Incidentally, such aggregates do form in the bacterial cell, and have often the role to protect DNA from detrimentals factors. In \cref{ch:naps}, I have investigated the formation of DNA hairpin loops by the H-NS protein. In the looped state, RNA polymerase cannot bind, resulting in the silencing of genes whose promoters are sequestered in these DNA loops. My findings suggest that that such hairpins are stable only when the length of the H-NS binding region is above a characteristic length. This gives credit to a conjecture proposing that genes silencing mediated by H-NS/DNA loops constitutes a mechanism for transcription regulation. Namely H-NS binding regions of intermediate lengths can lead to fragile DNA hairpins which can be easily perturbed, for instance by the binding of more dedicated transcription factors. In short, the formation of DNA hairpin loops by the H-NS protein may be seen as a mechanical switch for regulating the transcription.

In \cref{ch:ccc}, I have investigated the inverse problem which is to reconstruct the chromosome folding from chromosomal contacts measured in Chromosome Conformation Capture (CCC) experiments. Using analytical results and Brownian dynamics simulations, I have proposed reconstruction methods relying on a representation of the chromosome with an effective polymer model. The main achievement of these methods was to reproduce the experimental contacts. Although perfectible, these methods represent in my view an original departure from what have been proposed in the literature during the last decade. Namely, the effective model obtained may be used to perform Brownian dynamics simulations of the chromosome better rooted in biological data.

\paragraph{Value of the predictions\newline}
A common denominator of our approaches has been to base our investigation on experimental evidences. However, accurate \textit{in vivo} measurements are not always available. For example, measuring the transcription level of a pair of genes as a function of their spatial distance is an experimental challenge, and most often biologists must resort to indirect measures. Besides, even though high-throughput techniques relying of DNA sequencing and Polymerase Chain Reaction have produced a mine of experimental data, the relevant biological information can be hard to extract or it can be corrupted by noise. For these reasons, we are not yet in an era where established models can be confirmed by experiments to a quantifiable accuracy. This may explain partly at least why in this thesis I have remained at a rather qualitative level of comparison with experimental data. Maybe a fundamental limitation is the lack of a minimal biological system on which can be tested competing models. Indeed, bacteria are often considered as the simplest living system that can be investigated experimentally. Yet many biological processes in bacteria, like transcription regulation, replication or the cell-cycle control are very complex and not fully understood.

\paragraph{Link between architecture and transcription\newline}
The work presented in this thesis is a humble contribution to the broader scientific effort that has been undertaken to unveil the relation between chromosome architecture and genetic expression. Specifically, it has become clear that the interplay between chromosome folding and transcription regulation is highly dynamical and should be more generally considered as two related components of the cell physiology. Understanding the link between these two components is critical to decipher complex regulatory mechanisms of the genetic expression. For instance, chromosome folding seems to play an important role in still unresolved biological processes such as cell differentiation and cell senescence. More generally, it is widely assumed that a better understanding of chromosome architecture is a prerequisite for addressing modern challenges in biology such as conditional gene expression and epigenetics.

During my investigation, I have acquired the conviction that the chromosome should be envisioned as a cellular organ rather than as a mere carrier of the genetic information. In particular, the chromosome might provide a physical medium, or scaffold, for propagating genetic signals. Such signals can be for instance the transcriptional state (a gene is transcribed or not), or the existence of methylations in the context of epigenetics. Assuming that a genetic signal can propagate to nearest neighbors in space, the outcome of chromosome folding should determine the distribution of genetic signals on the chromosome. This schematic view illustrates the problem of context sensitivity. Transcription levels display uneven variations when considered as a function of the genomic coordinate. Yet this can be seen as the result of an unlucky projection from a three-dimensional to a one-dimensional space. Indeed, one can expect that the three-dimensional folding of the chromosome results in the genomic coordinates with the same transcription levels to be close in space. In short, I put forward the idea that the chromosome should be compared to a ``brain'' in which every locus is a ``neuron'' carrying a genetic signal. The particular folding of the chromosome results in contacts between loci that can be seen as synapses enabling the propagation of the genetic signal between neurons. If one associates a particular layout of synapses to a given physiological state, then adjusting chromosome folding can lead to the dynamical re-allocation of these synapses and may be interpreted as the transition to another physiological state.

\paragraph{Toward the design of synthetic gene networks?\newline}
In the classical view of the operon system, it is the affinity of a protein to a promoter that determines the efficiency of a transcription factor to repress or activate the transcription of a gene. Thus, chemical engineering might be used to produce a protein with a strong binding affinity to a promoter. An important message that I have tried to convey all along this manuscript is that the regulation of transcription can also be achieved by means of structural changes applied to the chromosome. For instance, DNA loops can be envisioned as mechanical switches, and similarly to protein folding, chromosome folding can result in active or inactive domains with a dedicated function.  Therefore, designing a regulatory mechanism to obtain a folding of the chromosome with mechanical switches and/or functional domains involves rather a structural and mechanical engineering approach. In short, there has been a shift in our conception of what constitutes a handle for regulating the transcription.

In terms of real applications, the design of a synthetic gene network would require for instance knowing how to position genes on the DNA sequence in order to achieve their co-expression. In particular this would require to have a deterministic knowledge of the functional structures formed. Despite the increasing number of physical models available, we are not able to achieve such a design yet.

\paragraph{Future research\newline}
My motivation for future research will be to obtain a better understanding of the connection between chromosome architecture and gene expression. Fundamentally, I would like to understand whether changes in the chromosome folding can be the driver of cell differentiation or cell senescence. In the short term, I would like to construct an empirical map that associates chromosome architecture to the physiological state of a cell. To serve that purpose, results obtained in the prediction of chromosome architecture from CCC data may be of precious help. Interesting outcomes of this mapping may be to provide novel diagnosis tools to detect cell deficiencies based on CCC assays.


\backmatter
\glsaddall
\printglossaries

\small
\clearpage
\printbibliography[heading=bibintoc, title=Bibliography]

\end{document}